\documentclass[authoryear]{elsarticle}

\usepackage[T1]{fontenc}
\usepackage{ae,aecompl}
\usepackage{graphicx}
\usepackage{amssymb}
\usepackage{amsmath}
\usepackage{mathptmx}
\usepackage{epstopdf}
\usepackage{booktabs}
\usepackage{float}
\usepackage{subfig}
\usepackage{epsfig,color,ulem}
\usepackage{tabularx}
\usepackage{deluxetable}
\usepackage{longtable}
\usepackage{multirow}
\usepackage{calrsfs}
\usepackage{hyperref}

\hypersetup{
    colorlinks
}

\def\apjl{ApJL }
\def\aplett{ApL }

\def\apj{ApJ }

\def\pasp{PASP }
\def\pasa{PASA }
\def\pasj{PASJ.}

\def\apjs{ApJS }
\def\araa{ARAA }
\def\aap{A\&A }
\def\jcap{JCAP }
\def\nat{Nature }
\def\nar{NewAR }
\def\mnras{MNRAS }
\def\prd{Phys. Rev. D. }
\def\prc{Phys. Rev. C. }
\def\physrep{Phys. Rep.}

\newcommand{\msun}{\,{\rm M_{\odot}}}
\newcommand{\cm}{\,{\rm cm}}

\newcommand{\degsq}{\,{\rm deg^2}}
\newcommand{\rp}{$r$-process }
\newcommand{\rpe}{$r$-process elements }
\newcommand\grays{$\gamma$-rays }
\newcommand\gray{$\gamma$-ray }
\newcommand\Ye{{\rm Y_e}}
\newcommand{\appropto}{\mathrel{\vcenter{
  \offinterlineskip\halign{\hfil$##$\cr
    \propto\cr\noalign{\kern1pt}\sim\cr\noalign{\kern-1pt}}}}}

\begin{document}

\begin{frontmatter}

\title{The electromagnetic counterparts of compact binary mergers}

\author{Ehud Nakar}

\address{School of Physics and Astronomy, Tel Aviv University, Tel Aviv 69978, Israel}

\begin{abstract}
Mergers of binaries consisting of two neutron stars, or a black hole and a neutron star, offer a unique opportunity to study a range of physical and astrophysical processes using two different and almost orthogonal probes - gravitational waves (GW) and electromagnetic (EM) emission. The GW signal probes the binary and the physical processes that take place during the last stages of the merger, while the EM emission provides clues to the material that is thrown out following the merger. The accurate localization, which only the EM emission can provide, also indicates the astrophysical setting in which the merger took place.  In addition, the combination of the two signals provides constraints on the nature of gravity and on the expansion rate of the Universe. The first detection of a binary neutron star merger by the LIGO-Virgo collaboration, GW170817, initiated the era of multi-messenger GW-EM astrophysics and demonstrated the great promise it holds. The event produced an unprecedented data set, and although it was only a single event, it provided remarkable results that revolutionized our knowledge of neutron star mergers. GW170817 is especially exciting since we know that it is not one of a kind and that many more events will be detected during the next decade.
In this review, I summarise, first, the theory of EM emission from compact binary mergers, highlighting the unique information that the combined GW-EM detection provides. I then describe the entire set of GW and EM observations of GW170817, and summarise the range of insights that it offers. This includes clues about the role that similar events play in the \rpe budget of the Universe,  the neutron star equation of state, the properties of the relativistic outflow that followed the merger, and the connection between neutron star binary mergers and short gamma-ray bursts. 
%I conclude by discussing some of the future prospects of this new window that has been opened. 

\end{abstract}


\end{frontmatter}

\tableofcontents

\section{Introduction}
For many decades, the mergers of two neutron starts or of a black hole and a neutron star, were considered the most promising candidates for a joint gravitational wave (GW) and electromagnetic (EM) detection. This promise was realized in August 2017 when the gravitational waves from a binary neutron star  merger were detected by 
the Laser Interferometer Gravitational Wave Observatory (LIGO) and Virgo \citep{abbott2017GW}. A short flash of \grays was detected independently by the Fermi Gamma-ray Burst Monitor \citep{goldstein2017} and by INTEGRAL gamma-ray observatory \citep{savchenko2017} only 1.7 s after the end of the chirping GW signal marked the merger of the two neutron stars. Later, EM emission was detected across the whole spectrum and GW170817 became one of the most studied astrophysical transients. 

The combination of GW and EM signals provides unique information. The GWs carry information about the binary at the last moments before it merges. This includes constraints on the masses, spins and tidal deformability of the binary members, as well as the orientation and eccentricity of the binary orbit, the exact time of the merger, and its luminosity distance from earth. The EM emission has the potential to provide a precise localization, which puts the merger in its astrophysical context, including a measure of its redshift as well as identification of the host galaxy and the specific environment in which the merger took place. In addition, the EM emission probes the outflow that the merger ejects into the circum-merger medium. By modeling this emission, we can learn about the various outflow components and their composition, geometry, and velocity. 

Further reward results from the combination of the two signals, which holds the key to a range of fundamental questions in physics and astrophysics. The difference in the arrival times of the GWs and the EM emission teaches us about the nature of gravity. The independent constraints on the luminosity distance and redshift provide a measure of the expansion rate of the Universe, a method which is independent of the assumptions used in all other methods. The precise measurement of the local merger rate, which can be achieved with a well-defined sample of merger events detected by their GW signal, together with the EM measurement of the mass and the composition of the ejecta, shed light on the origins of the heaviest elements.  GW information on the binary inspiral before and during the merger, together with EM mapping of the various outflow components, put unprecedented constraints on the equation of state of matter at supranuclear densities. Finally, the identification of the binary members and the emission of the outflow, may reveal the long-sought progenitors of short gamma-ray bursts (sGRBs).

There has been a significant progress in the theoretical study of compact binary mergers during the past decade. This progress was motivated by the continuous improvements in the sensitivity of GW detectors, and facilitated by growing computational power, which is needed for simulations of various phases of the merger. The breakthrough that came with the detection of GW170817 brought our understanding of compact binary mergers to a new level. It provided an exceptional observational data set, and put the study of binary mergers at the focus of numerous observational and theoretical efforts. In the past two years, about a thousand papers has been written on GW170817 and on various  issues that its detection raised. However, this is still a young field where most of the progress is yet ahead of us, anticipating the discovery of many more mergers during the next decade. This is, therefore, a good time to summarize the knowledge that had been accumulated to date and which will serve as a basis for future advancements.   

Naturally, the entire topic of compact binary mergers cannot be covered by a single review. Here I focus on the main sources of the EM emission from compact binary mergers and on the information that we can extract from their detection jointly with GWs. I pay special attention to the theory of the various components of the outflow that is ejected from the system following the merger and the physical processes that shape their EM emission. I then discuss the observations of GW170817 and the theoretical models of the event. Most of the review is a summary of published studies, but it includes also some novel derivations, synthesis and conclusions. 

Even within the narrow scope of this review, one cannot cover the entire relevant literature, and I apologize for all the important studies of the EM emission from binary mergers that are not discussed here. Below,  I provide, as an introduction, a very brief summary of the outflow from compact binary mergers and its EM emission, with references to the sections in the review that elaborate on each topic, followed by a description of the review layout.

\subsection{{\bf The outflow from compact binary mergers and its emission, in a nutshell}}

The outflows from a binary neutron star (BNS) merger and from a black hole-neutron star (BH-NS) merger are expected to have, broadly speaking, two components - a sub-relativistic and a relativistic one. The sub-relativistic outflow has three major sources: (i) tidal forces that operate during the final stages of the inspiral and the initial stages immediately after the merger; (ii) shocks driven by the collision between the two binary members in the case of a BNS; and (iii) winds from the accretion disk that forms following the merger (\S\ref{sec:BNSejecta} and \S\ref{sec:BHNSejecta}). All of the sources involves decompression of highly dense neutron rich material and therefore they are all prime sites for the nucleosynthesis of \rp elements (\S\ref{sec:rprocess}). In fact, BH-NS and BNS mergers were suggested already four decades ago as major sources of \rpe in the Universe \citep{lattimer1974,symbalisty1982}. The heavy nuclei are formed very far from the valley of stability and therefore they go through a chain of beta-decay, alpha-decay and nuclear fission on their way to stability. This radioactive decay provides a continuous source of heat, similarly to the case of Type Ia supernovae, which eventually escapes as a ultraviolet (UV), optical and infrared (IR) radiation that can be detectable for weeks and even months \citep{li1998} (\S\ref{sec:macronova}). In the case of compact binary mergers, this radiation has been called by various names, with kilonova and macronova being the most common ones. The properties of the sub-relativistic ejecta depend on the nature of the binary (i.e., BNS or BH-NS), the masses and spins of the binary members, and the NS equation of state (EOS). Observationally, the macronova/kilonova emission is sensitive to outflow properties such as its mass, velocity, composition, and geometry. Therefore, when the observations of the macronova/kilonova are combined with the information obtained from the GW signal, they provide a powerful tool to study the physical processes that take place during the merger. The sub-relativistic ejecta is expected to continue and radiate due to its interaction with the circum-merger medium.  This interaction is expected to produce a radio remnant, which may be detectable on a timescale of months to years, depending on the ejecta and the medium properties \citep{nakar2011}.

Within several dynamical times after the merger, some of the bound material settles into an accretion disk that surrounds a rapidly rotating central object. The central object may be a highly magnetized NS (magnetar) or a BH. In BNS merger, the nature of the central object depends on the total binary mass and on the NS EOS. If the merged central object is a NS which is supported by rotation, then it can collapse to a BH at any time after the merger. In BH-NS mergers the merged object is, naturally, a BH. Both types of systems (a rapidly rotating magnetar and a BH with a disk) are promising sources of ultra-relativistic jets, such as those that are present in gamma-ray bursts (GRBs). The jets are the second, relativistic, outflow component expected in a compact binary merger. And indeed,  the idea that BNS mergers are the progenitors of GRBs was suggested already three decades ago \citep{eichler1989}. Following the realization that GRBs are separated into at least to two sub-classes, short and long, the common expectation was that the short gamma-ray bursts (sGRBs) are generated by BNS and/or BH-NS mergers. The detection of sGRB afterglows in 2005 \citep{gehrels2005,fox2005,berger2005} provided strong support for this model, but the evidence was only circumstantial \citep[ and references therein]{nakar2007}. 

If compact binary mergers are sGRBs, then the jet that they launch must drill through the sub-relativistic ejecta that covers the polar region at that time, break out of the ejecta and release the intense burst of $\gamma$-rays. The following interaction of the jet with the circum-merger medium drives a blast wave that generates the long lasting X-ray, optical and radio afterglows. The emission from the jet is extremely bright and can be seen to high redshifts, but only if the observer is within the opening angle of the jet. Relativistic beaming renders the emission at large angles from the jet too faint for detection even at the distance where GWs from mergers are detectable. Since sGRB jets are most likely narrow, and the GW detection is not very sensitive to the inclination of the binary orbit, the expectation is that only rarely we will be able to see the jet directly in mergers that are detected by their GW signal. Moreover, the jet needs a significant power in order to cross the entire sub-relativistic ejecta successfully, at least in BNS mergers. Therefore, it is expected that in some mergers the jets will fail to do so and get choked while still within the ejecta, in which case the jets cannot be observed directly by any observer.

Luckily, the jet can leave an observable imprint also when it points away from us or when it is choked (\S\ref{sec:relativistic}). When a relativistic jet propagates through the sub-relativistic material it inflates a highly pressurise bubble known as the "cocoon" (\S\ref{sec:jet_propagation}). The jet and the cocoon interact with each other and evolve together within the ejecta. When (and if) the jet breaks out of the ejecta successfully, it is accompanied by the cocoon and together they form an outflow, which is spread over an opening angle that is much wider than the opening angle of the jet alone. The entire outflow is known as the "jet-cocoon" and it has a structure which is dictated by two factors, the structure with which the jet is launched from the central compact object and the interaction of the jet with the sub-relativistic ejecta. The expectation is that the jet-cocoon outflow will have  mostly an angular structure where,
along the jet axis, there is a narrowly collimated core with high isotropic equivalent energy and high Lorentz factor, while outside of the core the energy and Lorentz factor decrease with the angle. This type of outflow is often called a "structured jet". In this picture, a sGRB is produced by the narrow jet core and can be seen only by observers that look directly into its cone. Fainter, yet potentially observable emission (at GW detectable distances) is generated over a wider angle by the jet-cocoon  via several processes. First, the jet and the cocoon drive a mildly relativistic or even a fully relativistic shock into the ejecta. When such a shock breaks out it produces a short flash of \grays \citep{nakar2012} (\S\ref{sec:breakout}). 
\grays may be also emitted over a wider angle than the opening angle of the jet core, due to internal dissipation within the jet-cocoon outflow at regions that are outside of the core (\S\ref{sec:high_inclination_grays}). Second, the energy deposited by the shock in the cocoon diffuses as it expands and escapes to the observer, producing an X-ray, UV and optical cooling emission \citep{nakar2017,lazzati2017a} (\S\ref{sec:cocoon_cooling}). Third, radioactive decay of the \rpe in the cocoon can dominate the early macronova/kilonova emission \citep{gottlieb2018a} (\S\ref{sec:cocoon_macronova}). And, finally, the 
interaction of the jet-cocoon with the circum-binary medium produces an X-ray, optical and radio afterglow that can possibly be seen over a wide observing angle \citep{nakar2017,lazzati2017a} (\S\ref{sec:afterglow}). 

A choked jet may also produce observable emission . If the jet deposits enough energy in the cocoon before it is choked, then the cocoon can break out of the sub-relativistic ejecta. The cocoon outflow in that case is mildly relativistic and no sGRB is seen (by any observer). However, the cocoon produces emission that may be observable at a distance where GWs are detectable via most of the same processes as those of the successful jet-cocoon, i.e., shock breakout, cooling emission, radioactively powered cocoon emission and interaction with the circum-merger medium \citep{nakar2017}. Some of the properties of the emission in the case of a chocked jet are different than those of a successful jet-cocoon, allowing us to potentially distinguish between the two cases.

\subsection{{\bf Review layout}}
The goal of this review is to describe the current state of the theory and observations of the EM signal from compact binary mergers. It discusses separately the different components of the ejecta, where for each component I consider first the dynamics of the outflow, followed by a description of the resulting emission. Each of the sections and many of the subsections, are as self contained as possible, so that one can read only the section(s) of interest. The review is focused more on theory, but for each of the topics I describe the relevant observations and how can the observations constrain the theoretical models. I also try, to the degree possible, to explain what is the basis for each of the ingredients of our current theoretical views, paying special attention to how robust is each conclusion and how it can be tested in the future. Below I describe, briefly, the content of each of the sections.

In \S\ref{sec:massEjection} I review the properties of the sub-relativistic ejecta and the various physical processes that drive this outflow. Since the sub-relativistic ejecta is expected to be rich with \rp elements, I start (\S\ref{sec:rprocess}) with a brief review on the \rp and discuss observational constraints on \rp sources in the Universe. I Then discuss the various processes of mass ejection and the properties of the outflows that they drive from BNS mergers (\S\ref{sec:BNSejecta}) and from BH-NS mergers (\S\ref{sec:BHNSejecta}). 

The UV/optical/IR emission from the sub-relativistic ejecta is discussed in \S\ref{sec:macronova}. As already noted, different names are used in the literature for this emission, where the most common are kilonova and macronova. In this review I use the term macronova and I discuss shortly its etymology in \S\ref{sec:etimology}. I describe a simple model of the macronova emission (\S\ref{sec:MN_simple}), which depends on the radioactive heating rate (\S\ref{sec:RadioactiveHeat}) and on the opacity of \rp material (\S\ref{sec:opacity}). I then describe a robust method to estimate the mass of the ejecta (\S\ref{sec:KatzIntegral}). I conclude this section by discussing the first-day macronova emission and non-radioactive energy sources (\S\ref{sec:MN_nonRadioactive}).

I start the discussion of the relativistic outflow with a description of the interaction between the jet and the sub-relativistic ejecta, and the structure that this interaction induces on the relativistic outflow (\S\ref{sec:jet_propagation}). The \gray emission from the relativistic jet depends strongly on the observer viewing angle, where observers within the opening angle of the jet see, presumably, a GRB. Since GRB emission has been discussed extensively in a large number of reviews and it is most likely that the jet of a GW-detected merger will point away from us, I restrict the discussion here to the \gray emission seen by observers that are  away from the jet (\S\ref{sec:grays}). The discussion includes a generalization of the compactness limits to off-axis observers (\S\ref{sec:compactness}), shock breakout theory (\S\ref{sec:breakout}), jet off-axis emission (\S\ref{sec:off_axis_prompt}) and high latitude emission from the wings of a structured jet (\S\ref{sec:high_inclination_grays}). The interaction of the relativistic outflow with the external medium and the afterglow has been also discussed extensively in the literature. Here I provide only a partial description of this theory, which focuses on aspects and regimes that are important for the interpretation of the afterglow of GW170817, and are expected to be useful also for future events (\S\ref{sec:afterglow}). The discussion is separated according to the various phases of the afterglow (rise, peak and decay), where for each of these phases I focus on the information that we can extract from the observations on the structure of the outflow.
 
In \S\ref{sec:GW} I cover aspects of the GW emission that are either directly related to the physical interpretation of the EM emission, or that can provide unique information when combined with the EM signal. The former includes the GW-based measurements of properties such as the masses, spins and tidal deformability of the binary members as well the inclination of the orbital plane (\S\ref{sec:GWbinary}). The latter aspects of the GW emission that I discuss include the constraints that the arrival times of the GW and EM signals pose on the propagation of GWs (\S\ref{sec:GWspeed}), and the usage of compact binary mergers as standard sirens to measure the Hubble constant (\S\ref{sec:H0}).

Section \ref{sec:GW170817} reviews GW170817. First, I summarize the observations without any attempt to provide theoretical interpretations (\S\ref{sec:GW170817_observations}). Then, I describe what we can  learn from the observations on the properties of the binary (\S\ref{sec:GW170817_Binary}), the Hubble constant (\S\ref{sec:H0GW170817}), the sub-relativistic ejecta (\S\ref{sec:GW170817_SubRel}), the structure of the relativistic outflow (\S\ref{sec:GW170817relativistic}) and on the source of the observed \grays (\S\ref{sec:GW170817_grays}). I then discuss the implications of these results for the fate of the remnant (\S\ref{sec:remnant}), the NS equation of state (\S\ref{sec:GW170817_EOS}), and the origin of \rpe in the Universe (\S\ref{sec:GW170817_rpe}). 

In section \ref{sec:sGRBs}, I summarize our understanding of the connection between sGRBs and BNS mergers following GW170817, which contrary to the popular view is almost, but not entirely, secure. I review the various macronova candidates in sGRB afterglows (\S\ref{sec:sgrb_macronova}). I discuss the implications
of the jet seen in GW170817 on sGRB jets (\S\ref{sec:GW170817_sGRBjet}), and address the question of how many GRB170817A-like events hide in the current sample of sGRB prompt emission (\S\ref{sec:GRB170817A_like}). I compare the constraints on the rate of sGRBs to the rate of BNS mergers and discuss the implications of this comparison (\S\ref{sec:sGRB_rate}). I conclude this section by evaluating the current status of the possibility that BH-NS mergers are progenitors of sGRBs (\S\ref{sec:sGRB_BHNS}).

\section{Sub-relativistic mass ejection from compact binary mergers}\label{sec:massEjection}
A BNS merger ejects a considerable amount of mass at sub-relativistic velocities. Substantial theoretical effort has been invested over the past decades in order to predict the properties of the ejected material and its observational signatures. The general predictions about the mass of the sub-relativistic ejecta, before GW170817, varied by more than an order of magnitude centered around a canonical value of about $10^{-3}-10^{-2} \msun$ of \rp material, that is ejected at a velocity of about $0.1-0.3$ c, where c is the speed of light. This material was predicted to radiate with a peak luminosity of $\sim 10^{40}-10^{41} {\rm~erg~s^{-1}}$ in the optical on a timescale of a day and at a slightly fainter luminosity in the IR on a week timescale. The EM counterpart of GW170817 was brighter, bluer and evolved faster than those predictions. Nonetheless, given the uncertainty in the models, its general properties were in  remarkable agreement with the predictions.

The sub-relativistic ejecta originates from several different physical processes. Each process leads to different predictions with regard to the ejecta mass, velocity, composition and angular distribution, although there is some overlap among the processes. Thus, the emission carries a unique imprint of the physics of the merger (e.g., the NS equation of state, the fate of the remnant, etc.). However, deconvolving the observations to learn what is the contribution from each ejection process is and what are the physical conditions during and after the merger, is not a simple task, as evident from the case of GW170817, where there is still a debate regarding the interpretation of the UV/optical/IR emission. 

Mass ejection from coalescing neutron stars starts during the final stages of the inspiral. First, material is thrown out by tidal forces and by the shocks that form during the collision of the two NS cores and the bounce that follows. These mechanisms work on a dynamical timescale of a few orbits, namely milliseconds, and the matter they eject is called the {\it dynamical ejecta}. Within several orbits, the bound material that has sufficient angular momentum forms an accretion disk that surrounds the central merged compact object. The outflow from this point on is termed the {\it secular ejecta}. At first, if the central compact object does not collapse promptly into a black hole, neutrino- and magnetic-driven winds are expected to flow out from the surface of the merged neutron star and the inner disk. Later, on time scales of 100 ms to a few seconds,  disk winds, driven by viscosity (induced by magnetorotational instability), are expected to eject a non-negligible fraction of the initial disk mass. The mass ejection in this last phase is similar regardless of the remnant identity (NS or BH), although the electron fraction in the ejecta, and thus its final composition, does depend on the remnant identity. All processes are predicted to eject \rp rich material, but there are differences in the total mass, velocity, composition and direction of the outflow which are significant enough to affect the observed signal.

As explained above, the mass ejection can be strongly affected by the central compact object that forms during the merger. There are four possible outcomes:
(i) A prompt collapse to a BH; (ii) A {\it hypermassive neutron star} (HMNS) - a NS that is supported by differential rotation. The lifetime of a HMNS before it collapses to a BH depends on its mass as well as the angular momentum transport and the EOS, both of which are not well understood. Typical estimates are that a HMNS collapses to a BH within the first second, possibly even within a few ms \citep[e.g][]{duez2006,shibata2006}; (iii) A {\it supramassive neutron star} (SMNS) - a NS that is supported by rigid rotation. The lifetime of a SMNS depends on the time it takes to lose its angular momentum, which depends mostly on the large scale magnetic field. If the magnetic field is high, $\gtrsim 10^{16}$ G, the NS loses its supporting angular momentum and collapses to a BH within seconds, while for a lower field it can take considerably longer (see \S\ref{sec:central_engine}). In this process the remnant rotational energy, $\sim 5 \times 10^{52}$ erg, is released in the form of a magnetized wind; (iv) A stable NS.
The binary mass thresholds between the various end results depend on the EOS, were a softer EOS can support smaller masses of each object type. Observations of galactic NS masses show that neutron stars are stable at least up to $2 \msun$ \citep{antoniadis2013,linares2018,cromartie2019}. Simulations find typically that the threshold for a prompt collapse to a BH is roughly $2.8 \msun$ \citep[e.g.,][]{hotokezaka2011,bovard2017,radice2018}.

In the past few years, several comprehensive reviews were written on the theory of sub-relativistic mass ejection from compact binary mergers and on their observational signatures \citep{hotokezaka2015,tanaka2016,fernandez2016,metzger2017,shibata2019}, as well as on mergers and the production of \rp elements	 \citep{thielemann2017}. Most of these reviews were written before GW170817 and the significant theoretical progress that was achieved with the increased interest that followed it. In this section I summarize the theory of  the sub-relativistic ejecta, starting with a general discussion of \rp material, followed by a summary of the various mass-ejection processes and the expected ejecta properties. 

%The merger process of two neutron stars was studies by many authors, mostly numerically. Below I summarise the main conclusions drawn from these studies about the sub-relativistic outflow. It is important to bear in mind, though, that given the complexity of the processes and the limitations of the numerical simulations, there may be important ingredients that are missing and that some of the conclusions are likely to change in the future as our understanding of the system improves.

\subsection{{\bf  \rpe}}\label{sec:rprocess}
\begin{figure}
	\center
	\includegraphics[width=0.7\textwidth]{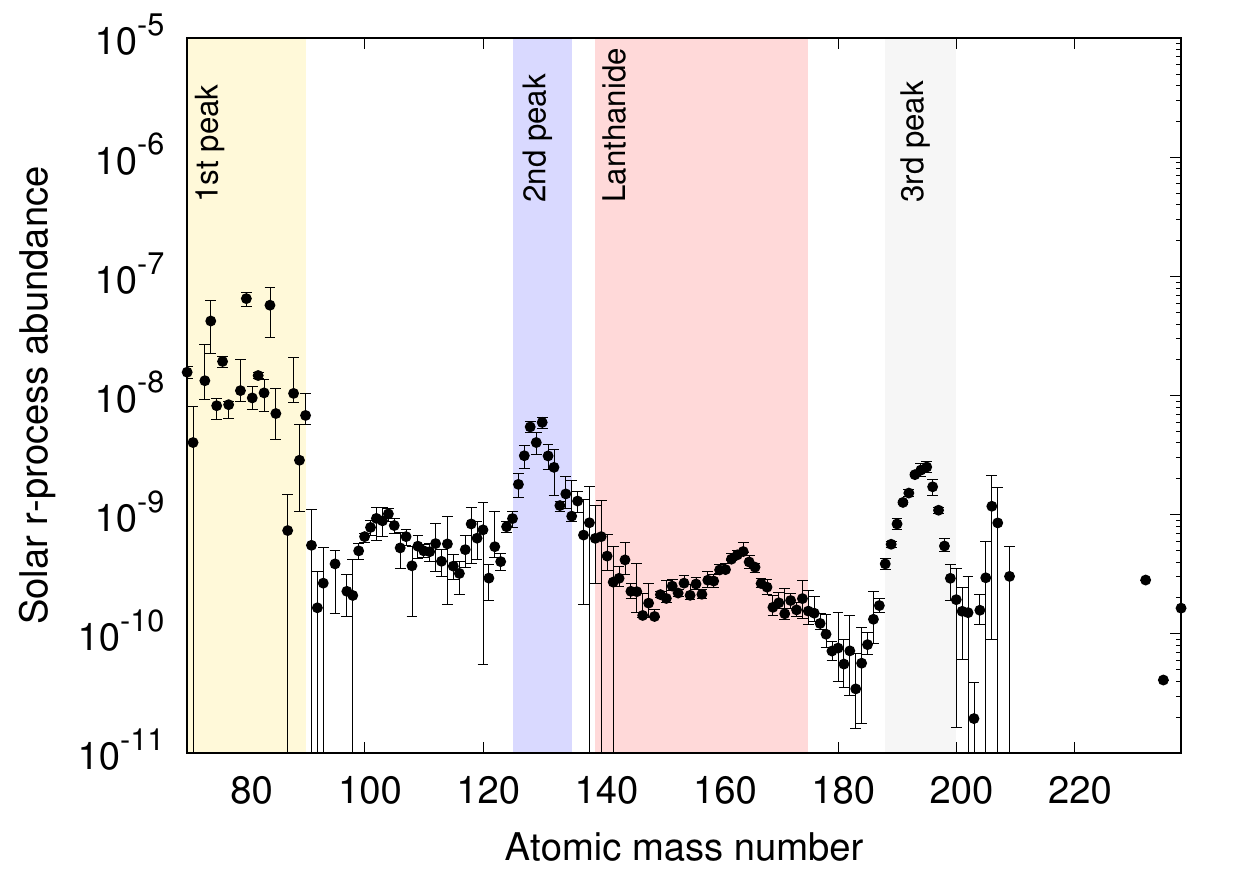}
	\caption{The solar abundance of r-process elements. The main source of uncertainty in the plotted error bars is the removal of the contribution of s-process elements. The figure is from \citet{hotokezaka2018a}, and is based on data from \citet{goriely1999} and \citet{lodders2003}. }%
	\label{fig:rprocess}
\end{figure}

Roughly half of the elements heavier than iron are \rpe  \citep{burbidge1957,cameron1957}, created primarily via rapid neutron-capture process, where rapid is compared to the radioactive decay time of the unstable nuclei that are formed by neutron capture \citep[For comprehensive reviews on \rpe see][]{sneden2008, cowan2019}. Nucleosynthesis of \rpe requires an extreme environment with density, pressure and neutron fraction that allow for such a rapid neutron capture rate. The two leading candidates for such an environment are core-collapse supernovae (cc-SNe) and compact binary mergers. Most of the other half of the heavy elements are formed mainly by slow neutron-capture (the $s$-process), which most likely takes place primarily in asymptotic giant branch stars. Most of the stable heavy elements have contributions from both the \rp and the $s$-process and it is not trivial to separate the contribution of each process. This introduces significant uncertainty when attempting to measure the \rp abundance of stars (including our own Sun). Some of the elements, however, are predominantly (or purely) $r$-process, and these can be used as reliable tracers of the $r$-process. Figure \ref{fig:rprocess} shows the solar abundance of \rp elements, after subtraction of the contribution from the $s$-process.  It shows three peaks at atomic masses $A \approx 80$, $A \approx 130$ and $A \approx 190$, which correspond to the magic neutron numbers 50, 82 and 126. The large uncertainties in the \rpe abundance seen in this figure are dominated by the uncertain subtraction of the  $s$-process contribution.

\cite{lattimer1974} were the first to suggest that decompression of NS material ejected during the merger of a BH and a NS is a promising  source of \rp elements. The same process in BNS mergers was discussed by \cite{symbalisty1982} and \cite{eichler1989}. Detailed calculations of nucleosynthesis in the ejecta of a BNS merger by \cite{freiburghaus1999}  and many following studies have shown that, indeed, these are promising sites. Nevertheless, for many years supernovae were considered the leading candidates for the dominant \rp production sites. The realization that \rp events must be rare compared to regular cc-SNe (see below) together with improved calculations of the properties of the ejecta from BNS mergers in the past decade, have shown that mergers are actually very promising candidates for being the major source of \rpe in the Universe. GW170817, provided the first direct observational test for this hypothesis.     

There are two important conclusions that can be drawn from various observations on the production of \rp elements. The first is that \rpe can be separated to two components, "light" and "heavy". The separation between the two components is at, or just above, the second peak $A \approx 130-140$. The difference between the two components is that the observed pattern of the heavy elements is largely universal while the abundance pattern of light component shows variability from one star to another. Also the ratio between the light and heavy \rpe varies between different stars. It is not clear if the two components are produced in the same sites, although there are some indications suggesting that they do. Theoretical considerations show that the heavy component is associated with nucleosynthesis in environments that are extremely neutron rich, while the light component is produced in environments that are only moderately neutron rich. The second important conclusion from observations is that the events that produce \rpe are relatively rare and that the mass that is generated in each event is relatively large. Quantitatively, observations suggest that the rate of \rp events is lower by a factor of $\sim 1000$ than that of cc-SNe and that the mass ejected in each event is $\sim 0.05\msun$. Below I describe the observations and the resulting conclusions in some detail.

\subsubsection{Two components: light and heavy}

Observations of \rp abundances in low-metallicity stars in the Milky way \cite[][ and references therein]{sneden2008, cowan2019} and in stars in ultra-faint dwarf (UFD) galaxies \citep[e.g.,][]{ji2016} suggest that there are several components to the \rpe production.  
Low-metallicity stars show a very large range in the total amount of \rp elements, yet in each star the relative abundances of different \rpe show similarities. First, the abundances of the lightest \rp elements, gallium (Ga; Z=31; $A=69,71$) and germanium (Ge; Z=32; $A=72-74,76$), are far below the best estimates of their solar abundance (although the uncertainty is large), implying that these elements are most likely not produced in the same sites as most of the \rp elements. Second, there seems to be a difference in the production of \rpe below the second peak ("light", $A \lesssim 130-140$) and above the second peak ("heavy", $A \gtrsim 130-140$). The total amount of \rpe in metal-poor stars varies by orders of magnitude, yet in extremely \rp rich stars 
the abundance pattern of the heavy elements seems to be universal and it resembles the one seen in the Sun\footnote{Note 
that the pattern of heavy elements is not similar to solar in all metal-poor stars. In some  stars, which are not the most \rp rich, the heavy elements pattern shows a decreasing trend as a function of the atomic number (compared to solar)  \citep[e.g.,][]{honda2006}.}. Light elements, on the other hand, show variance both in the abundance pattern of the light elements themselves and in the light to heavy abundance ratio. The exact atomic number that separates the two groups is not clear, where the  relative abundance of elements lighter than silver (Ag; Z=47; $A=107,109$) shows variability between different sites, while elements heavier than barium (Ba; Z=56; $A=135,137,138$) show a universal pattern. Between silver and barium there is not enough observational data. Similarly, it is unclear what the lightest nucleus in the "light" component is, which seems to be somewhere between germanium and strontium (Sr; Z=38; $A=88$), since between these two elements the abundances are not well constrained. Finally, we note that there are considerable uncertainties in the measurement of \rpe abundances in stars as well as the Sun, especially near the first peak (e.g., due to uncertainty in the amount of s-process contribution; \citealt{goriely1999}), and therefore the results with regard to the light elements, while being suggestive, should be taken with caution.

The different patterns of light and heavy \rpe seen in different stars suggests that there is either more than one type of sources of r-process elements (e.g., one type for heavy and one or more for light elements), or, if there is a single source then there is variability in the amount and in the abundance pattern of light elements that are produced in each event. The latter option is supported by observations of stars in the ultra-faint dwarf galaxy Reticulum II. These stars show both light and heavy \rpe which were probably all produced by a single event \citep{ji2016}. 

\begin{figure}
	\center
	\includegraphics[width=0.7\textwidth]{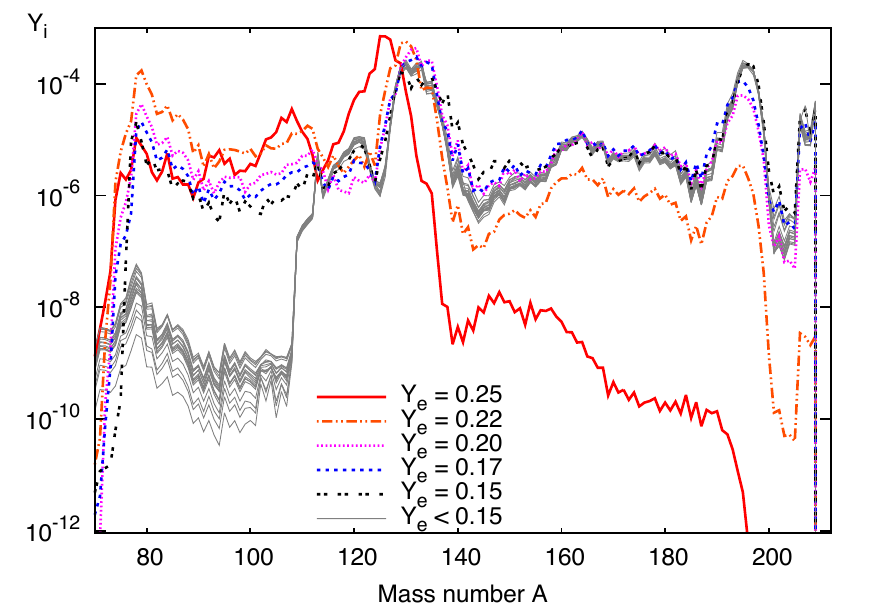}
	\caption{The composition of various fluid elements with a range of $\Ye$ values in a simulation of a BNS merger. Reproduced from \citet{korobkin2012}.}%
	\label{fig:Ye}
\end{figure}

The separation into light and heavy components is also guided by theoretical considerations. There are several parameters that determine the elements that are synthesized through rapid neutron capture when  neutron rich hot and dense material is decompressed. These include the entropy, the expansion time, and the ratio of the proton number to the total baryon number (protons+neutrons), defined as the electron fraction $\Ye$  \cite[e.g.,][]{hoffman1997}. Simulations of nucleosynthesis under the conditions expected in the ejecta from BNS mergers find that the value of $\Ye$ can vary significantly between different components of the ejecta, and that this is the main factor that determines which \rp elements are produced beyond the iron-peak \citep[e.g.,][]{goriely2011,korobkin2012,bauswein2013,perego2014,wanajo2014}. Figure \ref{fig:Ye}, from \cite{korobkin2012}, shows an example of the dependence of the final abundances of fluid elements ejected in a simulation of a BNS merger on their initial value of $\Ye$ (similar results are obtained by all studies). All the extremely neutron rich ejecta, $\Ye \lesssim 0.1$, produce efficiently only heavy elements with $A \gtrsim 130$. The final abundance pattern in such ejecta is weakly sensitive to the conditions of the decompressed ejecta. The reason is that the \rp beyond the third peak is terminated by fission and recycling of the heaviest nuclei into lighter seed nuclei, thereby closing a cycle with a robust abundance pattern\footnote{Note that while theory predicts a pattern of heavy elements that is insensitive to the exact conditions, the exact pattern predicted  depends on the nuclear model used, which is not well constrained.}. Moreover, the theoretical predictions of this pattern matches relatively well the observed solar abundance pattern (i.e., the same relative proportions) above the second peak. As the electron fraction increases, lighter elements starts being 
produced while heavier elements stop being produced. Ejecta with $\Ye \sim 0.2$ produce all three of the peaks, ejecta with $\Ye \sim 0.25-0.3$ produce all elements up to the second-peak but none at $A \gtrsim 130$ beyond the second peak, and ejecta with $\Ye \sim 0.35$ produce only first-peak elements up to $A \lesssim 100$. This is consistent with the observational separation into light and heavy \rp elements. Heavy elements are produced only by material with $\Ye \lesssim 0.2$ and in all this range the abundance pattern is  consistent with the observed one, while light elements are produced in material with $\Ye \gtrsim 0.15$ and its exact pattern depends on the exact value of $\Ye$ and the thermodynamic evolution of the ejecta. In almost all simulations, the lightest elements that are produced efficiently are with $A \approx 80$, while gallium and germanium are not produced efficiently. Note that the theoretical calculations of the synthesized composition are highly uncertain as they depend on the unknown properties of nuclei very far from the valley of stability. Thus, while the general trend of three peaks, which corresponds to the neutron magic number, and the general effect of $\Ye$ are robust, the abundances of specific elements predicted by theory are highly uncertain and can depend strongly on the specific nuclear model that is used.

\subsubsection{Constraints on the rates  and the yields of \rp sources }\label{sec:rp_rate}

One can obtain a rough estimate for the total production rate of \rpe in the Milky Way from the solar abundance, assuming that it is typical \citep[e.g.,][]{hotokezaka2018a}. The total mass of \rpe in the sun is dominated by the light component, in which the mass of elements with $A \gtrsim 80$ is about $2 \times 10^{-7} \msun$. The heavy component, $A>130$, carries about $25\%$ of this mass.
Taking the Milky-Way stellar mass in the disk,  $\sim 0.5-1 \times 10^{11} \msun$ \citep{licquia2015}, and assuming a constant production rate over the past 10 Gyr, we obtain an average production rate of the total \rpe (light and heavy) in the Milky Way of   $\sim 1~\msun {\rm~ Myr^{-1}}$.

This production rate can be achieved via common events, each of which produces a small amount of \rp elements, or by rare events where each event produces a large amount of \rp elements.  Several clues suggest that \rp sources are relatively rare (compared to normal cc-SNe) and that the yield of each event is relatively high. The first clue is the large variance in the amount of europium\footnote{Europium is considered as a reliable tracer of heavy \rpe. It is almost entirely produced by the r-process and it is the easiest heavy \rp element  to measure in optical spectra.} (Eu, Z=63, $A=151, 153$) seen in dwarf galaxies \citep{beniamini2016}. 
The most constraining results are from about half a dozen ultra-faint dwarf galaxies that contain a very small stellar population of $\sim 10^3 - 10^4 \msun$. Most ultra-faint dwarf galaxies seems to have a very low value of the ratio [Eu/Fe] while two, Reticulum II and Tucana III, show high [Eu/Fe] \citep{ji2016,hansen2017}. In Reticulum II, \cite{ji2016} measure the \rp abundances of 9 stars finding that  7 stars are \rp rich and that the two most metal poor stars (i.e., lowest [Fe/H]) have low abundance of \rp elements.  This suggests that in most ultra-faint dwarf galaxies there were no \rp events, while in Reticulum II there was a single event that produced a large amount of \rp material. This, in turn, suggests that the Hubble-time-integrated rate of events that make \rpe is significantly less (probably by about an order of magnitude) than one event per $10^4$ stars formed. For comparison, the corresponding number for cc-SNe is about one event per 100 stars formed, i.e. the ratio between the rate of events that produce \rpe,  ${\cal R}_{rp}$, and the rate of cc-SNe, ${\cal R}_{SN}$, is ${\cal R}_{rp}/{\cal R}_{SN} \sim  10^{-3}$.  Assuming a galactic rate of 3 SNe per century \citep{li2011}, this implies a Galactic rate of $\sim 20$ \rp events per Myr, each ejecting $m_{rp} \sim 0.05 \msun$ of \rp material into the interstellar medium.

\cite{tsujimoto2017a} examine the composition of  12 metal-poor stars in Draco, a dwarf galaxy with a stellar mass of about $3 \times 10^5 \msun$. They suggest that the pattern of Eu/Fe as a function of the metallicity (Fe/H) indicates two separate \rp episodes, where the second episode produced about an order of magnitude more \rpe than the first. This rate of 2 events per $3 \times 10^5 \msun$ is consistent with ${\cal R}_{rp}/{\cal R}_{SN} \sim 10^{-3}$. \cite{tsujimoto2017a} conclude that that the \rp pattern in Draco suggests that there is a large variability in the amount of \rpe produced by each event and that there are possibly different \rp production sites. 

A second clue to estimating ${\cal R}_{rp}$ and $m_{rp}$ comes from the abundances of radioactive elements in meteorites and in the deep-sea floor.  Meteorites hold record of the abundances of \rp elements in the vicinity of the solar system when it was formed. Therefore, a comparison of the abundances of elements with decay times of 0.01-5 Gyr  in meteorites constrain the delay between the last \rp event and the formation of the solar system. For example, the relatively large ratio of  $^{244}$Pu (half-life of 81 Myr) and $^{238}$U (half-life of 4.5 Gyr) in meteorites implies that the time delay between the \rp event and the formation of the solar system is $\sim 100$Myr  \citep[e.g.][]{wasserburg1969}.  The radioactive elements at the deep-sea floor are accreted continuously from the interstellar medium (ISM) and therefore their measurement constrain the current abundance of \rpe in the solar neighbourhood. \cite{wallner2015} measured the abundance of $^{244}$Pu in the deep sea floor, finding that it is lower than expected from continuous production in the Galaxy by about 2 orders of magnitude. \cite{hotokezaka2015a} conclude based on comparison of the abundance of $^{244}$Pu in the sea floor  and in meteorites that the abundance of \rpe in the vicinity of the solar system was much higher during its formation than it is today. They conclude that the large fluctuations in the \rpe abundance in the local ISM requires an \rp Galactic rate that is $\lesssim 90$ events per Myr and $m_{rp} \gtrsim 0.001 \msun$. \cite{tsujimoto2017} examined the implications of radioactive and stable \rpe in meteorites, obtaining similar results regarding the time between the last \rp event that affected the young solar system and its formation, and regarding the decrease in the \rp abundance in our vicinity since the formation of the solar system. They also conclude that ${\cal R}_{rp}/{\cal R}_{SN} \sim 10^{-3}$.

The conclusions of the studies described above are all based on multiple assumptions (as discussed in these works) and thus the uncertainty of each conclusion by itself is considerable. Nevertheless, the fact that independent studies based on different arguments all find ${\cal R}_{rp} \sim 10^{-3} {\cal R}_{SN} $ and $m_{rp} \sim 0.05 \msun$ provides  significant support to these conclusions. %Taking a volumetric cc-SN rate of $ {\cal R}_{SN} \sim 10^5 \,{\rm Gpc^{-3}\,yr^{-1}}$ \citep{strolger2015}), the corresponding volumetric rate of \rp events is ${\cal R}_{rp} \sim 100 \,{\rm Gpc^{-3}\,yr^{-1}}$. 
 
%\subsubsection{Constraints on the time delay distribution of \rp sources}
Another interesting clue can be drawn from the Galactic evolution of \rp element abundance with metallicity, usually represented by means of the iron to hydrogen ratio, Fe/H. It is instructive to compare the \rp evolution to the evolution of $\alpha$ elements, as represented, e.g., by the trends of stellar abundances seen in the plane of Mg/Fe vs. Fe/H. $\alpha$ elements show a complex evolution that may indicate various populations \cite[e.g.,][]{weinberg2018,helmi2018}, but there is a general trend of a decrease in the Mg/Fe ratio with increasing Fe/H above a metallicity of 0.1 solar (i.e., ${\rm [Fe/H]} \gtrsim -1$) \cite[e.g.,][]{nidever2014,ho2017}. The common explanation for this trend is a decrease of the ratio between the rate of core-collapse SNe, which are $\alpha$ rich, and the rate of Type Ia SNe, which are iron rich  \citep[e.g.,][ and references therein]{nomoto2013,maoz2017}. This is expected due to the different delay times distribution of core-collapse and Type Ia SNe. Core-collapse SNe explode promptly, within ~30 Myr, after the formation of their stellar progenitors, while type Ia SNe show a delay-time distribution that is roughly $\propto t^{-1}$ \citep[][ and references therein]{maoz2014}.

The chemical evolution of \rpe is traced readily via the abundance of Eu. There are hundreds of stars, almost all from the solar neighbourhood, with measured Eu/Fe \citep[e.g.,][ and references therein]{sneden2008,suda2008}. It is unclear how representative this sample is, since its selection effects are unknown (e.g., some of the surveys used criteria that targeted \rp rich stars) and it does not show the complexity of the evolution of the Mg/Fe ratio, which is seen in samples that probe also stars in different regions of the galaxy \cite[e.g.,][]{nidever2014}. Nevertheless, the Eu/Fe ratio does seem to show a similar trend to that of Mg/Fe, with its average value dropping by a factor of $\sim 3$ between ${\rm [Fe/H]}<-1$ and ${\rm[Fe/H]}\approx 0$. Two additional interesting features  are that the scatter in the Eu/Fe ratio is  large at  low metallically (much larger than seen in ${\rm \alpha/Fe}$ elements) and that there are extremely low metallically stars with high Eu/Fe. 

Translating these observations to properties of the \rp sources is not trivial. The large scatter at low metallically points most likely to rare sources, which is consistent with observations of \rpe in UFD galaxies and with the abundance of radioactive elements in the early solar system (see the discussion above). The simplest explanation for the Eu/Fe evolution is that the typical delay between the formation of the progenitor systems and the \rp events is shorter than that of Type Ia SNe and is more similar to that of core-collapse SNe \cite[e.g.][]{argast2004,wehmeyer2015}. This may suggest that either most of the \rpe where produced by events that took place promptly after star formation (similarly to cc-SNe) or, if there is some significant delay then the delay-time distribution falls faster than that of Type Ia SNe.  Simplified chemical evolution models that assume instantaneous mixing suggest that \rp events that have a delay-time distribution that drops faster than $t^{-1.5}$ are consistent with the Eu/Fe evolution  \citep{hotokezaka2018a,cote2018}. A prompt \rp source also makes it easier to explain the high Eu/Fe seen in some extremely metal poor stars. However, a prompt component may not be necessary to explain the observations. The chemical evolution is much more complex than that of a simple single zone model and models that include additional processes such as inhomogeneous mixing and/or hierarchical galaxy formation are able to explain the data also without a significant prompt component \citep[e.g.,][]{ishimaru2015,shen2015}. The conclusion is that the interpretation of the \rp chemical evolution is inconclusive, and while simplified models require a prompt \rp source (or a source with a rapidly declining time delay distribution), more detailed models suggest that a prompt component may not be necessary. A more detailed discussion of this topic can be found in \cite{cowan2019}.

\subsection{{\bf  Ejection processes and properties of the sub-relativistic ejecta from BNS mergers}}\label{sec:BNSejecta}

\subsubsection{Dynamical ejecta}\label{sec:dynamicalEjecta}
The dynamical ejecta is the first to be thrown out of the merging binary. It is driven by gravitational forces and by the hydrodynamics of the collision between the two binary members. The physics of these ejection processes is understood relatively well, and therefore the predictions for the dynamical ejecta are the most robust. Nevertheless, there are considerable uncertainties that are introduced mostly by the unknown NS equation of state (EOS). Due to the complexity of the problem, most estimates rely on numerical simulations and thus additional uncertainty is introduced by numerical limitations.

Numerical simulations of the dynamical ejecta from BNS mergers have been performed for over two decades \cite[e.g.,][]{ruffert1996,rosswog1999,rosswog2000,oechslin2002,korobkin2012,piran2013,rosswog2013,bauswein2013,hotokezaka2013,sekiguchi2016,foucart2016,radice2016,dietrich2017,dietrich2017a}. At first, simulations used Newtonian or other approximated gravity schemes as well as analytic approximations to the EOS and no treatment of neutrinos. The most recent simulations use full general relativity (GR), approximated neutrino transport schemes (at various levels of approximation) and a more realistic EOS. The different simulations agree about the total ejected mass to within about a factor of 10, and they provide a coherent picture of the processes that dominate the mass ejection, and of the ejecta's dependence on the binary parameters. Below, I describe the various processes and the effect of the system parameters on the outflow properties. Table \ref{table:BNSsubRel} summarizes the main results based on several  simulations that incorporate full GR, a microphysical equation of state, and some scheme of neutrino transport \citep{palenzuela2015,sekiguchi2015,lehner2016,bovard2017,radice2018}.

Mass ejection begins when two neutron stars approach the final stages of their inspiral. Tidal forces start ejecting mass, mostly along the equator, upon the last orbit before coalescence.  If the total binary mass is large enough ($\gtrsim 2.8 \msun$) then the first collision between the neutron stars leads to a prompt collapse to a BH and dynamical mass ejection stops. If not, tidal mass ejection from the central fast rotating compact object continues for several orbits while the collision of the two cores and the ensuing bounce drive shocks that eject more mass. Generally speaking, tidally ejected mass is concentrated towards the equator, it is slower than the shock driven ejecta, and its electron fraction is low ($\Ye \sim 0.1$). Shock driven ejecta is faster, more isotropic and its electron fraction is higher. Note that the tidal and shocked ejecta collide, interact and affect each other, and therefore the distinction between the two components is not well defined. Still, simulations find a clear correlation between the polar angle and the dynamical ejecta properties, where  material that is ejected closer to the equator is slower and more neutron rich.  The net result of the tidal and shock driven ejecta is that $\sim 90\%$ of the mass moves at a velocity in the range of $\sim 0.1c-0.3c$ for stiff EOS and $\sim 0.1c-0.4c$ for soft EOS. A small fraction of the mass moves very fast, $>0.6$c (the so-called "fast tail"). The outflow covers the whole $4\pi$ with more mass ejected near the equator and less towards the pole. 

The dynamical ejecta also has a wide range of electron fraction, with a roughly a uniform distribution of $\Ye$ in the range 0.1-0.4. The $ \Ye$ of the NS material before the merger is low. However, the temperature in the regions that are shocked by the collision rises to $\sim 10$ MeV or higher, leading to vigorous production of neutrinos and anti-neutrinos in the high density material (see e.g. \citealt{beloborodov2018}). The irradiation by neutrinos of neutron rich material increases its $\Ye$. Since the neutrino flux increases with the latitude,  the distribution of $\Ye$  depends on outflow angle above the equatorial plane, where higher latitude ejecta have larger $\Ye$.    

The dynamical ejecta properties depend strongly on the EOS. A soft EOS (i.e., a more compressible and compact NS) leads to the ejection of significantly more mass than a stiff EOS (by up to an order of magnitude), and at higher velocities. The reason for the higher velocities is that smaller NS radii lead to a faster collision, stronger shocks and a more compact faster rotating central object, with a higher escape velocity.

The binary mass ratio also affects the dynamical ejecta. When the mass ratio is large the lighter NS is partially disrupted during the last orbit, leading to stronger tidal forces and a weaker collision. Thus, unequal masses lead to enhanced tidal ejecta and reduce shock driven outflows.  The net result is ejecta that contain more neutron rich material and less material is ejected along the poles. 
Unequal masses produce a single tidal arm (from the lighter binary member), compared to two symmetric arms when the masses are equal. The result is a significant azimuthal anisotropy. A mild mass ratio of $q=0.85$ does not show a signifiant effect on the total ejected mass but \cite{dietrich2017a} find that a significant mass ratio, $q \approx 0.6$, can potentially enhance the total ejected mass by up to an order of magnitude. The mass ratio may also have some effect on the ejecta velocity, but such an effect does not always show a clear trend in simulations and it is much less significant than the effect of the EOS.

The main effect of the total binary mass is that it determines (together with the EOS) whether the central object collapses promptly to a BH or not. If it does, then material is ejected mostly tidally before the collapse and the total ejected mass is typically low. When there is no prompt collapse, the total mass does not have a clear and significant effect on the properties of the bulk of the ejecta.

Another possible source of dynamical ejecta is turbulent viscosity at the interface between the two neutron stars. A possible source for such viscosity is magnetohydrodynamic instabilities, but its level and exact effect is unknown. \cite{radice2018a}  introduce viscosity to their simulations, via a parametrized mixing length. They find that viscosity can increase the mass of the  dynamical ejecta by a factor of a few for a binary with a mass ratio $q=0.85$. For such a binary they find an even more significant enhancement in the ejection of fast, $>0.6$c, material (by up to four orders of magnitude). They find no significant effect of viscosity on the dynamical ejecta of equal-mass binaries.

The properties of the dynamical ejecta for the case of a prompt collapse to a BH have not been explored in detail. It is clear, however, that the amount of ejected mass is much lower in that case. For example, \cite{radice2018} find that in simulations where a collapse takes place within $\lesssim 1$ms of the merger, only $\sim 10^{-4}\msun$ are ejected with a relatively high average velocity of 0.2-0.3c and an average electron fraction $\Ye \approx 0.15$.\\

\noindent {\underline {The fast tail}}\\
While most of the dynamical ejecta move at a velocity $\lesssim 0.4$c, a small fraction of the mass is expected to reach hight, possibly even relativistic, velocities. Although the mass in this "fast tail" may be minute, it can have significant observational implications. First, its composition may be different than the rest of the ejecta. The short expansion time dictated by the fast velocity and the low density may lead to a nucleosynthesis freeze-out and a high fraction of free neutrons. If there are enough free neutrons, this can produce a detectable UV/optical signal on a time scale of hours \citep[e.g.,][]{kulkarni2005,metzger2015} (see section \ref{sec:neutrons}). Second, 
if the ejecta is shocked after expanding to large radii ($\gtrsim 10^{11}$ cm), e.g., by the launching of a relativistic jet, then the shock breakout from the fast tail may produce a detectable gamma-ray signal \citep[e.g.,][]{kasliwal2017,gottlieb2018b,beloborodov2018} (see section \ref{sec:breakout}).

Predicting the properties of the fastest ejecta is extremely difficult. The processes that accelerate it are hard to model accurately analytically, while the very small amount of mass that it contains makes it difficult to resolve numerically. Several possible processes that can accelerate a fast tail are discussed in the literature. The first is acceleration by means of a shock, and especially the breakout of this shock through the decreasing density near the edge of the merging binary material. The first to discuss this process were \cite{kyutoku2014}, who suggested based on an analytic toy model that $\sim 10^{-7} \msun$ are ejected at relativistic velocities. Note however, that the toy model used by \cite{kyutoku2014} assumes that the shock propagates in parallel to the density gradient in which it accelerates. In such a case the shock velocity is limited by the medium's optical depth \citep{nakar2012} and in principle gas can be accelerated to extremely high Lorentz factors, $\Gamma \gg 10$. A more realistic scenario is that the shock is oblique and that its maximal velocity is limited by its obliqueness \citep{matzner2013} so it is unlikely that the shock is accelerated to ultra-relativistic velocities, although it may be accelerated to mildly relativistic velocities. This process cannot be fully resolved by numerical simulations, yet simulations do find that the collision between the two neutron stars produces fast ejecta. First results of marginally resolved fast ejecta with $\sim 10^{-5}\msun$ moving at $v > 0.6$c were found by \cite{hotokezaka2013} and \cite{bauswein2013}. More recent simulations with much higher resolutions provide more reliable, although not conclusive, evidence for the existence of a fast tail \citep{hotokezaka2018b,radice2018}. 
They find typically $\sim 10^{-6}-10^{-5} \msun$ of material moving at $v >0.6$c.

\cite{radice2018} examined in detail the properties of the fast tail in their simulations and the processes that generate it. They find two distinct ejection episodes, with different properties. The first takes place upon the collision of the two stars as material is squeezed out from the contact interface. This material is ejected along the pole and at high latitudes. The second ejection episode takes place as the shock, which is driven by the first bounce of the remnant, breaks out of the cloud of slower ejecta material that surrounds the remnant. This second component is also found to have a highly anisotropic distribution, concentrated more towards the equator. The relative importance of the two components depends on the binary mass ratio. In an equal-mass binary a comparable amount of mass is ejected in each component. In an unequal-mass binary, the collision is not head-on, so the first squeezing episode is absent and practically all of the fast tail is ejected as the bounce-driven shock breaks out. The EOS also affects the fast tail, where a softer EOS results in a faster and more massive tail. The exact properties of the fast tail in  \cite{radice2018} still depend on the numerical resolution and therefore they cannot exclude that its origin is numerical. However,  the smooth velocity distribution that they find  for the fast-tail material supports a physical origin.

\cite{beloborodov2018} discuss two additional processes that can accelerate fast ejecta. First, they consider the ablation of material from the surface of the colliding neutron stars by heat deposited via neutrino-antineutrino annihilation. The entire process takes $\sim 10^{-5}$s and they estimate that it may be able to accelerate up to $\sim 10^{-6} \msun$ to relativistic velocities, where the fastest material, $\sim 10^{-10} \msun$, is accelerated to $\Gamma \sim 1000$. The second process that they suggest is the breakout of a shock driven by the central source, e.g., a magnetar burst, after the ejecta has already expanded to its homologous phase (see also \citealt{metzger2018}).

\subsubsection{Secular ejecta}
Part of the material that is torn from the neutron stars during the merger, and from the central remnant during the first $\sim 10$ ms, remains bound and has enough angular momentum to form a rotationally supported thick torus around the central remnant. Magnetorotational instability (MRI) is expected to generate viscosity in the disk and facilitate its accretion. The heat it generates near the plane of the disk is balanced at first by neutrino cooling and the accretion is efficient. During this phase mass is ejected as material above the disk plane is heated by neutrinos, MRI viscosity and nuclear recombination energy. After a significant fraction of the disk mass is accreted and angular momentum transport leads to its expansion, the cooling rate drops and accretion becomes inefficient. At this point, a thermally driven wind ejects what is left of the accretion disk at relatively low velocities. The entire process starts as soon as the material settles to form a disk, i.e., $\sim 10$ms after the merger. The transition from an efficient to an inefficient accretion takes place after $\sim 0.1-1$s, and the thermal wind ends after $\sim 10$s.

The total ejected mass depends mostly on the disk mass, which in turn depends mostly on the fate of the remnant during the first 10 ms. Most recent simulations find that if the remnant does not collapse to a black hole for longer than 10ms, then the disk mass is $\sim 0.1\msun$ \citep[e.g.,][]{hotokezaka2013,radice2016,sekiguchi2016,dietrich2017a,radice2018}. A prompt collapse to a BH can significantly reduce the disk mass. For example, \cite{radice2018} find disk masses of $\sim 10^{-4}-10^{-3}\msun$ when the collapse takes place within less than 1ms. The composition of the ejecta depends mostly on the level of neutrino irradiation, which in turn depends on the nature of the remnant as well.   Table \ref{table:BNSsubRel} summarizes the main results obtained by numerical simulations of the secular mass ejection from BNS mergers. Below, I discuss the different ejection phases in some detail. \\

\noindent {\underline {\it Neutrino driven wind}}\\
Neutrino flux is expected both from the disk and from the central object as long as it does not collapse to a BH. Energy deposition by the streaming neutrinos unbinds some of the material. In case the central object is a BH, mass ejection via this process is negligible \citep{fernandez2013,just2015}. A central NS (HMNS, SMNS or a stable NS) significantly enhances the neutrino emissivity, both by its own emission and by the accretion of the disk material on the NS surface. As a result, the mass ejection is also much stronger.  The neutrino driven wind from $\sim 0.1 \msun$ which is accreted onto a NS was studied numerically by several groups \cite[e.g.,][]{dessart2009,perego2014,martin2015,fujibayashi2017}. They find that the central NS  and its surrounding radiate $\sim 10-20$ MeV neutrinos at a luminosity of $\sim 10^{53} {\rm~erg~s^{-1}}$. Interaction of these neutrinos with the plasma and heat deposition via $\nu-{\bar \nu}$ annihilation ejects a wind into the funnel constructed by the torus shadow at angles $\gtrsim 30^\circ$ above the equatorial plane. The total neutrino driven wind mass is typically $\sim 10^{-3}-10^{-2}\msun$ and its velocity is 0.1-0.2c. The high neutrino irradiation results in $\Ye \gtrsim 0.2$, where higher latitude ejecta have higher $\Ye$. The resulting composition is mostly light \rp elements, where at high latitudes ejecta can be entirely free of heavy elements beyond the second peak (e.g,. the lanthanides).\\
 
\noindent {\underline {\it Viscosity driven wind}}\\
Viscous heating is expected to unbind a significant fraction of the disk mass. The source of viscosity in the disk, and the remnant if the latter is a NS, are  magnetic fields. These are amplified mostly via the MRI. Due to the huge dynamic range of the problem, numerical simulations cannot follow the entire evolution from before the merger and up to the end of the disk evolution. A numerical study of the disk wind is therefore typically done by a simulation of a disk that surrounds a central object, with initial conditions that are based (to a limited extent) on the end results of merger simulations. Some of these simulations consider accretion upon a NS \citep[e.g][]{metzger2014,martin2015,lippuner2017,fujibayashi2018} while others consider a central BH \citep{fernandez2013,fernandez2015,just2015,wu2016,siegel2018,fernandez2019,miller2019,christie2019}. Earlier studies used 2D hydrodynamics with parametrized viscosity, while more recent simulations use 3D GRMHD codes that resolve the MRI viscosity and use an approximated neutrino treatment \citep[e.g.,][]{siegel2018,fernandez2019,miller2019,christie2019}.

The results of these simulations show that viscous mass ejection can have up to three different phases. The first phase is expected only if the remnant is a differentially rotating NS. In this case, the magnetic field, which is built-up through differential rotation, trasports angular momentum from the inner parts of the remnant outward, driving a wind \cite[e.g.,][]{siegel2014,fujibayashi2018}. This wind is expected to last until the NS loses all of its differential rotation or until it collapses to a BH, whichever comes first, and it coincides to the most part with the neutrino driven wind discussed above, but enhances its mass ejection by up to an order of magnitude. \cite{fujibayashi2018} find this phase to last $\lesssim 0.1$s in their simulations and that its product is a wind that is similar to the one driven by neutrinos alone, with possible enhancement of the total mass. During the first 0.1s about  $\sim 0.01\msun$ is ejected at 0.1-0.2c with $\Ye=0.2-0.5$. This wind is distributed roughly isotropically above the disk shadow, i.e., at latitudes  $\gtrsim 30^\circ$, and its high $\Ye$ is due to the strong neutrino irradiation. 

The other two phases depend only on the disk evolution and therefore take place regardless of the central object's identity. At first,  the temperature and density in the disk plane are high enough that the accretion flow is neutrino dominated (NDAF), cooling is efficient and the accretion rate is high. The gas in the plane is in a mildly degenerate state and it is neutron-rich ($\Ye \approx 0.1)$ \citep[e.g.,][]{siegel2018}. Above the disk the density is lower and so is the neutrino cooling rate, which drops below the rate at which heat is deposited, mostly via magnetic viscosity and 
recombination of free nucleons into $\alpha$ particles. The result is a strong wind, mostly in a funnel of $\gtrsim 30^\circ$ above the plane, which is ejected at a velocity of $\sim 0.1-0.15$c. The composition of the outflow from this phase does depend on the central engine, as discussed below. All together, the efficient cooling NDAF phase lasts for $\sim 0.1-1$s. The total mass ejected during this phase differs among the different types of simulations. There are three recent 3D GRMHD simulations, all of which examine the evolution of a $0.03\msun$ torus around a $3\msun$ BH \citep{siegel2018,fernandez2019,christie2019}. These simulation find that during this phase up to 20\% of the disk mass can be ejected. It is not fully clear, at this point, how the mass ejection scales with the disk or the BH mass, or to what extent the uncertain initial conditions (e.g., magnetic field configuration) affect this result. For example, \cite{christie2019} examine the effect of the initial magnetic field configuration, finding that it can affect the mass ejection during this phase. The maximal mass ejection of 20\% is obtained for a strong poloidal magnetic field (similar ot the one used in  \citealt{siegel2018,fernandez2019}), while a weak initial  poloidal field or a toroidal field configurations reduce the total mass ejection and its velocity by a factor of a few. 
Simulations of similar setups that use parametrized $\alpha$-viscosity find much lower mass ejection during this phase both in 2D \citep{just2015,fernandez2015} and in 3D \citep{fernandez2019}. The reason for this difference is unclear at this point. There are no GRMHD simulations of a torus around a NS, and $\alpha$-viscosity simulations find low mass ejection during the NDAF phase, similar to simulations of BH systems. For example, \cite{fujibayashi2018} carry out a 3D simulation of a $0.2\msun$ torus around a $2.65\msun$ NS with parametrized $\alpha$-viscosity finding that following a short phase ($\sim 0.1$s) of mass ejection driven by the central engine, the mass ejection rate drops significantly during the NDAF phase. The conclusion is that it seems that a self-consistent treatment of the MRI during the NDAF phase enhances the mass ejection significantly and that it can get up to a non-negligible fraction of the total initial disk mass, although further exploration is still needed to verify this point for a range of system parameters and initial conditions.

The third phase starts as the disk expansion, due to the angular momentum transport, reduces the density in the disk plane to the point that neutrino cooling becomes inefficient. At this point, the accretion becomes advection dominated. The accretion rate drops and most of the mass that remains in the disk is expelled roughly isotropically within a few seconds (the viscous time scale) as a relatively slow, $\sim 0.05$c, wind \cite[e.g.,][]{fujibayashi2018,fernandez2019}. The total disk mass at the time that the accretion becomes cooling inefficient, and thus the mass ejected during this phase, was found to be about $20\%$ of the initial disk  mass in a range of simulations with different initial conditions and numerical schemes (various disk masses, BH/NS central object, 2D/3D, GRMHD/$\alpha$-viscosity, magnetic field configuration, etc.) \citep{just2015,fernandez2015,fujibayashi2018,siegel2018,fernandez2019,christie2019}, suggesting that this result is rather robust.

The electron fraction, and thus the composition of the disk wind, depends on the central object. As long as the accretion is onto a NS, there is a strong neutrino flux that increases $\Ye$ \citep[e.g.,][]{metzger2014}. \cite{fujibayashi2018} find ejecta with high electron fractions in all directions, with some angular dependence, $\Ye \approx 0.4-0.5$ at high latitudes and $\Ye \approx 0.3-0.4$ at low latitudes. The result is an outflow that contains almost only light \rpe with a very small fraction of elements above the second peak which are synthesized only in mass that is ejected along the equator. The effect of a central BH on the outflow composition is less certain. It is clear that the lack of strong neutrino irradiation reduces $\Ye$ in the outflow, but it is not clear by how much. Recent 3D GRMHD simulations find that during the cooling inefficient phase ($t \gtrsim 1$s), the outflow is highly neutron rich with an electron fraction in the range of $\Ye \approx 0.1-0.3$ \citep{fernandez2019,christie2019}. During the NDAF phase (efficient accretion; $t \lesssim 1$s), some of the simulations find no $\Ye>0.3$ ejecta \citep{siegel2018}, while other simulations find that about 20\% of the ejecta, which is concentrated at high latitudes, have $0.25<\Ye<0.4$ \citep{fernandez2019,christie2019,miller2019}. 2D simulations that use pseudo-Newtonian potentials and $\alpha$-viscosity find that during the entire evolution a non-negligible fraction of the ejected mass has $\Ye>0.3$ \citep{just2015,fernandez2015,fernandez2017}. The conclusion, assuming that the 3D GRMHD simulations are more accurate, is that most of the ejecta from a BH-disk system have $\Ye \approx 0.1-0.3$, which results in a composition that  includes  the whole range of \rp elements. It is possible that a small fraction of the ejecta at high latitudes has $\Ye \gtrsim 0.3$, and therefore it contains only light \rp elements (i.e., it is lanthanide free). 

\begin{table}
	\caption{BNS Sub-relativistic ejecta}
\begin{minipage}{1\textwidth}
\begin{center}
	\includegraphics[width=1\textwidth]{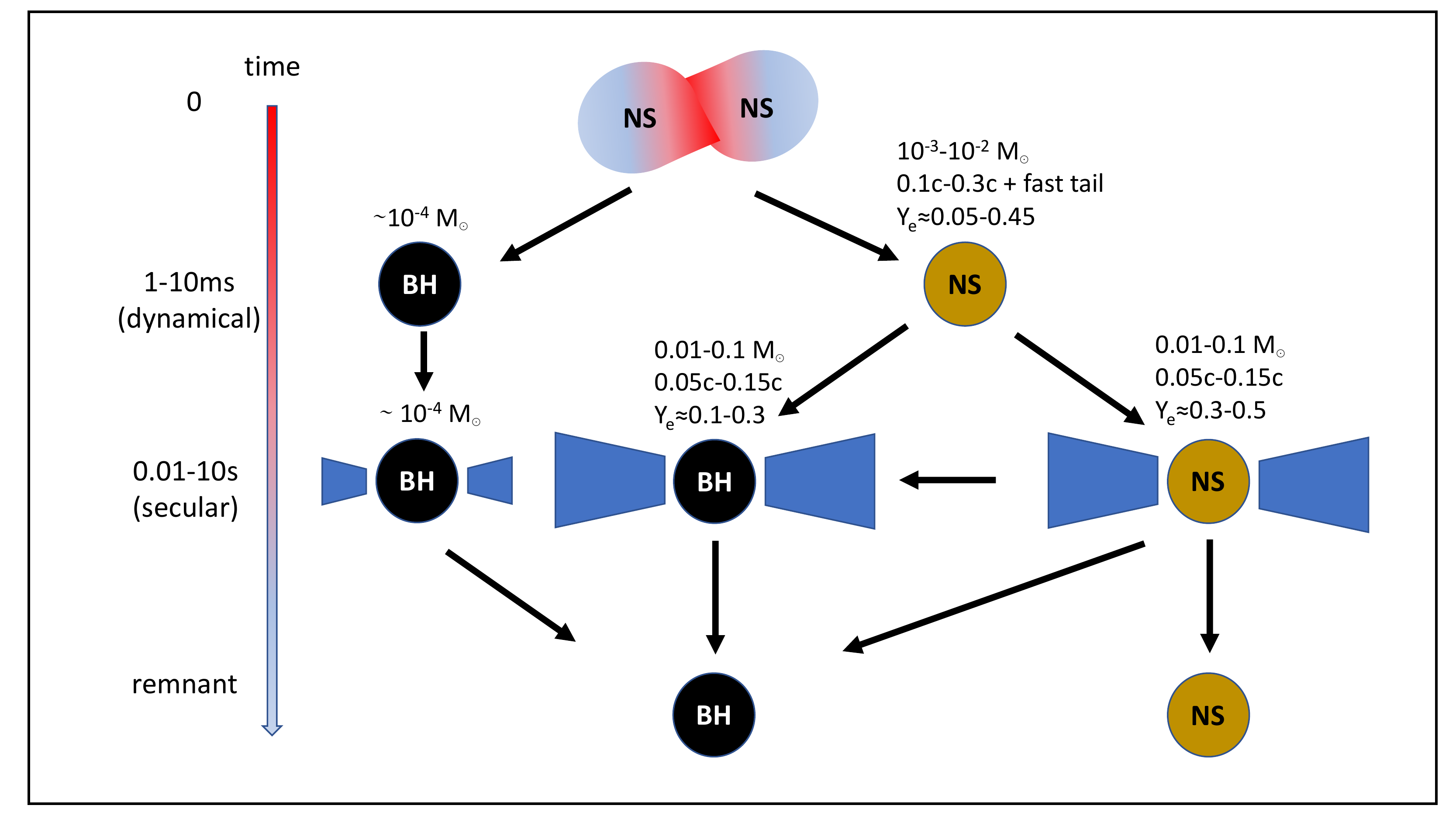}
	\includegraphics[width=0.95\textwidth]{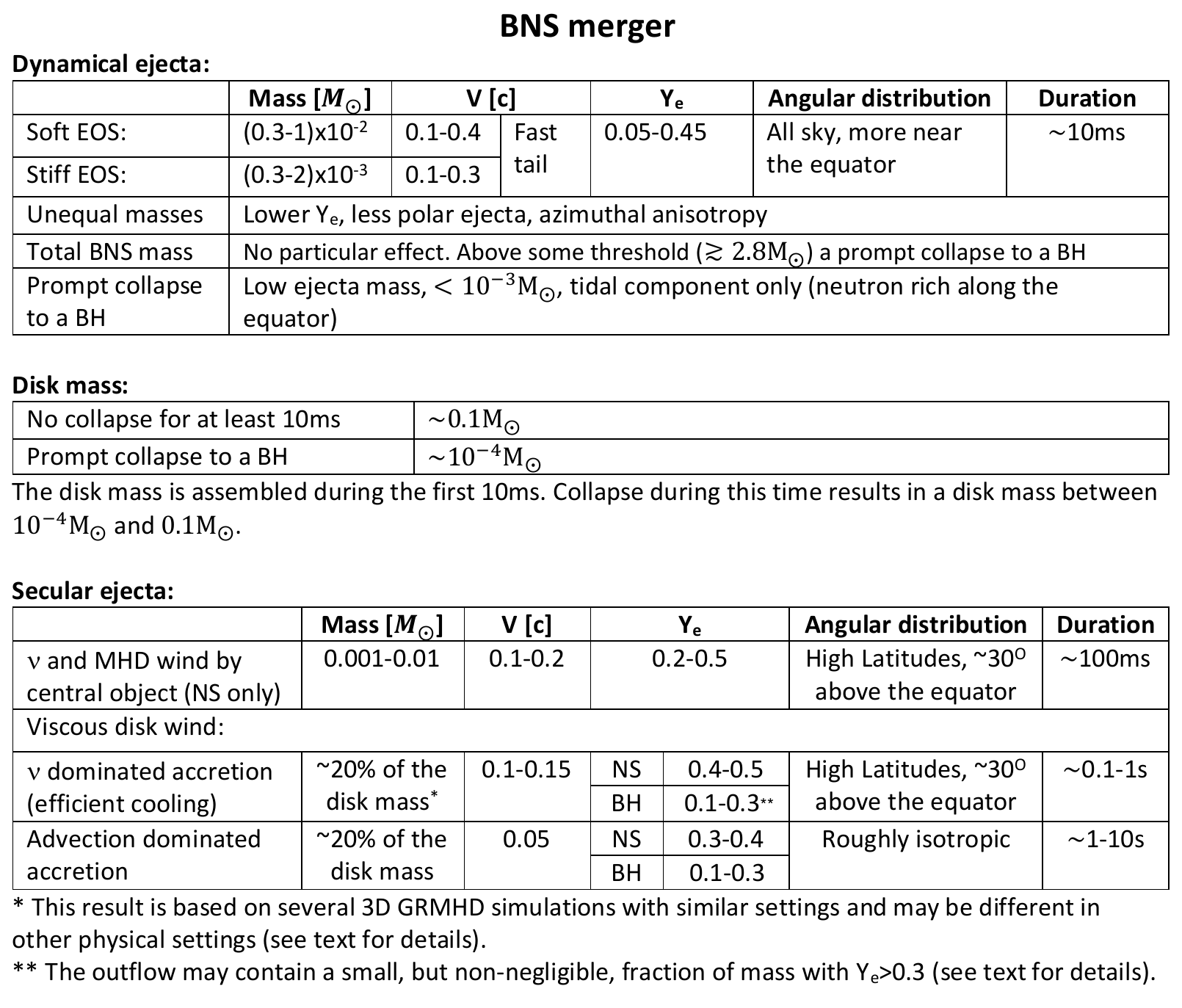}
\end{center}
\end{minipage}
\label{table:BNSsubRel}
\end{table}
 
\begin{table}
	\caption{BH-NS Sub-relativistic ejecta}
\begin{minipage}{1\textwidth}
\begin{center}
	\includegraphics[width=1\textwidth]{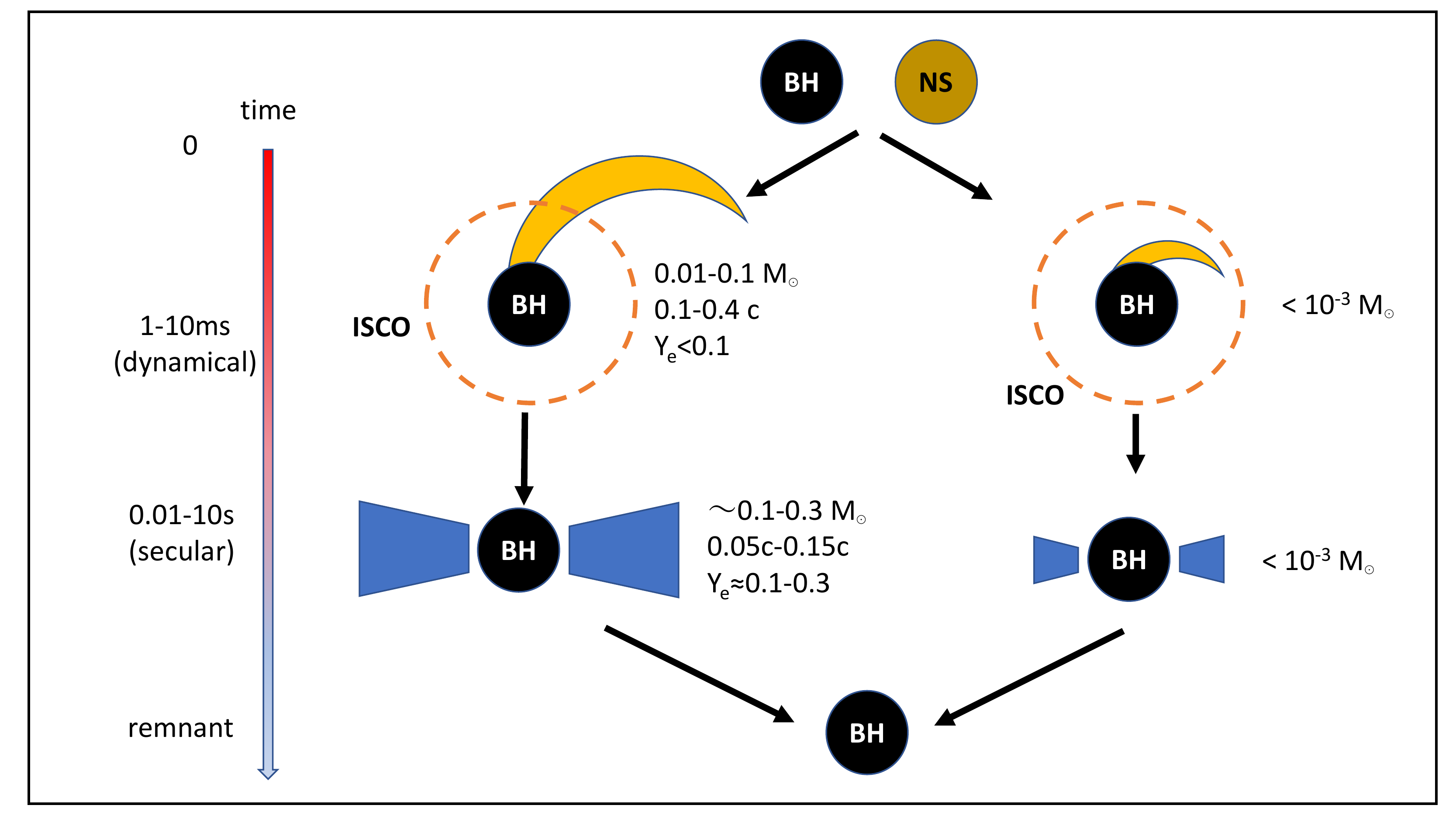}
	\includegraphics[width=1\textwidth]{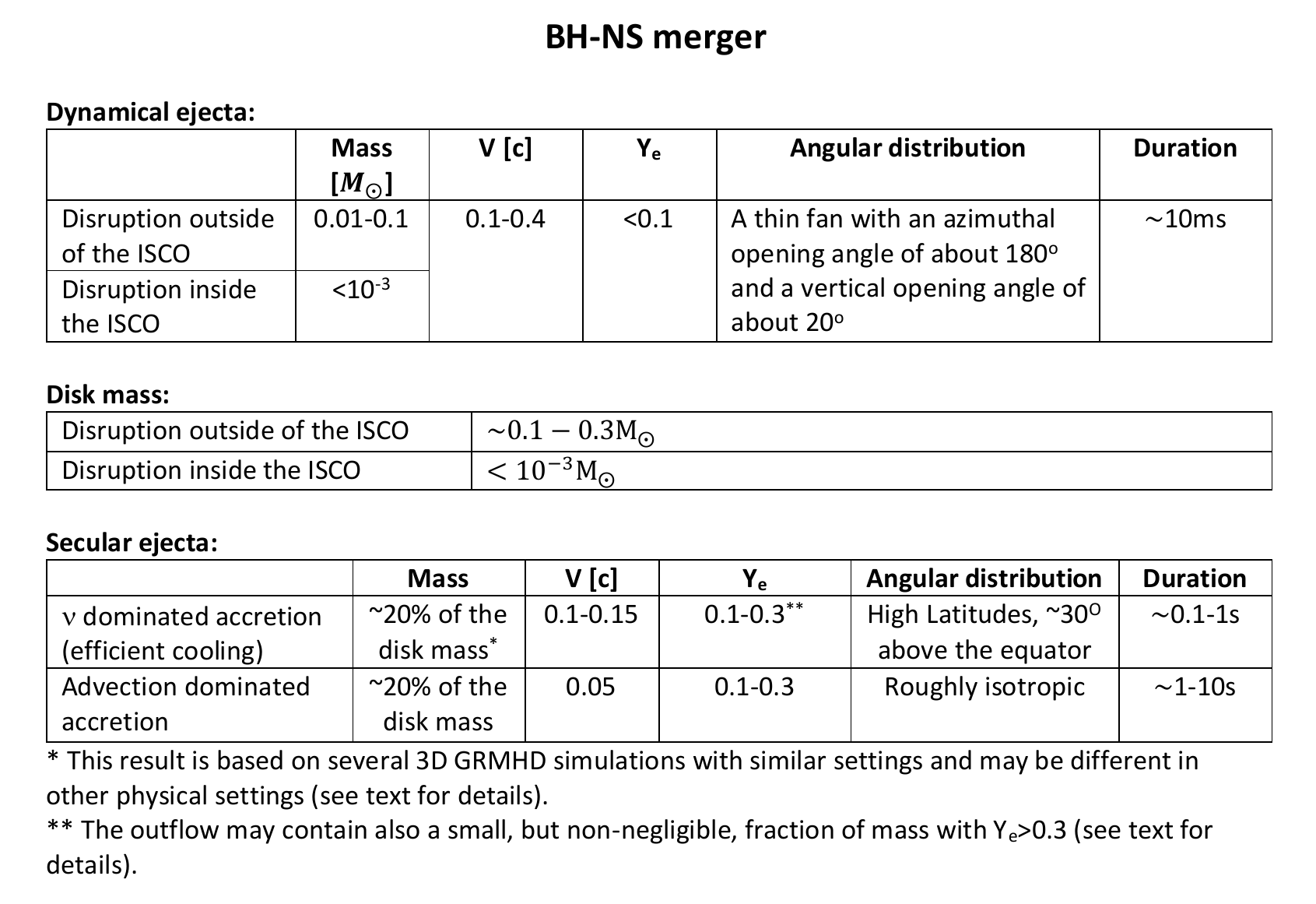}
\end{center}
\end{minipage}
\label{table:BHNSsubRel}
\end{table}

\subsection{{\bf  Ejection processes and properties of the sub-relativistic ejecta from BH-NS mergers}}\label{sec:BHNSejecta}
The dynamics of a BH-NS merger starts very differently than that of a BNS merger, with the following sequence of events. At the tidal radius of the BH, the NS is totally disrupted by tidal forces to form a single tidal arm, a small fraction of which is ejected. If the tidal radius is outside of the BH innermost stable circular  orbit (ISCO), then part of the ejected mass is unbound, and it forms the dynamical ejecta. The other part is sent into bound orbits and falls back after some time. Almost the entire NS material falls into the BH within a few ms and a small fraction of the material remains outside of the BH, forming a disk within $\sim 10$ ms. The BH-disk system and its secular evolution are rather similar to those of the disk that is formed following a BNS merger. 

\begin{figure}
	\includegraphics[width=0.47\textwidth]{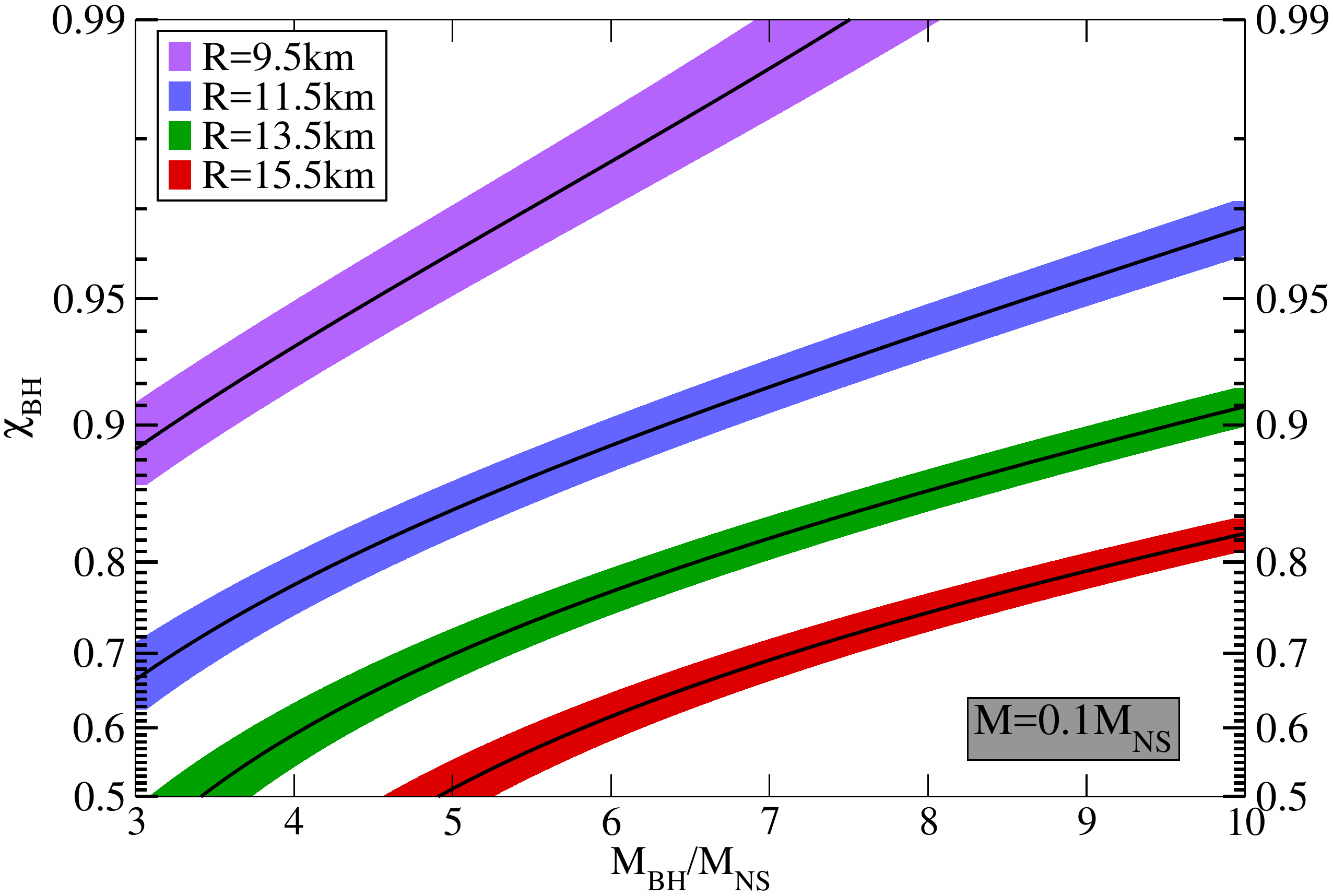}
	\includegraphics[width=0.445\textwidth]{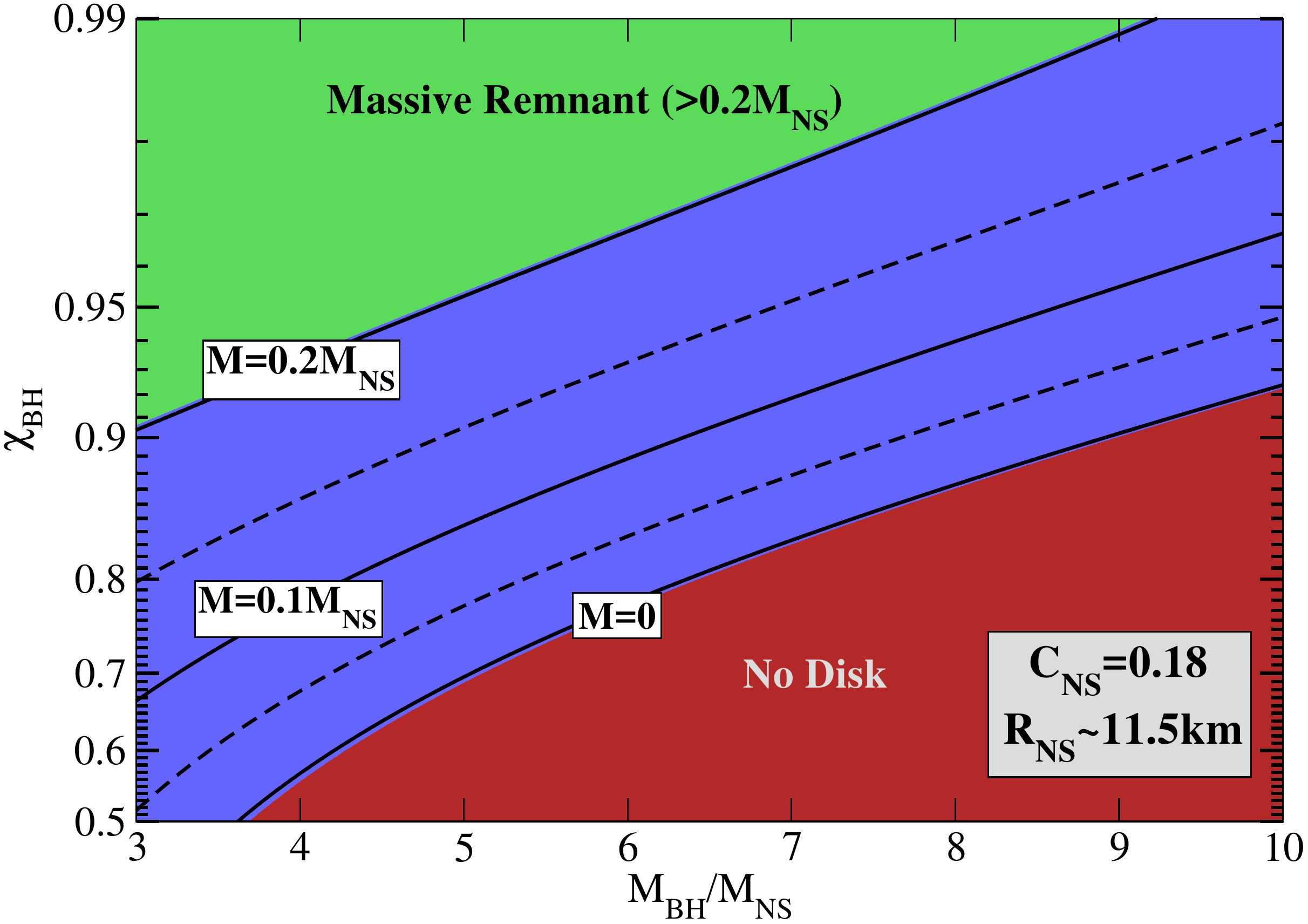}
	\caption{{\it Left}: The BH spin parameter and BH-NS mass ratio for which the remnant mass (i.e., disk mass + dynamical ejecta) is 10\% of the NS mass. The BH spin is fully aligned with the orbital angular momentum of the binary. Each line represents a different NS radius assuming a NS mass of $1.4\msun$.   {\it Right}: The mass of the remnant as a function of the BH spin parameter and the BH-NS mass ratio from a NS with a radius of 11.5 km (consistent with the constraints found based on observations of GW170817, see \S\ref{sec:GW170817_EOS}). From  \cite{foucart2012}. }%
	\label{fig:BHNS}
\end{figure}

The main parameter that determines the properties of the ejecta is the location of the tidal radius with respect to the ISCO. When the tidal disruption of the NS takes place within the ISCO almost all of its material falls directly into the BH and there is almost no dynamical ejecta and no disk. When the disruption takes place outside the ISCO, a significant amount of mass is ejected. The location of the tidal radius depends mostly on the NS radius (and thus on the NS EOS), where a larger NS radius implies a larger tidal radius (there is also a weak dependence on the BH/NS mass ratio). The location of the ISCO increases with the BH mass and decreases with its spin if it is aligned with the orbital angular momentum. Thus, for a given NS radius there is a maximal BH mass above which there is no significant ejecta. The larger the BH spin component that is aligned with the orbital angular momentum of the binary, the larger the value of this maximal mass. Figure \ref{fig:BHNS} shows that results of \cite{foucart2012} for the dependence of the total mass that remains outside of the BH after $\sim 10$ms  (i.e., tidal ejecta + disk)  as a function of the NS radius, BH spin and BH/NS mass ratio.    

The properties of the ejecta as a function of the NS EOS, the mass ratio, and the spin magnitude and orientation, were explored numerically in GRMHD simulations \citep[e.g.][]{foucart2012,foucart2013,foucart2014,kawaguchi2015,kawaguchi2016,kiuchi2015,kyutoku2015,foucart2018}. These simulations follow the merger and the post-merger evolution until the formation of the disk (at least 10-20 ms after the merger). The results of the various studies are in general agreement, probably because the evolution during the first 10 ms depends mostly on gravitational forces with minor effects of magnetic fields and neutrinos. The main result is that when the tidal disruption takes place well outside of the ISCO, the mass of the dynamical ejecta is $\sim 0.01-0.1\msun$ and the disk mass is $\sim 0.1-0.3\msun$. When the tidal  disruption takes place within the ISCO, both the dynamical ejecta and the disk mass are very small $<10^{-3}\msun$. 

The dynamical ejecta has a highly anisotropic distribution. It is thrown out in the shape of a thin fan concentrated along half of the equatorial plane, namely with an azimuthal opening angle of about $180^\circ$ and a vertical opening angle of about $20^\circ$. The ejected material is highly neutron rich, $\Ye \lesssim 0.1$ and therefore it contains almost only heavy \rp elements. The ejecta velocity depends on the depth of the potential well at the location of the disruption. Thus, high-spin, high-mass, BHs produce the fastest ejecta (as long as the disruption takes place outside of the ISCO). Typical velocities are in the range of $0.1-0.4$c.  

The secular evolution of the disk is expected to be similar to the one that follows a BNS merger when the central object collapses to a BH. At first, during the phase of efficient accretion, up to 20\% of the disk mass is ejected in a high latitude wind at a velocity of about $0.1-0.15$ c. Later, during the inefficient accretion phase, the mass that remains in the disk, which is about 20\% of the disk's original mass, is expelled as a roughly spherical wind at a velocity of about $0.05$ c. Simulations of a wind from a BH-disk system  typically find an outflow that is less neutron-rich than the dynamical tidal outflow, but the electron fraction is still low, $\Ye \sim 0.1-0.3$ \citep[e.g.,][]{siegel2018,fernandez2019,christie2019}, although some high $\Ye$ mass may be ejected at high latitudes \citep{miller2019,christie2019} (see the discussion in \S\ref{sec:rp_rate} for more details).

\subsection{{\bf  Summary}}
The current census of theoretical predictions for the properties of the various ejecta components and the evolutionary routes that a BNS merger may follow are summarized in a sketch and a table (Table \ref{table:BNSsubRel}). The most important ingredient that determines the nature of the outflow is whether the central object collapses to a BH, and if it does then at what time. A prompt collapse results in a very small ejecta mass of $\sim 10^{-4}\msun$ in both the  dynamical and the secular phases. Survival of the central NS for longer than 10 ms leads to mass ejection from various processes with a range of velocities, geometries, and compositions. These include:
(i) A dynamical ejecta of $\sim 10^{-3} \msun$ (stiff EOS) to  $\sim 10^{-2} \msun$ (soft EOS) that covers all $4\pi$ steradians at a velocity range of $\sim 0.1-0.3$c. The dynamical ejecta most likely include a low-mass tail ($\lesssim 10^{-6}\msun$) of fast material ($>0.6$c), that may even reach relativistic velocities. The dynamical ejecta have a wide range of electron fractions ($0.05 \gtrsim \Ye \lesssim 0.45$), where higher $\Ye$ material is ejected typically at higher latitudes. Thus, the dynamical ejecta is expected to include the whole range of \rp elements, where high-latitude material may contain no elements beyond the second peak (lanthanide-free), and material along the equator may be rich with heavy \rp elements;
(ii) Secular ejecta with a total ejected mass of $0.01-0.1\msun$. The outflow starts with a short phase of $\sim 100$ ms in which the central NS drives a wind of  $\sim 10^{-3}- 10^{-2} \msun$ at 0.1-0.2c with a high electron fraction ($0.2 \lesssim \Ye \lesssim 0.5$), and is followed by $0.01-0.1\msun$  of slower ejecta at 0.05-0.15c which are expelled within $\sim 10$s. This last component is expected to dominate the total ejected mass, and its electron fraction is high as long as the central object is a NS and lower if and when the central object collapses to a BH. 

The evolutionary path and the ejecta components of a BH-NS merger are also summarized in a sketch and a table (Table \ref{table:BHNSsubRel}). Here, the most important factor is the location of the tidal radius with respect to the ISCO. When the NS is tidally disrupted within the ISCO then practically all of its mass plunges into the BH, while if the disruption takes place outside of the ISCO then a significant amount of mass is thrown out dynamically by tidal forces, and even more mass remains outside of the BH to form an accretion disk. The locations of the tidal radius and the ISCO depend on the NS and BH parameters, where higher NS radius, lower BH mass and higher BH spin aligned with the orbital angular momentum of the binary, lead to a larger tidal radius relative to the ISCO. For example, assuming a NS radius of $11-13$ km and a BH mass of $5\msun$, then even a moderarte spin can lead to a 
significant ejecta mass. In contrast, a BH mass of $10\msun$ requires high spin, or otherwise there is practically no ejecta. When the NS disruption takes place outside of the ISCO, $0.01-0.1\msun$ are ejected dynamically by tidal forces. Part of this material forms a wide and thin fan of unbound material that expands along the orbital plane, while the rest moves in bound orbits and falls back on the accretion disk. This dynamical ejecta is highly neutron rich $\Ye<0.1$ and is, therefore, expected to contain only heavy \rp elements. The velocity of the dynamical ejecta is between 0.1c and 0.4c, where the higher range of the velocity is obtained for BHs with higher mass and high spin. The disk that is formed around the BH contains typically $0.1-0.3\msun$ and $\sim 20-40\%$ of this mass is ejected in a roughly spherical wind. The wind velocity is $0.05-0.15$c and its electron fraction is most likely moderate, $\Ye \approx 0.1-0.3$, so it probably contains both light and heavy \rp elements. There is a possibility that part of the disk wind has higher $\Ye$ and is therefore lanthanide-poor, but more theoretical work (or, better yet, observational data) is needed to test this option.

The observational manifestations of the various ejecta components, and how the observations can constrain the physics of the merger, are discussed in the following section.

\section{Theory of the UV/optical/IR emission (macronova/kilonova)}\label{sec:macronova}
A sub-relativistic $r$-process-rich outflow produces a radioactively powered UV/optical/IR transient in a manner that is very similar to Type Ia SNe. The basic physics is that radioactive heat is deposited into the expanding material, is thermalized to some extent, and leaks out towards the observer. The luminosity, spectrum and duration of the transient therefore depend on the mass, velocity, radioactive heating rate, and opacity of the outflow. The first to suggest that compact binary mergers will be followed by a supernova-like transient were \cite{li1998}. As an illustration, they calculated the predicted emission using a parameterized radioactive heating rate and opacity. The next to address the predicted signal was \cite{kulkarni2005}. He considered energy deposition by free neutrons and $^{56}$Ni and a parameterized grey opacity. A more realistic model was presented by \cite{metzger2010}. They used results from a numerical simulation of a merger \citep{rosswog1999} to calculate the nucleosynthesis of the \rpe (similar to \citealt{freiburghaus1999}), based on which they obtained realistic heating rates. They assumed an iron-peak-element opacity (similar to Type Ia SNe) to obtain the predicted light curve. The relatively low opacity resulted in a relatively bright and blue signal with a time scale of a day. The first to calculate a more  realistic opacity of \rpe  were \cite{tanaka2013} and \cite{barnes2013}. They showed that lanthanides (a group of heavy \rpe that are just above the second peak; A=140-176; Z=58-70) provide a much higher opacity than do iron-peak elements. The result is that merger ejecta that contain lanthanides produce a fainter transient, that peaks in the IR on a time scale of a week. Later, it was realized that some of the mass ejection processes may lead to lanthanide-free ejecta, which produce a brighter and bluer signal than the lanthanide-rich ejecta, and that peaks on a day time scale \citep{perego2014,metzger2014}. Over the past several years there have been many calculations, at various levels of approximation, of the radioactive-powered emission that follows BNS and BH-NS mergers. They all show a blue signal which peaks at $\sim 10^{41}$erg s$^{-1}$  after $\sim 1$d for lanthanide-free ejecta, and a red signal that peaks at  $\sim 10^{40}$erg s$^{-1}$  after $\sim 1$ week when lanthanides are present in the ejecta. 

The results of the previous section (\S\ref{sec:massEjection}) can be used to calculate the UV/optical/IR counterparts of BNS and BH-NS mergers. The initial conditions are the outflow properties (spatial distribution, velocity distribution, composition), from which the radioactive energy deposition (including thermalization) can be calculated. Then, calculation of the transfer of the deposited radiation through the ejecta provides the luminosity and the spectrum of the emission seen by an observer. An accurate calculation requires the exact mass, velocity and composition of each fluid element, a detailed account of the deposited energy, and a 3D radiative-transfer code that accounts properly for the frequency-dependent opacity of the \rp ejecta. When attempting to calculate the predicted signal from first principles, most of the uncertainty in the currently available models lies in the first and last steps, namely the ejecta properties and the radiative transfer, as described in some details in sections \ref{sec:massEjection} and \ref{sec:opacity}. The radioactive heating (for a given composition) is relatively well understood. Now that we have actual observations, we can try and work backwards, where the goal is to constrain the properties of the ejecta. Here, again, the large uncertainties of the opacity of \rpe is a major limitation. Thus, while the total mass and typical velocity of the ejecta can be estimated fairly well, it is more difficult to constrain the composition, as can be seen from the case of GW170817 (see section \ref{sec:GW170817_ejecta_property}).   

Radioactive decay of \rpe is a guaranteed source of energy in $r$-process-rich ejecta, and typically, it is also regarded as the dominant source of energy on time scales of a day and longer. However, it is plausible that there are additional energy sources, some of which are expected to dominate the UV/optical/IR  signal on time scales of hours or shorter, and some can possibly dominate at all times. The main energy sources that have been considered are a "central engine" (e.g., powered by  late accretion or by a magnetar), cooling emission from shocks that take place at large radii, and the decay of free neutrons. 

Below, I review the main physical processes and provide analytic and semi-analytic models of the UV/optical/IR emission from the sub-relativistic ejecta. I begin with  a simple analytic model of the emission from a spherical outflow with a given velocity, opacity, and heat rate,  that captures the main properties of the predicted light curve (\S\ref{sec:MN_Analytic}). I then describe a slightly more accurate semi-analytic model that can account for velocity and opacity gradients in the ejecta (\S\ref{sec:MN_semiAnalytic}). In \S\ref{sec:RadioactiveHeat} I discuss at some length the radioactive energy deposition and its thermalization, followed by a discussion of the opacity (\S\ref{sec:opacity}). In \S\ref{sec:KatzIntegral} I present a method to obtain a robust measure of the ejecta mass, which is independent of the highly uncertain radiative transfer. I end by discussing the emission during 
the first day, and non-radioactive energy sources (\S\ref{sec:MN_nonRadioactive}).

I begin, however, with a brief discussion of the origin of the names that are used for the UV/optical/IR supernova-like transient that follows BNS and BH-NS mergers (\S\ref{sec:etimology}). This is a possible source of confusion since different names have been used in the literature to describe exactly the same phenomenon. These include "Li-Paczynski nova", "merger-nova" and "r-process supernova", but the most popular names have been "kilonova" and "macronova", which are still  widely used by many authors. In this review, I use the name macronova, while emphasizing that all of these names are synonymous, and describe exactly the same class of transients. 

\subsection{{\bf  Etymology}}\label{sec:etimology}
I give here a short description of the history of the terms macronova and kilonova. \cite{li1998} did not give a name to the class of transients that they introduced in their paper, but only referred to it as being similar in nature to supernovae. \cite{kulkarni2005} dedicated part of his abstract to the naming of the new transient. He argued that since the physics is similar to that of a supernova (radiation from expanding gas that is powered by radioactive decay of unstable elements), but it is less luminous than a supernova, then the appropriate name would be "mini-supernova". However, in view of the oxymoronic nature of this term, he decided to use the term "macronova", to highlight the fact that it is related to supernovae, but fainter. The term was in sporadic use for several years, until \cite{metzger2010} renamed the phenomenon as a "kilonova". The rationale behind this name is that, according to the model presented in \cite{metzger2010}, the transient is expected to be 1000 times brighter than a typical nova. Since then, both names, as well as several others, have been used intermittently by different authors to describe the same phenomenon. According to NASA-ADS, at the time that this review is being written, there are about 1000 papers that include the term kilonova in their text and about 300 that include the term macronova. In this review, I will use the name macronova simply because this was the first to be used, and it is as good as any other name that relates the phenomenon to supernovae. Incidentally, 
the peak optical luminosity seen in GW170817, $\sim 10^{42}$erg s$^{-1}$, is similar to that of many core-collapse SNe (and brighter than some). However, due to its much shorter duration, the total energy of a macronova, $\sim 10^{47}$erg, is about two orders of magnitude lower than that of a SN.

\subsection{{\bf  Simplified macronova models}}\label{sec:MN_simple}
In terms of its physics, a macronova (similarly to Type Ia SNe)  is simply a flow of gas that expands homologously (i.e., at a given time $t$, each fluid element at radius $r$ has a constant velocity obeying $v(r)$ =r/t), and in which energy is deposited continuously in the form of radiation, and finally leaks out as observable emission. The key to estimating the resulting emission is to realize that, as long as the optical depth is large enough, the outflow can be generally divided into two regions--an external region where the diffusion time, $t_{\rm diff}$, is shorter than the dynamical time, and an internal region where it is longer. Homologous expansion implies that the dynamical time is also the time since the explosion, $t$, and the boundary between the two regions is $t_{\rm diff} \approx t$. The diffusion time of a photon from radius $r$ is $t_{\rm diff} \approx \tau r/c = \tau vt/c$, where $\tau$ is the optical depth to the observer and $v$ is the expansion velocity of gas at radius $r$.  Hence, the criterion $t_{\rm diff} \approx t$  can be translated  to a criterion on the optical depth to the observer $\tau \approx c/v$. Energy deposited at $\tau \lesssim c/v$ can be approximated as being radiated out immediately (within $t_{\rm diff} \lesssim t$), while energy deposited at $\tau \gtrsim c/v$ is largely trapped in the expanding gas, and thus suffers adiabatic loses. We define $r_{\rm trap}$ as the radius that separates these two regions, namely the radius where $\tau \approx c/v$. The expansion of the gas implies that the optical depth towards the observer drops fast enough such that increasing parts of the ejecta contribute to the luminosity with time, or in other words, $r_{\rm trap}$ moves inwards in Lagrangian coordinates (its motion in Eulerian coordinates depends on the velocity distribution). This continues up to the point where the optical depth of the entire ejecta drops below $c/v$. From this point on, the luminosity approaches the instantaneous  heating rate of the entire ejecta. 

\subsubsection{Analytic model}\label{sec:MN_Analytic}
The above principles can be quantified to obtain a very simple, yet useful, model for the luminosity. We consider a spherically symmetric ejecta with a total mass $m_{\rm ej}$, a velocity distribution in which mass is moving faster than $v$ is denoted $m(>v)$  or simply $m(v)$, a known energy deposition rate per unit of mass (including thermalization), $\dot{\epsilon}_{\rm heat}$, and a grey opacity $\kappa$. In the following, we use $m(v)$ as a Lagrangian coordinate. The optical depth from the mass $m(v)$ to the observer at any time is 
\begin{equation}\label{eq:tua_m}
	\tau(m) = \kappa \int_{vt}^\infty  \rho(r) dr \sim \kappa m/(4\pi v^2 t^2) ~.
\end{equation} 
At any given time, $t$, observed emission originates from the mass $m_{\rm obs}$ which is at $r>r_{\rm trap}$, where radiation diffuses to the observer over a dynamical time, namely $m_{\rm obs}=m(r>r_{\rm trap})=m(v_{\rm trap} \approx c/\tau)$, where $v_{\rm trap} = r_{\rm trap}/t$ is the velocity of the gas at the trapping radius. By requiring $\tau(m_{\rm obs})=c/v_{\rm trap}$, we obtain 
\begin{equation}\label{eq:mobs}
	m_{\rm obs}(t) \approx 5 \times 10^{-3} \msun~ C_{m} \left(\frac{v_{\rm trap}}{0.1c}\right) \left(\frac{\kappa}{1 {\rm~cm^2~ g^{-1}}}\right)^{-1}  \left(\frac{t}{{\rm day}}\right)^{2} ~~~;~~~ (m_{\rm obs}<m_{\rm ej}) ~.
\end{equation}
$C_{m}$ is a factor of order unity that depends on the exact velocity distribution. Equation \ref{eq:mobs} is implicit  since $m$ is a function of $v$ and therefore it should be solved simultaneously for $m_{\rm obs}$ and $v_{\rm trap}$. For example, there is an implicit dependence on $t$ via the time evolution of $v_{\rm trap}$, which can be solved for a given velocity distribution. Namely, in order to find $m_{\rm obs}(t)$ explicitly, one needs to plug $v(m)$ into equation \ref{eq:tua_m}. However, since $v$ in the ejecta varies by less than an order of magnitude, while $m$ can vary by many orders of magnitude, equation \ref{eq:mobs} provides an estimate of the observed mass even if the velocity distribution is not known exactly.   
As an example of an explicit solution of equation \ref{eq:mobs}, let us assume a power-law distribution  $m(>v) \propto v^{-k}$ for $v_{\rm min}<v$ where $k>0$. With this distribution, $v_{\rm trap} \propto t^{-\frac{2}{1+k}}$ and $m_{obs} \propto t^{\frac{2k}{1+k}}$. The coefficient $C_m$ for this power-law distribution is  $C_{m}=(k+2)/3k$, where we take the photon velocity in the radial direction as $c/3$, so $v_{\rm trap} = c/3\tau$. The observed mas  is growing with time until the entire ejecta mass, $m_{\rm ej}$, can be seen at time 
\begin{equation}\label{eq:tej}
	t_{\rm ej} \approx 3.5 {\rm~d} ~ C_{m}^{-1/2} \left(\frac{m_{\rm ej}}{0.05\msun}\right)^{1/2} \left(\frac{v_{\rm min}}{0.1c}\right)^{-1/2} \left(\frac{\kappa}{1 {\rm~cm^2 ~g^{-1}}}\right)^{1/2},
\end{equation} 
where $v_{\rm min}$ is the ejecta minimal velocity.
At $t>t_{\rm ej}$ the observed mass is $m_{\rm obs}=m_{\rm ej}$. 

Most of the internal energy that is deposited below the trapping radius ($r<r_{\rm trap}$) is lost to adiabatic expansion within one dynamical time. Therefore, we can approximate the emitted luminosity as the instantaneous energy deposition rate above the trapping radius, namely:
\begin{equation}\label{eq:MN_Lanalytic}
	L (t) \approx m_{\rm obs} \dot{\epsilon}_{\rm heat} \approx \left\{
	 \begin{array}{lr}
		 10^{41} {\rm~ erg~s}^{-1} ~ C_{\rm ad} C_{m} \left(\frac{v_{\rm trap}}{0.1c}\right) \left(\frac{\kappa}{1 {\rm~cm^2 ~g^{-1}}}\right)^{-1} \left(\frac{\dot{\epsilon}_{\rm heat}}{10^{10} \left(\frac{t}{{\rm day}}\right)^{-1.3} {\rm~\frac{\rm erg}{g\cdot s}}} \right) \left(\frac{t}{{\rm day}}\right)^{0.7} & t<t_{\rm ej}\\
		 &\\
		 8 \times 10^{40} {\rm~ erg~s}^{-1}~ \left( \frac{m_{\rm ej}}{0.05\msun}\right) \left(\frac{\dot{\epsilon}_{\rm heat}}{10^{10} \left(\frac{t}{{\rm day}}\right)^{-1.3} {\rm~\frac{erg}{g\cdot s}}} \right) \left(\frac{t}{{\rm week}}\right)^{-1.3}  & t>t_{\rm ej}
\end{array} \right.
\end{equation} 
where $C_{\rm ad}$ is a correction factor that accounts for energy that was deposited at time earlier than $t$, suffered some adiabatic losses, and was released toward the observer at time $t$. $C_{\rm ad}$ depends on the velocity distribution and varies slowly with time. An integration of equation \ref{eq:arnett} (see below) shows that for reasonable distributions $C_{\rm ad} \approx 2-3$ at $t<t_{\rm ej}$. Note that $\dot{\epsilon}_{\rm heat}$ in equation \ref{eq:MN_Lanalytic} is normalized by its analytic approximation, that is shortly presented below (equation \ref{eq:Qtot}), so  deviation of the heating rate from this approximation also affects the luminosity evolution with time. Here, again, at $t<t_{\rm ej}$ there is an implicit dependence on $t$ via the velocity $v_{\rm trap}(t)$.  For example, assuming a power-law distribution  $m(>v)\propto v^{-k}$ and $\dot{\epsilon}_{\rm heat} \propto t^{-1.3}$, the luminosity evolves as $L \propto t^{\frac{0.7(k-1.86)}{k+1}}$, so it is rising for $k>1.86$ and dropping otherwise.

As long as the ejecta is optically thick, $\tau(m_{\rm ej}>1)$, the observed spectrum is expected to be a modified blackbody, with a temperature $T \approx (L/4\pi\sigma v_{\rm ph}^2 t^2)^{1/4}$, namely:
\begin{equation}
	T \approx \left\{ \begin{array}{lr}
	 5,000{\rm~ K} ~ \left(C_{\rm ad}C_{m}\right)^{1/4} \left(\frac{v_{\rm trap}}{0.1c}\right)^{1/4}  \left(\frac{v_{\rm ph}}{0.2c}\right)^{-1/2} \left(\frac{\kappa}{1 {\rm~cm^2 ~g^{-1}}}\right)^{-1/4} \left(\frac{\dot{\epsilon}_{\rm heat}}{10^{10} \left(\frac{t}{{\rm day}}\right)^{-1.3} {\rm~\frac{erg}{g\cdot s}}} \right)^{1/4} \left(\frac{t}{{\rm day}}\right)^{-0.33}  & t<t_{\rm ej}\\
		 &\\
		 1,700{\rm~ K} ~ ~ \left( \frac{m_{\rm ej}}{0.05\msun}\right)^{1/4} \left(\frac{v_{\rm ph}}{0.2c}\right)^{-1/2} \left(\frac{\dot{\epsilon}_{\rm heat}}{10^{10} \left(\frac{t}{{\rm day}}\right)^{-1.3} {\rm~\frac{\rm erg}{g\cdot s}}} \right)^{1/4} \left(\frac{t}{{\rm week}}\right)^{-0.83}  & t_{\rm ej}<t<t_{\rm ph}
		 \end{array} \right.
\end{equation}
where $v_{\rm ph}=v(\tau=1)$ is the photospheric velocity and $t_{\rm ph}$ is defined below. The photospheric velocity is higher than $v_{\rm trap}$, typically by a small factor ($<2$).

The photospheric phase is expected to end when $\tau(m_{\rm ej}) \approx 1$, at: 
\begin{equation}\label{eq:tph}
	t_{\rm ph} \approx t_{\rm ej} \left(\frac{c}{v_{\rm min}}\right)^{1/2} \approx 10 {\rm~d} ~ C_{m}^{-1/2} \left(\frac{m_{\rm ej}}{0.05\msun}\right)^{1/2} \left(\frac{v_{\rm min}}{0.1c}\right)^{-1} \left(\frac{\kappa}{1 {\rm~cm^2~ g^{-1}}}\right)^{1/2}.
\end{equation} 
Around that time the spectrum is expected to make a transition from a modified blackbody to a nebular spectrum.

This simple model can be applied to any spherically symmetric mass distribution, grey opacity value and heat function. It is also simple to generalize it to a velocity-dependent opacity and heat deposition (i.e., non-uniform composition), and with some additional approximations also to anisotropic ejecta. It encompasses the essence of the physics involved and provides an order-of-magnitude estimate for the temporal evolution of the luminosity and the spectrum. The main approximations in this model, in addition to spherical symmetry, are the simplistic treatment of the frequency dependent opacity and of the energy deposited at $r<r_{\rm trap}$, which suffers adiabatic losses. The opacity cannot be treated properly without a sophisticated radiation transfer code, and even then our limited knowledge of \rpe opacity puts severe limits on the accuracy of the modeling. The adiabatic losses, however, can be accounted for rather simply, at the expense of making the model semi-analytic. I describe such a model next.

\subsubsection{Semi-analytic model}\label{sec:MN_semiAnalytic}
The following model is taken from \cite{hotokezaka2019a} and it is an extension of the Arnett model for SNe (\citealt{arnett1982}). Its advantage is that it can treat a flow with velocity and opacity gradients in a rather simple way, while taking into account the adiabatic losses and leakage of radiation from regions at $r<r_{\rm trap}$.
The basis for the model is the equation for the conservation of internal energy in a radiation-dominated gas that goes through a homologous expansion,
\begin{equation}\label{eq:arnett}
	\frac{dE}{dt}=-\frac{E}{t}-L(t)+\dot{Q}(t)
\end{equation} 
where $E$ is the internal energy within the outflow, $L$ is the bolometric luminosity and $\dot{Q}$ is the deposition rate of  energy that ends up as thermal gas and radiative energy. The first term on the right-hand-side accounts for adiabatic losses of the radiation (for a radiation dominated gas with an adiabatic index of 4/3). This equation is exact, and for a given composition we also know $\dot{Q}$ rather well. The main approximation is in the evaluation of $L$. If we are interested in a rough approximation of the outflow as a single zone, with typical $m$, $v$, $\kappa$ and $\dot{Q}$, then the luminosity can be approximated as $L \approx \frac{E}{t_{\rm diff}+vt/c}$ \citep[e.g.,][]{metzger2017}, where 
\begin{equation}
	t_{\rm diff} \approx 2 {\rm~d} ~ C_{m}^{-1} \left(\frac{m}{0.01\msun}\right) \left(\frac{v}{0.1c}\right)^{-1} \left(\frac{\kappa}{1 {\rm~cm^2 ~g^{-1}}}\right) \left(\frac{t}{{\rm day}}\right)^{-1}
\end{equation}
is the time it takes a photon to escape from the ejecta when the diffusion limit is applicable. When the optical depth drops below unity, the photon escape time is $vt/c$ and the sum $t_{\rm diff}+\frac{vt}{c}$ provides a smooth interpolation of the escape time from the diffusive regime ($\tau \gg 1$) to the free streaming regime ($\tau \ll 1$). 

The single-zone approximation does not work very well when we wish to account for the effect of the ejecta velocity structure on the light curve. The reason is that the diffusion time to the observer is different from different regions in the ejecta. A useful approximation that does account rather well for the ejecta velocity structure is to divide the ejecta into shells  according to the velocity distribution, where the j$^{\rm th}$ shell properties are $m_j$,  $v_j$, $\kappa_j$ and $\dot{Q}_j=m_j\dot{\epsilon}_{{\rm heat},j}$, and to integrate equation \ref{eq:arnett} for each of the shells. The total observed luminosity is then the sum of the shell luminosities. The diffusion of photons out of the   j$^{\rm th}$ shell is calculated by taking into account the optical depth of all the shells with velocity larger than  $v_j$, i.e. $\tau_j(t) = \int_{v_jt} ^\infty \kappa \rho(r) dr$. This optical depth is used when approximating the diffusion time, $t_{{\rm diff},j} \approx \tau_j v_jt/c$. The approximation to $L$ in this approach should take into account the fact that there are three different regimes in which: (i) the radiation is trapped, namely $t<t_{\rm diff}$; (ii) the radiation streams away over a diffusion time, $vt/c<t_{\rm diff}<t$, and (iii) the radiation streams freely, $\tau<1$.  In the first regime, where the radiation is trapped, only an exponentially small fraction of the internal energy escapes to the observer. A good approximation for the energy fraction that escapes from the   j$^{\rm th}$ shell over a dynamical time, when $t_{\rm diff} > t$, is \citep{piro2013}
\begin{equation}\label{eq:fesc}
	f_{{\rm esc},j} \approx {\rm erfc}\left[\sqrt{\frac{t_{{\rm diff},j}}{2t}}\right] ,
\end{equation}
where erfc is the complementary error function. When $t_{\rm diff} < t$ most of the energy escapes, namely $f_{\rm esc} \approx 1$ and this fraction escapes over a diffusion time as long as $\tau>>1$ , and over a light-crossing time when $\tau<<1$. Since, ${\rm erfc}\left[\sqrt{\frac{t_{\rm diff}}{2t}}\right] \approx 1$ when $t_{\rm diff} < t$, equation \ref{eq:fesc} provides a good approximation to $f_{\rm esc}$ at all times, and the escape time of the radiation can be approximated as
\begin{equation}
	t_{{\rm esc},j} \approx \min\{t_{{\rm diff},j},t\} + \frac{v_jt}{c} .
\end{equation}
The luminosity of the   j$^{\rm th}$ shell is then approximated as 
\begin{equation}
	L_j \approx \frac{f_{{\rm esc},j}E_j }{t_{{\rm esc},j}},
\end{equation}
and the total luminosity is the sum of the luminosity from all shells,
\begin{equation}
	L \approx \sum L_j
\end{equation}
%A python code that implement this semi-analytic approximation of the light curve was written by \cite{hotokezaka2019a} and can be found at 
\subsection{{\bf Radioactive heating}}\label{sec:RadioactiveHeat}
\rp nucleosynthesis, by its nature, produces nuclei that are far from the beta valley of stability. Therefore, as soon as nucleosynthesis ceases, the freshly formed nuclei start a chain of beta and alpha decays, and in some cases also fission, until stable (or very long-lifetime) nuclei are reached. Below, I discuss first the energy released by each of these processes, then I discuss the thermalization processes to find the radioactive energy that is deposited as heat. I conclude with a simple analytic model of the radioactive heating in $r$-process-rich ejecta. Note that if the composition is known, then both the radioactive decay and the thermalization processes can be approximated rather accurately since they depend mostly on experimental data of nuclei that are close to the valley of stability. In this subsection I follow mostly the discussion in \cite{hotokezaka2019a}, who provide a publicly available code (\url{https://github.com/hotokezaka/HeatingRate}) that calculates the heat rate numerically and the resulting light curve semi-analytically.

\subsubsection{Total radioactive energy release}
\noindent {\underline {\it $\beta$-decay:}}\\
Most of the unstable nuclei go through a chain of beta decays (keeping the mass number A constant while increasing the atomic number Z) on their way to the valley of stability. Each $\beta$-decay releases an energy of the order of 0.5-5 MeV in the form of an electron, an anti-neutrino, and gamma-rays (the gamma-rays are released when the excited nucleus decays to its ground state). The energy carried by the neutrinos escapes, while part of the energy released in the form of electrons and gamma-rays is absorbed in the ejecta and thermalized, providing the main energy source of the optical/IR counterpart. Each nucleus has its own half-life, $\tau_{1/2}$, which in general increases the closer it is to the valley of stability. As a result, at any given time after the merger, $t$, there are several elements with $\tau_{1/2}  \sim t$ which dominate the energy deposition. Along the decay chain, the number of elements with  $\tau_{1/2} \sim t$ is roughly equal per logarithmic interval of $t$ and therefore, to first order, the dependence of the energy deposition on time is $\dot{\epsilon}_{dep} \propto t^{-1}$ (as approximated, e.g., by \citealt{li1998}). At higher orders, the average energy per decay drops slowly with $\tau_{1/2}$ and therefore $\dot{\epsilon}_{dep}$ drops slightly faster than $t^{-1}$ \citep{colgate1966,metzger2010}. \cite{hotokezaka2017} show, based on analytic considerations, that as long as many elements contribute simultaneously, $\dot{\epsilon}_{dep}$ can be approximated as a power-law in the range between $t^{-6/5}$ and $t^{-4/3}$. 

Numerical simulations of nucleosynthesis in ejecta from BNS and BH-NS mergers show that on a time scale of seconds, the energy deposition depends on the exact conditions in the expanding material and the history of each fluid element. For typical fluid elements found in these simulations, a total energy of about 1 MeV per nucleon is released at roughly a constant rate during the first second\footnote{Most of the radioactive energy  is released during the first seconds. All of this energy, except for the part which is carried away by neutrinos, is converted via adiabatic losses to bulk motion. Note that 1 MeV per nucleon  corresponds to a bulk motion velocity of $\sim 0.05$c. This implies that the effect of radioactive power on the final velocity of ejecta moving at velocity $\gtrsim 0.1$c is negligible.}, i.e., a deposition rate per unit of mass of $\sim 10^{18}$ erg s$^{-1}$g$^{-1}$. At later times, the energy deposition decays roughly as $t^{-1.3}$ (with some dependence on the exact ejecta composition; see below). Thus, the total energy injected per unit of mass as electrons and gamma-rays (not including thermalization factors; see \S\ref{sec:thermalization}) is usually approximated as \citep[e.g.,][]{korobkin2012}:
\begin{equation}\label{eq:Qtot}
	\dot{\epsilon}_{\rm dep} \sim 10^{10} \left( \frac{t}{\rm day} \right)^{-1.3} {\rm erg ~s^{-1}~g^{-1}}.
\end{equation}   

\begin{figure}
	\includegraphics[width=0.5\textwidth]{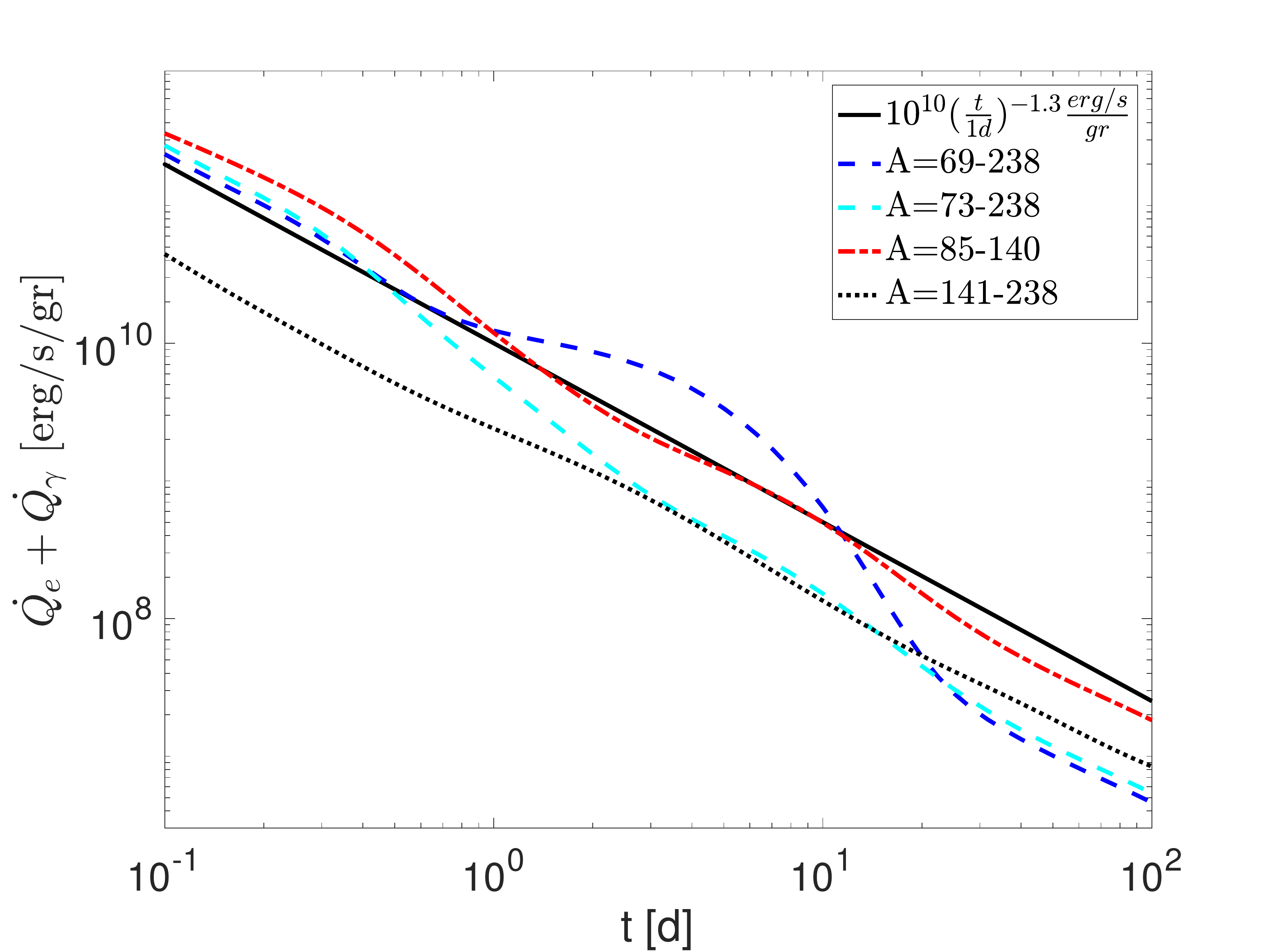}
	\includegraphics[width=0.5\textwidth]{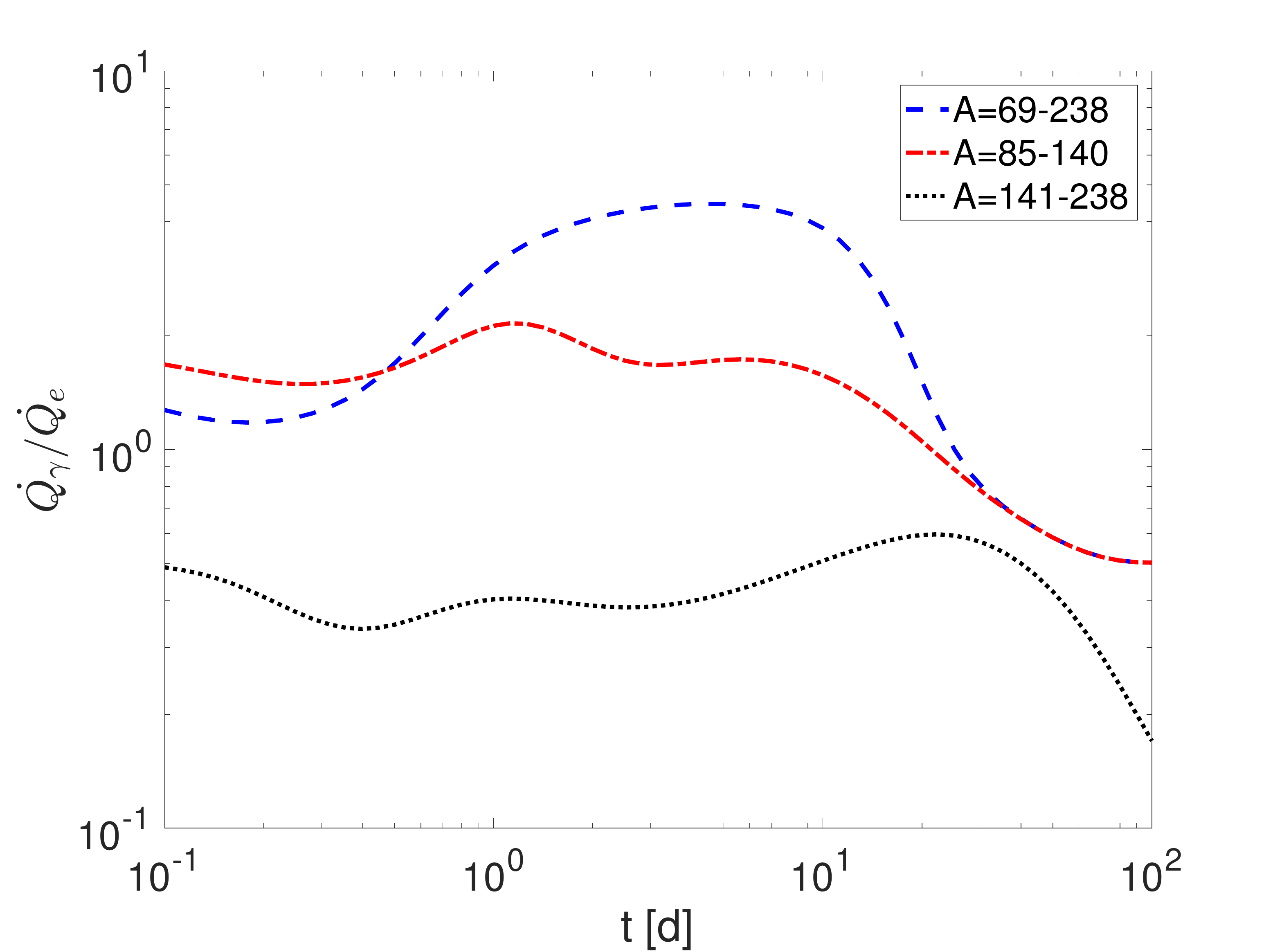}
	\caption{{\it Left:} The total radioactive energy deposition rate per unit of mass in the form of electrons and gamma-rays for different compositions. The different curves are for different ranges of atomic masses (as indicated in the legend), where the relative abundances are solar (taken from \citealt{sneden2008}). Also shown is the analytic approximation of equation \ref{eq:Qtot}. It shows that the analytic approximation is good to within an order of magnitude but that there are significant differences, both in normalization and in the decay rate, between different compositions and as a function of time. {\it Right}: Ratio of the energy deposited in gamma-rays and the energy deposited in electrons for three of the compositions shown in the left panel. This ratio affects the thermalization of the deposited energy. The figure shows that this ratio depends on the composition and that it can fluctuate with time.}%
	\label{fig:Qtot}
\end{figure}

On the timescales relevant for the observed emission, i.e., $t\gtrsim 10^3$ s, all of the unstable \rpe are very close to the valley of stability (e.g., one or two beta decays away), where there is accurate experimental data on the decay time scale and products\footnote{The data on unstable nuclei that are close to the valley of stability are typically known from experiments and are therefore quite accurate, while the properties of nuclei that are far from the valley of stability are based mostly on highly uncertain theoretical calculations. This is the reason why calculations of \rp nucleosynthesis, which takes place far from the stability valley, are  highly uncertain, while calculations of the heat deposition rate at late times ($\gtrsim 10^3$ s), are relatively accurate if the composition is known.} \citep[e.g.,][]{chadwick2011}. Since the atomic mass number, A, is conserved during $beta$-decay, for every stable element that is produced by $beta$-decay, there is only a single $beta$-decay chain that leads to this element.
Thus, for a given final (stable) ejecta composition, the total energy deposition can be readily calculated and it is probably the least uncertain ingredient in the calculation of the observed radiation. The main uncertainty is introduced by the unknown composition and by the thermalization, which depends on ingredients such as the ejecta mass and velocity (see \S\ref{sec:thermalization}). Figure \ref{fig:Qtot} depicts the total radioactive energy deposition rate by $\beta$-decay (in electrons and gamma-ray) per unit of mass for several compositions, which include different ranges of atomic masses with a solar abundance ratio as given by \cite{sneden2008} (note that for light \rpe there is a considerable uncertainty in the solar abundance; \citealt{goriely1999}). The compositions shown include the entire range of \rpe (A=69-238), light \rp elements, not including the first peak but including the second peak (A=85-140) and only heavy \rpe above the second peak (A=141-238). The figure demonstrates several important points. First, the normalization of the analytic approximation is accurate to within an order of magnitude. For example, the normalization of light elements is about a factor of 10 higher than that of the heavy \rp elements. Also the power-law index at any given time (i.e., the instantaneous luminosity logarithmic derivative), can deviate significantly from $-1.3$.  For example, the power-law index for all \rpe (A=69-238) varies between $-0.5$ at day 1.5 and $-3.8$ at day 15. When the first peak is excluded there are less fluctuations, but still, for A=85-140, the power-law index varies between $-2$ at day 1 to $-1$ at day 5. The reason for the significant differences between the energy depositions of different compositions, as well as the fluctuations of the decay rate, is that typically a small number of elements dominate at any given time, and in some cases only a single element dominates. For example, in the case of A=69-238 the emission between 1 and 10 days is fully dominated by the decay chain $^{72}$Zn$\rightarrow^{72}$Ge$\rightarrow^{72}$Ga. This is due to the half-life times of this chain (1.98 d and 0.59 d), the relatively high energy released per nucleus (3.5 MeV), and the relatively high solar abundance of $^{72}$Ga according to  \cite{sneden2008} (the actual amount of \rp $^{72}$Ga in the sun is highly uncertain). The contribution of this chain is the sole reason for the difference in the energy deposition of A=69-238 and A=73-238. 

The relative fraction of the deposited energy that goes to gamma-rays and electrons is also important. The reason is that \grays deposit their energy as heat only during the first days, while electrons thermalize efficiently up to weeks after the explosion (see \S\ref{sec:thermalization}). Figure \ref{fig:Qtot} shows the $\gamma$-ray to electron energy deposition ratio as a function of time for several compositions. It shows that for light \rpe the energy deposited in \grays is about twice the energy deposited in electrons (at least for the first several weeks), while for heavy \rpe it is the opposite.\\

\noindent {\underline {\it $\alpha$-decay:}}\\
Some of the decay chains of elements above the third peak include also a release of $\alpha$ particles. Each $\alpha$ particle carries 5-9 MeV, and for some elements the chain contains several such decays. If a significant amount of elements with A>220 is synthesized, then $\alpha$-decay can potentially provide a significant contribution to the heating of the sub-relativistic ejecta after about a week\footnote{On a time scale of a day, the $\alpha$-decay contribution is dominated by three chains that start at  $^{212}$Tl, $^{216}$Bi and $^{220}$Po. All chains end at $^{208}$Pb and they all  deposit about  $\dot{\epsilon}_{\alpha}(t=1d) \approx 2 \times 10^{11}$ erg s$^{-1}$ for each gram of these elements. For compositions that resemble solar, this contribution is insignificant compared to that of the  $\beta$ decay.}  \citep{barnes2016,hotokezaka2016b,wu2019}. This increased importance at late times is a result of a slower decline in the deposition rate from $\alpha$-decay (roughly as $t^{-1}$ compared to $t^{-1.3}$ for $\beta$-decay). The thermalization efficiency of $\alpha$ particles is comparable to that of electrons from $\beta$-decays. The $\alpha$-decay deposition on a time scale of days to weeks is dominated by four elements with A=222-225 ($^{222}$Rn, $^{223}$Ra, $^{224}$Ra, $^{225}$Ac) \citep{wu2019}. These elements have half-life of 3-10 days, after which they go through a sequence of 4-5 $\alpha$-decays and several $\beta$-decays, almost all with a relatively short half-life, until they reach stable (or very long lifetime) nuclei with A=207-209 ($^{207}$Pb, $^{208}$Pb, $^{209}$Bi). Each of these chains releases about 25 MeV per parent nucleus in $\alpha$ particles. This corresponds to a deposition rate of $\dot{\epsilon}_{\alpha}(t=10~{\rm d}) \approx 4 \times 10^{10}$ erg s$^{-1}$ for each gram of elements with A=222-225. On a timescale of 10 days, the energy deposition in electrons from $\beta$-decay is $\sim 2 \times 10^{8}$ erg s$^{-1}$ g$^{-1}$ (on this time scale gamma rays escape and do not contribute, for typical ejecta parameters). This implies that $\alpha$-decay is significant on that time scale, if the mass in those four elements is $\gtrsim 0.005$ of the total ejecta mass. The actual amount of elements with A=222-225 that are synthesized in various conditions is hard to estimate, as there are no direct observational constraints, and the theoretical predictions (for the same physical conditions) vary by orders of magnitude. The amount of the decay products of these elements ($^{207}$Pb, $^{208}$Pb and $^{209}$Bi) in the Sun suggests that if BNS and BH-NS mergers synthesize \rpe with a pattern that is similar to solar, then the fraction of elements with A=222-225 is probably too low for $\alpha$-decay to be the dominant heating source \citep{hotokezaka2019a}. Observations of late IR emission from BNS mergers can potentially constrain this contribution \citep{wu2019}.\\

\noindent {\underline {\it Fission:}}\\
Some very heavy elements with $A \gtrsim 250$ decay through spontaneous fission. The energy released in the process is roughly Mev per nucleon, namely $\sim 200$ MeV per parent nucleus, and the thermalization of the fission products is very efficient. Thus, a small amount of nuclei that go through fission on the relevant timescales can provide a significant heat source \citep{hotokezaka2016b}. Unfortunately, there is no reliable information on the half life and the decay chains for most of these elements. An exception is Californium-254, for which experimental data show that it almost always decays through  spontaneous fission  with a half life of 60.5 days and an energy release of about 185 MeV per nucleus \citep{zhu2018,wu2019}. This corresponds to an energy deposition rate of $\dot{\epsilon}_{\rm Cf254}(t=100d) \approx 2 \times 10^{10}$ erg s$^{-1}$ per gram of $^{254}$Cf. The contribution from $\beta$-decay on this time scale depends on the thermalization of the electrons, which is inefficient at this time, but for parameters similar to those observed in GW170817, it is roughly $\dot{\epsilon}_{\beta}(t=100~{\rm d}) \sim 2 \times 10^{6}$ erg s$^{-1}$g$^{-1}$. Thus, fission of $^{245}$Cf will dominate the energy deposition on time scales of months, if its mass fraction in the ejecta is $\gtrsim 10^{-4}$. Theoretically, we cannot estimate this fraction reliably, except for the fact that only ejecta with $Y_e \lesssim 0.15$ might, although not necessarily, produce a significant amount of $^{245}$Cf. Observations of late IR emission from compact binary mergers can potentially constrain the production of $^{245}$Cf in those sites \citep{zhu2018,wu2019}.

\subsubsection{Thermalization of the radioactive products}\label{sec:thermalization}
The energy released in the radioactive decay is converted to optical/IR emission only if it is thermalized and converted to heat. Neutrinos escape immediately and are therefore of no interest for the thermal emission. The different thermalization efficiencies of \grays, electrons, $\alpha$ particles and fission products have been discussed in \cite{hotokezaka2016b,barnes2016,waxman2018,waxman2019,kasen2019} and \cite{hotokezaka2019a}. Below, I describe the different thermalization processes and conclude by providing simple formulae that estimate the total energy deposition that goes into heat. The discussion follows, in most parts, the derivation of \cite{hotokezaka2019a}.  \\

\noindent {\underline {\it $\gamma$-rays:}}\\
Every  $\beta$-decay releases $\gamma$-rays. Most of the energy is carried by photons with $\sim 0.5-1$ MeV, but the spectrum ranges between about 50 keV and 3 MeV. The energy fraction carried by \grays out of the total $\beta$-decay energy
depends on the composition (e.g., figure \ref{fig:Qtot}). Considering light and heavy \rpe at solar abundance, \grays energy constitutes between 10\%-60\% out of the total energy (including neutrinos) and between 25\%-75\% of the energy that can thermalize (electrons and gamma-rays). Gamma-rays deposit their energy via several processes, depending on their energy. At low energies ($\lesssim 0.5$ MeV) the dominant process is photoelectric absorption, at intermediate energies ($\sim 1$ MeV) it is Compton scattering, and at high energies ($\gtrsim 5$ MeV) pair-production. The optical depth for gamma-rays can depend, to some degree, on the composition, mostly at lower energies and through the ionization energy of K-shell electrons ($\propto Z^2$). At photon energies $E_\gamma \gtrsim 1 {\rm MeV}$ the opacity is roughly constant, $\kappa_\gamma \sim 0.05 {\rm~ cm^2~g^{-1}}$, while at lower energies the opacity increases rapidly, roughly as $\kappa_\gamma \propto E_\gamma^{-2.7}$.   

The thermalization of radioactively deposited gamma-rays in expanding material has been studied in detail in the context of $^{56}$Ni decay in supernova ejecta \citep[e.g.][]{swartz1995,jeffery1999,wygoda2017}. These studies find that the fraction of the gamma-ray energy that is deposited at any given time can be approximated rather accurately by finding a single time scale, $t_0$, which is the time where the effective optical depth for gamma-rays is $\tau_{\gamma,{\rm eff}}=1$. The fraction of the energy deposited as a function of time is then 
$f_\gamma \approx 1-e^{-\tau_{\gamma,{\rm eff}}}=1-e^{-(t_0/t)^2}$. For a given outflow mass, velocity distribution and $^{56}$Ni distribution, $t_0$ is calculated as follows. The effective opacity, $\kappa_{\gamma,{\rm eff}}$, is found by averaging over the specific spectrum of gamma-rays emitted during the decay of 
$^{56}$Ni and $^{56}$Co. The effective column density, $\Sigma_{\gamma,{\rm eff}}$ is calculated by averaging over  all fluid elements and for each element averaging over all directions (see details in \citealt{wygoda2017}). Homologous expansion implies $\Sigma_{\gamma,{\rm eff}} \propto t^{-2}$ and the characteristic 
time for \grays thermalization is $t_0=\sqrt{\kappa_{\gamma,{\rm eff}}~ \Sigma_{\gamma,{\rm eff}}~t^2}$.
 
This calculation can be generalized to \rp ejecta, as shown in  \cite{hotokezaka2019a}. 
%Here we will assume a homogenous composition and spherical symmetry, but carrying out the calculation for any outflow distribution and geometry is straight-forward. 
For a given composition, $\kappa_{\gamma,{\rm eff}}$ can be calculated for each element based on the gamma-ray spectrum released upon its decay. If the ejecta is composed of light \rpe the typical $\kappa_{\gamma,{\rm eff}}$ values range between $0.03 {\rm~ cm^2~gr^{-1}}$ and $0.1 {\rm~ cm^2~g^{-1}}$. If the ejecta is composed mostly of heavy \rpe then $\kappa_{\gamma,{\rm eff}}$ of the various elements is in the range of $0.03-1 {\rm~ cm^2~g^{-1}}$. The reason for the difference is that the heavy \rpe have higher photoelectric absorption cross-sections (higher Z), and typically lower energies of their emitted gamma-rays. An accurate calculation requires using the specific $\kappa_{\gamma,{\rm eff}}$  for each element, however a reasonable approximation can be obtained by taking an average value of $\kappa_{\gamma,{\rm eff}}$ for all elements. \cite{hotokezaka2019a} find that for \rp material in the atomic mass range $A=85-209$ and solar abundance (i.e., dominated by light \rp elements) $\kappa_{\gamma,{\rm eff}} \approx 0.07{\rm~ cm^2~g^{-1}}$ while for $A=141-209$ (heavy \rp elements) $\kappa_{\gamma,{\rm eff}} \approx 0.4{\rm~ cm^2~g^{-1}}$.

The effective column density can be written as $\Sigma_{\gamma,{\rm eff}} = C_\Sigma m v^{-2} t^{-2}$ where $m$ is the total ejecta mass and $v$ is the characteristic velocity. $C_\Sigma$ is a constant of order $1/4\pi$ that depends on the velocity distribution of the ejecta and is calculated by carrying out a double integral over all fluid elements, and for each element finding the average column density over all directions \citep[e.g.,][]{wygoda2017}. For example, $C_\Sigma=0.18$ for a uniform density within a sphere with a radius $vt$. Below, I provide some useful analytic approximations, including a value of $C_\Sigma$ for a more realistic velocity distribution.\\

\noindent {\underline {\it Electrons:}}\\
The fraction of the $\beta$-decay energy that goes into electrons depends on the ejecta composition and on time (e.g., figure \ref{fig:Qtot}). 
Considering light and heavy \rpe at solar abundances, the energy in electrons constitutes between 20\%-30\%  of the total energy (including neutrinos) and between 25\%-75\% of the energy that can thermalize (electrons and gamma-rays). The typical energy of the $\beta$-decay electrons declines slowly with the half-life time of the element emitting those electrons. On time scales of 10-100 days, most of the electrons have energies in the range of 50-1000 keV.

The process of energy loss of electrons propagating through a dense medium is known from theory and experiments. The stopping power, $dE/dX$ (i.e., energy loss per unit of mass column density), at the relevant energies is described by the Bethe-Bloch formula, which includes a logarithmic term that is calibrated by experiments (see chapter "Passage of particles through matter" in \citealt{tanabashi2018}). Experiment-based tables for all stable elements can be found at \url{https://physics.nist.gov/PhysRefData/Star/Text/ESTAR.html}. The energy deposition rate of an electron with a velocity $\beta c$ is ${\dot E_e}=\rho c \beta~ dE/dX$, where $\rho$ is the ejecta density. The stopping power depends only weakly on the composition (it is slightly lower for heavier elements). The function $\beta~ dE/dX$ has a broad minimum around an electron energy of $\sim 0.5$ MeV, where $\beta dE/dX \approx 1 {\rm~MeV~ cm^2 ~g^{-1}}$, and it rises very slowly with $E_e$ at lower energies (roughly by a factor of 2 at $E_e=10$ keV). It is useful to define an effective opacity \citep{waxman2018} $\kappa_e=E_e^{-1} \beta~ dE/dX \approx 4.5 \left( \frac{E_e}{250 {\rm keV}}\right)^{-1} {\rm~cm^2 ~g^{-1}}$. Note that this definition already includes the electron velocity (in units of $c$), which facilitates comparing electron thermalization to that of $\alpha$ and fission particles. With this definition, an electron deposits a significant fraction of its energy after spending a time $t$ in matter with a density $\rho$ such that its effective optical depth is $\tau_e =\kappa_e \rho c t \approx 1 $.

${\dot E_e}$ depends linearly on the density, where in a homologous expansion,  for each fluid element, $\rho \propto r^{-3} \propto t^{-3}$. This  implies that the time it takes electrons to thermalize increases sharply with time. This sets a critical time for thermalization, $t_e(E_e)$, which is defined such that $t_e = \frac{E_e}{{\dot E_e}(t_e)}$. At $t \ll t_e$, an electron thermalizes within less than a dynamical time (recall that homologous expansion implies that the dynamical time is also the time since the explosion, t). Therefore, during this time the heating rate is equal to the radioactive deposition rate, namely ${\dot Q}_{e,{\rm heat}}(t < t_e)={\dot Q}_{e,{\rm dep}}$. At $t \gg t_e$ the electron energy loss to thermalization is insignificant compared to the electron energy, and electrons are cooling only adiabatically. Since  ${\dot E_e}$ is only weakly sensitive to $E_e$, the adiabatic cooling of electrons deposited after $t_e$  does not affect their energy deposition rate  significantly. Therefore, under the reasonable assumption that electrons do not escape from the ejecta (see below), all the electrons deposited at $t>t_e$ accumulate and contribute to the heating. Since, in material with a mix of \rp elements, the number of deposited electrons typically drops faster than $t^{-1}$, the number of emitting electrons at $t>t_e$ is approximately constant, and therefore ${\dot Q}_{e,{\rm heat}} \appropto t^{-3}$. Taking into account the very slow increase  of $\beta~ dE/dX$  as the electron energy drop adiabatically, one obtains approximately ${\dot Q}_{e,{\rm heat}}(t > t_e) \propto t^{-2.8}$. The transition between the two asymptotes (before and after $t_e$) takes roughly an order of magnitude in time, and its exact shape can be found numerically. 

The critical time $t_e$ depends on the density and the distance that an electron travels in the ejecta over a dynamical time. The density can be written as $\rho=C_\rho m v^{-3} t^{-3}$ where $C_\rho$ is a constant that depends on the velocity distribution of the ejecta. For example, for a uniform density within a sphere with a radius $vt$, then  $C_\rho=3/4\pi$. Below, I provide some useful analytic approximations, including a value of $C_\rho$ for a more realistic velocity distribution. The time that an electron spends within the ejecta depends on the magnetic field configuration. For any reasonable magnetic field strength, the electron Larmor radius is smaller by many orders of magnitude than the ejecta radius, implying that it is most likely trapped within the ejecta, roughly at the location where it was deposited\footnote{In the unlikely event that an electron escapes the ejecta directly, the time it spends within the ejecta is $vt/(\beta c)$ where $v$ is the ejecta velocity, $\beta c$ is the electron velocity, and $t$ the dynamical time. In that case $t_e=\sqrt{C_\rho (v/\beta) \kappa_e m v^{-3}}$. This may happen only if radial magnetic field lines thread the ejecta. This seems highly unlikely for two reasons. First, a large-scale field is expected only if it is  advected from a central compact object, in which case the radial field would decay faster than the toroidal one. Second, even if a minute fraction of the heat deposited by radioactive decay goes to turbulence,  then it will generate a small-scale field that is much stronger than anything that can be advected from the source out to a radius of $\sim 10^{14}$ cm. }. The critical electron thermalization time is then $t_e(E_e)=\sqrt{C_\rho c \kappa_e(E_e) m v^{-3}}$. \\

%For any reasonable field strength the Larmour radius is smaller by many order of magnitudes than the length scale of the ejecta $sim v t$, and it is confined to move along field lines. The field can be either advected from the merger site or generated by turbulent motions powered by the heat depotion 

\noindent {\underline {\it $\alpha$-particles:}}\\
The thermalization of  $\alpha$-particles is similar to that of electrons. Experiment-based tables of the stopping power of a representative set of elements can be found at
\url{https://physics.nist.gov/PhysRefData/Star/Text/ASTAR.html}.  The function $\beta~ dE/dX$ has a broad plateau for $\alpha$-particle energies around $5-9$ MeV, at a value of $\beta~ dE/dX \approx 20 {\rm~MeV~ cm^2 ~g^{-1}}$. This implies $\kappa_\alpha \approx 3  {\rm~cm^2 ~g^{-1}}$ and corresponds to $t_\alpha=\sqrt{C_\rho c \kappa_\alpha m v^{-3}}$. \\

\noindent {\underline {\it Fission:}}\\
The thermalization of  fission products is more efficient than that of electrons and $\alpha$ particles.  The function $\beta~ dE/dX$ in \rp material has been calculated by \cite{barnes2016}. At typical fission fragment energies of 100-200 MeV, they find $\beta~ dE/dX \sim 1500 {\rm~MeV~ cm^2 ~g^{-1}}$,  which corresponds to $\kappa_{\rm fis} \sim 10  {\rm~cm^2 ~g^{-1}}$, so $t_{\rm fis} \approx 2 t_{\alpha}$.

\subsubsection{A simple analytic  model for radioactive heating}
For a given ejecta composition and velocity distribution, the radioactive heating as a function of time can be calculated quite accurately based on a small number of assumptions. Here, I provide a simple analytic approximation for ejecta consisting of all \rpe above the first peak (A=85-238), with a solar abundance based on \cite{sneden2008}. Note that, for different compositions, the heating can vary by about an order of magnitude. For example, when only heavy elements with $A>140$ are considered, the gamma-ray heating is lower by an order of magnitude than the values given here. For more accurate heating rates for various compositions, one can make use of the code provided by \cite{hotokezaka2019a} at \url{https://github.com/hotokezaka/HeatingRate}. This code calculates the heat deposition and the thermalization numerically, and it also implements the semi-analytic light curve model given in \S\ref{sec:MN_semiAnalytic}.\\

\noindent {\underline {\it Velocity distribution:}}\\
 The effect of the velocity distribution on the thermalization appears in the coefficients $C_\Sigma$ and $C_\rho$. To evaluate their values, we consider  ejecta with a total mass $m_{ej}$ and a power law velocity distribution, $\frac{dm}{d\log(v)}\propto v^{-k}$ between $v_{\rm min}$ and $v_{\rm max}$, where $k>0$. For this distribution, both constants depend only on $k$ and on the ratio $w=v_{\rm min}/v_{\rm max}$.  $C_{\rho}$ is obtained by averaging over the outflow density profile, and we define $\Sigma_{\gamma,{\rm eff}}$ and $\rho$ using $v_{\rm min}$, namely $\Sigma_{\gamma,{\rm eff}} = C_\Sigma~ m_{\rm ej}~ v_{\rm min}^{-2}~ t^{-2}$, and $\rho=C_\rho ~m_{\rm ej}~ v_{\rm min}^{-3}~ t^{-3}$. Under this definition,
\begin{equation}
	C_{\rho} \approx \frac{k}{4\pi (2+3/k)(1-w^{k})^2},
\end{equation} 
and  
\begin{equation}
	C_{\Sigma} \approx 0.1w+0.003\frac{k}{w} ,
\end{equation} 
where the approximation for $C_{\Sigma}$ is good to within a factor of order unity for  $0 < k < 5$ and $0.1 < w < 0.5$.
The canonical values taken below are $k=2$ and $w=1/4$, for which $C_{\Sigma}=0.05$ and $C_{\rho}=0.05$.\\

\noindent {\underline {\it $\gamma$-rays:}}\\

\begin{equation}\label{eq:Qg}
	\dot{\epsilon}_{\gamma,{\rm heat}} \approx 8 \times 10^{9} \left( \frac{t}{\rm day} \right)^{-1.4} \left( 1-e^{-(t_\gamma/t)^2}\right) {\rm{erg~s}^{-1}{\rm g}^{-1}}
\end{equation} 
where, 
\begin{equation}\label{eq:tg}
	 t_{\gamma}=2.3  \left(\frac{C_\Sigma}{0.05}\right)^{1/2}  \left(\frac{m_{\rm ej}}{0.05\msun}\right)^{1/2} \left(\frac{v_{\rm min}}{0.1 c}\right)^{-1} \left(\frac{\kappa_{\gamma,{\rm eff}}}{0.07 {\rm cm}^2 ~{\rm g}^{-1}}\right)^{-1/2} {\rm~day~}.
\end{equation}
Note that the canonical value of $\kappa_{\gamma,{\rm eff}}=0.07 {\rm~ cm^2 ~g^{-1}}$ is appropriate for the composition considered here, A=85-238, which is dominated by light \rp elements. For a discussion of $\kappa_{\gamma,{\rm eff}}$ see \S\ref{sec:thermalization} and \cite{hotokezaka2019a}. \\

\noindent {\underline {\it Electrons:}}\\
\begin{equation}\label{eq:Qe}
	\dot{\epsilon}_{e,\rm heat} \approx 4 \times 10^{9} \left( \frac{t_e}{\rm day} \right)^{-1.3}  \left[ \left( \frac{t}{t_e} \right)^{1.3} + \left( \frac{t}{t_e} \right)^{2.8} \right]^{-1} {\rm erg~s}^{-1}{\rm g}^{-1} ,
\end{equation} 
where 
\begin{equation}\label{eq:t}
	 t_{e} \approx 55  \left(\frac{C_\rho}{0.05}\right)^{1/2}  \left(\frac{m_{\rm ej}}{0.05\msun}\right)^{1/2} \left(\frac{v_{\rm min}}{0.1 c}\right)^{-3/2} \left(\frac{E_e}{250 {\rm~ keV}}\right)^{-1/2} {\rm~day~} .
\end{equation}
Note that $E_e$ can vary by about an order of magnitude between different element decay products, and therefore $t_{e}$ can be a factor of $\sim 2$ larger or smaller than the one for the canonical value of $E_e=250$ keV.\\

\noindent {\underline {\it $\alpha$-particles and fission fragments:}}\\
As discussed above, there is a small number of elements that contribute to the heating via $\alpha$-particles at each time. On a time scale of 1-10 days, these are mostly $^{222}$Rn and $^{224}$Ra,  and on a time scale of 10-100 days these are $^{223}$Ra and $^{225}$Ac (see the heating rate of these elements in the discussion above). $\alpha$-particle heating therefore depends on the exact amount of each of the few contributing elements, which is unknown, and there is no point in providing an average heating rate, as for each assumed composition the heating has a different functional form. 
The radioactive energy deposition rate (without accounting for thermalization loses) of $\alpha$ particles of each decay chain is
\begin{equation}
	\dot{\epsilon}_{\alpha,dep} \approx 4 \times 10^8~e^{t/\tau} \left(\frac{Y_\alpha}{10^{-5}} \right)  
	\left(\frac{\tau}{10{\rm~day}} \right)^{-1}  \left(\frac{E_{\alpha,tot}}{30 {\rm~MeV}} \right) ~{\rm{erg~s}^{-1}{\rm g}^{-1}}~,
\end{equation}
where $\tau$ is the mean lifetime, $E_{\alpha,tot}$ is the total energy release per decay chain, and $Y_\alpha$ is the ratio between the initial number of 
parent nuclei and the total number of nucleons (e.g., the mass fraction of parent nuclei that have mass number $A$ is $A Y_\alpha$). The thermalization time of $\alpha$ elements is:
\begin{equation}\label{eq:talpha}
	 t_{\alpha} \approx 45  \left(\frac{C_\rho}{0.05}\right)^{1/2}  \left(\frac{m_{ej}}{0.05\msun}\right)^{1/2} \left(\frac{v_{min}}{0.1 c}\right)^{-3/2} {\rm~day~}~.
\end{equation}
At $t\lesssim t_\alpha$ there are no significant thermalization losses and the heat rate is  $\dot{\epsilon}_{\alpha,heat} \approx \dot{\epsilon}_{\alpha,dep}$. At $t > t_\alpha$ the heat rate should be integrated numerically for each decay chain in order to properly account for thermalization losses \citep[see e.g.,][]{hotokezaka2019a}.  Note that the abundance of elements with A=222-225 is practically unknown from observations, and cannot be estimated robustly from theoretical considerations. However, if BNS mergers produce heavy \rpe with abundances similar to solar, then $\alpha$-particles are not expected to contribute significantly to the heating of the ejecta.

The uncertainty relating to fission-fragment heating rates is even larger. The only element for which there are experimental data and that may contribute on relevant timescales is $^{245}$Cf. There may be other contributing elements, but theory cannot accurately predict the fission half-life of those very-heavy, unstable, elements. The radioactive deposition by spontaneous fission of a given parent nucleus is 
\begin{equation}
	\dot{\epsilon}_{sf,dep} \approx 3  \times 10^8~e^{t/\tau} \left(\frac{Y_{sf}}{10^{-6}} \right)  
	\left(\frac{\tau}{10{\rm~day}} \right)^{-1}  \left(\frac{E_{sf}}{200 {\rm~MeV}} \right)  ~{\rm{erg~s}^{-1}{\rm g}^{-1}} ~,
\end{equation}
where $E_{sf}$ is the energy release per spontaneous fission of this nucleus and  $Y_{sf}$ is its the initial number of parent nuclei per nucleon.
The thermalization time of fission particles is roughly:
\begin{equation}\label{eq:tfis}
	 t_{fis} \approx 85 \left(\frac{C_\rho}{0.05}\right)^{1/2}  \left(\frac{m_{ej}}{0.05\msun}\right)^{1/2} \left(\frac{v_{min}}{0.1 c}\right)^{-3/2} {\rm~day~}~.
\end{equation}

\subsection{{\bf  Opacity}}\label{sec:opacity}
The opacity of the expanding sub-relativistic outflow has a major effect on the optical/IR emission. The opacity of $r$-process-rich ejecta is strongly wavelength-dependent and it is dominated by line transitions (bound-bound) in the expanding material. Thus, the exact opacity depends on the specific lines of the different elements that compose the ejecta, which are not well known. However, the general evolution of the light curve and its temperature depends mostly on statistical properties of the lines, such as their density distribution in frequency space, on which there are better constraints. The line opacity of expanding ejecta ("line expansion opacity") that are rich with heavy elements has been studied in the context of Type Ia SNe, where the opacity is dominated by iron-peak elements \citep[e.g.][]{karp1977}. Here, I give only a brief and partial description of this process based on the approximation used by \cite{pinto2000}. I refer the reader to this paper (and references therein) for an excellent and much more detailed explanation of this topic. The basic principles for $r$-process-rich material are similar, although 
the different compositions can lead to very different opacities. This was realized  first by \cite{tanaka2013} and \cite{kasen2013}. Below, I start with a brief qualitative description of line expansion opacity and then discuss the estimates for the opacity of $r$-process-rich ejecta and its dependence on the composition.

The main characteristic of the opacity of Type Ia SN ejecta is the large number of optically thick lines in the UV and optical. When combined with expansion, this implies that a photon may be scattered by many different lines (via absorption and re-emission) as its rest frame frequency varies during its travel through the ejecta's velocity gradient. To see this, consider, for simplicity, a photon with a lab frame frequency $\nu$ that is emitted at the rear end of the ejecta, where the velocity is $\beta_{min}c$, and that makes its way, moving in the radial direction, to the front edge, where the velocity is $\beta_{max}c$. On its way, the photon sweeps the frequency range from $\nu(1-\beta_{min})$ to $\nu(1-\beta_{max})$, from blue to red. 
As it crosses a fluid element with a velocity $\beta c$, its frequency in the fluid rest frame is $\nu(1-\beta)$. If there is no line at this frequency, the photon propagates onward, since the continuum optical depth is negligible. If there is a line resonance at this frequency with a significant optical depth, then the photon is absorbed and re-emitted. In case that the line optical depth is  high, then the photon can be absorbed and re-emitted many times until its frequency either diffuses out of the line, or it is re-emitted as two or more photons at longer wavelengths (i.e., fluorescence). The photon then continues on its way through the ejecta while its frequency, as seen in the ejecta rest frame, is continuously redshifted, until its rest frame frequency "hits" another optically thick line. Thus, the optical depth for a photon with a lab-frame frequency $\nu$ is simply the total number of optically thick lines\footnote{Low-optical-depth lines may also contribute, if their number is  large enough. See \cite{pinto2000}.} between $\nu(1-\beta_{max})$ and $\nu(1-\beta_{min})$. Note that, unlike typical continuum opacity (e.g., electron scattering), the optical depth does not depend on the actual distance the photon travels, and its dependence of $\rho$ is not trivial. Instead, it is determined by the density of high-optical-depth lines within each frequency interval (which is high in the UV and drops sharply with increasing wavelength), and on the velocity gradient of the ejecta. Nevertheless, in Type Ia SNe, the light curve and color evolution can be reasonably approximated by using a grey opacity, with $\kappa \approx 0.1 {\rm~cm^2~ g^{-1}}$ \citep{pinto2000}. A similar process takes place when a photon travels through the ejecta of a compact binary merger, although the higher velocities of the ejecta and the larger number of lines in the \rp material may lead to some differences \citep[][ see discussion below]{fontes2017,wollaeger2018}.  

The consequence of these properties is that the line opacity depends on the number of atomic-level transitions that the various elements in the ejecta have in the UV/optical/IR. This, in turn, depends on the structure of the electron valence shell of each element, and specifically on the number of different ways that valence electrons can be  distributed within the open shell. Those ways depend on the number of magnetic sub-levels in the open shell, i.e.,  $g=2(2l+1)$, where $l$ is the orbital angular momentum of the shell (azimuthal quantum number). An element with more sub-levels in its open shell (i.e., larger $g$) and with roughly half of those levels occupied by electrons, will have more permutations possible with distinct energy levels, i.e. more lines, and thus presumably a higher opacity (see discussion in \citealt{kasen2013}). For example, carbon and oxygen are both p-block elements, i.e. elements with an open p-shell ($l=1$) which has $g=6$ sub-levels, and their line opacity is typically too low to have a significant effect on the opacity of partially ionized ejecta composed of these elements. Iron-group elements, on the other hand, are d-block elements, with a nearly half-filled d-shell ($l=2$) and with $g=10$ sub-levels. In these elements, line opacity dominates. 

\rpe are distinct from the elements that compose Type Ia ejecta. The most significant difference is that heavy \rpe contain f-block elements, where the open f-shell ($l=3$) has $g=14$ sub-levels and,  as a result, the number of line transitions is larger by about two orders of magnitude compared to iron-peak elements. This includes lanthanides, which are rare-earth elements that are just above the second peak (A=140-176; Z=58-70) and actinides (A=227-266; Z=89-103) that are beyond the third peak. Theoretical models cannot predict reliably the expected mass fraction of  actinides in the ejecta, but it may be significant \cite[e.g.,][]{mendoza-temis2015}. However, even if the ejecta contains a small amount of heavy \rpe, then lanthanides are expected to dominate the opacity. \cite{kasen2013} show that even a minute lanthanide fraction of $\sim 10^{-3}$ is enough to dominate the ejecta opacity. For comparison, the mass in lanthanides in the Sun is $\sim 10^{-8} \msun$, and their mass fraction out  of the entire \rpe ($A \geq 69$) group is $1\%$. Among the \rp elements beyond the first peak ($A \geq 85$), the lanthanide mass fraction is $7\%$ , and among the elements beyond the second peak ($A \geq 140$) it is  $30\%$. Thus, if the ejecta contains a significant fraction of heavy \rp elements, then its opacity should be significantly higher than ejecta that contains only light \rpe or iron-peak elements.

A major source of uncertainty in calculations of \rpe opacity is the atomic line data. The calculations described above need,  as their basic input, a list of all the frequencies and oscillator strengths of the relevant transitions. For \rp elements, little experimental data are available, and macronova opacity calculations have been based on line lists generated using an approximate  theoretical atomic structure model
(e.g. the Autostructure code; \citealt{badnell2011}). The accuracy of these line lists is limited, and different theoretical assumptions and approximations result in different lists. The fact that the various models and radiative-transfer approximations often result in roughly similar light curves does give some hope that the predictions for  the general light curve evolution in the various photometric bands may be reasonably reliable. However, the current models are certainly not accurate enough to predict the precise form of the spectrum, especially during the nebular phase. This point is demonstrated by \cite{kasen2013}, who show how several different optimization schemes, used in the code that generates the line list, result in very different spectral shapes (see their figure 13). On top of that, the high velocity of the ejecta smears the absorption and emission features considerably. Thus, while a comparison of observed spectra to theoretical models can be instructive, the information that it provides is limited. In particular, it is very hard to infer from macronova spectra the presence of specific elements in the ejecta, as is done routinely for SNe.

As explained above, the opacity is wavelength-dependent, and its dependence on the density and temperature is also not negligible. Nonetheless, several studies find that a grey opacity can be used as a rough, but reasonable, approximation for various \rp mixtures \cite[e.g.][]{tanaka2013,kasen2013,barnes2013,wollaeger2018,tanaka2019}. A typical grey-opacity value for lanthanide-rich ejecta is $\kappa \sim 10{\rm~cm^2~ g^{-1}}$, while for lanthanide-free ejecta it is $\kappa \sim 1 {\rm~cm^2 ~g^{-1}}$.  Ejecta that is dominated by light \rp elements, but has some small fraction of lanthanides, can be approximated by an intermediate value of $\kappa$ \citep{tanaka2019}. The reason that the opacity of light \rp elements, which are lanthanide-free,  is higher than that of Type Ia ejecta is that light \rpe contain more d-block elements than Type Ia ejecta. Since each element has its own set of lines, the line opacity of these elements simply adds up. Another important property of lanthanide opacity is that, unlike d-block elements, their opacity remains high also in the IR, while in the absence of lanthanides (and actinides) the opacity in the IR is expected to be low. Thus, even a small fraction of lanthanides can have a significant effect on the macronova IR spectrum at late times.

The rough calculation of the opacity, described above, uses the Sobolev approximation, and assumes that strong lines are well-separated. This approximation may not be fully justified for a fast expanding $r$-process-rich outflow. \cite{fontes2017} and \cite{wollaeger2018} use a different (and in some sense complementary) line-smearing approach, where lines are smeared over a frequency bin and the opacity of this bin is taken to be the sum of all the lines in that bin. This approach can, in principle, 
produce significantly higher opacity compared to the Sobolev expansion opacity approximation. \cite{wollaeger2018} find that the results of the two approaches (line smearing and line expansion) for the opacity of $r$-process-rich material in compact binary merger ejecta are in broad agreement, although there are some differences in the details.
  
%Finally, it is important to bear in mind that the opacity models of \rpe are highly uncertain. The reason is that the opacity depends on the line frequencies and, more importantly, on their oscillator strengths. There is almost no experimental information on that, and theoretical calculations are hardly reliable. The fact that different theoretical models and different radiative transfer approximations are often giving roughly similar light curves, give us the hope that the models predict the general light curve evolution in the various photometric bands reasonably well. However, current models are certainly not accurate enough to predict the exact shape of the spectrum, especially during the nebular phase. This point is demonstrated in \cite{kasen2013} that show how several different assumptions used in the model that generates the line list result in very different spectral shapes (see their figure 13). On the top of that, the high velocity of the ejecta smear the absorption and emission features considerably. Thus, while a comparison of observed spectra to theoretical models can be instructive, the information that it can provide may be limited. Particularly, it seems much harder to infer from macronova spectra the presence of specific elements in the ejecta, in the same way that it is done for SNe.

\subsection{{\bf  A robust measure of the ejecta mass based on time integrated energy deposition}}\label{sec:KatzIntegral}
\cite{katz2013} have shown that a clever manipulation of equation \ref{eq:arnett} provides a way to measure the total deposited energy, which is completely independent of the radiation transfer details. To do this, they multiply both sides of equation \ref{eq:arnett} by $t$ and integrate over time, to obtain $E(t)\cdot t=\int_0^t \dot{Q}(t')t'dt' - \int_0^t L(t')t'dt'$. At $t \gtrsim t_{ph}$ the radiation streams away rapidly from the ejecta, so $E(t)\cdot t$ becomes negligible. Furthermore, after that time $L \approx \dot{Q}$, and therefore at any $t > t_{ph}$,
\begin{equation}\label{eq:katz}
	\int_0^{t>t_{ph}} \dot{Q}(t')t'dt' = \int_0^{t>t_{ph}} L(t')t'dt'
\end{equation}
The right-hand-side of this equation is an observable that was measured rather accurately in GW170817, and is expected to be measured also in many future mergers. The left-hand-side depends on $\dot{Q}(t)$, which is known rather well for a given composition. Moreover, due to the statistical nature of the radioactive decay from a mixture of \rpe, $\dot{Q}(t)$ does not depend strongly on the composition, as long as it is a mix of many \rp elements. Thus, equation  \ref{eq:katz} provides tight constraints on the total ejecta mass. The main advantage of this constraint is its robustness, as  it is independent of the radiative transfer, including the highly uncertain opacity, as well as the unknown velocity distribution and outflow geometry\footnote{There is a weak dependence of $\dot{Q}(t)$ on the ejecta velocity and geometry via the efficiency of the thermalization (see \S\ref{sec:thermalization}).}. The key point is that, when constructing equation \ref{eq:katz}, the conserved quantity becomes $E\cdot t$, which accounts for the evolution of the radiative energy from the time that it is deposited as $\dot{Q}$,  until the time that it escapes the ejecta as $L$. This quantity is thus independent of the time that the radiation spends, and its route going through the ejecta.  

\subsection{{\bf  The first-day macronova emission and non-radioactive energy sources}}\label{sec:MN_nonRadioactive}
Radioactive decay of \rpe is an unavoidable macronova energy source. However, it is not the only possible energy source of quasi-isotropic UV/optical/IR emission.  The main additional energy sources that have been discussed in the literature are a central engine, decay of free neutrons, cooling emission from the ejecta, and emission from the cocoon that arises when a relativistic jet interacts with the sub-relativistic ejecta.  Below, I discuss, very briefly, each of these sources. All sources, except for the central engine, are expected to be short lived, and can possibly dominate the emission during the first several hours, or even several days. 

\subsubsection{Central engine}\label{sec:central_engine}
Continuous energy deposition from a central engine can have a significant effect on the macronova light and, in some scenarios, it can even be much more significant than radioactive heating. The two main central-engine energy sources discussed in the literature are late accretion of marginally bound material on the central object, and a magnetar wind. There are only loose constraints on the possible range of central engine activities, both in terms of the form in which the energy is carried (e.g., radiation, magnetic wind, baryonic wind, etc.,) and in terms of the amount of energy that the central engine deposits at any given time. Thus, a central engine can be invoked to explain almost any observation, and indeed it often is, in transients that cannot be explained easily by other sources. Given this freedom, the key to a credible model is to calculate how the deposited energy is converted to observed radiation, and to try to predict unique observational signatures of the model. Below, I discuss two possible power sources: a magnetized wind from a magnetar, and X-ray emission from the accretion onto the central object. 

If the central object is a magnetar with $B=10^{15}B_{15}$ G and a rotation period of $P=1ms \cdot P_{ms}$, then its considerable rotational energy, $E_{mag} \sim 5 \times 10^{52} P_{ms}^{-2}$ erg, is released within $t_{mag} \sim 500~ B_{15}^{-2}~ P_{ms}^2$ s in the form of a magnetic wind with a luminosity of $L_{mag} \sim 10^{50}B_{15}^{2} P_{ms}^{-4}$ erg/s \citep{usov1992,spitkovsky2006}. After $t_{mag}$ , the luminosity drops roughly as $t^{-2}$. The wind is terminated if and when the magnetar collapses to a black hole. \cite{metzger2014a} discuss the observational imprint of a stable magnetar that deposits all of its energy into the expanding ejecta. They find that, for canonical parameters, a bright UV/optical $\sim 10^{44}$erg/s signal is expected on a time scale of 1-10 days, and that in some cases non-thermal X-rays with similar luminosity and time scale may also be seen. In addition, most of the energy released by the magnetar's spin-down is deposited in the kinetic energy of the ejecta, which potentially leads to a bright radio remnant \citep[][]{nakar2011}. Limits on late radio emission from several sGRBs put limits on the energy of the ejecta and suggest that stable magnetars are not common remnants of these events \citep{metzger2014b,horesh2016,fong2016}.

An alternative energy source is X-rays that are emitted from the central object, which is most likely powered by accretion. The X-rays are then absorbed by the ejecta and reprocessed to produce UV/optical/IR emission \citep{kisaka2016,matsumoto2018}. The main motivation for this model is that sGRBs that show an excess of IR light that may be macronova emission (sGRBs 130603B, 050709, etc.) are almost always also accompanied 
by a simultaneous excess of X-ray emission (see \S\ref{sec:sgrb_macronova}).  \cite{kisaka2016} discuss the constraints that this model puts on the ejecta which, on the one hand, need to be opaque enough to absorb the X-rays and reprocess them, and, on the other hand, cannot be too opaque, so that the deposited X-ray energy can diffuse to the observer over a dynamical time scale. The main characteristic of this model (besides invoking an energy source) is that, unlike the case with radioactivity, the heat source is decoupled from the optical depth. While both the radioactive heating and the optical depth depend linearly on the ejected mass, the X-ray heat source is independent of the ejecta mass. As \cite{matsumoto2018} show, X-ray powered macronova models can produce similar macronova signals with a much broader range of ejecta masses than the radioactively powered models. This point is discussed below in some detail in the context of GW170817 (\S\ref{sec:GW170817_ejecta_property}). 

\subsubsection{Free neutrons}\label{sec:neutrons}
The composition of the ejecta depends, among other things, on the time available for nucleosynthesis and on the density of the neutron-rich material as it decompresses. Once the expansion time is short enough and the density is low enough, the interaction rates drop to the point that nucleosynthesis freezes out.  In very fast fluid elements, which are also usually less dense, freeze-out can take place at times such that some, and possibly even most, of the neutrons remain free. The exact neutron fraction in each fluid element depends on its history, but generally it is found in simulations that elements with a final velocity that exceeds about $0.5c$ may contain free neutrons \citep{metzger2015,ishii2018}. The heat deposited by the beta-decay of these neutrons can potentially lead to an observable early blue signal. The first to discuss free neutrons as a potential macronova energy source was \cite{kulkarni2005}. Later, \cite{metzger2015} used the results of a merger simulation by \cite{bauswein2013} to estimate the number of free neutrons and their velocity distribution, and to calculate the expected signal.

An approximation of the free-neutron macronova can be obtained using the same method as the one described above for the \rp powered macronova, simply by replacing the heat deposition rate in equation \ref{eq:arnett} by
\begin{equation}
	\dot{Q}_n(t) \approx 6 \times 10^{42} {\rm~erg~s^{-1}~}   \frac{m~X_n}{10^{-5}\msun} e^{-{t}/{900~{\rm s}}} ~~,
\end{equation}
where $m~X_n$ is the free-neutron mass. 

A simple and rather accurate approximation of the peak luminosity can be obtained when the energy deposited by the neutrons remains trapped for longer than 900s \citep{kasliwal2017}. Consider that the leading part of the ejecta has a mass $m$, velocity $v$, opacity $\kappa$ and its free-neutron mass fraction is $X_n$. The deposited heat is trapped in the mass $m$ up to (see equation \ref{eq:tej})
\begin{equation}\label{eq:tn}
	t_{tr} \approx 0.5 {\rm~hr} ~ \left(\frac{m}{10^{-5}\msun}\right)^{1/2} \left(\frac{v}{0.5c}\right)^{-1/2} \left(\frac{\kappa}{1 {\rm~cm^2~ gr^{-1}}}\right)^{1/2},
\end{equation} 
We see that for $m \gtrsim 10^{-6}-10^{-5}\msun$ , the trapping time is $t_{tr} > 900$ s. In such a case, for $t<t_{tr}$, equation \ref{eq:arnett} can be integrated while neglecting radiative losses, obtaining that for $900 <t<t_{tr}$, the total  trapped energy is $E \approx 6 \times 10^{43} {\rm~erg}~ X_n~  \frac{m}{10^{-5}\msun} \left(\frac{t}{{\rm day}}\right)^{-1}$.  The peak luminosity, seen at $t_{tr}$, is then 
\begin{equation}
	L_{n,p} \approx \frac{E(t_{tr})}{t_{tr}} \approx 10^{42} {\rm~erg~s^{-1}}~ X_n \left(\frac{v}{0.5c}\right) \left(\frac{\kappa}{1 {\rm~cm^2 ~gr^{-1}}}\right)^{-1}.
\end{equation}
Note that the peak luminosity in this case ($t_{tr}>900$s) is independent of the total mass in the regions that contain free neutrons, and is therefore not sensitive to the exact mass distribution of the ejecta. Instead, the bolometric luminosity depends mostly on $X_n$ and $\kappa$. Finally, the observed temperature can be approximated by the effective blackbody temperature, 
\begin{equation}\label{eq:Tn}
	T_n \approx 35,000 {\rm~K}~ X_n^{1/4} \left(\frac{m}{10^{-5}\msun}\right)^{-1/4} \left(\frac{\kappa}{1 {\rm~cm^2~ gr^{-1}}}\right)^{-1/4}.
\end{equation}

Examination of equations \ref{eq:tn}-\ref{eq:Tn} shows that, while the bolometric luminosity can be high if the fast ejecta is rich in free neutrons, the small emission radius dictates a very blue signal that peaks in the far UV (unless $\kappa$ is very large and then the luminosity is low). As a result, the optical signal is expected to be rather faint and the best place to search for free neutron emission is the near UV (e.g., by the proposed ULTRASAT mission; \citealt{ganot2016}). Moreover, a detectable signal requires  rather optimal conditions in which an extremely neutron rich mass of $10^{-5}-10^{-4}\msun$ is ejected at $\gtrsim 0.5$c. Nonetheless, it is worth looking for such a signal on a timescale of hours or less after the merger, as it may be detectable.

\subsubsection{Cooling emission}\label{sec:cocoon_cooling}
When a fluid element crosses the trapping radius, all of the internal energy that it contains diffuses out and is observable. This includes any recently deposited radioactive energy, as well as heat that remains in the element from the last time that it was shocked. The latter is called "cooling emission" and it is the main luminosity source in some supernovae (e.g., Type IIP). At the time that the ejecta is shocked, the internal energy can be very high (as high as the bulk-motion kinetic energy), but as long as it remains trapped it suffers adiabatic losses. The radiated energy is thus suppressed by a factor of $R_{sh}/R_{tr}$ (assuming homologous expansion), where $R_{sh}$ is the radius at which the energy was deposited by shocks, and $R_{tr}=v t_{tr}$ is the radius at which the energy is released. Thus, the emission generated by the energy $\sim E_{sh}$ that was deposited at $R_{sh}$ peaks at $t_{tr}$, and its peak luminosity is $L_{cool,p} \sim E_{sh} R_{sh}/(v t_{tr}^2)$.

The cooling emission thus depends on the specific model, which dictates how much energy is deposited, and at which radius. However, an approximation for the cooling emission can be obtained for the case where $E_{sh} \sim m v^2/2$, which is the deposited energy for the case that the shock itself is the mechanism that accelerates the mass $m$ to its velocity $v$. In this case, the peak luminosity is independent of the mass $m$ \citep[e.g.,][]{nakar2014}:
\begin{equation}
	L_{cool,p} \approx 10^{41} {\rm~erg~s^{-1}}~ \frac{R_{sh}}{10^{10}{\rm~cm}} \left(\frac{v}{0.3c}\right)^{2}  \left(\frac{\kappa}{1 {\rm~cm^2 gr^{-1}}}\right)^{-1} .
\end{equation}
The time of the peak luminosity is $t_{tr}$, given by equation \ref{eq:tn}, implying that, for typical parameters, cooling emission may be important during the first day, or at most for several days. Moreover, a detectable signal is expected only if the mass is shocked at a large radius, $R_{sh} \gtrsim 10^{10}$ cm. Thus, the internal energy that is deposited during the mass ejection, at a radius of $10^6-10^7$ cm, is irrelevant, and the ejecta must go through a strong shock after it has expanded significantly, at least $1$s after the merger. One possible source of such shocks is a central engine that works for seconds after the merger. Such an engine can drive shocks via a delayed launching of a relativistic jet \citep[][]{nakar2017,gottlieb2018a,piro2018}, or by a more spherical magnetar wind \citep{metzger2018,beloborodov2018}. Another suggestion has been the interaction of the ejecta with a companion in a mass-transferring triple system \citep{chang2018}.

\subsubsection{Cocoon emission} \label{sec:cocoon_macronova}  
The above discussion of the macronova emission focuses on the emission of the sub-relativistic ejecta. During the first hours, the cocoon, that arises from the interaction of the relativistic jet with the sub-relativistic ejecta, can be a significant source of UV/optical emission. The physics that governs the cocoon formation and evolution is discussed in section \ref{sec:jet_propagation}. The important point for the macronova discussion is that jet-ejecta interaction can bring a non-negligible amount of $r$-process-rich material to mildly relativistic velocities ($\gtrsim 0.5$c). This mass can radiate both by radioactive heating and by cooling emission. The cocoon macronova emission was explored analytically by \cite{nakar2017} and \cite{piro2018}  and  numerically by \cite{gottlieb2018a,gottlieb2018b,kasliwal2017} and \cite{nakar2018}.

\newpage
\section{The Relativistic outflow --- propagation, \gray emission, and the afterglow}\label{sec:relativistic}

BNS mergers launch relativistic jets. This was a long-standing expectation based on the conjectured association of short GRBs with BNS mergers \citep{eichler1989}, and was confirmed by the afterglow observations of GW170817. The engine of the jet is presumably the central compact object formed following the merger, either a magnetar or an accreting BH. The launched jet needs to propagate through the sub-relativistic ejecta that covers the polar region at that time. The jet propagation drives a cocoon that engulfs the jet and affects its propagation, forming a structured relativistic outflow known as the jet-cocoon.  EM emission  begins upon the breakout of the shock, driven by the jet-cocoon, from the ejecta. It is followed by emission from internal dissipation in the jet and cooling emission from the cocoon. At larger radii, interaction of the relativistic outflow with the circum-merger medium generates a long-lasting non-thermal afterglow that can be seen over the entire spectral range from radio to X-rays for a long time, possibly even years. 
	
A BH-NS merger is also expected to launch a relativistic jet, in the case that the tidal disruption of the NS takes place outside of the ISCO. The accreting BH then serves as the engine for the jet. The sub-relativistic dynamical ejecta from a BH-NS merger is concentrated along the equatorial plane, while the sub-relativistic secular wind from the accretion disk covers the polar regions as well. Therefore, if there is some delay between the formation of the disk and the launching of the jet, then the jet must propagate through the disk wind, forming a cocoon that is rather similar to the one that is formed in a BNS merger. If, however, the jet is launched with no delay, then the front-end of the jet propagates into a relatively clean environment. There may still be interaction between the jet and the disk wind, although this type of interaction has never been explored in the context of GRB jets. In any case, the jet emission from a NS-BH merger can possibly generate the same types of EM emission as a BNS merger jet (prompt $\gamma$-rays, afterglow, etc.), with the caveat that, if there is no delay in the jet launching, then there is no shock breakout, and possibly also no cocoon emission.

Due to the relativistic and anisotropic nature of this outflow, the observed EM signal depends strongly on the observer's line-of-sight viewing angle with respect to the jet axis.  An observer who looks down into the jet's opening angle sees, presumably, a short gamma-ray burst (sGRB) -- a bright burst of gamma-rays lasting about 1 s or less, followed immediately by a bright X-ray, optical, and radio afterglow (see \citealt{nakar2007,berger2014} for reviews). An observer positioned away from the jet's opening angle sees much fainter emission, which may be dominated by the cocoon during a large fraction, and possibly all, of the time. The physical processes that are most relevant in shaping the emission toward a line of sight within the opening angle of the jet have been studied over the last several decades in thousands of papers that discuss GRBs. These processes include the launching of the jet, the effect of its propagation through the ejecta on the jet itself, and the emission of the prompt gamma-rays and the afterglow, as seen along the jet axis. I do not cover these topics here as they are treated
in a large number of reviews on GRBs \citep[e.g.,][]{piran1999,piran2004,meszaros2002,meszaros2006,nakar2007,lee2007,gehrels2009}. Moreover, the jet is narrow, and therefore in the vast majority of BNS and BH-NS events that are discovered via their GW signal, the jet will be pointing away from us.  I therefore focus here on the processes that are most relevant for the emission from these events  as seen by observers who are not looking into the jet opening angle. These processes include the cocoon evolution, the shock breakout, and the prompt and afterglow emission at large viewing angles. 

In the flood of papers that followed GW170817, the terminology was sometimes ambiguous. There were cases where different terms were used to describe a single physical phenomenon, while in other cases similar terms were used to describe different physical phenomena. For example, the terms "structured jet" and "cocoon" where discussed in many papers as two distinct and competing models, when in fact they are often the same thing---whenever there is a successful jet, the jet itself and the cocoon, that must accompany it, constitute together a structured jet. To clarify the terminology I provide in an inset, below, a list of short definitions of the main terms that I use in this section.

\begin{minipage}{1\textwidth}
\begin{center}
	\includegraphics[width=1\textwidth]{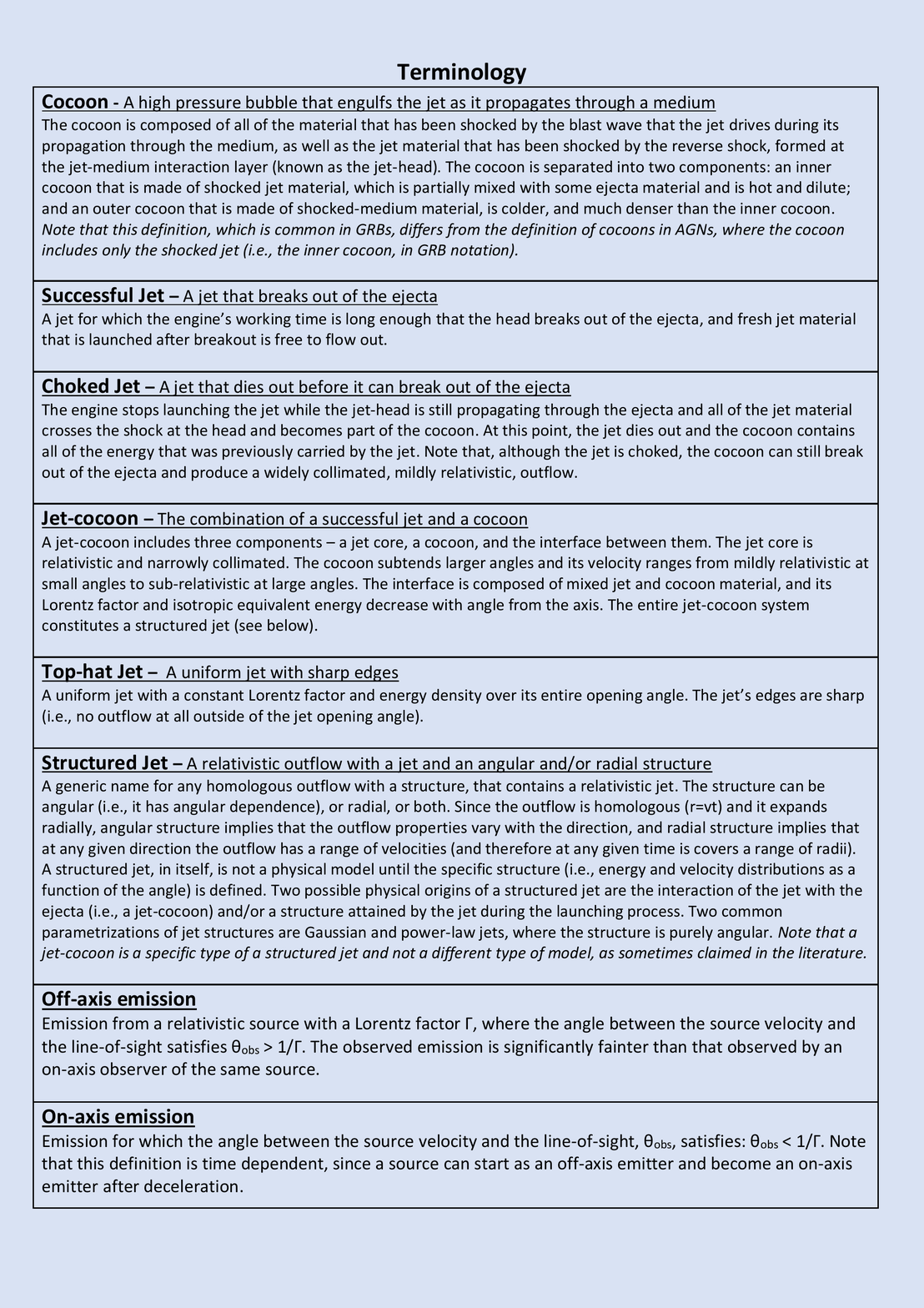}
\end{center}
\end{minipage}

\subsection{{\bf  Propagation of a relativistic jet in sub-relativistic ejecta}}\label{sec:jet_propagation}
\begin{figure}
	\center
	\includegraphics[width=0.7\textwidth]{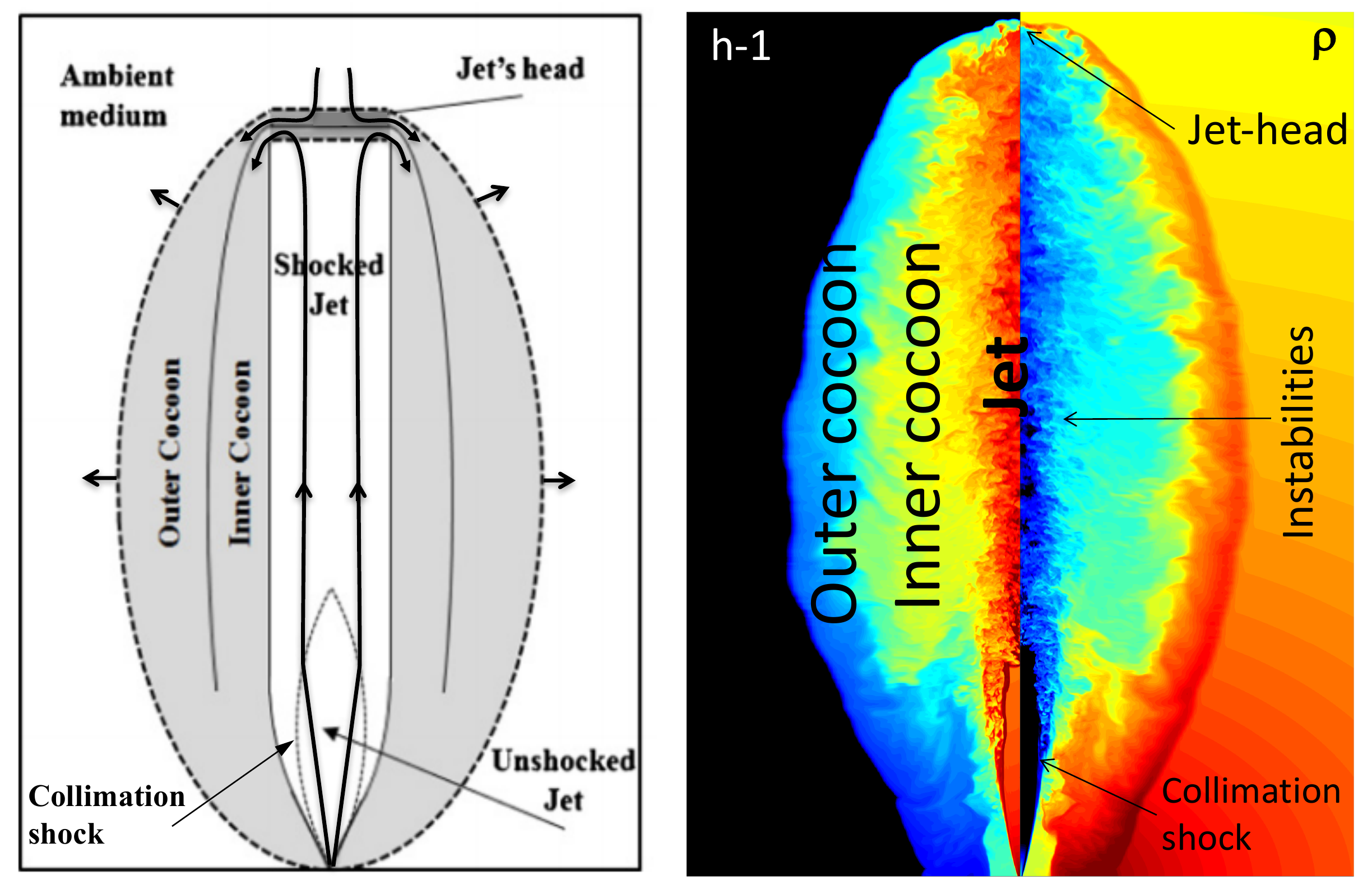}
	\caption{Illustration (left) and 3D RHD simulation (right) showing the structure of the jet and the cocoon during the propagation of a relativistic jet through a dense medium. The right half of the simulation shows mass density, $\rho$, and the left shows internal energy per baryon, $h-1$, where $h$ is specific enthalpy. Color palettes in both halves are logarithmic with blue for low values and red for high. The collimation shock and jet-head are seen clearly, as well as the instabilities along the interface between the jet and the cocoon, that lead to some mixing. The low-density, hot, inner cocoon and the high-density cold outer cocoon are also seen to be well-separated.}%
	\label{fig:Jetsketch}
\end{figure}

A relativistic jet that propagates in a dense medium (the sub-relativistic ejecta, in our case) drives a strong forward bow shock. At its tip, the jet develops a slowly moving head with a reverse shock that separates the head from the rest of the jet. Ambient medium that crosses the forward shock, and jet matter that crosses the reverse shock, spill sideways and form a hot enveloping cocoon. Depending on the jet and medium properties, the cocoon may collimate the jet, accelerating its propagation. If not, the jet remains roughly conical and its propagation is not affected by the cocoon. The cocoon is composed of two parts, an inner cocoon that is composed mostly of shocked jet material and an outer cocoon that is composed of shocked medium material. Figure \ref{fig:Jetsketch} shows a schematic illustration of the jet and the cocoon during the jet propagation (i.e., before they break out of the medium). As long as the jet head is within the ejecta, its propagation is driven by fresh jet material that crosses the reverse shock and spills into the cocoon. Thus, most of the jet energy that is injected during the time that the jet drills its way through the ejecta is deposited into the cocoon. Moreover, if the jet injection stops too early and jet material stops crossing the reverse shock before the head successfully crosses the entire ejecta, the jet is choked and all of its energy is deposited into the cocoon.

%A jet that propagates in dense media inflates a high pressure cocoon, which in turn applies pressure on the jet, influencing its propagation. 
The propagation of a relativistic jet and its interaction with the surrounding media  has been studied mostly numerically, using relativistic hydrodynamic (RHD), or relativistic magneto-hydrodynamic (RMHD) simulations. 
Jet simulations that were carried out in the context of GRBs typically considered the case of long GRBs, where the jet propagates in an approximately static and spherically symmetric stellar envelope \citep[e.g.,][]{aloy2000,macfadyen2001,zhang2004,morsony2007,mizuta2009,mizuta2013,lopez-camara2013,lopez-camara2016,harrison2018}. These simulations were also limited to unmagnetized jets, due to numerical difficulties in simulating RMHD jets in 3D over the required dynamical range\footnote{The propagation of highly magnetized jets requires a 3D  study for reliable results, due to the growth of instabilities that can be seen only in 3D \citep{bromberg2016}. These simulations are expensive numerically, and currently there are no 3D RMHD simulations of a magnetically dominated jet that crosses the ejecta of a BNS merger or the envelope of a collapsar. There are studies of magnetized jets in 2D that propagate in the ejecta from a BNS merger \citep[e.g.,][]{kathirgamaraju2018,bromberg2018} and a study of a magnetic jet in 3D, which ignores the interaction with the ejecta \citep{kathirgamaraju2019}, but no 3D simulations of a magnetized jet that interacts with merger ejecta.}. 
\cite{bromberg2011a} derived an analytic solution to the propagation of a hydrodynamic jet in a static and spherically symmetric medium, as long as the jet is within the medium (i.e., before breakout). The solution, which describes the self-consistent coevolution of the jet and the cocoon, was later verified and calibrated numerically by \cite{harrison2018}. These studies are not applicable directly to compact binary mergers where the jet propagates into a medium that is expanding at high velocities and that has, most likely, an angular structure. Numerical simulations of jets in expanding and/or aspherical media were carried out prior to GW170817  by only a handful of studies \citep{nagakura2014,murguia-berthier2014,murguia-berthier2017a,duffell2015,gottlieb2018a}, and after GW170817 almost only in the context of finding a fit to the afterglow data (see section \ref{sec:GW170817relativistic}). 

Currently there is no analytic model for the propagation of a jet in an expanding and/or aspherical medium. There is also no quantitative understanding of the dependence of the jet propagation on various parameters such as the ejecta velocity and angular structure. Below I use the analytic model of a static, spherically symmetric, medium (as calibrated by \citealt{harrison2018}), with a minor necessary adjustment, to approximate the jet propagation velocity and the time it takes the jet to break out of the merger ejecta. This approximation is reasonable if the jet propagation is significantly faster than the ejecta expansion velocity, and if the ejecta properties do not vary significantly over the jet opening angle. 

In order to estimate the time it takes the jet to cross the ejecta and break out, we need to compare the ejecta velocity to the jet-head velocity. Note that the head may be sub-relativistic even though the jet itself is ultra-relativistic.
The propagation velocity of the jet head is determined by the ram-pressure balance of the jet and the medium, as seen in the head's rest frame. Thus, the head velocity can be approximated by the propagation in a static medium \citep{harrison2018}, but the velocity is measured in the ejecta rest frame:
\begin{equation}\label{eq:vh}
	v_h-v_{\rm ej} \approx 0.2 ~c~ \left(\frac{L_{\rm iso}}{10^{51}{\rm~erg~s^{-1}}}\right)^{1/3} \left(\frac{\theta_{j,0}}{0.1{\rm~rad}}\right)^{-2/3} \left(\frac{m_{\rm ej}}{0.01\msun}\right)^{-1/3} \left(\frac{R_{\rm ej}}{10^{10}{\rm~cm}}\right)^{1/3} ,
\end{equation}    
where both $v_h$ and $v_{\rm ej}$ are measured in the merger rest frame, $L_{\rm iso}$ is the jet isotropic equivalent luminosity, and $\theta_{\rm j,0}$ is the jet half-opening angle upon injection. $m_{\rm ej}$ is the isotropic-equivalent ejecta mass, as seen by the jet (i.e., $m_{ej}$ is the ejecta mass within the jet opening angle, times $2/\theta_{\rm j,0}^2$). If, for example, the ejecta is anisotropic, then $m_{\rm ej}$ is the isotropic-equivalent ejecta mass along the poles at the time that the jet is launched. $R_{\rm ej}$ is the ejecta radius at the time that the jet crosses it, which is roughly $v_{\rm ej} t \approx v_h (t-t_{\rm delay})$, where $t$ is the time since the merger and $t_{\rm delay}$ is the delay time between the merger and the jet launch. Equation \ref{eq:vh} provides a reasonable approximation, as long as the head is sub-relativistic, and only if the jet crosses the ejecta before it doubles its radius, namely $v_h \gtrsim 2 v_{\rm ej}$. If those criteria are satisfied, the jet breaks out of the ejecta roughly at time $t_b$ after the merger where 
\begin{equation}\label{eq:tb}
	t_b-t_{\rm delay} \approx 1 ~s~ \left(\frac{L_{\rm iso}}{10^{51}{\rm~erg~s^{-1}}}\right)^{-1/3} \left(\frac{\theta_{j,0}}{0.1{\rm~rad}}\right)^{2/3} \left(\frac{m_{\rm ej}}{0.01\msun}\right)^{1/3} \left(\frac{R_{\rm ej}}{10^{10}{\rm~cm}}\right)^{2/3} .
\end{equation}   

Equation \ref{eq:tb} shows that there are necessary conditions that are required for the jet to break out of the ejecta. If the work-time of the engine that launches the jet is shorter than about $t_b-t_{\rm delay}$, then the jet is choked\footnote{The minimal engine work-time for breakout, $t_b-t_{\rm delay}$, is an approximation for a Newtonian head. If the head is  relativistic, moving with a Lorentz factor $\Gamma_h$, then the minimal engine work-time for breakout is  $t_b-t_{\rm delay}-R_{\rm ej}/2c\Gamma_h^2$, where the last term accounts for the time it takes the last jet element that the engine launches to reach the jet head.} . Given that the typical work-time of sGRB jets is $\lesssim 1$s, equation \ref{eq:tb} implies that the jet must have relatively high $L_{\rm iso}$ and a narrow opening angle to be able to break out. It also shows that a longer delay makes it harder for the jet to break out because  $R_{\rm ej}$ is larger. The delay may also affect $m_{\rm ej}$ to some extent, since sub-relativistic mass ejection continues during the first second, or even longer (see \S\ref{sec:massEjection}).

\subsubsection{Successful jets---the jet-cocoon structure}\label{sec:JetCocoon}
\begin{figure}
	\center
	\includegraphics[width=0.42\textwidth]{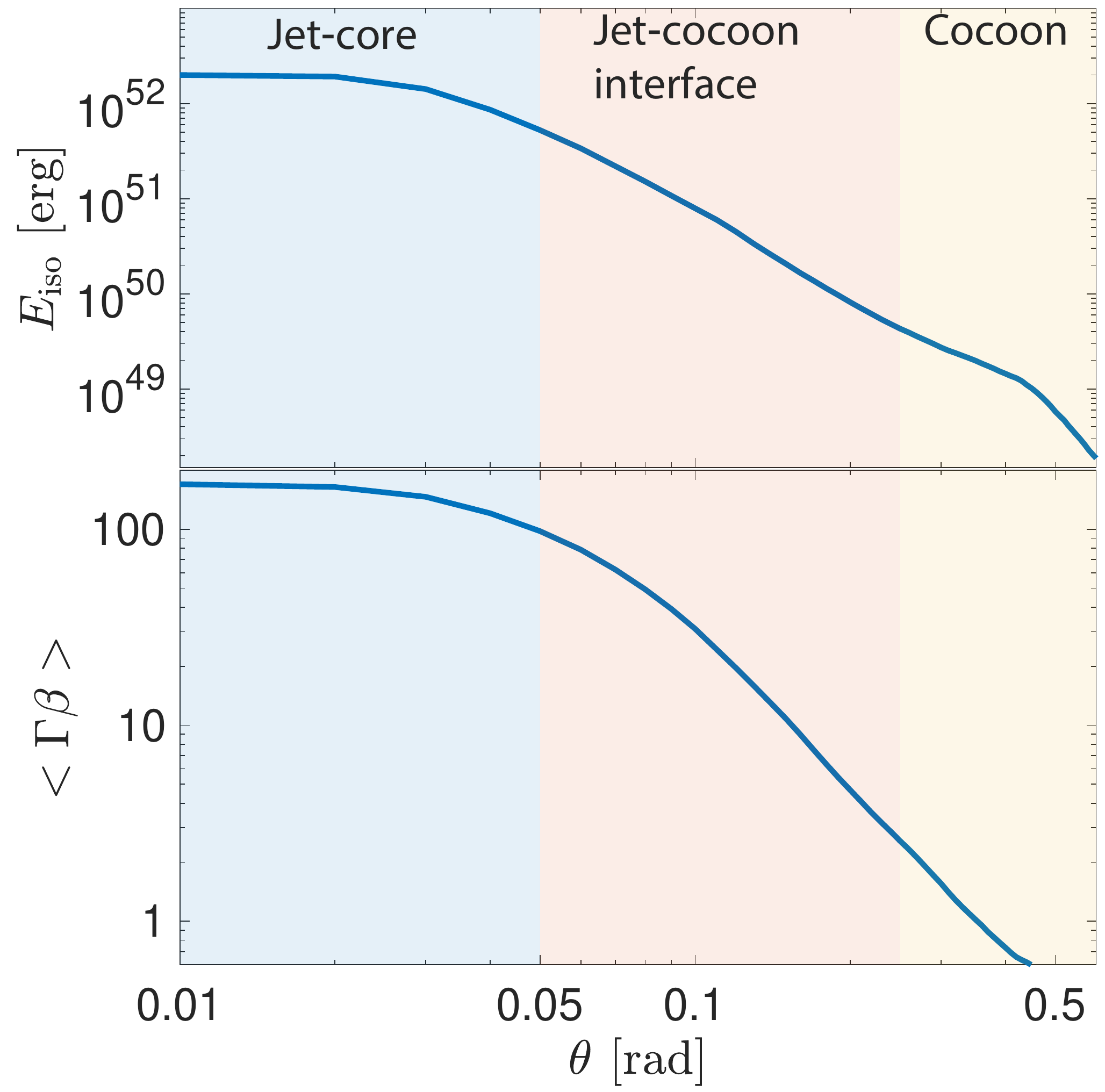}$~~~$
	\includegraphics[width=0.45\textwidth]{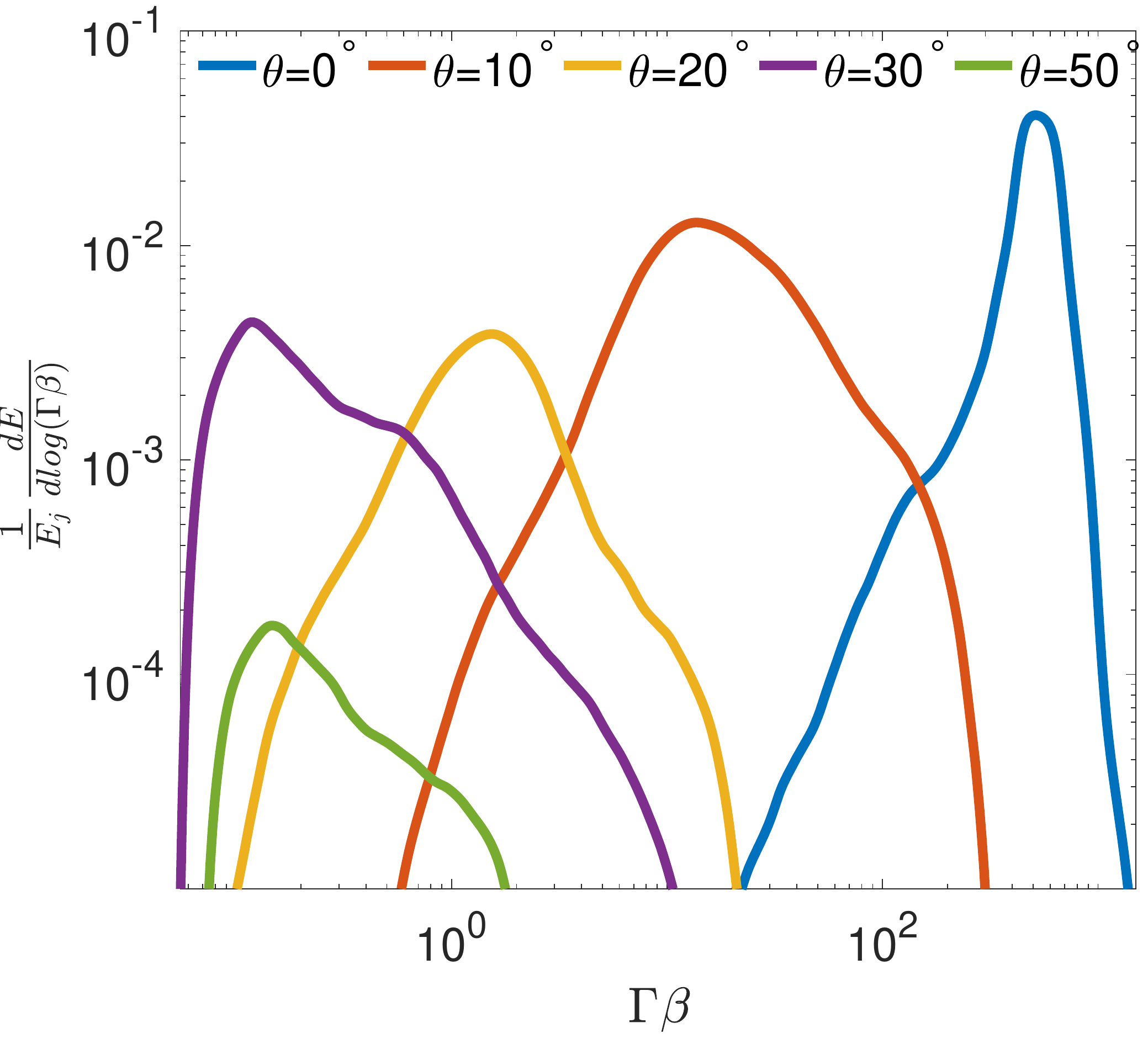}
	\caption{The jet-cocoon structure as obtained from a 3D RHD simulation of a successful jet that punches through the ejecta of a BNS merger (this jet was used in \citealt{mooley2018b} to fit the afterglow of GW170817; see \citealt{gottlieb2019b} for details of the structure). {\it Left: }The angular structure of the isotropic equivalent energy, $E_{\rm iso}$, and the average four velocity $\Gamma\beta$. $\theta$ is the angle from the jet axis. The different background colors show the three components: (i) jet core; (ii) jet-cocoon interface; and (iii) cocoon.  {\it Right:} The radial structure---the energy distribution as a function of four-velocity along a radial direction, for several different angles. At each angle, the material has a characteristic four-velocity, implying that the structure of the outflow is mostly angular (i.e., depends mostly on $\theta$).}%
	\label{fig:successful}
\end{figure}

RHD simulations follow the jet as it propagates in the sub-relativistic ejecta, breaks out of the ejecta, and expands up to the phase where the entire outflow (jet+cocoon) becomes homologous. All simulations show a similar general outflow structure---a relativistic core of relatively pure jet material, surrounded by cocoon material that arises from the jet-ejecta interaction \citep{gottlieb2018a,gottlieb2018b,nakar2018,lazzati2017b,lazzati2018,xie2018,zrake2018}. 
During the propagation of the jet, most of the injected energy is deposited in the cocoon. After the breakout, both the cocoon and the jet are free to expand and the outflow transitions to a structure that is composed of three components: (i) jet; (ii) cocoon;and (iii) the interface between them. Figure \ref{fig:successful} shows the structure of a jet-cocoon as found in 3D RHD numerical simulation by \cite{gottlieb2019b}. The figure shows the angular structure of the isotropic equivalent energy, $E_{\rm iso}$, and the four-velocity $\Gamma\beta$, where the three different components are highlighted with different colors. It also shows the radial structure along several angles. At each angle the material has a characteristic four velocity, implying that the structure of the outflow is mostly angular (i.e., $E_{\rm iso}$ and $\Gamma\beta$ depend mostly on the angle from the axis and at each direction there is almost no radial structure).
Below I discuss the properties of each of the three components of the outflow.\\

\noindent \underline{Cocoon:}\\ 
The cocoon is composed of all the shocked ejecta and all the jet material that was shocked within the jet-head during the  head's propagation through the ejecta. The total energy of this material is roughly the energy launched into the jet before it breaks out, which can be approximated as \citep{harrison2018},
\begin{equation}
	E_{\rm cocoon} \approx  L_{\rm j}(t_b-t_{\rm delay}) \approx 5 \times 10^{48} {\rm~erg} \left(\frac{L_{\rm iso}}{10^{51}{\rm~erg~s^{-1}}}\right)^{2/3} \left(\frac{\theta_{j,0}}{0.1{\rm~rad}}\right)^{8/3} \left(\frac{m_{\rm ej}}{0.01\msun}\right)^{1/3} \left(\frac{R_{\rm bo}}{10^{10}{\rm~cm}}\right)^{2/3} ,
\end{equation}
where $L_{\rm j}=L_{\rm iso}\theta_{j,0}^2/2$ is the total jet luminosity and $R_{\rm bo}$ is the radius of the ejecta at the time that the jet breaks out. This estimate is applicable for head velocities that are not relativistic\footnote{Only jet material that was shocked within the head deposits its energy into the cocoon. Hence,  jet  material that was launched between $t_b-R/c$ and $t_b$ does not contribute to the cocoon energy. Thus, a more accurate estimate of the cocoon energy is $E_{\rm cocoon} \approx L_{\rm j} (t_b-t_{\rm delay}-R/c)$. This correction is of order unity for a sub-relativistic head, but it can be significant for a relativistic head.}. 

Upon breakout, the cocoon expands and a rarefaction wave causes it to accelerate and to spread sideways. 
%according to the  Lorentz factor of its material at the time of the breakout, $\Gamma$, i.e., to an angle of $\sim 1/\Gamma$} \note{why does it accelerate? and why does it spread sideway a 1/Gamma? I understand if emission from the structure is beamed as 1/Gamma, but not this}. 
The Lorentz factor of the cocoon depends sensitively on the mixing between the ejecta and the shocked jet material in the inner cocoon, which can be explored only by 3D numerical simulations\footnote{Obtaining reliable estimates of the mixing is a hard task. Mixing cannot be studied analytically due to its highly nonlinear nature, and in 2D, mixing is affected strongly by the artificially enforced symmetry. Even in 3D simulations, it is hard to be confident that the mixing that is observed is physical, rather than numerical, even after numerical convergence is obtained. Thus, results based on the mixing seen in simulations should be treated with caution.} (as shown by e.g., \citealt{harrison2018,gottlieb2018a}). 
The terminal angular and radial distributions of the cocoon, after acceleration and spreading is completed, has been studied in 3D RHD simulations by \cite{gottlieb2018a,gottlieb2019b}. An example is shown in figure \ref{fig:successful}, where the cocoon can be seen at angles that are larger than 0.25 rad. \cite{gottlieb2018a} find that, in their simulations, the shocked jet (inner cocoon) accelerates to Lorentz factors of 2-4 and spreads over an angle of $\sim 0.25-0.4$ rad ($\sim 15^\circ-25^\circ$). The outer cocoon, which is made of shocked ejecta, has more mass. It accelerates to a velocity of $\sim 0.5$c, and it spreads out to an angle of $\sim 0.6$ rad ($\sim 35^\circ$). The result is a relatively wide-angle outflow with mostly an angular structure where the four-velocity, $\beta \Gamma$, drops from $\sim$4 to 0.5 between $\sim 15^\circ$ and $35^\circ$. \\

\noindent \underline{Jet-cocoon interface:} After breakout, freshly launched jet material has a clear path out of the ejecta through the cavity carved out during the propagation of the jet-head. However, this does not mean that the jet stops interacting with the ejecta. It takes a relatively long time for the cocoon, parts of which are sub-relativistic, to escape out of the unshocked ejecta, and therefore the cocoon continues to apply pressure on the jet long after the breakout. This pressure drives a collimation shock into the jet, and more importantly, Rayleigh-Taylor, Richmeyer-Meshkov and Kelvin-Helmholtz instabilities develop along the interface between the jet and the cocoon  \citep[e.g.,][]{matsumoto2013,matsumoto2017}. These instabilities, which can be seen only in 3D simulations, mix some of the jet material with cocoon material along its path. This mixing generates an interface layer between the ultra-relativistic jet and the mildly relativistic cocoon. The energy source of this layer is jet material that was launched after the breakout. Since the level of mixing drops rather slowly with time, the energy in the interface is roughly proportional to the energy launched into the jet after it breaks out, i.e., $E_{\rm interface}\appropto L_{\rm j} (t_e-t_b)$, where $t_e$ is the time at which the engine stops injecting the jet. The fraction of the jet energy that goes to the interface layer increases with the mixing, which, in turn, is stronger for lower-luminosity and wider jets, and for denser ejecta. The entire outflow distribution is continuous and the mixing is weak near the jet axis, and gets stronger away from the axis. The interface therefore has an angular structure where both $E_{\rm iso}$ and $\Gamma$ drop with angle from the axis.  \cite{gottlieb2019b} carried out several simulations of jets with various properties, that are launched into expanding merger ejecta. They find that the distribution of $E_{\rm iso}$ in the interface can be well fitted by a power-law $E_{\rm iso} \propto \theta^{-\delta}$ , were $\theta$ is the angle from the jet axis and the value of $\delta$ depends on the level of mixing. In their simulations,  they find $\delta \approx 3-4$. An example of the interface distribution in one of the simulations is shown in the top-left panel of figure \ref{fig:successful} at angles $0.05-0.25$ rad. The Lorentz factor distribution that they find in the interface falls sharply with angle, from the high Lorentz factor in the core of the jet to that of the cocoon within $\sim 5\theta_j$ (see bottom-left panel of figure \ref{fig:successful}), where $\theta_j$ is the opening angle of the jet core (see below).

Finally, it is important to note that the mixing is sensitive to the magnetization of the outflow, and even a weak magnetic field can suppress the mixing and could affect the interface layer. Currently there are no studies that explore this effect in detail. \\

\noindent \underline{Jet (also known as jet-core):} 
The parts of the launched jet that are closer to the symmetry axis are less susceptible to mixing. Thus, near the axis, the jet properties, $\Gamma$ and $E_{\rm iso}$, are set by the engine at the launching site. This part of the outflow is called the core of the jet and it has the highest $\Gamma$ and $E_{\rm iso}$ in the entire outflow. The total energy in the jet-core is proportional to the energy launched into the jet after it breaks out $E_{\rm j} \propto L_{\rm j} (t_e-t_b)$, implying that $E_{\rm interface} \appropto E_{\rm j}$. The angular size of the jet-core, $\theta_{\rm j}$, is typically smaller than the angle at which the jet is launched, $\theta_{\rm j,0}$. For example  \cite{gottlieb2019b} find in their simulations $\theta_{\rm j} \approx \theta_{\rm j,0}/2$. The jet core can be seen in the left top panel of figure \ref{fig:successful} at angles smaller than $0.05$ rad.

\subsubsection{Choked jet}
As evident from equation \ref{eq:tb}, the durations, luminosities and opening angles inferred from  observations of sGRB jets \citep[e.g.,][]{berger2014} are roughly at the levels needed to successfully break out of the sub-relativistic ejecta of BNS mergers. This implies that, if these mergers are the progenitors of sGRBs, then it is possible that, in many mergers, the jets are choked and there is no sGRB. In other words, the engine dies out too early and the jet dissipates all of its energy into the cocoon before it is able to break out of the ejecta. Interestingly, there is observational evidence, based on the distribution of sGRB durations, that suggests the existence of a significant population of choked sGRB jets \citep{moharana2017}. While choked jets are not expected to produce sGRBs, they are expected to generate a mildly relativistic outflow. If this outflow breaks out of the ejecta, it produces \gray and afterglow emission that may be detectable at the distances of GW events.
% (see section \ref{sec:breakout}). 

A choked jet produces a significant observational signature only if the cocoon driven by the jet is able to break out. For that to happen, the jet must be choked only after crossing a significant fraction of the ejecta. Given that many jets do break out successfully, if there are choked jets then at least some, and possibly many, of those jets are choked at a location where the cocoon breaks out.  The structure of the cocoon outflow, after it breaks out, is quite different than that of a successful jet. First, there is no narrow core of ultra-relativistic material. The typical opening angle of the outflow is $\sim 0.5$ rad, and the fastest ejecta is, at most, mildly relativistic (Lorentz factor of 2-3). Second, the outflow has both an angular and a radial structure. The exact structure, however, can vary significantly from one case to another, and depends mostly on properties such as where the jet was choked, and on whether or not the jet was collimated before it was choked.

%out depends on whether the jet was collimated or not, which in turn depends mostly on its opening angle \citep{bromberg2011a}.  The outflow from a narrowly collimated jet that is choked just before it breaks out is similar to the cocoon from a successful jet without the jet-core and the interface. Namely an outflow with an opening angle of $\sim 45^\circ$ and an angular dependent Lorentz factor, where up to an angle of $\sim 30^\circ$  the outflow is mildly relativistic and at larger angles it is subrelativistic. An example of the structure from a collimated choked jet is shown in figure ???. 

%An uncollimated wide angle choked jet must have a relatively high energy in order for the cocoon to break out of the ejecta \citep[e.g.][]{kasliwal2017,duffell2018}. The reason is that the uncollimated jet has to shock all the ejecta within its opening angle. Thus, for the shock to break out and accelerate the ejecta that it shocks to a four velocity $\gamma \beta \gtrsim 1$ the energy in the jet should be $E_j \approx  10^{51} \gamma (m_{\rm ej}/0.01\msun) (\theta_{j,0}/0.4{\rm~rad})^2  {\rm~erg}$. The resulting outflow has a radial structure dictated mostly by the initial velocity profile of the ejecta, before it was shocked by the jet. An example of the structure from a a wide uncollimated jet is shown in figure ???. 

\subsection{{\bf Prompt \grays}}\label{sec:grays}
The relativistic outflow is expected to produce \grays from at least two (and possibly more) different sources. The initial \gray signal is produced by a (mildly) relativistic shock breakout as the cocoon's forward shock emerges from the sub-relativistic ejecta \citep{nakar2012}. Later an additional signal is emitted by the jet, if it emerges successfully. The source of jet emission is internal dissipation within the jet (e.g., shocks, magnetic reconnection) and this jet emission is presumably what we see as the prompt emission of sGRBs. It is much brighter than the shock-breakout emission, but is expected to be narrowly collimated around the opening angle of the jet. The cocoon breakout emission releases much less energy, but this energy is emitted over a wider opening angle and therefore can dominate the signal seen by observers that are not within the jet's opening angle. Finally, the jet may have an angular structure, with a narrow core at the center and wider wings, where the energy and the Lorentz factor  drop with the angle. Such a structure is inferred from the afterglow observations of GW170817 (\S\ref{sec:GW170817relativistic}) and is also seen in simulations of successful jets (the wings are the jet-cocoon interface \S\ref{sec:JetCocoon}). In the case of such a structured jet, it is possible that \grays are generated by dissipation in the wings. There has been no direct evidence that emission is generated outside of the jet core, but if it is, then it is expected to be fainter than that of the jet. However, similarly to the case of the cocoon breakout, the radiation is emitted over a wider angle and may compete with the cocoon emission in some directions. Below, I first discuss at some length the theory of shock breakout (\S\ref{sec:breakout}). I then discuss the emission from the jet (\S\ref{sec:off_axis_prompt}). I focus only on the off-axis effects on the observed jet emission, and do not discuss at all the mechanisms that generate the jet emission, as they have been addressed in detail in the GRB literature (even though much about them remains unknown). Finally, I  discuss briefly the high inclination emission from the jet's wings, which is the least explored among the three sources of \gray outlined above (\S\ref{sec:high_inclination_grays}).  However, before discussing the various \gray sources, I summarize the constraints that compactness poses on the Lorentz factor and on the observing angle of the sources of the \grays that we observe. These limits are useful for identifying the sources and constraining their properties.

\subsubsection{Compactness limits on the Lorentz factor and observing angle of the \gray source}\label{sec:compactness}
The first limit on the Lorentz factor of GRB jets, which remains the most stringent limit, was set by a compactness argument, namely, that the source is not too optically thick to the observed $\gamma$-rays\footnote{The source of the observed \grays can have only a limited optical depth, $\tau$. When the spectrum is non-thermal (i.e., it includes a high-energy tail), then it is clear that $\tau \lesssim 1$, but also when the spectrum is Comptonized (i.e., it has an exponential cut-off) $\tau$ cannot be too large (see \citealt{matsumoto2019} for a discussion).} \citep{ruderman1975,schmidt1978}. Until recently, compactness was explored only for an on-axis observer, and it was then realized that there  are at least three different sources of opacity that should be considered, where each depends on different assumptions, and each provides a different lower limit on the source Lorentz factor, $\Gamma$ \citep[e.g.,][]{lithwick2001}. Recently, these limits have been generalized also for an off-axis observer, and it was shown that the compactness argument simultaneously places a lower limit on the Lorentz factor for all observers, $\Gamma_{\rm min}$, as well as an upper limit on the angle between the source and the observer, $\theta_{\rm max} \approx 1/2\Gamma_{\rm min}$ \citep{matsumoto2019}. 
The three different causes of optical depth for \grays are (using the notation of \citealt{lithwick2001}): {\bf a.} production of $e^+e^-$ pairs by scattering of the \grays on lower-energy photons; {\bf b.} Thomson scattering of the \grays on electrons and positrons produced by photons with energy higher than the electron rest-mass, as seen in the source rest-frame; {\bf c.} Thomson scattering on electrons that accompany the baryons in the outflow. The first two sources of opacity depend on the \gray  spectrum while the third one depends on the outflow composition (i.e., its baryonic fraction). In GW170817, as well as in many sGRBs, the spectrum is Comptonized \citep{veres2018,ghirlanda2004}, and the most stringent constraints are obtained from limits b and c. I therefore summarize below the relevant equations only for these two cases,  and refer the reader to \cite{matsumoto2019} for a thorough discussion of all the limits and the various caveats.

Consider a source of \grays  at redshift $z$, moving with Lorentz factor $\Gamma$ (corresponding to velocity $\beta c$), at an angle $\theta$ with respect to the observer. The Doppler factor of the source is
\begin{equation}\label{eq:doppler}
	\delta_D=\frac{1}{\Gamma(1-\beta\cos(\theta))} \approx \frac{2\Gamma}{1+(\Gamma\theta)^2} \approx 
	\left\{ 
	\begin{array}{lr}
	\Gamma & \theta \lesssim \Gamma^{-1}  \\
	2 \Gamma^{-1}\theta^{-2} & \theta \gg \Gamma^{-1}  
	\end{array} 
    ,
	\right.
\end{equation}
where the approximated expressions are for $\Gamma \gg 1$ and $\theta \ll 1$. Assume that the observed \grays have an isotropic equivalent luminosity $L_{\gamma,\rm iso}$, a pulse duration $\delta t$, and a Comptonized spectrum $dN/dE \propto E^{\alpha}\exp[-E(\alpha+2)/E_p]$, where $N$ is the photon number, $E$ is the photon energy, $\alpha$ the low-energy power-law index, and $E_p$ the peak energy. The compactness limit set by these observables depends on two dimensionless parameters, $\mathcal{L}$ and $\mathcal{E}$:
\begin{equation}
	\mathcal{L} \equiv \frac{\sigma_T}{\pi c^2} \frac{L_{\rm \gamma,iso}}{E_p\delta t} 
	\approx 1.5 \times 10^{12} \left(\frac{L_{\rm \gamma,iso}}{10^{50}{\rm~erg~ s}^{-1}}\right) 
	 \left(\frac{E_p}{100{\rm~keV}}\right)^{-1}\left(\frac{\delta t}{0.1{\rm~s}}\right)^{-1}~~~;~~~{\rm limits~b~\&~c}~,
\end{equation}
where $\sigma_T$ is the Thomson cross-section, and
\begin{equation}
	\mathcal{E} \equiv \left\{ 
	\begin{array}{lr}
	\frac{E_p(1+z)}{m_e c^2(\alpha+2)} & {\rm limit~b} \\
	&\\
	\frac{E_p(1+z)}{m_p c^2} & {\rm limit~c} 
	\end{array}
	\right. ~.
\end{equation}
Note the dependence on $m_e$ in limit b, and on $m_p$ in limit c.

Limit b is obtained by considering the optical depth for Thomson scattering on pairs at the source, while limit c considers the optical depth to electrons in a baryonic outflow with a total mass $m$, assuming that the observed \gray energy cannot be larger than $mc^2\Gamma^2$ (this is the internal energy, in the observer frame, if the mass $m$ is shocked to a Lorentz factor $\Gamma$). For each of these limits, the  minimal optical depth that is consistent with the observables is:
\begin{equation}\label{eq:compactness}
	\tau_{\rm min} \approx 
    \left\{ 
   	\begin{array}{lr}
    \frac{\mathcal{L}}{\cal{E}^\alpha}\delta_D^{\alpha-4}\exp\left[ -\frac{\delta_D}{\mathcal{E}}\right] & {\rm limit~b} \\
   	&\\
    \frac{\mathcal{LE}}{\Gamma\delta_D^5}& {\rm limit~c} 
   	\end{array} 
   	\right.
\end{equation}
The compactness limit is obtained by requiring\footnote{The strong dependence on $\delta_{D}$, and the typical value of $\mathcal{L}$ ($\gtrsim 10^{10}$), imply that the exact value of $\tau_{\rm min}$ does not have a significant effect on the result.} $\tau_{\rm min} \lesssim 1$. 
Equation \ref{eq:compactness} shows that limit b sets, in turn, a lower limit on the value of the Doppler factor, $\delta_{D,\rm min}$, which can be found by solving for it numerically. Limit c sets a lower limit to the value of $\Gamma \delta_D^5$. Equation \ref{eq:doppler} shows that, for both limits, this implies a lower limit on the source Lorentz factor, 
\begin{equation}
	\Gamma_{\rm min} \approx 
    \left\{ 
   	\begin{array}{lr}
    \delta_{\rm D,min}/2 & {\rm limit~b} \\
   	&\\
    \left(\frac{\mathcal{LE}}{32}\right)^{1/6}& {\rm limit~c} 
   	\end{array}
	\right.
\end{equation}
For both limits, there is a corresponding upper limit on the viewing angle, 
\begin{equation}
	\theta_{\rm max} \approx 1/2\Gamma_{\rm min}
\end{equation}

It is interesting to apply these limits to several observed events. For example, the best-fit parameters of the gamma-ray flare in GW170817 were $L_{\rm \gamma,iso} \approx 2 \times 10^{47}$ erg s$^{-1}$, $\delta t \approx 0.2$ s, $E_p \approx 520$ keV, and $\alpha=-0.6$ \citep{veres2018}, which corresponds to $\mathcal{L} \approx 3 \times 10^8$ and $\mathcal{E} \approx 0.7[5 \times 10^{-4}]$ for limit b[c]. Solving equation \ref{eq:compactness} for limit  b with these values and $\tau_{\rm min} = 1$ results in $\delta_{\rm D,min} \approx 8$. This implies that the source of the observed gamma-rays must have had a Lorentz factor larger than $\Gamma_{min} \approx 4$, and the maximal angle between the direction of its velocity and the line-of-sight was smaller than $\theta_{\rm max} \approx 0.13$ rad ($7^\circ$). Limit c provides similar constraints. 

Another interesting event is sGRB 150101B, that showed some similarities to the prompt emission and the afterglow of GW170817 \citep{burns2018,troja2018a}. It has been suggested that this is due to a \gray source (relativistic jet) seen off-axis at an angle of $\theta \approx 13^\circ$ \citep{troja2018a}. The characteristics of the prompt emission of sGRB 150101B are $L_{\rm \gamma,iso} \approx 10^{51}$ erg s$^{-1}$, $\delta t \approx 0.01$ s, $E_p \approx 1300$ keV and $\alpha=-0.8$ \citep{burns2018}.  For both limits b and c,  the corresponding minimal Lorentz factor is $\Gamma_{\rm min} \gtrsim 20$, and the maximal viewing angle is $\theta \lesssim 3^\circ$. This implies that the prompt emission of sGRB 150101B was seen on-axis (i.e., within an angle of $1/\Gamma$ from the \gray source), or very close to on-axis. The limits for sGRB 150101B are similar to the limits obtained for many sGRBs for which the prompt emission spectra are well fit by a Comptonized model \citep{nakar2007}.

\subsubsection{Shock breakout}\label{sec:breakout}
The propagation of the jet through the ejecta drives strong shocks into the ejecta, both by the jet-head and by the expanding cocoon. Strong shocks may also be driven into the ejecta by an uncollimated wind, e.g., via a magnetar's spin-down. As further discussed below, these shocks are radiation-mediated and radiation-dominated, in the sense that the dissipation within the shock transition layer is mediated by radiation, and the downstream energy and pressure are dominated by radiation as well. Once a shock, for instance the forward shock driven by the cocoon, breaks out of the ejecta, the photons that are within the shock transition layer are released to the observer, followed by photons that diffuse from behind the shock. For a shock velocity greater than about $0.5$c, these photons are $\gamma$-rays. Thus, when the shock driven by the jet and/or cocoon breaks out of the ejecta, it emits a flash of $\gamma$-rays. 

The breakout emission is determined by the interaction with the ejecta of photons released from  the shock transition layer and from the shock downstream, as they stream towards the observer. Especially in the relativistic case, this is a dynamical process, far from thermal equilibrium, that involves different species (photons, electrons, positrons and baryons) that interact on very different scales.  Currently, there is no theoretical calculation of this interaction in detail, and the models provide only order-of-magnitude estimates of the main observables---energy, duration and typical photon frequency---of the signal from a breakout  of a spherical shock that propagates in a plasma with no free neutrons. Below, I discuss these predictions for the main properties of the shock breakout emission, and under what conditions it is expected to be detectable. With this in mind,  I first give a brief overview of the structure of radiation-mediated shocks, and then discuss the hydrodynamics and the emission from various stages of the shock breakout and the ensuing cooling emission.\\

\noindent \underline{\textbf{The structure of radiation-mediated shocks:}} Since the breakout photons are emitted from the shock transition layer and the immediate shock downstream, their nature depends strongly on the structure of the radiation-mediated shock (RMS). This structure depends, in turn, on the shock velocity and on the conditions in the unshocked ejecta. In recent years, there has been significant progress in the theoretical understanding of RMS structure, especially in the mildly relativistic and ultra- relativistic regimes. Below, I provide a short description of the main RMS properties that are most relevant to the shock breakout emission from compact binary merger ejecta, and I refer the reader to \cite{levinson2019} (and references therein) for a comprehensive review on radiation-mediated shocks and their observational signatures. 

\begin{figure}%[ht]
\centering
\vspace*{-25mm}
\includegraphics[width=9.8cm]{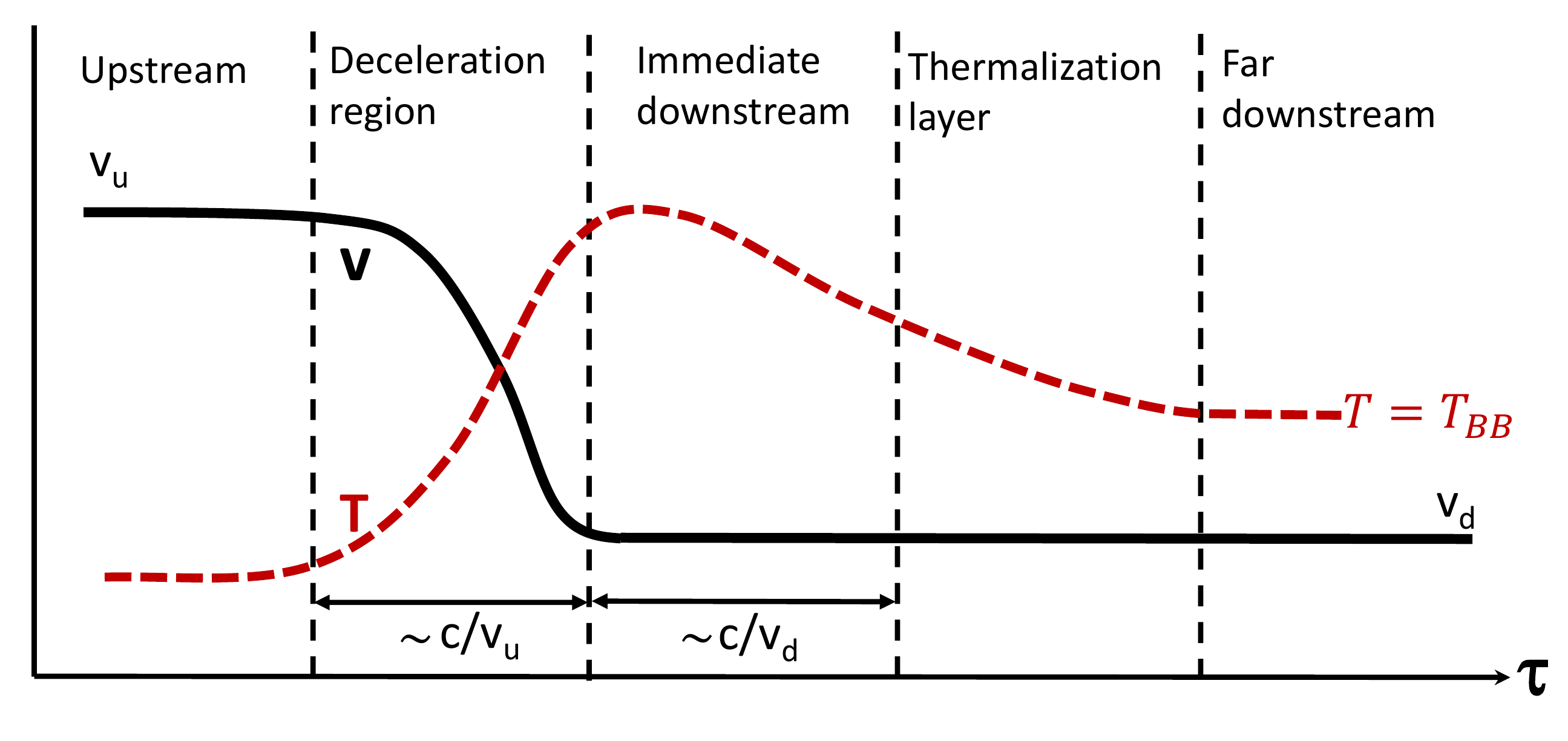} %\vspace*{-5mm}
\includegraphics[width=10cm]{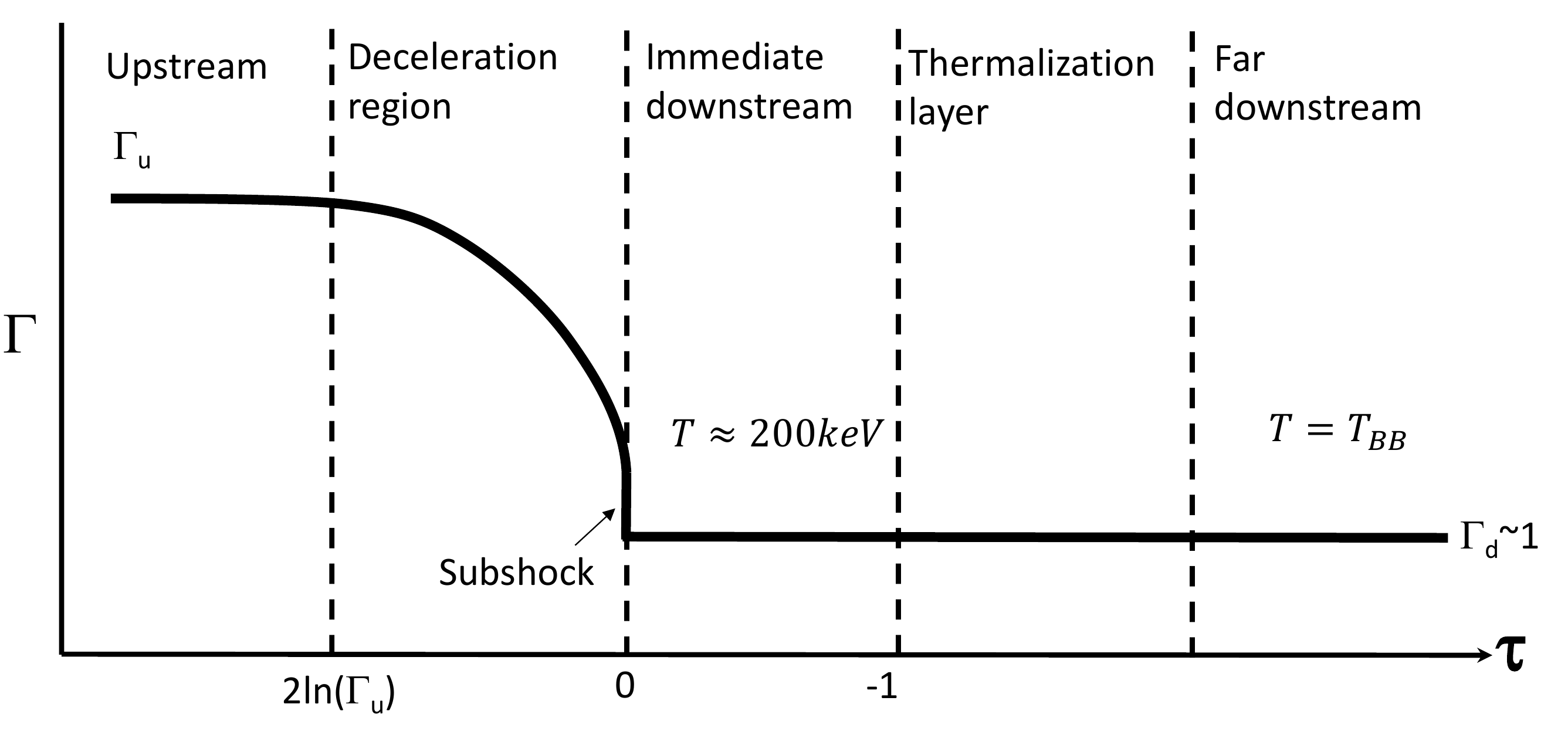} \vspace*{-5mm}
\caption{.Schematic illustrations of the structure of radiation-mediated shocks that propagate in a photon-starved unmagnetized fluid, as seen in the shock rest frame. The horizontal axis gives the optical depth traversed by a photon moving upstream. {\it Upper panel}: a fast sub-relativistic RMS, where the temperature in the immediate downstream is out of thermal equilibrium. The solid-black  and dashed-red curves delineate the 
velocity and temperature profiles, respectively. {\it Lower panel}: relativistic RMS. The solid black line is the flow's Lorentz factor. Five distinct regions are indicated in each of the RMS types: far-upstream, where the shock has no effect; deceleration region, were the fluid decelerates from the upstream velocity to the downstream velocity; immediate-downstream, where the temperature is considerably higher than the blackbody temperature; a thermalization zone, where the temperature drops by production of photons; and far-downstream, where the radiation reaches thermodynamical equilibrium. From \cite{levinson2019}.}
\label{fig:RMS_sketch}
\end{figure}

It is most convenient to describe the shock structure as seen in the shock frame. In this frame, the shock is stationary. An unshocked plasma streams into the shock at a velocity $\beta_s c$ and corresponding Lorentz factor $\Gamma_s$. The plasma, which arrives from the upstream, is decelerated and heated in the shock transition layer, and then flows out of the transition layer towards the downstream at a velocity $\beta_d c$. A radiation-mediated shock is defined as one in which the momentum of the upstream incoming plasma is carried away by photons that stream back into the shock transition layer from the downstream. It can be shown that this requirement dictates that the optical depth of the shock transition layer, as seen by a photon that heads from the downstream towards the upstream, is\footnote{ In relativistic shocks, the optical depth for photons depends on their direction relative to the shock. Also, the optical depth as seen by a photon that heads from the downstream towards the upstream increases logarithmically with the shock Lorentz factor $\tau_s \approx \ln (\Gamma_s)$ \citep{budnik2010,nakar2012,granot2018}.}  $\tau_{\rm s} \approx 1/\beta_{\rm s}$ (see figure \ref{fig:RMS_sketch}).

The optical depth of the shock transition layer dictates that for the formation of a RMS, the shock must propagate in a medium with optical depth in the upstream direction, $\tau$, that is larger than $1/\beta_s$. This condition is always satisfied at the relevant radii within the bulk of the merger ejecta. A second condition for the formation of a RMS, is that the energy density in the downstream is dominated by radiation. For an upstream with a mass density $\rho_u$, composed of heavy ions with atomic mass $A$, this condition is satisfied for $\beta_s \gtrsim 0.03 (A/100)^{2/3} (\rho_u/10^{12} {\rm~g~cm}^{-3})^{1/6}$. Thus, fast shocks, with $\beta_s \gtrsim 0.1$, that are driven into the merger ejecta are always radiation-mediated and radiation-dominated. Note that, as the downstream energy density is dominated by radiation, the fluid has an adiabatic index of 4/3.

The shock breaks out of the ejecta once the optical depth ahead of the shock cannot sustain the shock anymore, namely when $\tau \approx 1/\beta_s$. At this point, all the photons that are within the shock transition layer are free to stream towards the observer, and the shock makes a transition into a collisionless shock. Following the shock breakout, photons that were behind the shock begin diffusing out of the shocked ejecta, forming the so called "cooling emission". 

The RMS structure depends strongly on the shock velocity and on the conditions in the shock upstream (e.g., photon number, magnetization, composition). In our case, the shock velocity in the ejecta is $\gtrsim 0.1$c, and the upstream is most likely unmagnetized. The main composition of the ejecta is \rpe although the fastest parts of the ejecta may be rich with free neutrons and light elements (see Section \ref{sec:neutrons})\footnote{In this review I do not consider the effect of free neutrons, owing to the lack of a proper theory for RMS with free neutrons.}.  An important property of a RMS in these conditions is that the radiation in the immediate downstream is most likely dominated by photons that were emitted in the downstream itself ( the shock is photon poor)\footnote{The photon-to-baryon ratio in the ejecta following its ejection from the vicinity of the central compact object is too low to affect the shock structure. Photon-rich shocks can occur only if the ejecta is shocked repeatedly at large radii, where one shock generates the photons and successive shocks propagate thorough a photon-rich plasma. Such a scenario may take place under some conditions for the case of variable energy injection from a long-lasting central engine \citep{beloborodov2018}. }. This implies that the temperature in the downstream, which strongly affects the shock breakout signal, depends on the number of photon that are produced in the layer of the downstream from which photons can diffuse back into the shock transition layer. The width of this layer is $\Delta_{ph} \sim (3 \beta_d \kappa \rho_d)^{-1}$, where $\rho_d$ is the downstream mass density. This layer is denoted the "immediate downstream", and its temperature during the shock breakout determines the observed temperature. 

A rough estimate of the temperature in the immediate downstream can be obtained by dividing the energy density, $\epsilon$, by the number density of photons that are generated over the advection time, $\Delta_{ph}/\beta_d c$. The main photon-generation mechanism is free-free emission, and the photons that determine the downstream temperature, $T_d$, are those that are emitted with temperature $\sim T_d$, together with photons that are inverse-Compton up-scattered to this temperature. Thus, the generation rate of photons with temperature $\sim T_d$ is 
\begin{equation}\label{eq:dot_nff}
	\dot{n}_{ff} \approx 4 \times 10^{36} {\rm~s^{-1}~cm^{-3}} \frac{\left<z\right>\left<z^2\right>}{\left<A\right>^2} \rho_d^2 T_d^{-1/2} \Lambda_{ff} ,
\end{equation}
where $\rho_d$ and $T_d$ are in c.g.s units, and the averages are over the atomic fraction, e.g.,  $\left<z^2\right> = \Sigma x_j Z_j^2$ where $x_j$ is the atomic fraction of element $j$.  The factor $\Lambda_{ff}(\rho_d,T_d)$ accounts for photons up-scattered by inverse Compton, and it can be approximated \citep{weaver1976,nakar2010,sapir2013} as
\begin{equation}\label{eq:Lambdaff}
	\begin{array}{l}
	\Lambda_{ff} \approx \max\left\{ 1,\frac{1}{2}\ln(y)[1.6+\ln(y)] \right\},\\
	\\
	y=500  \left(\frac{\left<z^2\right>}{\left<A\right>}\right)^{-1/2}\left(\frac{\rho_d}{10^{-9} {\rm~g~cm^{-3}}}\right)^{-1/2}\left(\frac{T_d}{\rm{keV}}\right)^{9/4}.
	\end{array}
\end{equation} The immediate downstream temperature, $T_d$, is then found by solving the equation 
\begin{equation}\label{eq:Td}
	3k_B T_d\approx  \frac{ \epsilon }{\frac{\Delta_{ph}}{\beta_d c} \dot{n}_{ff}(\rho_d,T_d)} =\epsilon \frac{ 3 \beta_d^2 \kappa \rho  c }{\dot{n}_{ff}(\rho_d,T_d)}
\end{equation}
where $\epsilon$, $\beta_d$ and $\rho_d$ are determined by the upstream density and shock velocity (e.g., for Newtonian high-Mach-number shocks, $\beta_d=\beta_s/7$, $\rho_d=7\rho_u$, and $\epsilon=\frac{18}{7}\rho_u \beta_s^2 c^2$). If the solution of equation \ref{eq:Td} results in a temperature that is lower than the blackbody temperature $T_{BB}=(\epsilon/a_{BB})^{1/4}$, where $a_{BB}$ is the radiation constant, then photon generation is rapid enough to maintain thermal equilibrium in the shock transition layer, and $T_d=T_{BB}$. If, however, the solution of this equation provides a temperature that is higher than $T_{BB}$, then the immediate downstream is out of thermal equilibrium, and its temperature is roughly the one that solves the equation. In that case, photon generation continues to reduce the temperature as the fluid is advected away from the shock, reaching thermal equilibrium only far from the shock transition layer, at the far downstream.

\begin{figure}
	\center
	\includegraphics[width=0.65\textwidth]{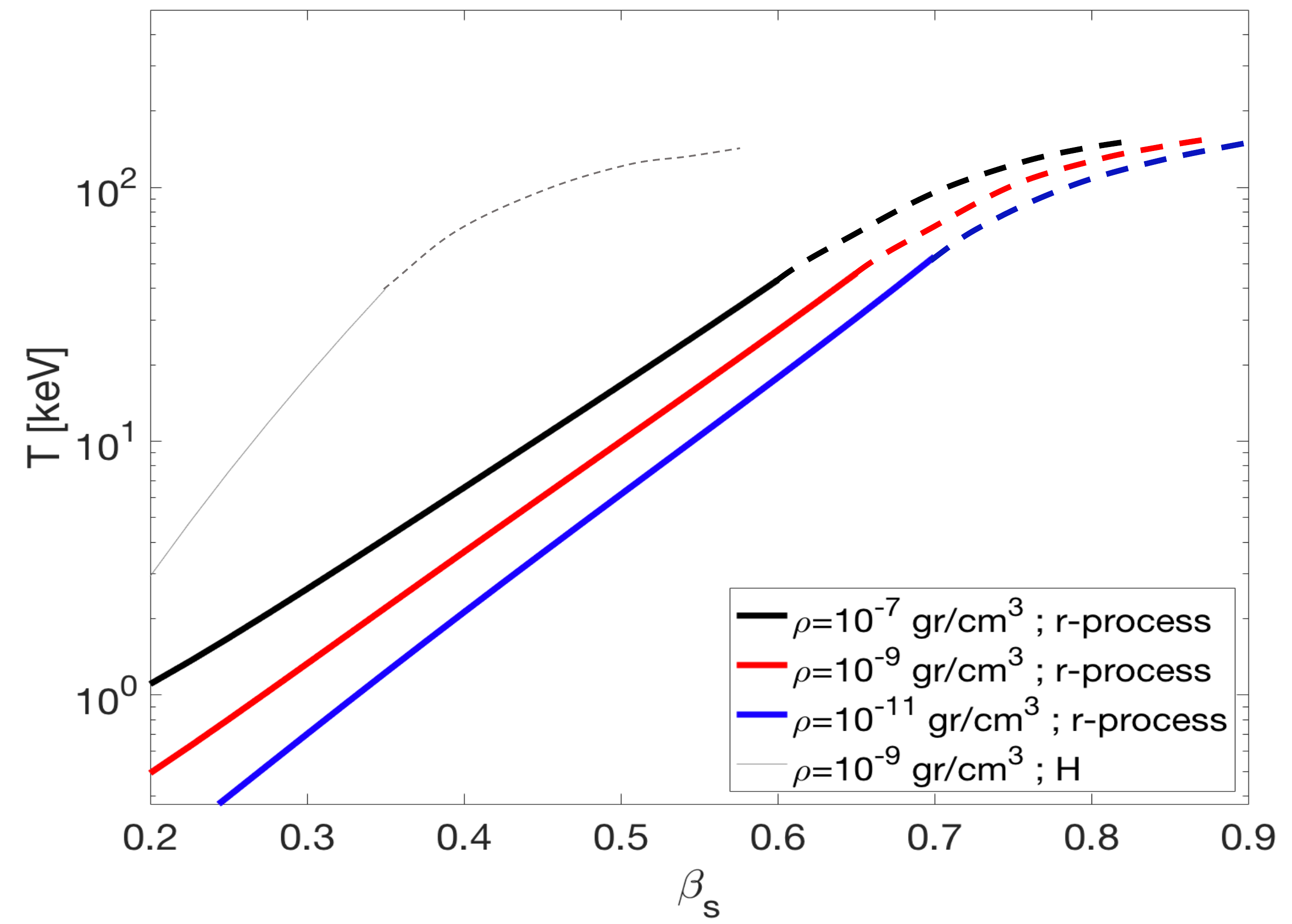}
	\caption{The temperature in the immediate downstream of a radiation-mediated shock, as a function of the shock velocity. The solid curve is calculated by solving numerically equations \ref{eq:dot_nff}-\ref{eq:Td}. For \rp [H-rich] composition we use $\left<z\right>\left<z^2\right>/\left<A\right>^2 = 10 ~[1]$, $\left(\left<z^2\right>/\left<A\right>\right)^{1/2} = 5~ [1]$, and $\kappa= 0.16~ [0.34] {\rm ~cm^2~g^{-1}}$. This calculation is applicable for $T \lesssim 50$ keV. At higher temperatures, vigorous pair production leads to increased photon generation that mitigates the rise in the temperature, bringing it to $100-200$ keV, almost independent of the shock's Lorentz factor. The dashed lines are illustrations of the temperature behavior in this regime.  From \cite{levinson2019}.}%
	\label{fig:T_RMS}
\end{figure}

Equations \ref{eq:dot_nff} and \ref{eq:Lambdaff} show that the photon generation rate depends on the composition via averages on $Z$ and $A$. For fully ionized \rpe with solar abundance and $A>85$, the values of these averages are ${\left<z\right>\left<z^2\right>}/{\left<A\right>^2} \approx 10$, $\left({\left<z^2\right>}/{\left<A\right>}\right)^{1/2} \approx 5$ and $\kappa=\left({\left<z\right>}/{\left<A\right>}\right)\frac{\sigma_T}{m_p} \approx 0.16 {~\rm cm^2~g^{-1}}$. These values depend only weakly on the exact composition, as long as it is dominated by \rp elements. Plugging these values into equations \ref{eq:dot_nff}-\ref{eq:Td}, we find that the downstream radiation falls out of thermal equilibrium, once the shock velocity exceeds $\beta_s > 0.12 \left(\rho_d/10^{-9} {\rm~g~cm^{-3}}\right)^{1/30}$,
where the weak dependence on $\Lambda_{ff}$ is neglected. Figure \ref{fig:T_RMS} depicts the temperature in the immediate downstream at higher velocities (obtained by solving equations \ref{eq:dot_nff}-\ref{eq:Td}) for \rp material at several representative densities, and for one density of H-rich plasma.  The figure shows that, for \rp material,  the temperature rises sharply from $0.1-1$ keV at $\beta_s=0.2$ to $50$ keV at $\beta_s = 0.6-0.7$.  Once the downstream temperature exceeds  $50$ keV, electron-positron pair production starts playing a role and the shock structure changes significantly. Pairs have an impact on practically all aspects of the shock structure, but the effect that probably has the largest influence on the observed signal is the regulation of the photon temperature. Pairs serve as a thermostat that regulates the temperature in the immediate downstream to $100-200$ keV \citep{katz2010,budnik2010}. This happens because the temperature depends on the photon production rate, which is a function of the number of pairs, and which, in turn, increases exponentially with the temperature. This effect, which, by coincidence, becomes important once the shock velocity approaches the speed of light (i.e., $\Gamma_s \beta_s \gtrsim 1$), implies that the downstream temperature of relativistic RMS, as measured in the downstream restframe, is always  $100-200$ keV, independent of the shock Lorentz factor. This is illustrated schematically by the dashed lines in figure \ref{fig:T_RMS}.

Pair production within the shock transition layer has another important aspect, which may affect the breakout signal. The produced pairs dominate the optical depth within the transition layer and, since the optical depth of relativistic RMS is $\tau_s \sim 1$, this implies that pairs significantly reduce the physical width of the shock \citep{nakar2012,granot2018,lundman2018}. Putting it differently, a relativistic RMS continues to propagate in a medium that has an pair-unloaded optical depth (i.e., without pairs) much smaller than unity, by generating its own optical depth via pair production. As I explain below, while this effect may be important in some cases (see footnote \ref{note:pairs}), in most scenarios this does not have a strong effect on observables such as the total breakout energy or duration, since the breakout emission in these cases is dominated by the region were the pair-unloaded optical depth is unity. \\

\noindent \underline{\textbf{Shock dynamics:}}
The dynamics of the shock that crosses the ejecta depends on the force that drives it (e.g., a jet), and on the ejecta density and velocity profiles. An important point to realize is that, since the shock breakout takes place when the optical depth ahead of the shock is about unity or less, its properties are dictated mostly by a minute amount of mass that is moving at the front of the ejecta. For example, if the breakout takes place at a radius of $10^{12}$ cm, then $\tau=1$ corresponds to a mass of $4 \times 10^{-8} \msun$ (taking $\kappa= 0.16 {\rm~cm^2~g^{-1}}$), namely a fraction of $\sim 10^{-6}$ out of the total ejecta mass. Thus, the breakout signal depends mostly on the fast tail that leads the ejecta,  at velocities that may be significantly higher than those of the bulk of the ejecta, possibly mildly or even ultra relativistic (see section \ref{sec:dynamicalEjecta} and \citealt{kasliwal2017,gottlieb2018b,beloborodov2018,hotokezaka2018b,radice2018}). It is difficult to predict theoretically the properties (e.g., velocity and density distribution) of this leading part of the ejecta which contains so little of the mass. Numerical simulations cannot be trusted to resolve such a minute mass fraction, and the applicability of analytic considerations is limited, given the complex non-linear hydrodynamics that launches this mass. Nevertheless, simulations \citep[e.g.,][]{hotokezaka2018b,radice2018} and analytic arguments \citep[e.g.][]{kyutoku2014,beloborodov2018} both suggest that the leading parts of the ejecta are launched by deposition of energy near the outer parts of the neutron stars at the time of the merger, and that this type of deposition typically leads to a steep velocity profile. For example,  \cite{hotokezaka2018b} find that the fast tail reaches at least mildly relativistic velocities (i.e., $\gamma \beta > 1$), and that the density distribution at $\beta>0.5$ is often steeper than $\rho \propto (\gamma \beta)^{-10} $.
%Another implication of the low amount of mass that is involved in the shock breakout is that if, indeed, the  leading edge of the ejecta is mildly or even ultra relativistic, then its velocity is very important in shaping the breakout signal. The reasons are that, first, the shock must "chase" the high velocity ejecta before catching it. This leads to a larger breakout radius. Second, there are three restframes that should be considered when calculating the signal, the observer frame the unshocked ejecta frame and the shocked ejecta frame. Below we show how these considerations affect the breakout emission.

An important question is, under which conditions is a shock breakout expected, and if there is a breakout, then what are its radius and velocity. The answers to these questions depend on the ejecta structure and on the force that drives the shock. Motivated by numerical and analytical considerations described above (see also \S\ref{sec:dynamicalEjecta}), I consider below ejecta which is composed of a slow massive bulk with a shallow density distribution, and a low-mass fast tail with a steep density distribution. If the jet is successful, then there must be a shock breakout. The velocity of the jet-head within the bulk of the ejecta is mildly relativistic for typical parameters, and it accelerates further within the fast tail. The cocoon, which is sub-relativistic and narrowly collimated while the jet-head is within the ejecta, expands sideways and accelerates to mildly relativistic velocities in the fast tail. As a result, upon the breakout from the fast tail, the shock is mildly relativistic where the cocoon breaks out, away from the jet axis. Near the jet axis, along the jet-head,  the shock is faster, possibly even ultra-relativistic.

Choked jets may also lead to a shock breakout, but not always. If a jet is choked after crossing a significant part of the bulk of the ejecta, then the cocoon is expected to drive a shock into the fast tail. Also, in the case that the jet is choked deep within the bulk of the ejecta, it may still drive a strong shock into that tail, but only if the energy deposited into the cocoon is larger than the energy in carried by the ejecta within the jet opening angle, i.e., the cocoon's isotropic equivalent energy is larger than about $10^{51}$ erg. In fact,  a very-wide-angle outflow or even a spherical energy injection from the central engine can drive a strong shock through the bulk of the ejecta, if it is energetic enough. The dynamics of the shock, once it starts crossing the tail, depends on the steepness of the density gradient, as I discuss below. A spherical Newtonian RMS (adiabatic index 4/3)  that propagates in a {\it static} medium with a power-law density $\rho \propto r^{-\alpha}$, accelerates if $\alpha>3.13$ \citep{waxman1993}. The propagation of a shock in {\it expanding} material behaves very differently. A shock that propagates in expanding ejecta with a density profile  $\rho \propto v^{-\alpha}$ and with $\alpha>8.2$ can accelerate not only with respect to the observer, but also compared to the ejecta, i.e. the shock velocity in the upstream frame increases and the shock becomes stronger. If, however, $\alpha<8.2$, the shock may accelerate compared to the observer, but it must decelerate compared to the ejecta, becoming weaker and weaker with time, until it dies eventually \citep{govreensegal2019}. This implies that, if the density gradient in the fast tail is steep enough, then any strong shock that is driven into the tail accelerates as it crosses the tail, at least as long as the shock is Newtonian, and eventually it is expected to break out. If the density gradient in the tail is too shallow, then the shock always becomes weaker during its propagation, and it may die without breaking out.  

The post-breakout hydrodynamic evolution is also important in shaping the observed signal. This evolution depends mostly on whether the shock (as measured in the upstream frame) is relativistic or not, and on whether pairs are produced. If the shock is relativistic, then most of the shocked gas energy is in the form of internal energy. As a result, after the shock crossing, the hot gas rarefies, converting internal energy to bulk energy, and the gas accelerates significantly. A fluid element that has been shocked to a Lorentz factor $\gamma_s$ accelerates to a final Lorentz factor that ranges from about $\gamma_s^2$ to $\gamma_s^{1+\sqrt{3}}$ (as measured in the upstream frame), depending on details such as the density profile of the unshocked gas \citep{johnson1971,pan2006,yalinewich2017}. The dynamics of  relativistic RMS is also affected by the production of pairs that dominate the optical depth. This enables the shock to continue propagating also beyond the point where the optical depth of the unshocked gas without pairs,  $\tau_{\rm unloaded}$, drops below unity. After the shock crossing, as the shocked fluid rarefies and accelerates, the rest-frame temperature drops, and so does the number of pairs, up to the point were the rest-frame temperature is $\sim 50$ keV and pair opacity stops playing a role \citep{nakar2012}. At this point, all the photons in regions where $\tau_{\rm unloaded} \lesssim 1$ are released and can travel to the observer. Photons from regions where  $\tau_{\rm unloaded} > 1$ cannot stream out directly, even after the pairs disappear, and therefore escape only at larger radii.

If the shock is sub-relativistic, $\beta_s\gamma_s \lesssim 1$, then the post-breakout evolution is much simpler. Current models of RMS structure suggest that sub-relativistic shocks do not produce pairs. Thus, the shock propagates up to the point where $\tau \approx 1/\beta_s$, and from this point on the photons are free to stream toward the observer. Again, photons from deeper regions diffuse out over longer timescales. Also, in sub-relativistic shocks, the post breakout acceleration is not significant (acceleration by a factor of 2, at most; \citealt{matzner1999}).\\

\noindent \underline{\textbf{The shock breakout signal:}}
I provide here a rough estimate of the main breakout observables under the approximation of a spherical shock that propagates in an expanding ejecta. I denote the Lorentz factor and velocity of the shock by $\gamma_s$ and $\beta_s$, and those of the unshocked ejecta by $\gamma_e$ and $\beta_e$, both measured in the observer frame. The shock quantities, as seen in the unshocked ejecta (upstream) frame, are  marked with "$'$". Thus,
\begin{equation}\label{eq:beta_s_tag}
	\beta_s'=\frac{\beta_s-\beta_e}{1-\beta_s\beta_e} ,
\end{equation}  
and
\begin{equation}
	\gamma_s'=\gamma_s\gamma_e(1-\beta_s\beta_e),
\end{equation}  
are, respectively, the velocity and the Lorentz factor of the shock, as measured in the upstream frame.

The breakout of a relativistic shock from the ejecta releases a short flare of \grays towards the observer from the breakout layer, where $\tau_{\rm unloaded} \sim 1/\beta_s'$. Immediately after, photons from deeper layers begin diffusing out of the expanding gas, producing the cooling emission. The cooling emission can be divided into (at least) two phases, planar and spherical. The planar phase lasts for as long as the hydrodynamics of the emitting region can be approximated as planar, specifically, until the expanding material roughly doubles its radius. The duration of the signal from the breakout layer is dictated by the difference in the light-travel time of photons emitted at different angles with respect to the line-of-sight, known as the {\it angular time}. The duration of the planar phase is dominated by the difference in the arrival times of photons emitted at different radii, known as the {\it radial time}. If the shocked material is relativistic (in the observer frame) then the radial time of the planar phase is comparable to the angular time of the breakout layer, and therefore photons from the breakout layer and from the planar phase arrive at the observer simultaneously, comprising together a shock-breakout \gray flare. Below, I derive the properties of the emission from the breakout layer under some simplifying assumptions, followed by a discussion of the planar phase emission. I then estimate in which scenarios the breakout layer dominates the emission, and in which ones the planar phase dominates.\\

\noindent \underline{\textit{The emission from the breakout layer:}}\\
As discussed above, the RMS propagates at least up to the point were $\tau_{\rm unloaded} \approx 1/\beta_s'$. If $\gamma_s'\beta_s' \lesssim 1$, then pairs in the shock are negligible, and the shock breaks out at this point.  If $\gamma_s'\beta_s' \gtrsim 1$, then self-generated pairs dominate the opacity, and the shock continues in the ejecta beyond this point. Nevertheless, even though the shock continues to propagate, in most cases the photons in the layer  where $\tau_{\rm unloaded} \approx 1/\beta_s'$ are released and dominate the emission from the entire region were $\tau_{\rm unloaded} \lesssim 1/\beta_s'$. This happens after the shocked gas accelerates, rarefies and cools down to $\sim 50$ keV. Thus, the properties of the radiation emitted from the layer where $\tau_{\rm unloaded} \approx 1/\beta_s'$  provide an approximation of the shock breakout signal in both the Newtonian and the relativistic regimes, and therefore we denote this region as the "breakout layer". The relation between the unloaded optical depth and the mass of the shock transition layer is $\tau_{\rm unloaded} \approx \kappa m/(4\pi R^2)$, and the corresponding mass of the breakout layer is roughly
\begin{equation}
	m_{\rm bo} \approx \frac{4\pi R_{\rm bo}^2}{\beta_s'\kappa} = 4 \times 10^{-8} \beta_{\rm s,bo}'^{-1} \left(\frac{R_{\rm bo}}{10^{12}{\rm~cm}}\right)^2 \left(\frac{\kappa}{0.16 {\rm~cm^2~g^{-1}}}\right)^{-1} \msun ,
\end{equation} 
where $R_{\rm bo}$ is the radius at the shock location where $\tau_{\rm unloaded} = 1/\beta_s'$.
The internal energy in the breakout layer right after it is shocked, as seen in the observer frame, provides a rough estimate of energy in the emission from the breakout layer,
\begin{equation}\label{eq:Ebo}
	E_{\rm bo} \sim m_{\rm bo} c^2 \gamma_{\rm s,bo} (\gamma_{\rm s,bo}'-1) \sim 7 \times 10^{46} {\rm~erg~} \frac{\gamma_{\rm s,bo} (\gamma_{\rm s,bo}'-1)}{\beta_{\rm s,bo}'}\left(\frac{R_{\rm bo}}{10^{12}{\rm~cm}}\right)^2 ,
	%\left(\frac{\kappa}{0.16 {\rm~cm^2~g^{-1}}}\right)^{-1} {\rm~erg}
\end{equation}
where $\gamma_{\rm s,bo}$ is the shock Lorentz factor at the breakout layer, and the dependence on $\kappa$ is omitted ($\kappa=0.16 {\rm~cm^2~g^{-1}}$ is assumed). Note that the total breakout signal may contain also contributions from layers deeper than the breakout layer, that radiate during the planar phase. In such a case, the breakout energy can be larger (see discussion of the planar phase, below).

Following shock crossing, the breakout layer accelerates. If $\gamma_s'\beta_s' \gtrsim 1$, then the acceleration is significant and, due to pair opacity, the acceleration occurs before the photons are released. The amount of acceleration depends on the density distribution of the ejecta (see discussion above) and I will approximate here the final Lorentz factor, as seen in the ejecta frame, by $\gamma_s'^2$, which in the observer frame is roughly $\gamma_f \approx \gamma_s\gamma_s'$. Note that, since acceleration is negligible when $\gamma_s'\beta_s' \lesssim 1$, this approximation is valid also if the shock is sub-relativistic. Assuming a spherical breakout, the duration of the breakout signal is dominated by the difference between the light travel time of photons emitted along the line of sight and that of photons emitted an an angle of $\sim 1/\gamma_f$ with respect to the line of sight. Thus, the duration of the breakout signal is  roughly
\begin{equation}\label{eq:tbo}
	t_{\rm bo} \sim \frac{R_{\rm bo}}{2c\gamma_{\rm f,bo}^2} \approx 16 {\rm~s~}\left(\frac{R_{\rm bo}}{10^{12}{\rm~cm}}\right) \left(\gamma_{\rm s,bo} \gamma_{\rm s,bo}'\right)^{-2}     ~.
\end{equation}

The breakout temperature is roughly the immediate downstream temperature of the breakout layer, as seen in the observer frame.  As evident from figure \ref{fig:T_RMS}, this temperature depends strongly on whether the shock is relativistic or not. If $\gamma_s'\beta_s' \lesssim 1$, then the rest-frame temperature should be calculated using equations \ref{eq:dot_nff}-\ref{eq:Td} (where $\beta_d \approx \beta_s'/7$), and then multiplied by $\gamma_s$ for transformation to the observer frame.  If $\gamma_s'\beta_s' \gtrsim 1$, then the rest-frame temperature at the time that the photons are released is about $50$ keV, and the Lorentz factor of the breakout layer is $\gamma_{\rm f,bo}$. The observed temperature is therefore roughly,
\begin{equation}\label{eq:Tbo}
	T_{\rm bo} \sim 50 \gamma_{\rm f,bo} {\rm~keV} \sim 50 \gamma_{\rm s,bo} \gamma_{\rm s,bo}' {\rm~keV}~~~;~~~ \gamma_{\rm s,bo}'\beta_{\rm s,bo}' \gtrsim 1 ~.
\end{equation}

Equations \ref{eq:Ebo}-\ref{eq:Tbo} show that when the shock is relativistic and the emission from the breakout layer is comparable to, or larger than, that of the planar phase, then the three main breakout observables, $E_{\rm bo}$, $t_{\rm bo}$ and $T_{\rm bo}$, depend on two physical parameters, $R_{\rm bo}$ and $\gamma_{\rm f,bo}$. In this regime, the observables provide a direct measure of $\gamma_{\rm f,bo}$ and $R_{\rm bo}$, where the former is obtained trivially from equation \ref{eq:tbo} and the latter is
\begin{equation}
	R_{\rm bo} \sim 2.5 \times 10^{11} \left(\frac{t_{\rm bo}}{1{\rm~s}}\right)^{-1}  \left(\frac{T_{\rm bo}}{100{\rm~keV}}\right)^2 {\rm~cm} ~~~;~~~ \gamma_{\rm s,bo}'\beta_{\rm s,bo}' \gtrsim 1 ~.
\end{equation}
Moreover, since the three observables depend on two physical parameters, equations \ref{eq:Ebo}-\ref{eq:Tbo} define  a "closure relation" that must be satisfied,
\begin{equation}\label{eq:closure}
	t_{\rm bo} \sim 1 \left(\frac{E_{\rm bo}}{10^{46}{\rm~erg}}\right)^{1/2} \left(\frac{T_{\rm bo}}{100{\rm~keV}}\right)^{-2.5} {\rm~s} ~~~;~~~ \gamma_{\rm s,bo}'\beta_{\rm s,bo}' \gtrsim 1 ~.
\end{equation}
I stress that the observables of a relativistic shock breakout that is dominated by the breakout layer need agree with this relation only to within an order of magnitude. First, the estimate of the breakout energy is quite rough, and the actual energy can be larger by some factor, due to contribution from the planar phase. Second, and more importantly, the relation depends sensitively on the temperature, which is not well defined, because the spectrum is not expected to be a black body (see below). Thus, one cannot measure the value of $T_{\rm bo}$ simply by taking the peak of the specific flux $F_\nu$ and applying the blackbody relation $F_{\nu,\rm peak}=2.8~kT_{\rm bo}$. A better estimate of $T_{\rm bo}$ for the case of a spectrum with an exponential cut-off, $F_\nu \propto e^{-h\nu/E_0}$ is probably $kT_{\rm bo}=E_0$. In any case, an uncertainty by a factor of 2-3 in the measurement of $T_{\rm bo}$ from the observed spectrum translates to an uncertainty of a factor of $\sim 10$ in the closure relation (eq. \ref{eq:closure}).\\

\noindent \underline{\textit{The delay between the GW and the shock breakout signal:}}\\
An observable that is unique to compact binary mergers is the delay between the merger time, as marked by the end of the chirping GW signal, and the $\gamma$-rays, $\delta t_{\scriptscriptstyle GW,\gamma}$. If the \grays originate from a shock breakout, then since the fast-tail of the ejecta is most likely launched immediately after the merger (within less than $10$ ms), the breakout radius is related to the observed delay via the velocity of the ejecta at the breakout, 
\begin{equation}\label{eq:R_Dt}
	R_{\rm bo}=\frac{\beta_{\rm e,bo}}{1-\beta_{\rm e,bo}}c \delta t_{\rm \scriptscriptstyle GW,\gamma}. 
\end{equation}    
Solving this equation together with equations \ref{eq:Ebo} and  \ref{eq:tbo} provides an estimate of the shock and the ejecta velocities and Lorentz factors. If the shock is relativistic, $\gamma_{\rm s,bo}'\beta_{\rm s,bo}' \gtrsim 1$, then the observed temperature is predicted by equation  \ref{eq:Tbo}. If $\gamma_{\rm s,bo}'\beta_{\rm s,bo}' \lesssim 1$, then the temperature in the shock rest frame is obtained by solving equations \ref{eq:dot_nff}-\ref{eq:Td}, and a boost by $\gamma_{\rm s,bo}$ provides a prediction to the observed temperature\footnote{Equations \ref{eq:dot_nff}-\ref{eq:Td} depend also on the composition and density of the shocked gas. The dependence on the composition may be significant, so for every assumed composition, there will be a different solution. The dependence on the density 
is extremely weak, as evident from figure \ref{fig:T_RMS} (T varies by a factor of 
order unity when $\rho$ varies by four orders of magnitude).  A lower limit on the density of the gas in the shock upstream is $\rho_u >\frac{ 1}{\kappa \beta_{\rm s,bo}' R_{\rm bo}} \sim 10^{-11}\left(\frac{R_{\rm bo}}{10^{12}{\rm~  cm}}\right)^{-1} {\rm~  g~cm^{-3}}$.}. A comparison of the predicted temperature to the observations provides a test of whether the observed emission is from the breakout layer of a shock breakout. This test is similar to the closure relation, but it is valid also in the case that the shock is not relativistic. 

The observed delay also constrains the delay between the merger and the time that the shock is driven into the ejecta. In the case that a jet is driving the shock, this is the time, after the merger, at which the jet launching starts, which as discussed in section \ref{sec:remnant}, may mark the delayed collapse of the central compact object to a BH. Under the assumption that the shock velocity does not vary significantly over most of the way to the breakout point, the delay between the merger and the jet launching is 
\begin{equation}
	\delta t_{\rm jet} \approx \frac{\beta_{\rm s,bo}-\beta_{\rm e,bo}}{\beta_{\rm s,bo}(1-\beta_{\rm e,bo})} \delta t_{\rm \scriptscriptstyle GW,\gamma}. 
\end{equation}

\noindent \underline{\textit{The Planar phase:}}\\
Following shock breakout, photons start diffusing from the shock downstream towards the observer. Photons that are able to diffuse, before the expanding shocked ejecta doubles its radius, contribute to the planar phase. 
If the shocked gas is relativistic (i.e., $\gamma_{\rm s,bo}\beta_{\rm s,bo} \gtrsim 1$), then these photons arrive to the observer over a duration of $\sim R_{\rm bo}/(2c\gamma_{\rm f,bo}^2)$ after the first breakout photon is observed. As noted above, this time is similar to the duration of the breakout layer signal (eq. \ref{eq:tbo}) and therefore photons from both phases are observed simultaneously.

The region that contributes to the planar phase, marked with subscript "pl"", has a diffusion time that is comparable to the dynamical time.  This condition is $t_{\rm diff,pl}''=\tau_{\rm pl} \Delta_{\rm pl}''/c \sim R/(c\gamma_f)$, where $"''"$ denotes quantities measured in the frame of the shocked fluid after it ends accelerating, $\Delta''$ is the width of a layer at the end of the planar phase (the layer may be spreading during the planar phase) and $R/(c\gamma_f)$ is the dynamical time as measured in the shocked fluid frame. Note that $\tau_{\rm pl}$ is the optical depth of the layer after the acceleration and cooling of the gas, and therefore it does not have a contribution from pairs. Next, we estimate the ratio between the energy in the planar-phase emission, which is roughly the energy deposited by the shock in the mass contained by $\Delta_{\rm pl}''$, and the energy of the breakout-layer emission. This ratio depends  on the ratio between the  width of the breakout layer, $\Delta_{\rm bo}''$, and the width of the layer over which the Lorentz factor of the shocked gas varies, namely the dynamical length scale $\Delta_{\rm dyn}''$. The width of the dynamical layer at the end of the planar phase is $\Delta_{\rm dyn}'' \sim R/\gamma_f$. Under the reasonable assumption that the density in the ejecta varies on the same scale as the Lorentz factor of the shocked gas, the density in the dynamical layer is roughly uniform, and the optical depth of the planar layer is $\tau_{\rm pl} \sim \Delta_{\rm pl}''/\Delta_{\rm bo}''$ (we used here the approximation $\tau_{\rm bo} \approx 1$). Plugging these relations into the condition that the diffusion time  is comparable to the dynamical time, one obtains $\Delta_{\rm pl}'' \sim \sqrt{\Delta_{\rm dyn}''\Delta_{\rm bo}''}$. Thus, under the assumption of a uniform density dynamical layer, the mass in $\Delta_{\rm pl}''$ is $m_{\rm pl}=m_{\rm bo} \sqrt{\frac{\Delta_{\rm dyn}''}{\Delta_{\rm bo}''}}$ , and the amount of energy radiated during the planar phase is $E_{\rm pl} \sim E_{\rm bo} \sqrt{\frac{\Delta_{\rm dyn}''}{\Delta_{\rm bo}''}}$.

The value of $\Delta_{\rm dyn}''/\Delta_{\rm bo}''$ depends mostly on the density distribution of the unshocked ejecta at $R_{\rm bo}$. 
One possibility is that the breakout takes place before the shock reaches the edge of the ejecta \citep[e.g.,][]{beloborodov2018}. There is, then, ejecta material with velocity larger than $\beta_{\rm e,bo}$, but the optical depth of this material is too low to sustain an RMS. In such a case, the width of the breakout layer is roughly $\Delta''_{\rm bo}  \sim R/\gamma_{\rm f,bo}$, and therefore $\Delta_{\rm dyn}'' \sim \Delta_{\rm bo}''$, and the planar phase contains only emission from the breakout layer
\footnote{There may be an important subtlety if the shock reaches the point in the ejecta were $\tau_{\rm unloaded}=1/\beta_s'$ while there is still-faster unshocked ejecta at larger radii. If the shock is relativistic, then the production of pairs provides opacity and the RMS continues to propagate beyond the point were $\tau_{\rm unloaded}=1$. In the discussion above, we assume that the breakout radiation is emitted from the region where $\tau_{\rm unloaded}=1/\beta_s'$, which we denote as the breakout layer. The validity of this assumption depends on the velocity of the shock as it continues beyond the breakout layer, to larger radii and lower optical depth. If the density profile is steep enough, then the shock accelerates. If the acceleration is fast enough so that the shock does not interact with photons that are released from the breakout layer, then it 
does not have a strong effect on the emission from the breakout layer or on the total energy of the breakout signal, and the discussion above is valid. If, however, the shock does not accelerate fast enough, or in the unlikely case that it is decelerating, then it can have a strong effect on the breakout signal, as discussed for example in \cite{granot2018} for the case of a breakout from a stellar wind. I do not discuss this case here. \label{note:pairs}}. 

Another possibility is that the breakout takes place because the shock reaches the edge of the ejecta, i.e. the maximal velocity of the fast tail is $\beta_{\rm e,bo}$. In that case, the breakout layer may be comparable to the dynamical layer, but it may also be much smaller, $\Delta''_{\rm bo} \ll \Delta_{\rm dyn}''$, and the contribution from the planar phase may dominate over the emission from the breakout layer. The exact value of $\Delta''_{\rm bo}/\Delta_{\rm dyn}''$ may depend on the exact structure of the ejecta near its leading edge, but even without information about that structure, this ratio can be constrained. In general, larger breakout radii and maximal ejecta velocities imply a smaller contribution of the planar phase. The reason is that a larger radius implies higher $m_{\rm bo}$, and a higher maximal ejecta velocity implies lower $m_{\rm dyn}$. For example, consider ejecta with a maximal velocity of $0.9$c and a breakout radius of $\sim 5 \times 10^{11}$ cm, which gives a delay of $2$ s between the merger and shock breakout $\gamma$-rays (similar to the case of GW170817). Only a small fraction of the mass is expected to have a velocity that is close to $0.9$c. If we take, for example, the results of \cite{radice2018} who find typical masses of $10^{-5}-10^{-6} \msun$ moving faster than $0.6$ c, then $m_{\rm dyn} < 10^{-6} \msun$, while at this radius, $m_{\rm bo}  \approx 10^{-8}\msun$, implying $m_{\rm pl} \sim (m_{\rm dyn}/m_{\rm bo})^{1/2} \lesssim 10$. Therefore, for these parameters, the planar phase increases the breakout energy by, at most, an order of magnitude, and it is likely that its contribution is not very significant. At the other extreme, consider ejecta without any fast tail, where the fastest ejecta velocity is $0.5$c and the breakout radius is $\sim 5 \times 10^{10}$ cm (also here, the delay is similar to that in GW170817). In this case, 
$m_{\rm bo}  \approx 10^{-10}\msun$, while the dynamical mass near the breakout can be a non-negligible fraction of the entire ejecta mass, say, $10^{-4} \msun$. In this case, $m_{\rm pl} \sim (m_{\rm dyn}/m_{\rm bo})^{1/2} \sim 10^3$, the planar phase completely dominates the breakout emission, and the total shock breakout signal has 
significantly more energy than the one given in equation \ref{eq:Ebo}, which includes only the contribution from the breakout layer.\\

\noindent \underline{\textit{The spectrum of the breakout signal:}}\\
While there are robust predictions for the typical photon energy of the breakout emission, the theoretical predictions concerning the spectral shape of the breakout emission are quite limited. Unlike the total energy or the duration, the spectrum depends on dynamical evolution of the shock transition layer as it breaks out, and the spectral shape of the breakout emission is currently unknown. Nevertheless, there are several predictions that seem rather robust.

First, the spectrum in the shock transition layer is neither a blackbody nor a Wien  spectrum, and therefore the shock breakout signal is not expected to have either of these forms. Second, RMSs are not expected to accelerate particles, and therefore the spectrum is not expected to have a high-energy power law tail that extends for many orders of magnitude above the peak of $\nu F_\nu$, such as the one often seen in the prompt emission from GRBs. A power-law above the spectral peak is possible over a limited spectral range (one or two orders of magnitude). Light-travel-time effects lead photons from different radii and different angles to arrive together at the observer. Therefore, in some scenarios, photons from shocks with different velocities, and from regions with different Lorentz boosts, are seen simultaneously. This implies that, even if the spectrum in the shock had a blackbody or a Wien form, the observed spectrum would be the sum of spectra of different temperatures. Thus, the observed breakout spectrum is not a blackbody, and it may appear approximately as a power-law below, and possibly also above, the peak of $\nu F_\nu$ (see, for example, the case of a relativistic breakout from a star; \citealt{nakar2012}).

A second prediction is that the spectrum at each location in the shock's downstream most likely has an exponential cut-off corresponding to its rest-frame temperature. The observed spectrum therefore also has an exponential cut-off, that corresponds to the observer-frame temperature of the layer with the highest post-shock velocity. In many, but not all, scenarios, this is  the breakout layer (i.e., where $\tau_{\rm unloaded}=1/\beta_{\rm s,bo}$).

A third prediction is that, if the breakout layer dominates the emission, then the typical photon energy is given by equation \ref{eq:Tbo} for a relativistic shock, and equations \ref{eq:dot_nff}-\ref{eq:Td}, for a Newtonian one. 
If the planar phase dominates the breakout emission, then this may affect the peak of $\nu F_\nu$, since the radiation in the layers that contribute to the planar phase have a longer time to produce photons, and thus their temperature may be significantly lower than that of the breakout layer. This is especially true if the shock does not produce pairs and then, as can be seen from equation \ref{eq:Td}, the generation of new photons, during the long time that it takes  the radiation to diffuse through the planar layers, reduce
the radiation's temperature significantly. If the shock is relativistic, then the effect on the temperature is much smaller. The reason is that after the annihilation of the pairs at $T \sim 50$ keV, production of new photons becomes much less efficient than it was right after shock crossing. Nevertheless, if $E_{\rm pl}$ is larger than $E_{\rm bo}$ by several orders of magnitude, then this will most likely result is a breakout temperature that is much lower than that of the breakout layer (equation \ref{eq:Td}). Thus, breakout emission that is softer and with a higher energy compared to the expectation from a breakout layer (e.g., the closure relation in equation \ref{eq:closure}) indicates a large contribution from the planar phase.\\ 

\noindent \underline{\textbf{The Spherical phase:}}
The spherical phase starts after the breakout layer doubles its radius. During this phase, pairs (if they previously existed) are typically gone, and the optical depth of the expanding gas drops rapidly with the radius as $\tau \propto r^{-2}$, and therefore photons diffuse to the observer from deeper and deeper layers, as time progress. The deeper layers contain more mass and, initially, after the shock crossing and before expansion's beginning, they contain also more energy. During the expansion, part of this energy is lost to adiabatic expansion. Since the more massive layers are shocked at smaller radii and radiate at larger radii, they suffer more-severe adiabatic losses. The spherical emission is calculated by taking into account all of these factors, and it depends also on the density distribution of the ejecta and on the velocity evolution of the shock. Typically, the luminosity during the spherical phase drops with time, but the integrated total emitted energy may be increasing with time. In addition to the drop in the luminosity, the emission also becomes softer with time. The reason for this is a combination of three factors: (i) deeper layers have more adiabatic loses; (ii) if the shock accelerates before it breaks out, then the initial temperature in the deeper layers can be smaller; (iii) the radiation spends more time in the deeper layers, so there is more time to generate photons and get closer to thermal equilibrium.   

A calculation of the spherical phase from a relativistic shock has been done to date only in the context of a breakout from a static stellar envelope \citep{nakar2012}. The exact luminosity and temperature evolution from this solution are not applicable here, but the general picture is similar. The luminosity drops with time, but not very fast (not much faster than $t_{\rm obs}^{-1}$, and possibly slower). As for the spectrum, during the spherical phase, we see at any given time only a single layer, and therefore the observed spectrum during this phase should be much more similar to a Wien spectrum than the spectrum of the breakout emission. The temperature during the spherical phase drops with time, but the rate depends strongly on whether the layer that releases the photons was shocked by a Newtonian or by a relativistic shock. As long  as we see layers that were shocked relativistically, the temperature drops rather slowly (slower than  $t_{\rm obs}^{-1}$), but when we see Newtonian layers, the drop in the temperature is sharp. The reason is that layers that are shocked relativistically are dominated by pairs, and they all have the same rest frame temperature immediately after being shocked. After the gas expands and the pairs annihilate, the production of new photons stops, and the evolution of the temperature with time is determined only by adiabatic losses and by different boost from the fluid frame to the observer. In the Newtonian shocked layers, the initial temperature depends strongly on the velocity. Since we see slower layers at later times, the difference in the temperature immediately after the shock dictates a fast drop in the observed temperature.

Finally, if the breakout layer is only a small fraction of the dynamical layer, $\Delta_{\rm bo}'' \ll \Delta_{\rm dyn}''$, then
at the beginning of the spherical phase it may have a comparable luminosity to that of the breakout. The reason is that the entire dynamical layer has the same initial conditions (Lorentz factor, radius, etc.) and, it turns out, that as long as we see photons released from the dynamical layer, the adiabatic losses are compensated by the growing visible mass, while the luminosity stays roughly constant. 
The emission, however, does become softer with time. After all the photons from the dynamical layer are released, and we start seeing the deeper layers, the luminosity starts to fall.\\

\noindent \underline{\textbf{Summary}}\\
Every BNS merger with a successful relativistic jet (such as GW170817) produces a flare of gamma-rays that accompany the breakout of the shock that the jet-cocoon drives into the ejecta. The same applies also to mergers with choked jets, in which only the cocoon breaks out. A BH-NS merger may generate shock breakout emission as well, if there is a delay between the merger and the launching of the jet, so that the jet needs to propagate through a sub-relativistic disk wind (see \S\ref{sec:BHNSejecta}).
The shock breakout emission can have a range of properties, depending on the various details (e.g., shock velocity, ejecta density and velocity profile, etc.) but there are several common properties to all the shock breakout signals,  which I summarize below (many of these properties are common also to many other types of shock breakout emission, including those from SNe and low-luminosity GRBs, as discussed in \citealt{nakar2012}).\\

\noindent General properties of shock breakout emission:
\begin{itemize}
\item \textbf{Low energy:} The energy released in the shock breakout is always a small fraction of the total energy released in the explosion. The reason is that the breakout emission is generated by energy deposited by the shock into a small fraction of the total mass. 

\item \textbf{Smooth light curve:} The breakout signal is not highly variable. It may have temporal structure, e.g. due to inhomogeneities in the ejecta, but large variability such as seen in the prompt emission of many GRBs is not expected. 

\item \textbf{hard-to-soft evolution:} The spectrum of the breakout emission and the ensuing cooling emission shows a hard-to-soft evolution. The spectrum of the breakout emission, which is composed of photons emitted from the breakout layer and from the planar phase, is harder and does not have a blackbody or Wien spectrum. The emission from the spherical phase, which follows the breakout emission, is softer (and continues to soften with time) and its spectrum is more akin to a Wien or a blackbody spectrum. 

\item \textbf{Delay between the GW signal and the \grays:} The energy of the breakout emission depends sensitively on the breakout radius. Assuming a mildly relativistic breakout velocity, a detectable signal at a distance of $\sim 100$ Mpc requires a breakout radius of $\gtrsim 10^{11}$ cm (equation \ref{eq:Ebo}). This radius corresponds to a delay of order one second or longer between the merger time, as indicated by the GW signal, and the gamma-rays emitted by the shock breakout (equation \ref{eq:R_Dt}).

\item \textbf{Relatively wide angle:} The emission from a cocoon breakout spreads over a much larger angle than the jet. Due to the strong suppression of the jet emission when seen off-axis, it is very plausible that at these angles the cocoon breakout emission dominates over the jet's off-axis emission.

\item \textbf{Closure relation:} The three main observables from a relativistic shock breakout---its energy, duration, and typical photon energy---are over constrained, and therefore predicted to satisfy a closure relation (equation \ref{eq:closure}). If a delay between the GW signal and the gamma-rays is measured, then also the breakout of a sub-relativistic shock is over-constrained. Note that the the closure relation is accurate only to an order of magnitude (see discussion). Note also that, if the emission from the breakout layer is negligible compared to  the planar-phase emission, then the closure relation is not expected to be satisfied.
 
\end{itemize}

\subsubsection{Off-axis jet emission}\label{sec:off_axis_prompt}
Off-axis jet emission can be related to the emission seen by an on-axis observer, via the dependence of the Lorentz boost on the angle between the source and the observer. Consider a top-hat jet with an opening angle $\theta_j$ and a Lorentz factor $\Gamma$, that emits \grays with a total  isotropic equivalent energy $E_{\rm iso}$ and a typical photon energy $E_p$, as measured by an on-axis observer. What does an off-axis observer, positioned at an angle $\theta_{\rm obs}$ with respect to the jet axis, where $1/\Gamma \ll (\theta_{\rm obs}-\theta_j) \ll 1$,  measure? The ratio between the on-axis observables and the off-axis observables depend on the factor $q \equiv \Gamma(\theta_{\rm obs}-\theta_j)$, where for $q \gg 1$, the
ratio of the Doppler factors between the on-axis and off-axis observers is approximately $q^2$. The relation between the observed peak energies is simply
\begin{equation}
	\frac{E_p}{E_p^\prime} \approx q^2 ,
\end{equation}
where we denote off-axis observables with "$^\prime$". The isotropic energy transformation depends on how far the observer is from the edge of the jet, compared to the jet opening angle \citep{kasliwal2017,granot2017,ioka2018,matsumoto2019a}:
\begin{eqnarray}
\frac{E_{\rm iso}}{E_{\rm iso}^\prime} \approx \begin{cases}
q^4&\text{;\,$\theta_{\rm obs}-\theta_j\ll\theta_j$\,},\\
q^6(\Gamma \theta_j)^{-2}&\text{;\,$\theta_{\rm obs}-\theta_j\gg\theta_j>1/\Gamma$\,},\\
q^6&\text{;\,$\theta_j<1/\Gamma$\,}.
\end{cases}
.
   \label{e_gamma_iso}
\end{eqnarray}

The duration of the emission depends on the radial structure of the jet. The prompt emission of GRBs is often highly variable, with the burst lasting much longer than the duration of the individual pulses. The duration of each pulse can be as short as $\delta t \approx R/(2c\Gamma^2)$, where $R$ is the emission radius. Such a pulse will have a much longer duration when observed by an off-axis observer, $\delta t^\prime \approx q^2 \delta t$. The duration of the entire burst, $T$, is usually attributed to the radial width of the outflow, $\Delta$, such that $T \approx \Delta/c$. This duration is not affected significantly by the viewing angle, i.e. $T' \sim T$, as long as $T' > q^2 \delta t$. Thus, the duration of single pulses is much longer for an off-axis observer, but the total duration of the burst is not necessarily longer. The result is that the time-separated pulses seen by an on-axis observer appear overlapping to an off-axis observer, and the entire off-axis light curve is less variable.  

\subsubsection{High inclination emission from a structured-jet}\label{sec:high_inclination_grays}
As we have learned from observations of the afterglow of GW170817, BNS mergers launch relativistic jets which, after they break out of the ejecta, have an angular structure (see section \ref{sec:GW170817relativistic}). The origin of the structure can be the interaction with the ejecta (i.e., jet-cocoon) or it may be induced at the launching site (near the central object). Regardless of its origin, the observations indicate that the jet in GW170817 has a narrow core ($<0.1$ rad) with high isotropic equivalent energy,  and outside of the core $E_{\rm iso}$ drops with the angle. The jet-cocoon model of a structured jet posits that the Lorentz factor also falls as the angle from the axis increases (see section \ref{sec:JetCocoon}), although there is no direct observational constraint on this from GW170817. Under the most reasonable assumption, that the core of the jet produces the bright \gray emission seen in sGRBs, it is reasonable to expect that there is also some high-inclination emission that originates from regions that are outside of the core. The rest-frame emissivity from this region is assumed to be significantly lower than that of the core. Nevertheless it may dominate at large viewing angles, due to the effect of the Lorentz boost. 

Observationally, there are no direct constraints on the high-inclination (i.e., outside of the jet core) \gray emission. A comparison of the prompt \gray and afterglow energies in GRBs, as well as the light curve shapes of many GRB afterglows, provide some indirect evidence. It suggests that, in regular GRBs, we observe the core of the jet, and that efficient \gray production is confined to a narrow region around the core, where the Lorentz factor is high, $\gtrsim 50$, while the isotropic equivalent energy is not much lower than that of the jet core \citep{beniamini2019}. The implication is that the \gray efficiency (i.e., the fraction of the local isotropic equivalent energy that is emitted as $\gamma$-rays) is not constant. Instead, it drops significantly at low Lorentz factors and at low isotropic equivalent energies. 

Theoretically, there are almost no constraints on the high-inclination \gray emission. Similar to the prompt GRB emission, the high-inclination emission also requires a mechanism that dissipates the energy of the outflow, and a radiation process that produces the observed $\gamma$-rays. We do not know what these processes are in the core of the jet (which produces the sGRBs), and if similar, or other, mechanisms take place outside of the jet core. In general, there have been only few attempts to calculate the high-inclination emission based on any type of physical model. Instead, typically,  some {\it ad-hoc} angular dependence of the gamma-ray emissivity is assumed \citep[e.g.,][]{troja2017,ioka2019}. Another common assumption, which seems to be in tension with the results of \cite{beniamini2019}, is that the \gray efficiency is constant, namely that the gamma-ray emissivity is a constant fraction of $E_{\rm iso}$, although in some cases the efficiency is taken as constant in the outflow rest frame \citep[e.g][]{kathirgamaraju2018,kathirgamaraju2019} while in others it is taken as constant in the observer frame \citep[e.g][]{troja2018}. These models often have almost no predictive power, as they can explain almost any observed signal. The unknown dissipation mechanism allows for almost any emission radius, implying no prediction of the duration or the temporal structure of the \gray emission, nor of its delay with respect to the end of the GW signal. The emission mechanism is unknown and therefore there are no predictions with respect to the observed \gray spectrum. Finally, the free parameters of the model often allow for a large range of luminosities at any observing angle.

The only attempts, that I am aware of, to calculate the high inclination \gray emission based on an actual physical model have been made in the context of numerical simulations of photospheric emission in un-magnetized (or weakly magnetized) jets \citep{parsotan2018,parsotan2018a,ito2019,gottlieb2019a}. These works carry out RHD simulations of jet propagation in dense media (stellar envelope or merger ejecta) to find the dissipation at the collimation shock that arises due to the interaction between the jet and the medium. Then they follow the radiation generated at the collimation shock all the way to the photosphere, in order to determine its efficiency.  \cite{parsotan2018,parsotan2018a} and \cite{ito2019} carried out RHD simulations of lGRB jets (propagating through a stellar envelope) and  employ a Monte-Carlo radiative transfer code for  post-processing of the RHD results, which provides light curves and spectra.  They all find a sharp drop of the gamma-ray luminosity (by orders of magnitude) when the observing angle is about twice the jet core opening angle. \cite{gottlieb2019a} provide an analytic estimate of the photospheric emission efficiency, and verify its applicability using 3D RHD simulations of lGRB and sGRB jets. They show that the efficiency of the photospheric emission depends sensitively on the baryonic loading of the ejecta, and thus on the outflow's final Lorentz factor. For jet parameters similar to those seen in GW170817, they find high efficiency for $\Gamma \gtrsim 300$, and a sharp drop of the efficiency, as $\Gamma^{8/3}$, at lower Lorentz factors. As discussed in section \ref{sec:JetCocoon}, the Lorentz factor drops sharply outside of the jet core (in the jet-cocoon interface), and therefore they find, similarly to the other studies, that the gamma-ray efficiency drops by orders of magnitude already when the observing angle is $6^\circ$ (twice the jet core opening angle). When combined with the drop in $E_{\rm iso}$, the resulting gamma-ray luminosity at larger angles is very faint. The conclusion of these studies is that, if photospheric emission, which is powered by dissipation of the jet energy at the collimation shock, is the main source of the \gray energy, then high-inclination emission is very faint and  unlikely to be observed at large angles.

\subsection{{\bf  The afterglow}}\label{sec:afterglow}
The outflow from a merger interacts with the circum-merger medium. This interaction generates a long-lasting, non-thermal, emission,  known as the afterglow. Detailed observations and a comprehensive theoretical modeling of afterglow has been carried out in the context of GRB afterglows. These studies have led to the so-called 'standard' afterglow model, where the emission is generated by synchrotron emission of electrons that are accelerated in the shock that was driven by the outflow into the surrounding medium. The model makes many simplifications.  Naturally, it cannot explain all the various observations of GRB afterglows, and it is clear that there are parts of the picture that are more complex. However, given that the observations span more than five orders of magnitude in time and nine orders of magnitude in frequency, and that the model explains the observations over such a vast dynamical range rather well, it is considered successful. The same model can be used to predict the afterglows of compact binary mergers and, in the case of GW170817, also to model the observations. Since the afterglow model was developed for GRBs, in almost all studies the observer is positioned within the opening angle of the jet. In the few studies in which the observer is outside of the jet opening angle, the jet is typically assumed to be a top-hat jet \citep[e.g.][]{nakar2002,granot2002,granot2005,van-eerten2010} and only rarely is a more complicated jet structure considered \cite[e.g,][]{granot2006,lamb2017}. However, in GW detected mergers, the observer is almost never within the jet opening angle and, as we learned from GW170817, the jet structure is not top-hat. Given the breadth of literature on GRB afterglow theory and the large number of reviews that deal with it, I will not cover the entire model here. To that end, I refer the reader to the many GRB reviews \citep[e.g.,][]{piran1999,piran2004,meszaros2002,meszaros2006,nakar2007,gehrels2009} and the references therein. Here, I only touch upon some aspects that are important for the interpretation of the afterglow of GW170817, and which are expected to be useful also for future events. After a brief general description of the afterglow model, I discuss the various phases of an afterglow produced by a structured jet, as seen by an observer who is away from the jet core. I highlight observational signatures that permit us to distinguish between on-axis and off-axis emission and to learn about the structure of the outflow (e.g., a successful vs. a choked jet).  I advise the reader to take a look at the terminology inset at the beginning of this section before proceeding, especially at the definitions of on-axis and off-axis emission, since they are essential for the following discussion.

As the outflow from the merger interacts with the surrounding medium, it drives a shock into it, known as the forward shock, and the medium drives a shock into the outflow, known as the reverse shock. After the outflow deposits all of its energy into the medium, the reverse shock dies and the forward shock continues to propagate in the medium, decelerating as it accumulates a growing amount of mass. The shocks are collisionless, so they accelerate particles efficiently. It is assumed that the shocks also generate magnetic fields, so the relativistic electrons gyrate along the generated field in the shocks downstream, and radiate synchrotron emission. Since we have only a limited understanding of particle-acceleration and magnetic-generation processes, the simplified model parametrizes these microphysical processes into three parameters. It assumes that a fraction $\epsilon_e$ of the internal energy deposited by the shock goes into acceleration of the entire electron population to a power-law energy distribution with index $p$, and it assumes that a fraction $\epsilon_B$ is deposited in the magnetic field. The values of $\epsilon_e$, $\epsilon_B$ and $p$ are typically taken to be constant in both time and space. Now, for a given outflow and circum-merger density distribution, the hydrodynamics of the shocks can be solved and, with the parametrization of the microphysics, the emission can be calculated. 

The synchrotron spectrum has three critical frequencies, which correspond to the synchrotron frequencies of the electrons that are accelerated to the minimal Lorentz factor, $\nu_m$, the electrons that cool down over a dynamical time scale, $\nu_c$, and the self absorption frequency, $\nu_a$ \citep[e.g.,][]{granot2002a}. In this review, I will consider only slow cooling cases, for which $\nu_a,\nu_m<\nu_c$, and I will deal only with the emission at frequencies that satisfy $\nu_a,\nu_m<\nu<\nu_c$. The radio to X-ray afterglow of GW170817  satisfies this criterion, although in events with much higher circum-merger densities, it is possible to have the X-ray band above the cooling frequency and the radio band below the self-absorption frequency (the latter is less likely). The main difference between the afterglow from a successful jet, as seen by an observer within the jet-core opening angle, as opposed to one outside the opening angle, is that the former sees bright emission that peaks on a very short time scale (typically minutes, or less), while the latter sees a much fainter signal that rises for a long time, possibly years \citep[e.g.,][]{granot2002,nakar2002}. The afterglow of a choked jet is more similar to a successful jet seen away from the jet core, as it is faint and rises for a long time \citep{gottlieb2018b}. Finally, even if there is no jet, or if the jet is choked deep within the ejecta, the fast-tail of the sub-relativistic ejecta can also generate an afterglow, which is, again, faint and with a long rise time \citep{nakar2011}. 

In what follows, I describe four phases of the light curve of the afterglow generated by the forward shock in the frequency range $\nu_a,\nu_m<\nu<\nu_c$ (for a discussion of the contribution of the reverse shock see \citealt{lamb2019}): (i) the rising light curve from an off-axis source; (ii) the rising light curve from an on-axis source; (iii) the properties of the peak of the light curve (flux and time) from a jet that points away from the observer; and (iv) the declining light curve. In each of these phases, I focus on the information that we can extract from the observations on the structure of the outflow.

\subsubsection{Off-axis emission}
Afterglow emission is defined as off-axis when the observed emission is dominated by a region with a Lorentz factor $\Gamma$, seen at an angle $\theta$, that satisfy $\Gamma > 1/\theta$. The most notable property of such emission is that the observed flux, $F_\nu$, must be rising sharply with time, at least as $t^3$ \citep{nakar2018a}.  The reason is that off-axis emission is strongly suppressed by the Lorentz boost to the observer frame. A shock with a constant velocity produces a light curve that rises as $t^3$ when seen off-axis. If the shock decelerates, then due to the weaker Lorentz boost suppression, the observed flux rises even faster. Since the shock is not expected to accelerate, if $F_\nu$ does not rise as $t^3$ or faster, then the region of the shock that dominates the emission is on-axis, namely $\Gamma < 1/\theta$. A top-hat jet that is observed outside of the jet's initial opening angle is initially seen off-axis, and therefore the rising phase of its light curve is faster than $t^3$. 
The light curve of a top-hat jet peaks when the blast wave decelerates enough for the beam of the radiation (with angle $\sim 1/\Gamma$) to includes the observer, i.e. when it becomes on-axis.\\

\subsubsection{The rising phase of on-axis emission}
A light curve that rises more slowly than $t^3$ must be from an on-axis source, i.e, the emission is seen 
from an angle  $\lesssim 1/\Gamma$ relative to the line of sight. Note that the definition of on-axis emission depends on the angle between the observer and the source of the emission, and not the angle between the observer and the jet axis, if there is one. Thus, the afterglow of a successful jet, which is seen by an observer that is outside of the jet core,  can also be dominated by on-axis emission, if the source of the emission is outside of the jet's core as well.

\cite{nakar2018a} have shown that as long as the blast wave is relativistic, the observed flux from an on-axis emission, as a function of time, provides constraints on the evolution of the observed Lorentz factor,
\begin{equation}
	\Gamma (t) \approx 5 ~ \left(\frac{F_\nu}{100\mu {\rm Jy}} \right)^{\frac{1}{6+2p}} 
	\left( \frac{\nu_{\rm obs}}{3 {\rm~GHz}}\right)^{\frac{p-1}{12+4p}} 
	\left(\frac{t}{10 {\rm~d}}\right)^{-\frac{3}{6+2p}} 
	\left( \frac{n}{10^{-3}{\rm~cm^{-3}}}\right)^\frac{-(p+5)}{24+8p} 
	\epsilon_{e,-1}^{-\frac{p-1}{6+2p}} \epsilon_{B,-3}^{-\frac{p+1}{24+8p}}\left(\frac{d}{\rm 100~Mpc}\right)^\frac{1}{3+p} \ ,
	\label{eq:gammaon}
\end{equation}
and on the isotropic equivalent energy of the observed region,
\begin{equation}
E_{\rm iso}(t) 
 \sim   2 \times 10^{49}~{\rm erg}~ \left( \frac{F_\nu}{100~\mu {\rm Jy}}\right)^\frac{4}{3+p} 
 \left( \frac{\nu_{\rm obs}}{3 {\rm~GHz}}\right)^{\frac{2(p-1)}{3+p}}
\left(\frac{t}{10 {\rm~d}}\right)^{\frac{3(p-1)}{3+p}} 
 \left( \frac{n}{10^{-3}{\rm~cm^{-3}}}\right)^{-\frac{2}{3+p}} %
	\epsilon_{e,-1}^{-\frac{4(p-1)}{3+p}} \epsilon_{B,-3}^{-\frac{p+1}{3+p}} \left(\frac{d}{\rm100~Mpc}\right)^\frac{8}{3+p}\ .
\label{eq:Eoniso}
\end{equation} 
Here, $\nu_{\rm obs}$ is the observed frequency, $t$ is the observer time, $n$ is the circum-merger density (assumed to be constant), $d$ is the distance to the merger, and $\epsilon_e$ and $\epsilon_B$ are in units of $10^{-1}$ and $10^{-3}$, respectively. These estimates assume that the entire observed region (within an angle of $ 1/\Gamma$) is radiating roughly uniformly. If this is not the case, and only a small, but non-negligible, fraction of the observed region contributes, then the Lorentz factor can be higher by a factor of order  unity, and the energy may be larger by up to an order of magnitude. Note that, since the angular area of the observed region is roughly $2\pi/\Gamma^2$ (the factor of 2 is to account for the counter jet under the assumption that the outflow is two sided), the total energy which is actually contained in the observed region is $E_{\rm obs} \sim E_{\rm iso}/(2\Gamma^2)$. 

Equation \ref{eq:Eoniso} shows that rising light curves (and also light curves that decline more slowly than $t^{-4(p-1)/3}$) require an increase with time in the isotropic equivalent energy. The source of this energy is must be a structure of the outflow, either angular, or radial, or both. In the case of an angular structure, the increase is due to high energy regions that become on-axis as the blast wave decelerates. Consider, for example, a successful jet with an angular structure  similar to the jet-cocoon structure shown in figure \ref{fig:successful}, where the the isotropic equivalent energy and the Lorentz factor decrease with angle outside of the jet core.  The initial Lorentz factor of the core is high, and therefore, if the observer's line of sight is outside of the core, the afterglow is dominated at first by on-axis emission of material closer to the line of sight, with lower energy and lower Lorentz factor (e.g. the cocoon or the jet-cocoon interface in figure \ref{fig:successful}). As the blast wave decelerates, the observer sees regions that are progressively closer to jet axis, where the isotropic equivalent energy is higher. If the angular structure is such that the isotropic equivalent energy increases fast enough towards the jet axis, then the result will be a rising light curve. The peak, in this case, will be seen when the core itself decelerates enough so that its beam of radiation includes the observer. 
%\begin{equation}
%E_{\rm obs}(t) 
% \approx  10^{48}~{\rm erg}~ \left( \frac{F_\nu}{100\mu Jy}\right)^\frac{3}{3+p} 
% \left( \frac{\nu_{\rm obs}}{3GHz}\right)^{\frac{3(p-1)}{2(3+p)}}
%\left(\frac{t}{10d}\right)^{\frac{3p}{3+p}}
% \left( \frac{n}{10^{-3}{\rm~cm^{-3}}}\right)^{-\frac{3-p}{4(3+p)}} %
%	\epsilon_{e,-1}^{-\frac{3(p-1)}{3+p}} \epsilon_{B,-2}^{-\frac{3(p+1)}{4(3+p)}} \left(\frac{d}{100Mpc}\right)^\frac{6}{3+p}\ .
%\label{eq:Eonobs}
%\end{equation}

A radial structure, where slower material carries more energy than faster material, also leads to an increase in the blast wave energy.  In this case, the slower material catches up with the forward shock as it decelerates, and injects energy into the shock. This may happen, for example, in the case of a choked, uncollimated, jet, in which the cocoon has a strong radial structure.

\subsubsection{The peak of the light curve from a succesful jet}
When a successful jet with an energetic core points away from the observer, the light curve rises until the core decelerates enough, such that its beam of emission includes the observer. The time and flux of the peak, in such an event, is independent of the entire jet structure, and only the properties of the core play any role. Therefore, as long as the jet energy is dominated by the jet's core, the main properties of the peak of the light curve, produced by a jet with an arbitrary structure, are similar to those of a top-hat jet, seen from an observing angle that is larger than the jet opening angle. The time and flux of the peak from such a top-hat jet were first calculated analytically \citep{granot2002,nakar2002}. \cite{gottlieb2019}  calibrated the coefficients of the analytic solution numerically, and verified that, indeed, the same equations are valid for a range of succesful jets, each with a different cocoon structure. They find that the time of the peak is:
\begin{equation}
t_{p} \approx 130 \rm{~d~}   \left( \frac{E}{10^{50}{\rm~erg}}\right)^{1/3} \left( \frac{n}{10^{-3}\cm^{-3}}\right)^{-1/3} 
 \left(\frac{\theta_{\rm{\rm obs}}-\theta_j}{15^\circ}\right)^2  ~,
 \label{eq:peak_time}
\end{equation}
and the peak flux is 
\begin{equation}
\label{eq:peak_mag}
F_{\nu,p} \approx 140  ~\mu\rm{Jy~}
\left(  \frac{E}{10^{50}{\rm erg}} \right) 
\left( \frac{n}{10^{-3}\cm^{-3}}  \right)^{\frac{p+1}{4}} 
 \left(\frac{\nu}{3\rm{GHz}}\right)^{\frac{1-p}{2}}\left(\frac{\theta_{\rm{\rm obs}}}{20^\circ}\right)^{-2p}
 \epsilon_{e,-1}^{p-1} \epsilon_{B,-3}^{\frac{p+1}{4}} \left(\frac{D}{100\rm{Mpc}}\right)^{-2}  \ ,
\end{equation}
%\end{eqnarray}
where $E$ is the jet energy, $\theta_j$ is the opening angle of the jet-core, and $\theta_{\rm obs}$ is the angle between the observer and the jet axis.

\subsubsection{The decline}
During the decline, the emission is always dominated by an on-axis source. The decline rate is a powerful diagnostic to distinguish between different possible outflow structures \citep{nakar2018a,lamb2018}. In all of the discussion above, I have considered only emission in the frequency range $\nu_a,\nu_m<\nu<\nu_c$. Here, I provide further, in square brackets, the decline rates above the cooling frequency, i.e., $\nu_a,\nu_m,\nu_c<\nu$, which may be relevant for the X-ray afterglow. If the decline rate is slower than $t^{-3(p-1)/4}$ [$t^{-(3p-2)/4}$], then energy injection  continues during the decline as well. A decline at a rate of $F_\nu \propto t^{-4(p-1)/3}~ [t^{-(3p-2)/4}]$ is consistent with  a constant isotropic equivalent energy in the observed region, while the blast wave is still relativistic. This, in turn, suggests that the emission is coming from a region that is quasi-spherical (i.e., roughly uniform over an opening angle that is larger than $1/\Gamma$), which indicates a radial structure. Such a structure is expected if the emission is dominated by a cocoon, either because the jet is choked, or because the jet energy is low compared to the cocoon.  A faster decline is expected if the peak is dominated by the core of a successful jet. In such a case, the expected decline rate is $F_\nu \propto t^{-p}$, both above and below the cooling frequency (similar to a GRB afterglow after the jet-break; \citealt{sari99}). Finally, once the blast wave becomes Newtonian, the predicted decay rate is $F_\nu \propto t^{-(15p-21)/10} ~[t^{-(3p-4)/2}]$ \citep{gao2013}.

\subsubsection{Radio afterglow from the merger sub-relativistic ejecta}
At late times, radio emission can also be generated by the ejecta itself \citep{nakar2011}, which is mostly sub-relativistic, but possibly with a mildly relativistic fast tail. This emission is on-axis at all times, and its evolution depends on the ejecta velocity profile, as calculated in \cite{piran2013}.	
Assuming that an energy $E$ is carried by material with velocity $\beta c$, the time scale of the peak of the afterglow that it generates is \citep{nakar2011}
\begin{equation}
	t \approx 400 {\rm~d~}  \left( \frac{E}{10^{50}{\rm~erg}}\right)^{1/3} \left( \frac{n}{0.1{\rm~cm^{-3}}}\right)^{-1/3} 
	 \left( \frac{\beta}{0.5}\right)^{-5/3} ~,%
\end{equation}
at an observed flux,
\begin{equation}
	F_\nu \approx 20 {\rm~\mu Jy}  \left( \frac{E}{10^{50}{\rm~erg}}\right) \left( \frac{n}{0.1{\rm~cm}}\right)^{\frac{p+1}{4}}
	 \left( \frac{\beta}{0.5}\right)^{\frac{5p-7}{2}}
	 \left(\frac{\nu_{\rm obs}}{3{\rm~GHz}}\right)^{-\frac{p-1}{2}}
	  \epsilon_{e,-1}^{p-1} \epsilon_{B,-3}^{\frac{p+1}{4}} 
	 \left(\frac{d}{100{\rm~Mpc}}\right)^{-2} ~.
\end{equation}
This emission is roughly isotropic and can be seen by all observers. It requires a density $\gtrsim 0.01 {\rm~cm^{-3}}$ in order be detectable at the typical distances of GW-detected mergers.

\section{Information carried by the gravitational waves}\label{sec:GW}
The combination of GW and EM signals makes BNS and BH-NS mergers unique. The focus of this review is on the EM emission, and here I give only a brief description on what can be learned from the GW signal alone about the properties of the binary, and from the combination of GW and EM emission on some fundamental questions in physics and cosmology. I refer the reader to  \cite{sathyaprakash2009} and \cite{baiotti2017} for reviews that discuss the GW emission from compact binary mergers in more detail.

\subsection{{\bf  Binary properties}}\label{sec:GWbinary}
The GW signal from the inspiral of a compact binary in a circular orbit, especially a BNS, can be seen for many orbits before the merger. Before relativistic effects becomes important, the frequency evolution depends to the leading order only on the redshifted chirp mass, ${\cal M}(1+z)$, where  ${\cal M}=(m_1m_2)^{3/5}/(m_1+m_2)^{1/5}$ is the chirp mass, $z$ is the redshift and $m_1$ and $m_2$ are the binary's member masses. As a consequence, this is the most robust binary quantity that can be measured. The effect of the mass ratio and the spins becomes increasingly important as the orbit shrinks and relativistic effects become significant. Their effects are degenerate to some extent, and therefore it is harder to constrain the mass ratio and the spins separately. Finally, the internal structure of the binary members becomes important as the separation between them becomes comparable to the tidal radius, and therefore the last few orbits before the merger provide constraints on the tidal deformability of the binary members.

GWs provide also a unique opportunity to measure the inclination angle between the binary orbital plane and our line of sight, $i$ (defined as the angle between the direction 
of the binary's orbital angular momentum and the line of sight). This definition distinguishes between binaries with an angular momentum that points towards us ($i<90^\circ$) and away from us ($i>90^\circ$). Note that, for many purposes, the important quantity is the viewing angle, defined as $\theta_{\rm obs}=\min[i,180-i]$.
However, measuring the inclination angle using the GW signal alone is difficult. It can be done, in principle, since the information on the inclination is encoded in the GW polarization. A fully face-on [face-off] binary produces a fully circular anti-clockwise [clockwise] polarization, i.e. when $\theta_{\rm obs}=0$ the signals in the two measurable polarizations have the same amplitude and a phase difference of $90^\circ$.
For larger viewing angles, there is a difference in the amplitudes of the two polarizations. Thus, a measurement of the amplitude ratio of the two polarizations provides a measure of the inclination.  However, the amplitude ratio is sensitive enough to be measured only when $\theta_{\rm obs} \gtrsim 50^\circ$. This can be seen from the leading order of the amplitudes of the two polarizations \citep[e.g.,][]{sathyaprakash2009}:
\begin{align}\label{eq:GWamplitude}
h_+ &\propto 2(1+\cos^2 i) \frac{\left[{\cal M}(1+z)\right]^{5/3}(\pi f)^{2/3}}{D_L}  \nonumber \\
& \\
h_x &\propto 4\cos i  \frac{\left[{\cal M}(1+z)\right]^{5/3}(\pi f)^{2/3}}{D_L} \nonumber 
\end{align}
where $D_L$ is the luminosity distance and $f$ is the GW frequency. The two amplitudes are practically identical at small inclination angle. Comparison of the two amplitudes shows that differences of $1\%$, $10\%$ and $25\%$, correspond to  $\theta_{\rm obs}=30^\circ,\, 50^\circ$ and $60^\circ$, respectively. Given that the error on the amplitude is roughly the inverse of the signal-to-noise ratio (SNR), 
a GW signal can measure the inclination relatively accurately only for events with $\theta_{\rm obs} \gtrsim 50^\circ$, assuming that both 
polarization amplitudes are measured with SNR larger than 10. For binaries with a lower inclination, the GW signal will usually provide only an upper limit on the viewing angle, of $\lesssim 50^\circ$. Note, however, that for many future mergers, only one of the polarizations will be measured to an accuracy of $10\%$ or better. The reason is that both of the LIGO detectors, which are currently the most sensitive detectors, are almost aligned, so their sensitivity to polarization differences is low, and thus a high SNR signal must be seen also by a third detector in order to measure the second polarization. \cite{chen2019} carried out an analysis that estimates the sensitivity to  inclination angle of the LIGO-Virgo network at design sensitivity, finding that, typically, it can constrain the viewing angle only if it is larger than about $70^\circ$. 

As evident from equation \ref{eq:GWamplitude} the measurement of the inclination can be significantly improved if there is additional information on the distance to the binary, such as the one obtained by the EM counterpart with the identification of the host galaxy.  Since ${\cal M}(1+z)$ is measured accurately from the chirp, once the distance is known the inclination can be determined. Note also that, when $D_L$ is known from the EM emission with some accuracy, the constraints on larger inclination angles are tighter, since at low inclination, for a given measured amplitude, $\cos(i) \propto D_L$, and thus any error in $D_L$ translates to a similar error in $\cos(i)$. 
 \cite{chen2019} estimate that, for BNS events detected by the LIGO-Virgo network at design sensitivity, redshift information will constraint the viewing angle to $\leq 7^\circ$ ($1\sigma$).

\subsection{{\bf The propagation speed of gravitational waves}}\label{sec:GWspeed}
Combined GW+EM observations have far-reaching implications for fundamental physics, particularly to the theory of gravity, through a measurement of the speed at which GWs propagate from the source to the Earth. Einstein's General relativity (GR) predicts that the velocity at which gravitational waves propagate in vacuum is identical to that of light. It also predicts that the effect of a gravitational potential well on  EM and GW propagation times is the same (i.e., Shapiro delay; \citealt{shapiro1964}), as long as the GWs effect on the metric is negligible. In contrast, many modified gravity theories predict different speeds for GWs in flat and/or curved spacetime.
Thus, the main constraint on the nature of gravity is derived from the difference in the arrival times of the GWs and the first observed photons. For example, in GW170817, the GWs and the \grays crossed a distance of about $130$ million light years, but arrived only 1.7 s apart. As discussed below, this single observation places the tightest limits ever on the speed of gravitational waves. The constraints that these limits pose on alternative gravity theories are beyond the scope of this review, but are discussed in numerous papers, such as \cite{ezquiaga2017,creminelli2017,langlois2018,kase2018}, and more. 

The most stringent constraint posed by the difference in the arrival times of GWs and photons is on the difference between the average velocities of the GW and EM signals along the way. Define $t_{\rm travel}=D/c$  as the travel time of a signal that traverses the distance to the merger\footnote{We assume here that the GWs and the gamma-rays were emitted at the same location, whereas in reality it is most likely that they were emitted at a separation of at least a few light seconds (see \S\ref{sec:relativistic}). This assumption has no significant impact on the conclusions.}, $D$,  moving at the speed of light in vacuum, $c$.  Next, define $\Delta t_{e}$ and $\Delta t_{\rm obs}$ as the differences in the emission and observation times of the GW and EM signals, respectively.  The difference in the photon and GW average velocities, $v_{\gamma}$ and $v_{\rm GW}$ respectively, can then be written as $(v_{\gamma}-v_{\rm GW})/c \approx (\Delta t_{\rm obs}-\Delta t_{e})/t_{\rm travel}$. We can measured $\Delta t_{\rm obs}$  accurately, and if the redshift to the source is measured, then its distance is known to better than  $\sim 10\%$, and therefore the main uncertainty lies in $\Delta t_{e}$. Theoretical models do not provide much useful information on $\Delta t_{e}$, since there are scenarios in which there is a significant delay between the merger and the EM emission (possibly even days or years), and there are even scenarios where the EM emission comes slightly before the merger \citep[e.g.,][]{tsang2012}. However, it is highly improbable that  $|\Delta t_{e}| \gg |\Delta t_{\rm obs}|$ and the difference between the velocities conspired such that the two signals arrived at a separation $\Delta t_{\rm obs}$. Thus, while the limit of the velocity difference depends on the assumptions about $\Delta t_{e}$, it is extremely constraining for any reasonable value taken. For example, in the case of GW170817, assuming $|\Delta t_{e}|<10$ s, one obtains $|v_{\gamma}-v_{\rm GW}|/c \lesssim 3 \times 10^{-15}$  \citep[e.g.,][]{abbott2017grays} and even under the extreme assumption of $|\Delta t_{e}|<1$ d,  we get that  $|v_{\gamma}-v_{\rm GW}|/c \lesssim 2 \times 10^{-11}$. 

An almost simultaneous arrival of the GWs and the gamma-ray photons also provides an observational test of the weak equivalence
principle between gravitational waves and photons---that photons and GWs propagate along the same path in the presence of a gravitational potential (both can be treated as test particles that do not affect the spacetime curvature along their path). A consequence of this is that both the photons and the GWs experience the same Shapiro time delay on their way to Earth. 
The accuracy with which the observations constrain the weak equivalence principle depends on the estimate of the Shapiro time delay relative to the difference in the travel time. Shapiro time delay can be written in terms of the parameterized post-Newtonian (PPN) parameter $\gamma$ \citep[e.g.,][]{krauss1988}: $\Delta t_s = \frac{1+\gamma}{c^3}\int_{t_e}^{t_{\rm obs}} U[r(t)] dt$, where the signal is emitted at time $t_e$, observed at time $t_{\rm obs}$, and propagates through the gravitational potential $U(r)$. The weak equivalence principle dictates that the $\gamma$ parameters for photons and GWs are the same. In GR, for example, $\gamma_{\rm EM}=\gamma_{\rm GW}=1$, where current observational constraints for EM waves are of the order $|\gamma_{\rm EM}-1| \lesssim 10^{-4}$ \citep{bertotti2003,lambert2011}. A constraint on violation  of the weak equivalence principle can then be expressed as $\Delta \gamma \equiv |\gamma_{\rm EM}-\gamma_{\rm GW}|$. With this parametrization the observational constraint is $\Delta \gamma < 2 (\Delta t_{\rm obs}-\Delta t_{e})/\Delta t_{s}$  where $\Delta t_s$ here refers to the photon Shapiro time delay (and we use $|\gamma_{\rm EM} -1| \ll 1$). For example, using the observations of GW170817, various authors have estimated the time delay due to the propagation through the potentials of various objects along the way to the source, to obtain limits on $\Delta \gamma$ in the range of $\Delta \gamma \lesssim 10^{-7}-10^{-10}$, depending on the potential considered (the Milky Way, the Virgo cluster, or the large-scale structure;  \citealt{abbott2017grays,wei2017,wang2017,boran2018,shoemaker2018}).

\subsection{{\bf The Hubble constant}}\label{sec:H0}
Currently, the Hubble parameter, $H_0$, is estimated by a variety of methods that are broadly divided into two categories, depending on the basic assumptions that underlie each method. The first category includes methods that are based on a locally calibrated distance ladder. Namely, a primary local distance estimator (e.g., Cepheids, RR Lyrae variables, the tip of the red-giant branch) is tied to a secondary distance estimator that can be observed at larger distances (e.g, Type Ia  supernovae, the Tully-Fisher relation, surface brightness fluctuations, etc.) \citep[e.g][]{freedman2001,freedman2012,riess2016,beaton2016,riess2019,freedman2019}. Currently, the tightest constraint on $H_0$ for this type of 
measurement is $H_0 =  74.03 \pm  1.42 {\rm ~km~s^{-1}~Mpc^{-1}}$ (1$\sigma$), based on a Cepheid-SN Ia distance ladder (the SH0ES team; \citealt{riess2019}). Another recent constraint using a local distance ladder, which is based on the tip of the red-giant branch and Type Ia SNe, is $H_0 =  69.8 \pm  1.9 {\rm ~km~s^{-1}~Mpc^{-1}}$ (1$\sigma$) (the Carnegie-Chicago Hubble Program; \citealt{freedman2019}).
The second category of $H_0$ measurments is based on observations at high redshifts (e.g., the cosmic microwave background, gravitational lenses, baryon acoustic oscillations), and requires assuming a cosmological model in order to constrain the local expansion rate of the Universe \citep[e.g.,][]{hinshaw2013,aubourg2015,planck-collaboration2016,bonvin2017,vega-ferrero2018}.  Most of the constraints of the second type are based on the CMB, and assume a standard flat $\rm{\Lambda CDM}$ 
cosmological model. The most recent constraint obtained based on the Planck data\footnote{ \cite{planck-collaboration2016} reports several slightly different constraints on $H_0$, that are based on different components of the data that are  included in the analysis. The limits reported here are based on Planck temperature data combined with Planck lensing data.} \citep{planck-collaboration2016} is $H_0 = 67.8 \pm 0.9 {\rm ~km~s^{-1}~Mpc^{-1}}$ (1$\sigma$), which is in tension with the SH0ES result at a level of about $4 \sigma$.  Currently, it is unknown what is the source of the possible inconsistency between the Planck team and the SH0ES  results, and especially whether there are some inaccuracies in one or both of the measurements, or whether the cosmological model needs revision. 

BNS and BH-NS mergers offer a unique opportunity to obtain a measurement of $H_0$ that is based on assumptions  orthogonal to all other current measurement methods. Specifically, it provides a {\it local} measure of $H_0$ (i.e., independent of a cosmological model), but which does not rely on a cosmic distance ladder. The basic idea is to use the GW signal as a standard siren (the analog to the electromagnetic standard candle) in order to determine the luminosity distance to the source, $D_L$, and the EM counterpart to measure the redshift \cite[e.g.,][]{schutz1986,holz2005,nissanke2010}. The basic measurement of $D_L$, and some of its limitations, can be understood from equation \ref{eq:GWamplitude}. The GW strain amplitude depends on ${\cal M}(1+z)$, $D_L$, and the inclination angle, $i$. The redshifted chirp mass, ${\cal M}(1+z)$, can be measured accurately from the chirping signal. Thus, the strain amplitude measures a function of $D_L$ and $i$, where for both polarizations when  $i \lesssim 50^o$, we can approximate $h \propto \cos(i)/D_L$. This degeneracy between $D_L$ and $i$ is often the main source of error in the measurement of $H_0$, especially if there is no sensitive polarization information (such as in GW170817). Additional information on $i$ from modeling of the EM counterpart can significantly improve the measurement of $D_L$, and thus of $H_0$, although it introduces systematic errors that are hard to quantify (see below). Additional sources of errors are the GW amplitude measurement and the peculiar-velocity of the host. Each of these error sources behaves differently with the distance. The statistical uncertainty in the GW amplitude is roughly the inverse of the SNR\footnote{There is also a small systematic error in the detector calibration, which may become important when the sample of events will be large enough. The current error is at a  level of about  $1\%$, but it is likely to improve in the future \citep{karki2016,chen2018}.}. Thus, the error in the amplitude increases with the distance and decreases with detector sensitivity. The peculiar velocity is independent of the distance. Its fraction relative to the Hubble flow velocity therefore decreases with increasing distance, and it is independent of the detector sensitivity. Thus, for each detector there is an optimal distance where the errors are comparable. For the LIGO-Virgo O2 run, it was $\sim 30$ Mpc, while for the detector's design sensitivity, it is $\sim 50$ Mpc \citep{chen2018} (this is assuming no EM information on the inclination angle is included). 

In the case of GW170817, the Hubble constant was measured with an error of $\sim 15\%$ based on the GW signal and the source redshift measurements  \citep{abbott2017H0}, were the dominant source of error in this case was the unknown inclination (see section \ref{sec:H0GW170817}). These observations are useful as a proof of concept, and in order to estimate how many future BNS mergers are needed in order to measure $H_0$  at a level of accuracy that may resolve the discrepancy between the Cepheids+SNe and the CMB measurements. \cite{chen2018} estimate that the error on $H_0$ from an individual future event, where the luminosity distance estimate is based on the GW signal only, will be $\sim 15\%$ (similar to the one obtained for GW170817), and that for $N$ events
the error will scale roughly as $1/\sqrt{N}$. Therefore, they estimate that about 50 GW-EM events are needed to reduce the error on $H_0$ to the level of 2\%.  Similar results are obtained by \cite{feeney2018}. This number of mergers may well be observed within the next 5-10 years, making GW standard sirens a promising tool for obtaining an independent, high precision, measurement of $H_0$ in the near future.

The EM signal can potentially provide tight constraints on the inclination angle, $i$. Thus, including the information from modeling of the EM emission can significantly improve the constraints on $H_0$ from individual events \citep{hotokezaka2018b,dobie2019}.
In GW170817, \cite{hotokezaka2018b} used the EM modeling to find $H_0$ to an accuracy of 7\%. Assuming that the error drops as $1/\sqrt{N}$, about 15 GW170817-like events may be sufficient to bring the error on $H_0$ to the level of 2\%. 
However, tighter constraints on the inclination angle require VLBI measurements, which were indeed available for GW170817 (see \S\ref{sec:GW170817_Afterglow}).  GW170817, it is unclear in what fraction of future BNS mergers EM observations will allow us to obtain similar constraints. Moreover, a major advantage of using GWs alone to determine the luminosity distance is that it is 'clean' and systematic errors are relatively easy to quantify. Using EM modeling to constrain the inclination introduces a dependence on the modeling that is hard to quantify. A better understanding of the systematic errors introduced by the EM modeling is required before it can be useful for high-precision measurements of $H_0$.

%\section{GW170817}
	%\input{Observations}
\section{GW170817}\label{sec:GW170817}
\subsection{{\bf  Observations}}\label{sec:GW170817_observations}
\subsubsection{Gravitational Waves}\label{sec:obsGW}
GW170817 was the fifth confirmed GW event detected by LIGO and the second joint detection with Virgo. But, unlike all earlier events, the mass of the binary, as deduced from the GW signal, implied that the binary members are most likely neutron stars rather than black holes. The main details of the GW signal detection are given in \cite{abbott2017GW} and references therein. Here I give a brief summary. 

The merger was detected on August 17, 2017, towards the end of the advanced LIGO O2 run (from November 30, 2016 to August 25, 2017, consisting of 117 days
of simultaneous LIGO-detector observing time) and shortly after Virgo joined the O2 run on August 1, 2017. At the time of detection, both LIGO sites and the Virgo site were in observing mode. The detectors' horizons (i.e., the maximal distance for a BNS merger detection at optimal sky location and binary orbital inclination) at the different sites at that time were 218 Mpc, 107 Mpc and 58 Mpc (LIGO-Livingston, LIGO-Hanford, and Virgo, respectively). The range distance averaged over the entire O2 run, defined as the radius of a sphere that contains the actual volume surveyed (taking into account all sky locations and binary orbital inclinations) for a 1.4$\msun$-1.4$\msun$ binary merger was about 80 Mpc \citep{chen2017}. 

\begin{figure}
	\includegraphics[width=0.45\textwidth]{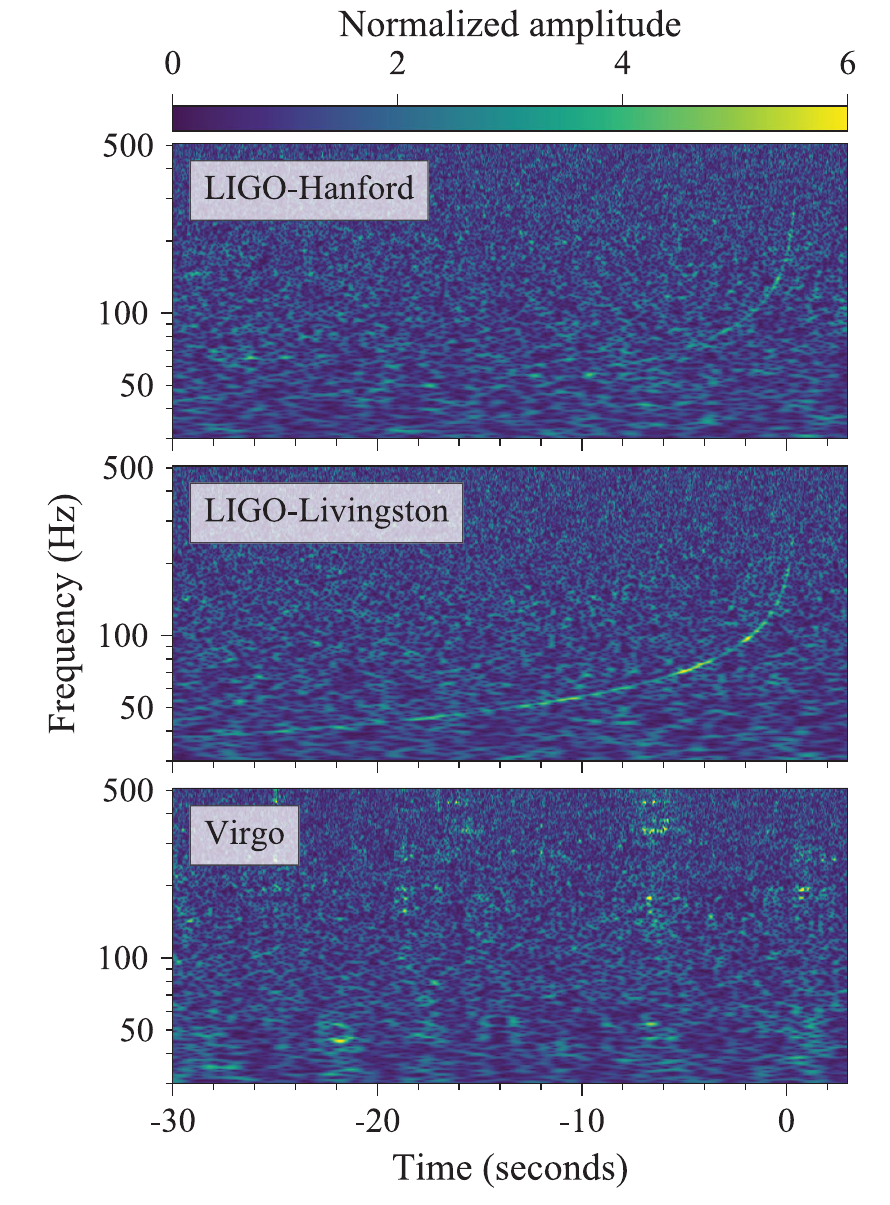}
	\includegraphics[width=0.45\textwidth]{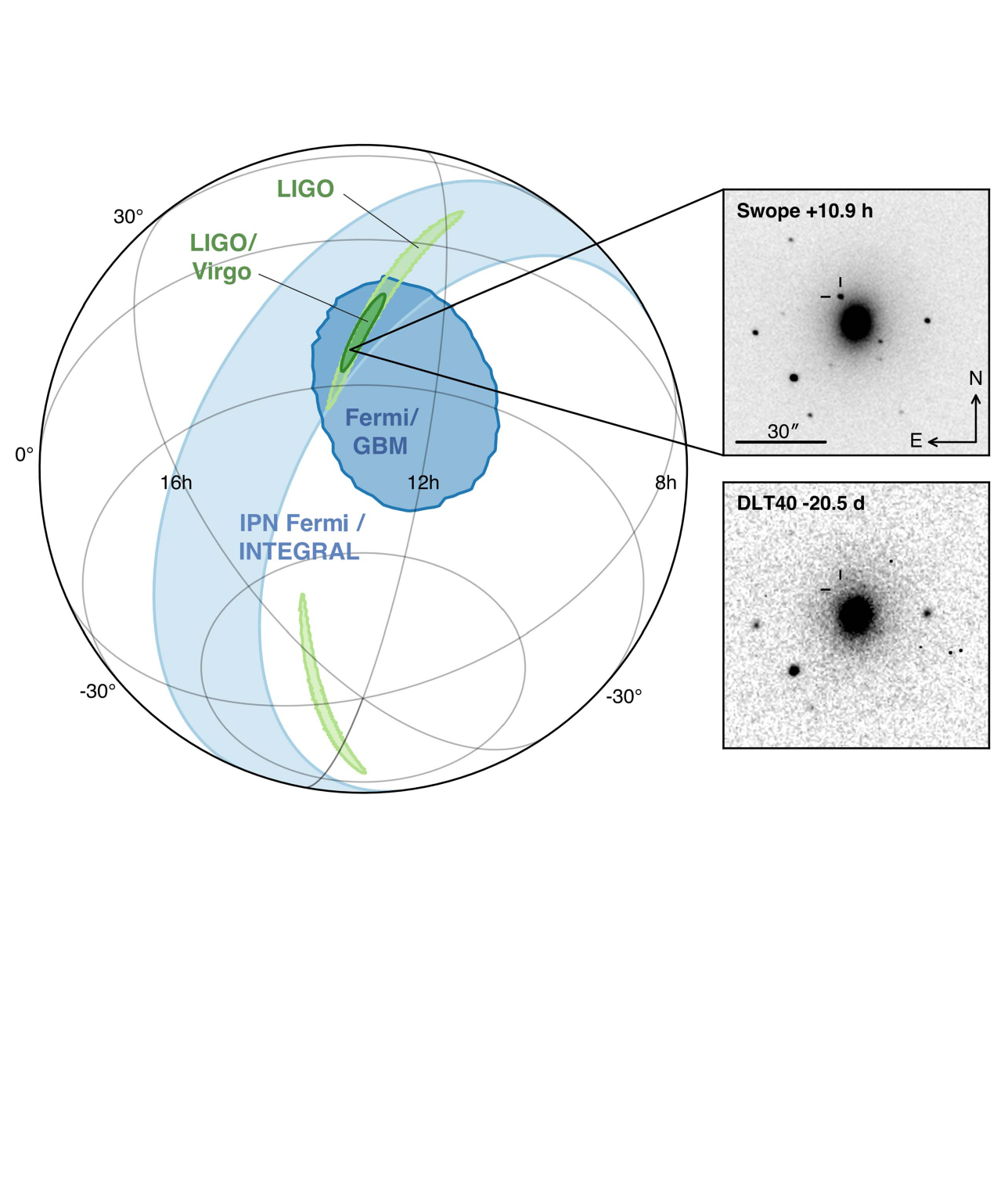}
	\caption{{\it Left:} Time-frequency representations of LIGO-Virgo data containing the gravitational-wave signal from GW170817. From 
 \cite{abbott2017GW}. {\it Right:} The localization (90\% error regions) of the gravitational-waves, \grays and optical signals.  From  \cite{abbott2017Multi}.}
	\label{fig:ligo_signal}
\end{figure}

The gravitational wave signal from GW170817 was loud. It was observed at a combined signal-to-noise ratio of 32.4, and was clearly identified for $\sim 75$ s by the LIGO detectors as its chirping signal crossed their sensitivity band in frequency, until  the inspiral signal ended at 12:41:04.4 UTC (see figure \ref{fig:ligo_signal}). Although the signal was very bright in both LIGO detectors, it was  initially identified as a single-detector event with the LIGO-Hanford detector by the low-latency binary-coalescence search. The reason was that the LIGO-Livingston detector suffered from a brief instrumental noise transient (a "glitch") just 1.1 s before the coalescence, and the glitch caused a temporary detector saturation. A first sky map of the GW signal, localizing it to a region of about 30 deg$^2$, was distributed about 5 hr after the merger, based on an offline analysis of the signal in all three LIGO/Virgo detectors (see figure \ref{fig:ligo_signal}; the localization was improved following a detailed analysis that was carried out several months after the merger to 16 deg$^2$, \citealt{abbott2019Binary}). The luminosity distance was constrained by the initial analysis to $40^{+8}_{-14}$ Mpc (90\% confidence). Together with the sky localization, the merger was constrained to be within a volume of about 380 ${\rm Mpc^3}$.
An interesting anecdote is that the signal in the Virgo detector was barely detected, since the source location on the sky was in a "blind spot" in the detector's antenna pattern. As a result, the contribution of the Virgo data to the analysis of the binary properties was limited, but it was essential for improving the source localization. 

%\begin{figure}
	%\includegraphics[width=1\textwidth]{GW170817_localization.pdf}
%	\caption{LIGO signal and localization map figs 1 
%	\label{fig:localization}
%\end{figure}

\subsubsection{Gamma-Rays}
About 2 s after the chirping GW signal ended, the gamma-ray detectors \textit{Fermi-GBM} and \textit{INTEGRAL} were triggered (independently of the GW signal and of each other) by a short flash of gamma-rays, named GRB 170817A \citep{abbott2017grays,goldstein2017,savchenko2017}. The Fermi-GBM localization was consistent with the refined GW localization and the estimated probability of the near-simultaneous detection and the spatial agreement of GRB 170817A and GW170817 occurring by chance is $5.0\times {10}^{-8}$, implying a secure association \citep{abbott2017grays}. 

The soft gamma-ray emission was seen in more detail by Fermi-GBM \citep{goldstein2017}, which is more sensitive than INTEGRAL at lower frequencies. The GBM analysis revealed that the emission is composed of a smooth pulse with a duration of $\sim 0.5$ s followed by a weaker and softer tail that lasted another $\sim 1$ s. The  beginning of the gamma-ray emission, defined as the first time the signal reaches 10\% of the peak rate,  was $1.74 \pm 0.05$ s after the time of the BNS merger. Its total fluence in the 1keV-10MeV range was $(2.4\pm 0.5)\times {10}^{-7}\ \mathrm{erg}\ {\mathrm{cm}}^{-2}$ divided into $(1.8\pm 0.4)\times {10}^{-7}\ \mathrm{erg}\ {\mathrm{cm}}^{-2}$  during the main pulse and $(0.61\pm 0.12)\times {10}^{-7}\ \mathrm{erg}\ {\mathrm{cm}}^{-2}$ during the tail.  At a distance of 40 Mpc, these measurements correspond to a total isotropic equivalent gamma-ray energy  $E_{\gamma,iso}=(4.6\pm 1.1)\times {10}^{46}\ \mathrm{erg}$  ($3.5 \times {10}^{46}\ {\rm erg}$ and $1.1 \times {10}^{46}\ {\rm erg}$ in the two components). This isotropic equivalent gamma-ray energy is about three orders of magnitude fainter than the faintest sGRB seen to date.

The average spectrum of the first pulse was consistent with a Comptonized function, i.e., $dN_\gamma/dE \propto E^\alpha \exp[-E/E_0]$, where $\alpha=-0.62 \pm 0.4$ and the peak of $\nu F_\nu$ is at $E_p=(2+\alpha)E_0=185 \pm 62$ keV. The time-averaged flux of the first pulse was $(3.1\pm 0.7)\times {10}^{-7} {\rm~erg\ s^{-1} cm^{-2}}$. The tail emission was spectrally soft, consistent with a blackbody spectrum with temperature of ${k}_{{\rm{B}}}T=(10.3\pm 1.5)$ keV, and a time-averaged flux of $(0.53\pm 0.10)\times {10}^{-7}{\rm~erg\ s^{-1} cm^{-2}}$. 

\cite{veres2018} carried out a time-resolved spectral analysis of the first pulse. The pulse rise time is about 0.1 s, after which the luminosity peaks and the flux starts to decay. The spectral hardness seems to be correlated with the flux, becoming hardest with the peak of the flux and then softening. They find that the peak energy and the luminosity (using a time bin of $0.064\,\rm{s}$ and assuming a distance of 40 Mpc) at maximum are $E_{\rm{p}}\simeq520_{-290}^{+310}\,\rm{keV}$ and $L_{\rm{\gamma,iso}}\simeq2.0_{-0.6}^{+0.6}\times10^{47}{\,\rm{erg\,s^{-1}}}$. 

\subsubsection{UV/Optical/IR}\label{sec:GW170817_obsMN}
Theory had predicted that BNS mergers would have optical/IR counterparts that peak on time scales of a day to a week, possibly detectable to a distance of $\sim 100$ Mpc, and named macronovae/kilonovae (see \S\ref{sec:macronova}). The most probable energy source of this emission is the radioactive decay of $r$-process elements that were synthesized during the merger \citep{li1998}. Following these predictions, the astronomical community prepared, during the last several years, both theoretically and observationally, to carry out a rapid follow-up campaign in a search for such a counterpart, once a BNS merger were detected via its GW signal. The main challenge is the relatively large error regions that GW detectors report, of order tens to hundreds (and possibly even thousands) of square degrees. The two main search techniques that have been employed are wide-field tiling of the entire error region, and a targeted search of galaxies within the volume consistent with the GW signal \citep[e.g.,][]{nissanke2013,gehrels2016}. The advantage of the blind tiling is, of course, the completeness, but the disadvantage is the time it takes and the depth it reaches, as well as a large number of false-positive transients. Teams that carried out a targeted search used catalogs of nearby galaxies \citep[e.g.][]{white2011,dalya2016,cook2019} and prioritized the search sequence based on various criteria, such as stellar mass, star formation rate, etc. Before GW170817, the community had several "training" opportunities in GW alert followups, including low signal-to-noise ones, of several BH-BH mergers, and even a  "blind injection" test alert \citep[e.g.,][]{evans2012}. Thus, at the time that the alert from GW170817 was announced, the community was ready and waiting to launch the largest-ever follow-up campaign dedicated to a  single astrophysical event. The campaign was hardly coordinated, so almost every interested astronomer with an available facility was trying to search for, or observe, the electromagnetic counterpart. This led, on the one hand, to an inefficient use of resources, but on the other hand it resulted in what is one of the most detailed data sets we have on a single extragalactic transient to date. 

The first alert of near simultaneous  GW and gamma-ray detections was announced about an hour after the merger. However, at that time, only the $\sim 1000 \degsq$ error circle obtained by the GBM was available \citep{goldstein2017}. An area that proved to be too large for rapid identification of the optical counterpart. Only 5 hours after the merger, the LIGO/Virgo team circulated an improved localization of approximately $30 \degsq$ (see \S \ref{sec:obsGW}). This limited the search to a volume of $\approx 380 {\rm ~Mpc^3}$, which contains only $\sim 100$ galaxies of significant size. Unfortunately, by that time there was no available observatory that could search for the transient, which was located in the southern hemisphere and visible for only 1-2 hours at the beginning of the night, due to its proximity to the Sun. The search was therefore delayed by almost another 6 hours, until the Sun set in Chile. An optical transient, previously unseen, was imaged not long after that by the Swope Supernova Survey, 10.9hr after the merger, and announced about an hour later \citep{coulter2017}. The transient was coincident with NGC 4993, an early-type galaxy at a distance of about 40 Mpc.  NGC 4993  was at the top of the galaxy search priority lists of several teams, and therefore images of the transient were taken by several teams as a part of their galaxy-targeted searches before the Swope team announced its detection. As a result, we have a detailed multi-color light curve of the GW170817 optical counterpart starting about 11 hr after the merger. A detailed account of the entire sequence of observations during the first 12 hr can be found in \cite{abbott2017Multi}.

\begin{figure}
	\center
	\includegraphics[width=0.7\textwidth]{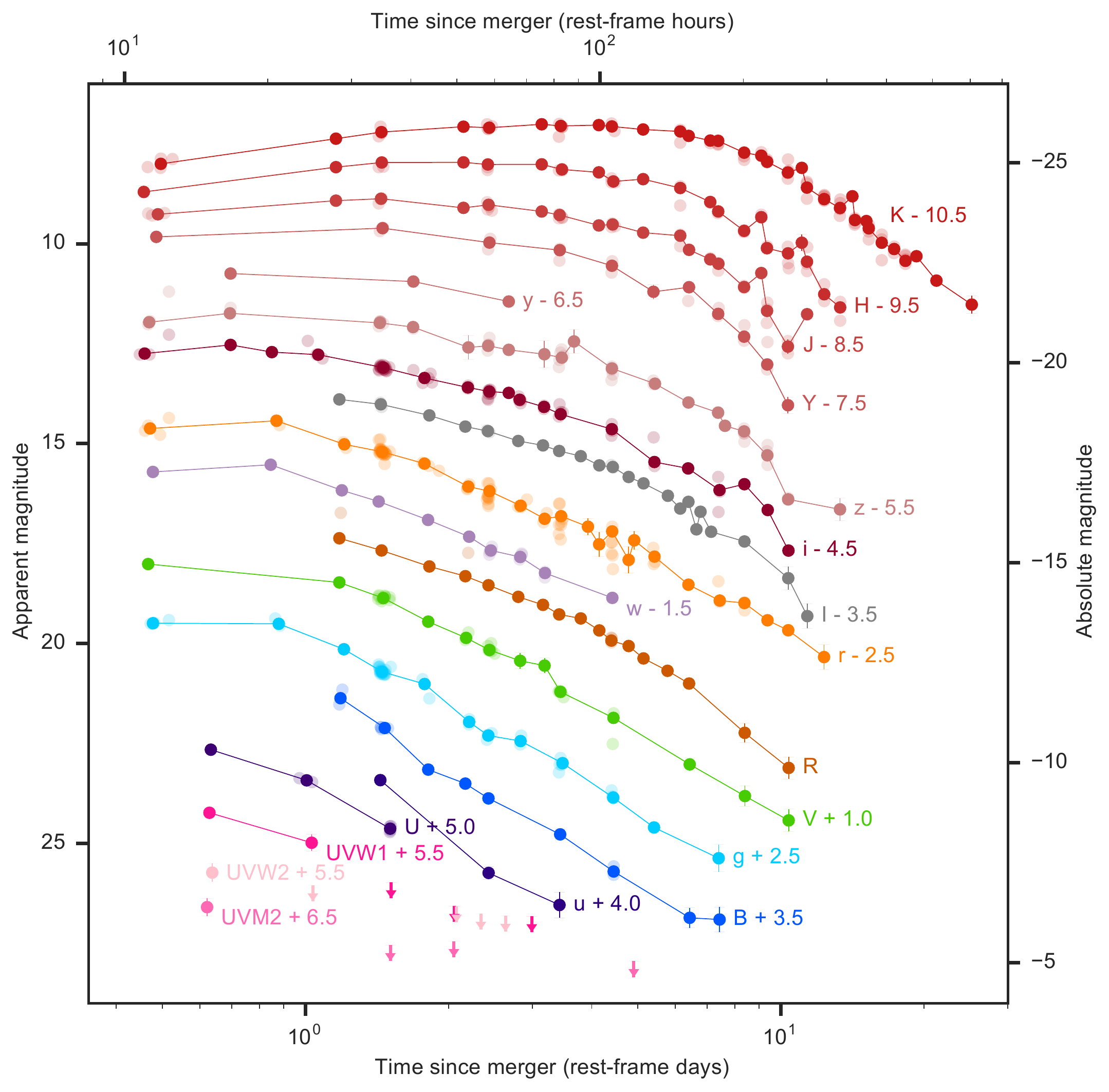}
	\caption{A compilation of the multi-wavelength UV/Optical/IR transient that followed GW170817  \cite[][ and references therein]{arcavi2018}. From  \cite{arcavi2018}}
	\label{fig:opt_phot}
\end{figure}

\begin{figure}
	\center
	\includegraphics[width=0.7\textwidth]{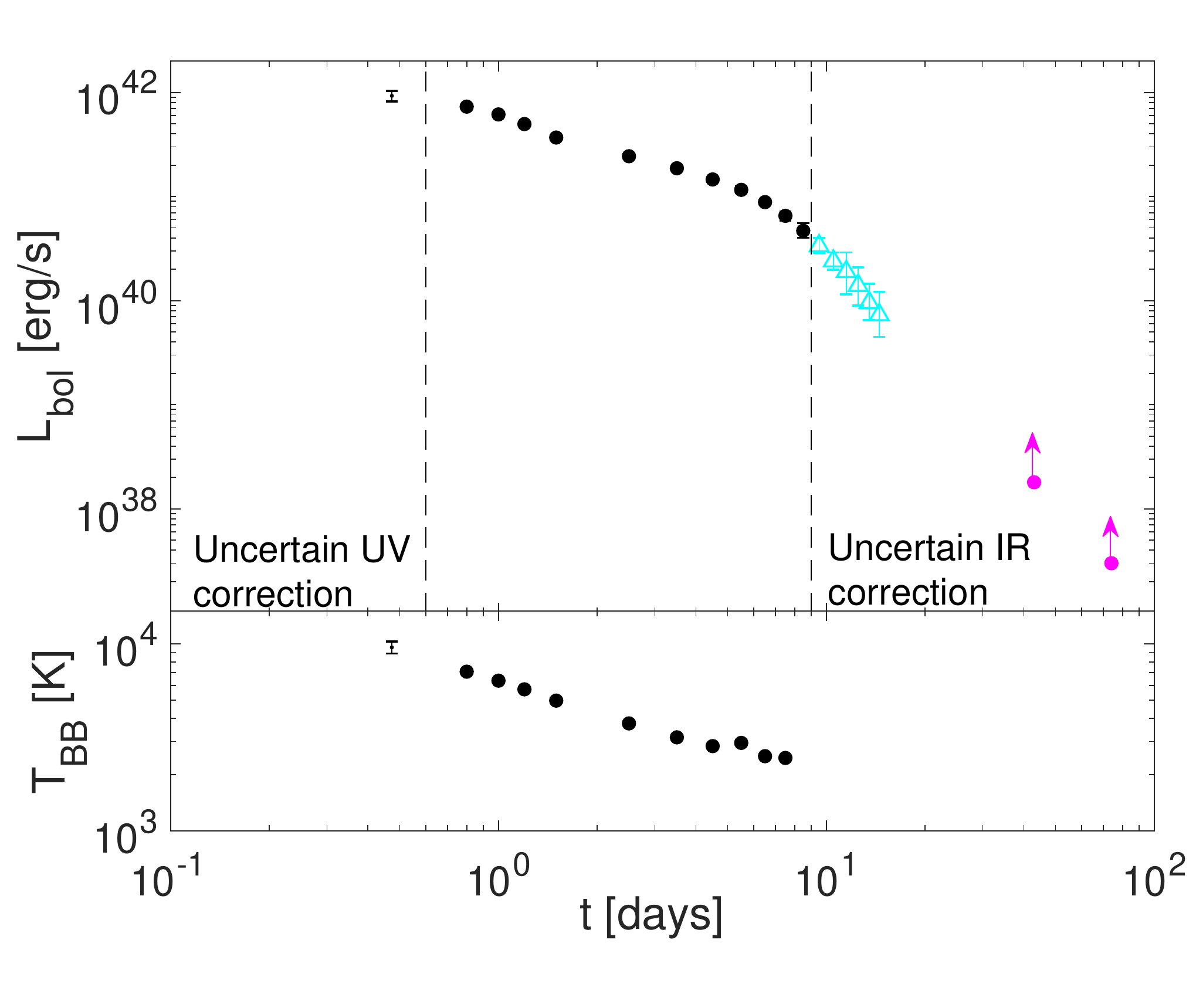}
	\caption{Bolometric luminosity and blackbody temperature of the UV/Optical/IR transient that followed GW170817. Between day 0.6 and day 9, the various estimates of the bolometric luminosity (by different authors and using different methods) are all in good agreement.  Here, the data during this period are taken from \cite{waxman2018} (their photometric integration estimate of $L_{bol}$) and is marked with filled black circles. The temperature is given only where the blackbody spectra provide a reasonable description of the data, which is until day 8. The contribution of UV light to the first observation at 0.47 day is uncertain. Here, the data is taken from \cite{arcavi2018}, who obtained two estimates for $L_{bol}$ and $T$ at this epoch. The error bars mark the differences between the two estimates. Between day 9 and day 16, a significant fraction of the luminosity is missing in the IR. The data here are taken from \cite{waxman2018} (their photometric integration estimate of $L_{bol}$) and should be considered a lower limit. The data on days 43 and 74 are taken from \cite{kasliwal2019} and are based on a detection in a single band, $4.5\mu$, by Spitzer. The spectrum at this time is clearly not thermal and cannot be used for a reliable estimate of $L_{bol}$.  We therefore include here only the actual luminosity that was observed within the Spitzer $4.5\mu$ band, and which is a strict lower limit.}
	\label{fig:opt_lc}
\end{figure}

\begin{figure}
	\center
	\includegraphics[width=0.6\textwidth]{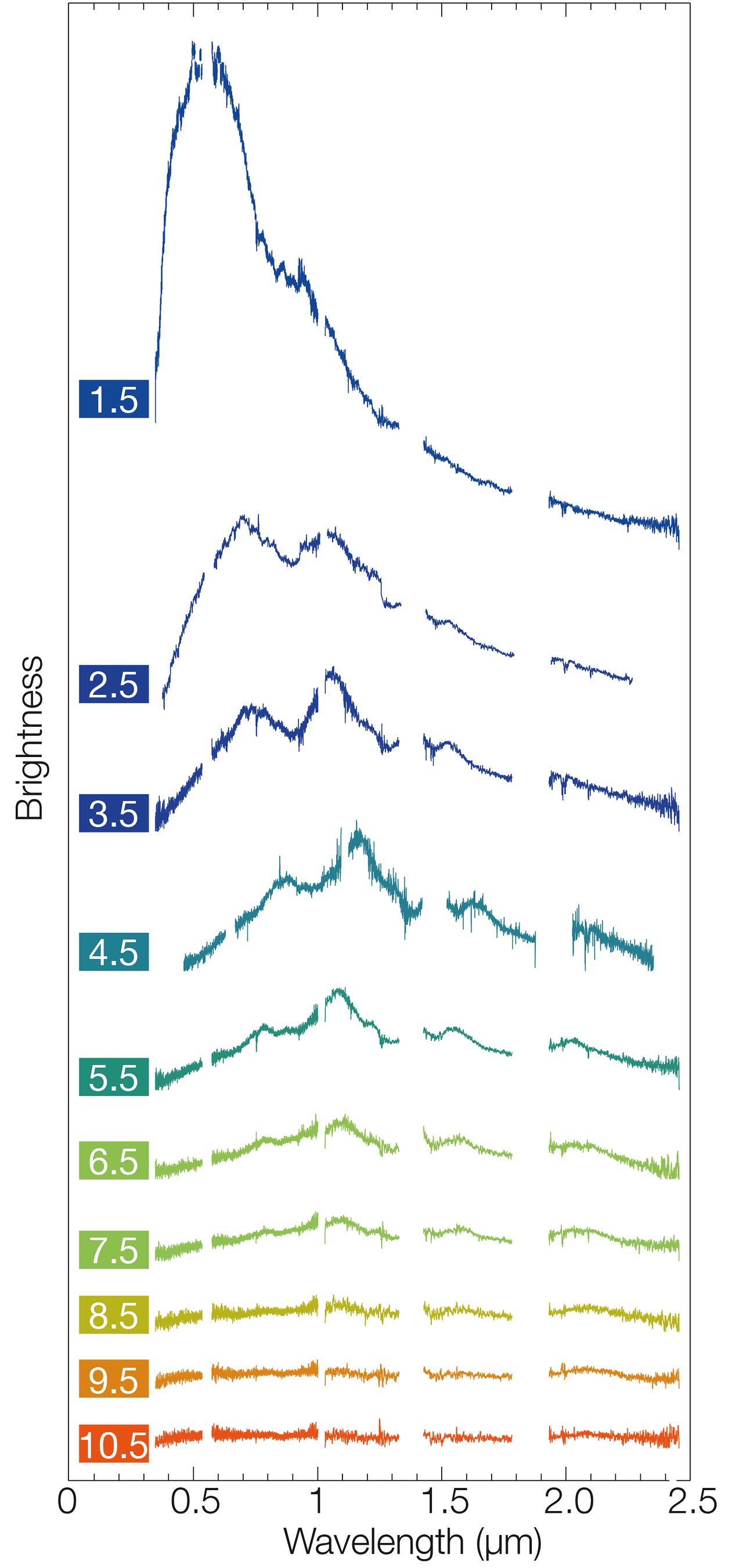}
	\caption{A series of spectra of the GW170817 UV/optical/IR transient obtained with the VLT X-shooter \citep{pian2017,smartt2017}. 
	From \url{https://www.eso.org/public/usa/images/eso1733j/}.  Reprinted with permission from European Southern Observatory.}
	%Credit:ESO/E. Pian et al./S. Smartt \& ePESSTO % 
	\label{fig:opt_spectra}
\end{figure}

The UV/optical/IR counterpart of GW170817 was initially given various names, including SSS17a, DL17ck, J-GEM17btc, and was later designated as astronomical transient AT2017gfo.  Its photometric and spectral observations were reported in a large number of papers: \citet{andreoni2017,arcavi2017a,arcavi2017b,buckley2018,chornock2017,coulter2017,cowperthwaite2017,diaz2017,drout2017,evans2017,hu2017,kasliwal2017,kilpatrick2017,lipunov2017,mccully2017,nicholl2017,pian2017,pozanenko2018,shappee2017,smartt2017,soares-santos2017,tanvir2017,tominaga2018,troja2017,utsumi2017,valenti2017,villar2018}. Many of these papers also offer interpretations of their observations, which will be discussed in \S \ref{sec:GW170817_SubRel}. Several studies compiled the data from these papers and reanalyzed them in a relatively homogeneous way:  \cite{villar2017,arcavi2018} and \cite{waxman2018}. Figure \ref{fig:opt_phot}, reproduced from  \cite{arcavi2018}, shows the compiled photometric data, and figure \ref{fig:opt_lc} shows the resulting bolometric luminosity and blackbody temperature. Figure \ref{fig:opt_spectra} shows a series of spectra taken by  the X-shooter spectrograph on the ESO VLT, every night between days 1.5 and 10.5  \citep{pian2017,smartt2017}. The photometric and spectral evolution of AT2017gfo is unique and had never been observed before in any other astrophysical transient. The association with GW170817 is considered secure \citep[e.g.,][]{siebert2017}.  Since AT2017gfo is just one part of the much broader-band multi-wavelength electromagnetic emission that followed GW170817, I will not use this (or any other) name and will refer to it as the optical/IR counterpart of GW170817. Below, I give a brief summary of its main observational properties.

The GW170817 optical counterpart was brighter and bluer than predicted by most of the macronova models. Its bolometric luminosity upon detection was $\sim 10^{42} {\rm~erg~s^{-1}}$, and a fit to a blackbody spectrum yielded a temperature of $\approx 10,000$ K. The earliest observations (at 11 hr) covered only the optical and IR bands. The first UV measurements were made by Swift, 14 hr after the merger \citep{evans2017}. The photometric data at 14 hr provide a reasonable fit to a blackbody, with some deviations.  Assuming blackbody emission from a spherical source, the luminosity and temperature after 12 hr correspond to a photospheric radius of $\approx 4 \times 10^{14}$ cm and a velocity of $\approx$ 0.3c. 

The evolution of the luminosity during the first hours after optical detection is unclear, due to the lack of UV data during that time \citep{arcavi2018}. Specifically, we do not know if the luminosity was on the rise or whether it was already decaying when detected. This is unfortunate, since different models have different predictions for the early evolution (see \S \ref{sec:macronova}).  After 14 hr, it is clear that both the luminosity and the temperature were falling rapidly. By the end of the first day after the merger, the bolometric luminosity had dropped to $\approx 6 \times 10^{41} {\rm~erg~s^{-1}}$,  and the 
temperature to about $7,000$ K. The luminosity and the temperature continued to decay roughly as power-laws during the first week,  $L \propto t^{-1}$ and $T \propto t^{-0.5}$, where $t$ is the time since the merger. By the end of the first week, the luminosity had fallen to $\approx  10^{41} {\rm~erg~s^{-1}}$,  the temperature to $2,000$ K, and the photospheric velocity to $0.1$c. After day 10, there are no reliable estimates of the bolometric luminosity and the
temperature for two reasons. First, the spectrum had  become highly non-thermal, and could no longer be fitted reasonably with a blackbody. Second, the light had become so red that it quickly disappeared from all the optical bands, and from most IR bands. By day 10, it was detected only in H and K, and after day 14 only in K. Thus, we do not know how much luminosity we missed in the IR at these epochs. Nevertheless, some attempts to estimate the luminosity after day 10 suggest that it started declining more rapidly than $t^{-1}$ \citep[e.g.,][]{waxman2018}. The latest detections of IR emission were obtained by {\it Spizer} at 4.5 $\mu m$ on days 43 and 74 \citep{villar2018,kasliwal2019}. The bolometric luminosity cannot be determined based on a single band detection, but the luminosity in this band alone sets a lower limit of $\sim 10^{38} {\rm~erg~s^{-1}}$ at that time. This luminosity is brighter than expected by the afterglow emission  (see \S \ref{sec:GW170817relativistic}) and it suggests that significant radioactive energy deposition continued at least up to that time.

While the rapid luminosity decay and the fast redenning of the light curve are unique, the spectral evolution is even more interesting.
The first spectra were obtained with Magellan/LDSS-3 and Magellan/MagE 12 hr after the merger \citep{shappee2017}. These spectra showed a smooth blue continuum in the optical that could be fit by part of a blackbody spectrum, with no evident spectral features. During the following 10 nights, spectra of the event were taken by many observatories. Figure \ref{fig:opt_spectra} shows X-shooter spectra that were taken nightly 
between 1.5 and 10.5 days \citep{pian2017,smartt2017}. The spectrum taken at 1.5 d is fit by a blackbody quite well, with a minor absorption feature around 0.8 $\mu m$, and a possible emission near 1 $\mu m$. On the following days, these features became more prominent (the emission feature was seen clearly) while moving slowly redward. Around day 3.5, additional broad features appear at 1.5 $\mu m$ and 2.1 $\mu m$ (hints of these features can be seen in earlier spectra). These features became more dominant with time, so that by the end of the first week most of the emission was in three broad peaks around 1.1, 1.6 and 2.1 $\mu m$. At day 8, a blackbody no longer provides a reasonable fit to the data, and most of the emission is seen in the broad 2.1 $\mu m$ peak. The observed spectral features reflect the outflow's composition, but unlike normal supernova, it is hard to attribute them to any specific chemical element, although the 1 $\mu m$ feature may correspond to P Cygni profile from near-IR lines of singly ionized strontium (Sr; Z=38, A=88) , centered in the restframe around 1,050 nm \citep{watson2019}. Nevertheless, similar spectra have never been seen in any other transient, and they are broadly consistent with the rough theoretical predictions of $r$-process elements. In \S \ref{sec:GW170817_SubRel}, the implications of the observed light curve and spectra on the outflow properties are discussed in detail.

Finally, \cite{covino2017} reported optical polarization measurements that were taken  between 1.5 and 9.5 days. They find linear polarization at a level of 0.5\%, which is at the level driven by the photometric uncertainties and the spread in polarizations seen in  field stars. This level is therefore an upper limit, and the observations are consistent with the optical light being unpolarized.

\subsubsection{Afterglow - radio to X-rays}

%\subsubsection{Light curve and spectrum}
\noindent \underline{{\it Light curve and spectrum}}\\
In typical GRBs, the prompt \grays are  followed by an afterglow emission seen in X-ray, optical, and radio bands. The common wisdom is that the source of this non-thermal emission is the blast wave that the relativistic jet, which generates the $\gamma$-rays, drives into the circum-burst medium. An X-ray afterglow, already in decline, is typically detected as soon as an X-ray telescope is pointed to the location of the burst. In GW170817 the optical macronova emission was clearly thermal and therefore it could not have been generated by the same mechanism that produces non-thermal GRB afterglows. Since no macronova emission is expected in X-ray or radio, those were the natural bands to search for nonthermal afterglow emission. Indeed, X-ray and radio searches began as soon as possible after the identification of the optical counterpart. However, unlike in normal GRBs, no X-ray or radio emission was detected for over a week following the prompt \grays. \\

\noindent The limits during the first week are as follows:

\noindent {\it $\gamma$-rays}: INTEGRAL observations set limits on continuous and bursting gamma-ray activity between 19.5 hr and 5.4 days after the merger \citep{savchenko2017}. They placed a limit of $<7.1 \times 10^{-11}  {\rm~erg~s^{-1}~cm^{-2}}$ on continuous emission at 80-300 keV. During the same period, bursting activity on time scales of 1s [0.1s] was limited to $<1.4 [4.5] \times 10^{-8} {\rm~erg~s^{-1}~cm^{-2}}$ in the 20-80 keV energy range and $<7.8 [24] \times 10^{-8} {\rm~erg~s^{-1}~cm^{-2}}$ at 80-300 keV.

\noindent {\it X-rays}: The earliest X-ray observations were carried out by MAXI  4.6 hr after the GW trigger, yielding an  upper limit of $8.6 \times 10^{-9} \,{\rm~erg~s^{-1}~cm^{-2}}$ in the 2-10 keV band \citep{sugita2018}. Much deeper limits were obtained a few hours later by $Swift$ and $Nu$STAR \citep{evans2017}, $<1.9 \times 10^{-13} {\rm~erg~s^{-1}~cm^{-2}}$ at 0.3-10 keV 15 hr after the merger ($Swift$) and  $<2.6 \times 10^{-14} {\rm~erg~s^{-1}~cm^{-2}}$ at 3-10 keV 20 hr after the merger ($Nu$STAR). A yet-deeper limit of $<1.7 \times 10^{-15} {\rm~erg~s^{-1}~cm^{-2}}$ at 0.3-10 keV was obtained by the Chandra X-ray Observatory 2.3 d after the merger \citep{margutti2018}.

\noindent {\it Radio}: Upper limits were obtained during the first two weeks (starting 14 hr after the merger) over a wide range of radio bands (1-100 GHz). Observations were carried out almost nightly and the limits were at the level of 10-100 $\mu$Jy
  \citep{hallinan2017,alexander2017}.\\

The first afterglow light was detected in X-rays 9 days after the merger \citep{troja2017}, and in the radio 16 days after the merger \citep{hallinan2017}. Optical emission that was attributed to the afterglow was detected first by the Hubble Space Telescope at day 110 \citep{lyman2018}.  Following the first detection, both the radio and the X-ray emission showed the same light curve, which rose continuously, reaching a peak around $\approx 130$ days. The peak was relatively narrow, and it was followed by a rapid decay. The entire afterglow radio observations were reported in \cite{hallinan2017,alexander2017,mooley2018a,margutti2018,alexander2018,dobie2018,mooley2018c,hajela2019,troja2019a}. Optical observations that were attributed to the afterglow were reported by \cite{lyman2018,margutti2018,alexander2018,lamb2019b,fong2019}. X-ray observations were reported in \cite{troja2017,haggard2017,margutti2017,ruan2018,pooley2018,davanzo2018,troja2018,troja2019a,margutti2018,nynka2018,hajela2019}.

\begin{figure}
	\center
	\includegraphics[width=0.7\textwidth]{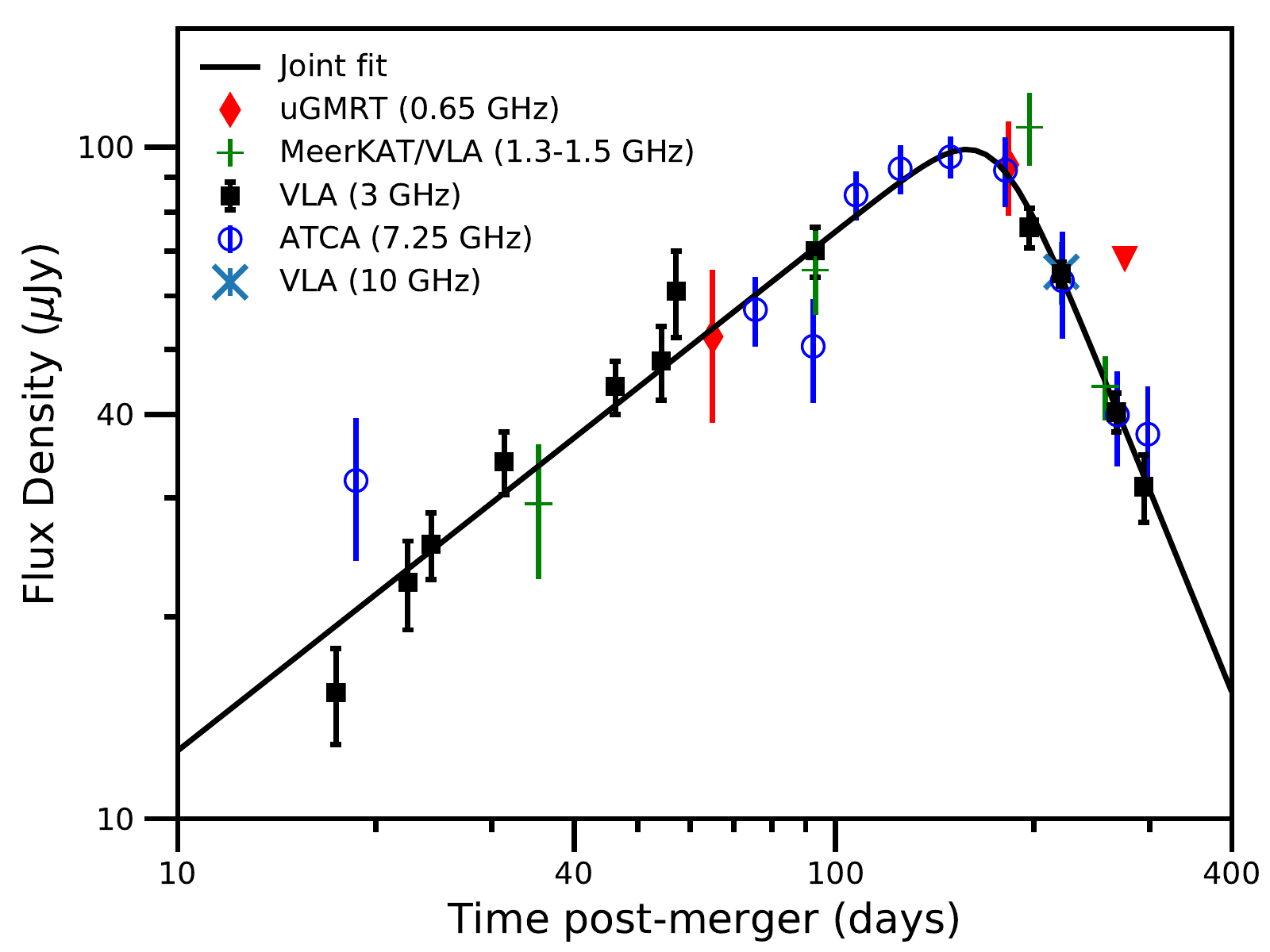}
	\caption{The radio light curve of GW170817 spanning multiple frequencies, and scaled to 3 GHz assuming a power-law spectrum. The optical and X-ray afterglows, being on the same spectral power-law segment as the radio, show similar light curves. From \citet{mooley2018c}.}%
	\label{fig:radio}
\end{figure}

Figure \ref{fig:radio} shows the multi-wavelength radio afterglow of GW170817. THe afterglow first displays a continuous rise that is consistent with a smooth power-law, $F_\nu \propto t^{0.78}$. The rise starts leveling off around 120 d, reaching a peak before 150 d, and then turning rapidly into a decline which, after day 200, is consistent with roughly $F_\nu \propto t^{-2}$. This decline continues at least up to 700 days after the merger (beyond that there are no data yet). Most interesting is the multi-wavelength spectrum, which shows that, during the entire afterglow evolution, all the radio, optical and X-ray measurements fall on a single power-law $F_\nu \propto \nu^{-0.58}$ (thus, the radio and X-ray light curves show exactly the same evolution). Note, that there is no sign for any evolution in time of the radio to x-ray spectral index, which is determined accurately from all of the nearly simultaneous radio and X-ray observations. 

\cite{corsi2018} report a linear polarization measurement taken at 2.8 GHZ on day 244, not long after the peak. They do not find any linear polarization, with an upper limit of 12\% (at 99\% confidence). \\

%\subsubsection{Radio image superluminal motion}

\begin{figure}
	\center
	\includegraphics[width=0.7\textwidth]{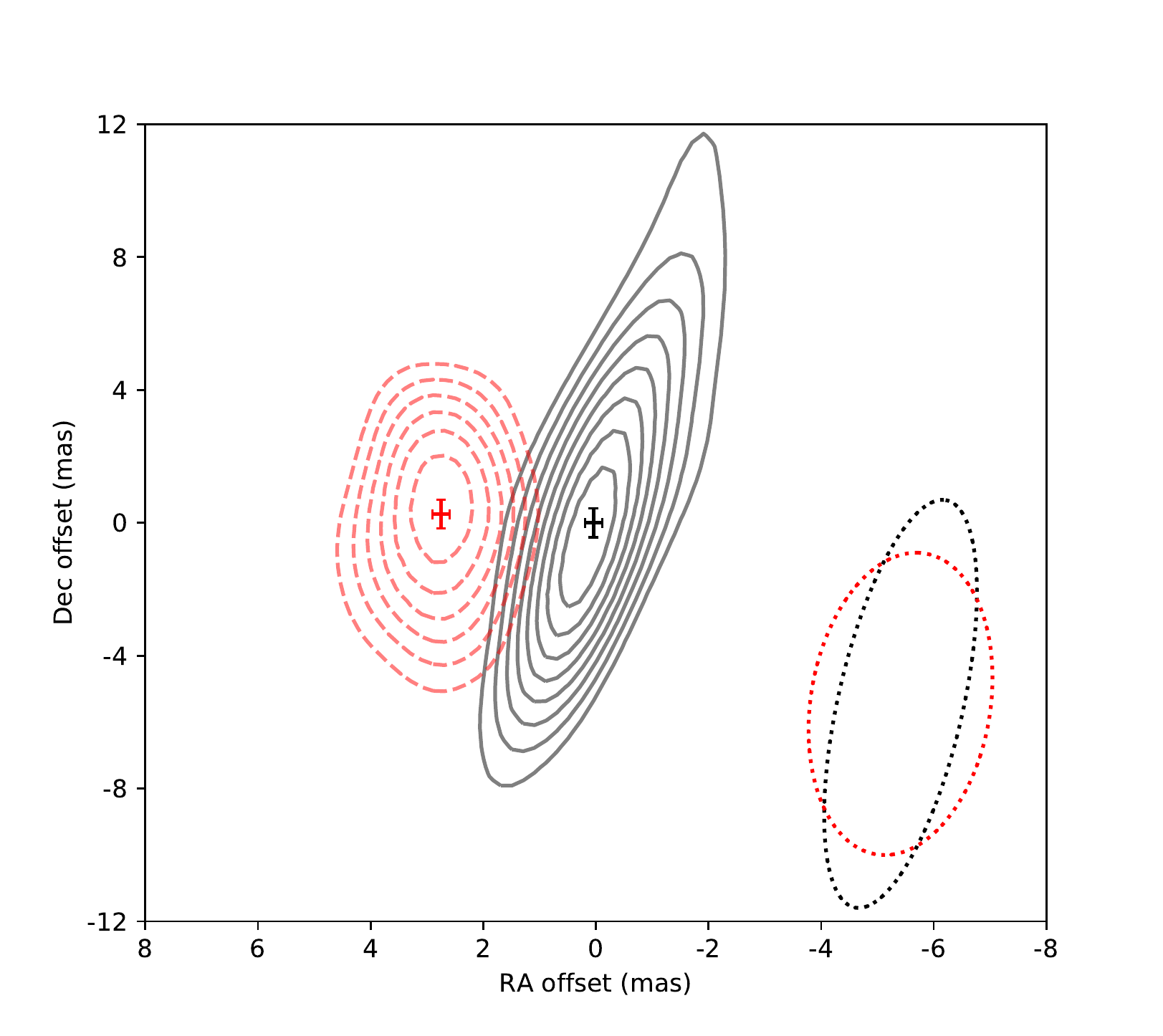}
	\caption{The VLBI radio images of GW170817 taken on days 75 (black) and 230 (red). The centroid offset between the two images is $2.7 \pm 0.3$mas.   The
shapes of the synthesized beam for the images are shown as dotted ellipses at the
lower right corner. From \citet{mooley2018b}.}%
	\label{fig:VLBI}
\end{figure}

\noindent \underline{{\it Superluminal motion of the radio image}}\\
The radio source was observed by the Very Long Baseline Interferometer (VLBI) at three epochs. \cite{mooley2018b} observed it on days 75 and 230 with the High Sensitivity Array (HSA). In both epochs, they find an unresolved source, were the centroid of the two epochs is displaced  by $2.7 \pm 0.3$mas (figure \ref{fig:VLBI}). Accounting for the uncertainty in the distance to NGC 4993, this displacement is translates to a super-luminal motion with an average apparent velocity of $4.1\pm 0.5$ c ($1\sigma$ error). As can be seen in figure \ref{fig:VLBI}, the shape of the beam was highly asymmetric, such that the narrow part of the beam was almost parallel to the centroid motion and the wide part was perpendicular to the centroid motion. As a result, limits on the size of the unresolved source are asymmetric, where the image size in the direction parallel to the centroid motion is $\lesssim 1$ mas, while in the perpendicular direction it is much less constrained, $\lesssim 10$mas. Additional observations on day 207 were obtained by \cite{ghirlanda2019} using the global VLBI. They find a displacement of the centroid by $2.44 \pm 0.32$mas compared to the HSA observation on day 75, fully consistent with the results of \cite{mooley2018b}. The source was unresolved also in this observation, but  the array configurations of the global VLBI allow  \cite{ghirlanda2019} to limit the size of the source to less than $2.5$ mas.

\subsubsection{Neutrino non-detection}
A search for a neutrino signal associated with GW170817 was done using ANTARES, IceCube,  the Pierre Auger Observatory \citep{albert2017}, and Super-Kamiokande \citep{abe2018}. The search was done in two time windows: $\pm 500$ s around the gravitational wave detection time, and during a 14-day period after the merger. The reported upper limits for the neutrino fluence are: 

\noindent {\it IceCube}: In both searches ($\pm 500$ s and 14 days) the limits in 10 TeV to 100 PeV are similar. Assuming a power-law neutrino spectrum  $\propto E^{-2}$, the spectral fluence limit is $F(E) \lesssim  0.2 (E/{\rm Gev})^{-2} \rm{~GeV^{-1} cm^{-2}}$.

\noindent {\it  Pierre Auger Observatory}: Assuming a power-law neutrino spectrum  $\propto E^{-2}$ the single-flavor upper limits to
the spectral fluence are  $F(E) \lesssim  0.77 (E/{\rm Gev})^{-2} \rm{~GeV^{-1} cm^{-2}}$ over a time window of $\pm 500$ s, and $F(E) \lesssim  25 (E/{\rm Gev})^{-2} \rm{~GeV^{-1} cm^{-2}}$ during 14 days after the merger, both over the energy range from $10^{17}$ eV to $2.5 \times 10^{19}$ eV.

\noindent {\it Super-Kamiokande}: Assuming a power-law neutrino spectrum  $\propto E^{-2}$, the limit on a time scale of $\pm 500$ s is $F(E) \lesssim  10^3 (E/{\rm Gev})^{-2} \rm{~GeV^{-1} cm^{-2}}$ over the energy range from 1.6 GeV to 100 PeV.

Altogether, at a distance of $\sim 40$ Mpc these limits corresponds to a maximal isotropic equivalent energy of $E_{{\rm iso},\nu}\lesssim 10^{50}$ erg carried by neutrinos in the energy range $10^{12}$ eV to $2.5 \times 10^{19}$ eV and  $E_{{\rm iso},\nu}\lesssim 10^{53}$ for 1-1000 GeV neutrinos. 
\subsubsection{Host galaxy}
NGC 4993, the host galaxy of GW170817, is otherwise and unremarkable early-type galaxy. It is at a redshift z = 0.0098 and its distance is $\approx 40$ Mpc (see \S\ref{sec:distance} for various constraints on the host and binary distance). The detailed properties of NGC 4993 in general, and at the (projected) location where the merger took place, are discussed in detail by \cite{blanchard2017,im2017,kasliwal2017,levan2017,palmese2017,pan2017,contini2018,ebrova2018} and \cite{wu2018}. Here, I briefly summarize these properties.

NGC 4993 has been classified as an E-S0 galaxy with a morphological T-type of -3 \citep{capaccioli2015}. Its effective radius is $r_{eff} \sim 3$ kpc and its Sersic index is $n \sim 4$. The total stellar mass is $\sim 3 \times 10^{10}\, \msun$   and its current star formation rate is low, $10^{-2}-10^{-3} \,\msun/{\rm yr}$. Most of the stellar population is very old ($\gtrsim 5$ Gyr), and the youngest significant population is $\sim 2-3$ Gyr old, where only a very small fraction ($\ll 1\%$) is younger than  500 Myr. The mass-weighted mean stellar metallicity is roughly solar. 

Optical line ratios, as well as radio and X-ray emission from the nucleus, show a weak activity of a low-luminosity AGN. The estimated mass of the super-massive black hole  is $\sim 10^8\,\msun$, and it accretes at $\sim 10^{-6}$ of its Eddington rate. 

Deep images reveal the presence of concentric shells, as well as a large ($\sim 10$ kpc) face-on spiral structure. A complex structure of dust lanes is seen near the nucleus. These features are most likely the results of a recent ($\lesssim 1$ Gyr) dry merger which has not fully relaxed yet.

The EM counterpart of GW170817 is located at an offset of $10.3^{''}$ from the nucleus, which corresponds to a projected distance of  $\sim 2$ kpc. There is nothing special at this location, i.e., no evidence for recent star formation or a point source that would suggest a globular cluster (down to a cluster mass of $\sim 10^4\msun$; \citealt{fong2019}), and no indication of dust or significant extinction. 

\subsubsection{Distance}\label{sec:distance}
The luminosity distance to the merger can be measured directly from the amplitude of the GW signal only, but not very accurately. The reason is that there is a degeneracy between the distance and the binary inclination (see \S\ref{sec:GWbinary}). The initial GW analysis  yielded $40^{+8}_{-14}$ Mpc (90\% confidence), as reported by \cite{abbott2017GW}, and  $43.8^{+2.9}_{-6.9}$ Mpc (1$\sigma$;  note the different confidence interval) as reported by \cite{abbott2017H0}. A more detailed analysis \cite{abbott2019Binary} gave a similar range (24-46 Mpc at 90\% confidence interval). The EM signal significantly improves the constraints, via the association of the merger with NGC 4993. Following the merger, the distance to NGC 4993 was measured using several different methods. \cite{hjorth2017} combined the measured redshift and the estimated peculiar velocity with the location of NGC 4993 on the Fundamental Plane of E and S0 galaxies to find a joint constraint on the distance of $41.0 \pm 3.1$ Mpc ($1 \sigma$). This estimates assumes the SH0ES measurement of the Hubble constant, $H_0 = 73.24 \pm 1.74\, {\rm~ km~s^{-1}~Mpc^{-1}}$ \citep{riess2016}. If instead the Planck value of
$H_0 = 67.8 \pm 0.9\, {\rm~ km~s^{-1}~Mpc^{-1}}$ is taken \citep{planck-collaboration2016} the best fit distance increases to about 44 Mpc (the error remains roughly the same). \cite{cantiello2018} used the surface brightness fluctuation method to find a distance of $40.7 \pm 2.4$ Mpc ($1 \sigma$). Finally, \cite{lee2018} used the luminosity function of globular clusters that are candidate members of NGC 4993, to estimate a distance of $41.65 \pm 3$ Mpc ($1 \sigma$).

\subsection{{\bf  The binary}}\label{sec:GW170817_Binary}
\subsubsection{Masses, spins, deformability and inclination}
An initial analysis of the properties of the binary was given by \cite{abbott2017GW}, and was followed by a more detailed study in \cite{abbott2019Binary}. The analyses were based mainly on the GW emission, but the distance derived from the EM counterpart improved some of the constraints (mostly on the inclination). The analysis is Bayesian and therefore depends on the assumed priors, which are described in \cite{abbott2019Binary} in detail. An important prior is the range of the binary members' spins. \cite{abbott2019Binary} consider two priors, where in both cases the spins of the two members are assumed to be uncorrelated in magnitude or orientation. The first prior on the spin magnitude is uniform  in the range $\chi=0-0.89$, where $\chi=cJ/Gm^2$ is the dimensionless spin parameter and $J$ and $m$ are a neutron-star's angular momentum and mass, respectively. This prior assumes that we have no previous knowledge on the spin of each of the binary members. The second prior is uniform in the range $\chi=0-0.05$, which is based on Galactic BNS systems that will merge within a Hubble time. All of these binaries  show relatively low spins, which at the time of merger are predicted to be $\chi \lesssim 0.05$ \citep{stovall2018}. Table \ref{table:binary}, reproduced from \cite{abbott2019Binary}, summarizes the main binary parameters derived under each of these priors. Below, I discuss some of the parameter derivations. \\

%\subsubsection{Masses}
\noindent {\underline {\it Masses:}}\\
The most robust binary quantity measured by the long-lived chirping signal is the chirp mass,${\cal M}=(m_1m_2)^{3/5}/(m_1+m_2)^{1/5}$. In GW170817, it was measured to be ${\cal M}=1.186 \pm 0.001 \msun$. For an equal masses binary, this chirp mass corresponds to $m_1=m_2=1.36 \msun$, which is similar to many observed Galatic BNS systems. The mass ratio is much harder to extract from the waveform. It is also partially degenerate with the spins and, therefore, it is less constrained when there is no prior knowledge of the spins. Nonetheless, under either prior (high or low spins) an equal mass binary (mass ratio $q=1$) is fully consistent with the data. The low-mass limit of the light member is $1 \msun$  ($q=0.53$) when a high-spin prior is assumed, and $1.16 \msun$  ($q=0.73$) for a low-spin prior.\\

%\subsubsection{Spins}
\noindent {\underline {\it Spins:}}\\
Spins affect the GW signal through several different effects (see \citealt{abbott2019Binary} for a brief discussion). In general, it is easier to constrain the spin components that are aligned with the orbital angular momentum, compared to those that are perpendicular to it. In GW170817, the effective spin, defined as $\chi_{eff}=(m_1 \chi_{1,\parallel}+m_2 \chi_{2,\parallel})/(m_1+m_2)$, where $\chi_{\parallel}$ is the spin component aligned with the orbital angular momentum, is constrained to the range $\chi_{eff}=0-0.1$, for the high spin prior. Also, the individual spin component of each of the members cannot be high in the direction aligned or anti-aligned with the orbital angular momentum. The perpendicular spins, however, are not well constrained, and both low spins and spins as high as 0.5-0.6 for each of the members are consistent with the data.\\

\begin{table}
	\caption{Binary properties of GW170817}
\begin{minipage}{1\textwidth}
\begin{center}
	\includegraphics[width=1\textwidth]{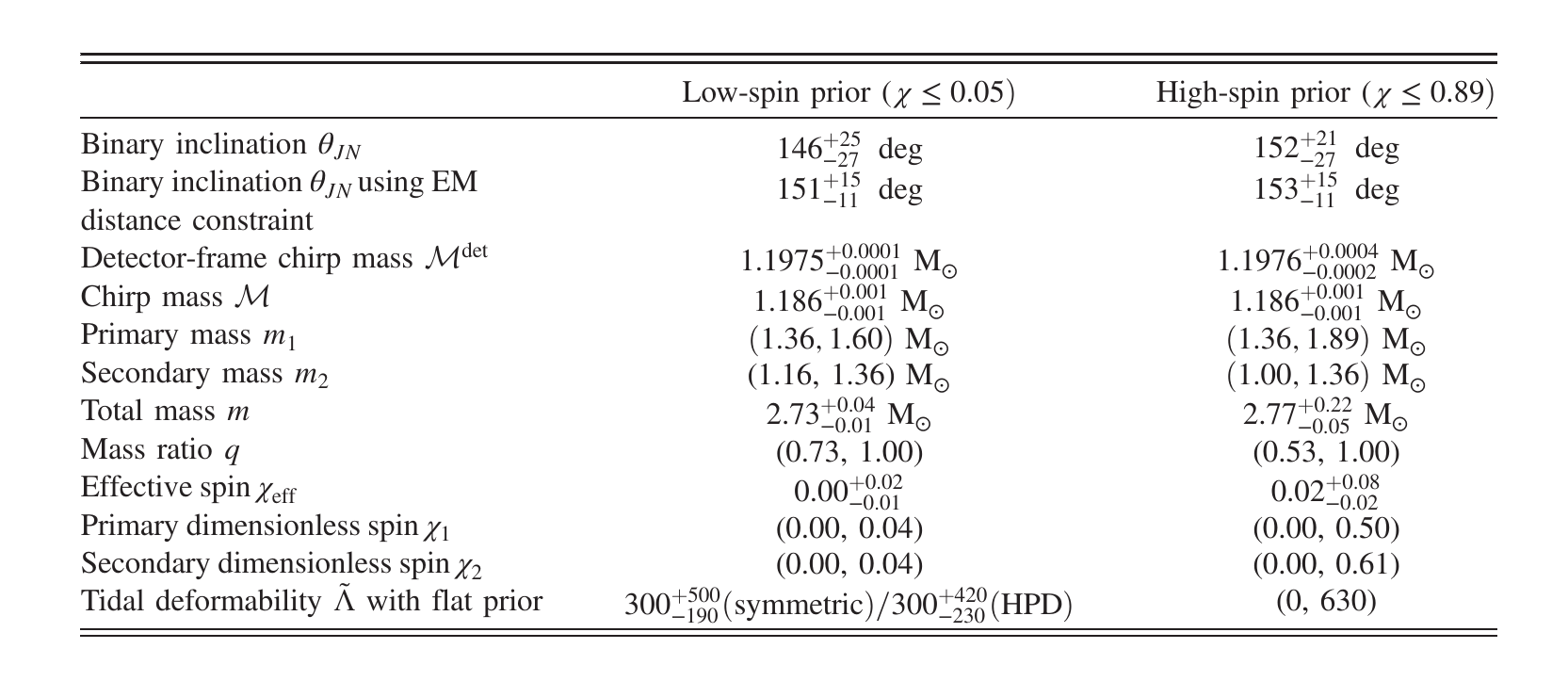}
\end{center}
\end{minipage}
\label{table:binary}
{\small The main properties of the binary of GW170817 as derived from the GW signal. From  \citealt{abbott2019Binary}}.
\end{table}

\noindent {\underline {\it Deformability:}}\\
The tidal deformation of each of the binary members is measured using the dimensionless tidal deformability parameter $\Lambda$. A larger value of $\Lambda$ corresponds to a stronger tidal deformation and thus a larger NS radius, where for a BH $\Lambda=0$. The leading term of the tidal deformation effect on the GW signal is a combination of the dimensionless tidal deformability parameters of the two binary members, $\tilde{\Lambda}$ (see  \citealt{abbott2019Binary} for a brief discussion). The deformability of the binary members is constrained  to $\tilde{\Lambda}<630$ for high-spin prior, and $\tilde{\Lambda}=300^{+420}_{-230}$ for a low-spin prior (90\% highest posterior density interval). Note that the GW signal is consistent with one of the binary members having $\Lambda=0$ also when taking the low-spin prior. Thus, we cannot exclude based on the GW observations alone, that one of the binary members of GW170817 was a BH \citep[e.g.,][]{hinderer2018}. \\

%inclination
\noindent {\underline {\it Inclination:}}\\
The GW signal by itself provided only poor constraints on the inclination of GW170817. The reason is that inclination can be measured from GWs alone only by comparing the two polarizations (see \S\ref{sec:GWbinary}). In GW170817, a high SNR signal was seen only by the two LIGO detectors, which are almost aligned, and therefore have high sensitivity only to one of the polarizations. \cite{abbott2019Binary} set constraints of $i=119^\circ-171^\circ$ (90\% confidence), i.e., $\theta_{obs}=9^\circ-61^\circ$, assuming a low spin prior (a similar constraint is obtained for the high-spin prior).

The measurement of the inclination is significantly improved when additional information on the distance to the binary is included. The best contraint on the distance to GW170817 is based on the EM identification of the host galaxy. Several studies have used it to constrain the inclination, with results that depend on the estimated distance to NGC 4993. \cite{abbott2017GW} used the estimates of the Hubble flow velocity near NGC 4993, and measurements of $H_0$, to estimate the distance. They find, assuming the Planck measurement of $H_0 = 67.9 \pm 0.55\, {\rm~ km~s^{-1}~Mpc^{-1}}$ \citep{planck-collaboration2016}, that the viewing angle is $\theta_{obs}<28^\circ$, while if the Sh0ES measurement of  $H_0 = 73.24 \pm 1.74\, {\rm~ km~s^{-1}~Mpc^{-1}}$ \citep{riess2016}  is used, then $\theta_{obs}<36^\circ$ (at 90\% confidence). \cite{mandel2018} obtained similar constraints assuming a similar value of $H_0$. \cite{finstad2018} used, instead, the distance estimate of \cite{cantiello2018}, $D=40.7 \pm 2.4$ Mpc, based on surface-brightness fluctuations of the host galaxy (i.e. independent of the assumed value of $H_0$). They find $\theta_{obs}=32^{+10}_{-13}$ deg (90\% confidence).

The afterglow observations can be used to constrain the observing angle with respect to the jet axis, which is typically assumed to be aligned with the orbital angular momentum. The best constraints are obtained by combining the VLBI radio imaging with the afterglow light curve\footnote{There are number of papers that attempt to constrain $\theta_{obs}$ based on the afterglow light curve alone, ignoring the available VLBI data. Some of these studies find that the light curve can be fitted with observing angles $>30^\circ$ that are inconsistent with \cite{mooley2018b}. However, at such angles the motion of the radio source would also be inconsistent with the VLBI data.}. \cite{mooley2018b} used this information to constrain the viewing angle to the range $\theta_{obs}=14^\circ-29^\circ$  with a most likely value of $\theta_{obs}=20^\circ$.

\subsubsection{Merger Rate}\label{sec:merger_rate}
Although GW170817 is only a single event, it still provides a constraint on the BNS merger rate that is more accurate than all previous methods, as it is based on direct observation with few assumptions. A detailed analysis by the LIGO/VIRGO collaboration find a merger rate of ${\cal R}_{BNS}=1540_{-1220}^{+3200} \,{\rm Gpc^{-3}\,yr^{-1}}$ at $1\sigma$ \citep{abbott2017GW}. The error in naturally dominated by the Poisson statistics of a single event. Given that this error is highly asymmetric, it is useful to estimate also the 90\% confidence interval of the rate. To do so, I first roughly recover the result of \citep{abbott2017GW}, simply by considering the duration of O2 (117 days of simultaneous twin-LIGO-detector observing time) and its range distance ($\approx 80$ Mpc; see \S \ref{sec:obsGW}). Finding one event within this time and volume implies an event rate of  $\approx 1500 \,{\rm Gpc^{-3}\,yr^{-1}}$, where the Poisson statistics $1\sigma$ interval is ${\cal R}_{BNS}=250-5000 \,{\rm Gpc^{-3}\,yr^{-1}}$, similar to the result of \cite{abbott2017GW}. With this estimate, I find that the 90\% interval is ${\cal R}_{BNS}=75-7100 \,{\rm Gpc^{-3}\,yr^{-1}}$. A more detailed analysis of the full runs O1 (in which there were no events) and O2 finds a somewhat narrower 90\% confidence interval, of ${\cal R}_{BNS}=110-3840 \,{\rm Gpc^{-3}\,yr^{-1}}$ \citep{the-ligo-scientific-collaboration2018}.

%\subsection{Formation channels and implications for binary evolution}

\subsection{{\bf The Hubble constant}}\label{sec:H0GW170817}

\begin{figure}
	\center
	\includegraphics[width=0.7\textwidth]{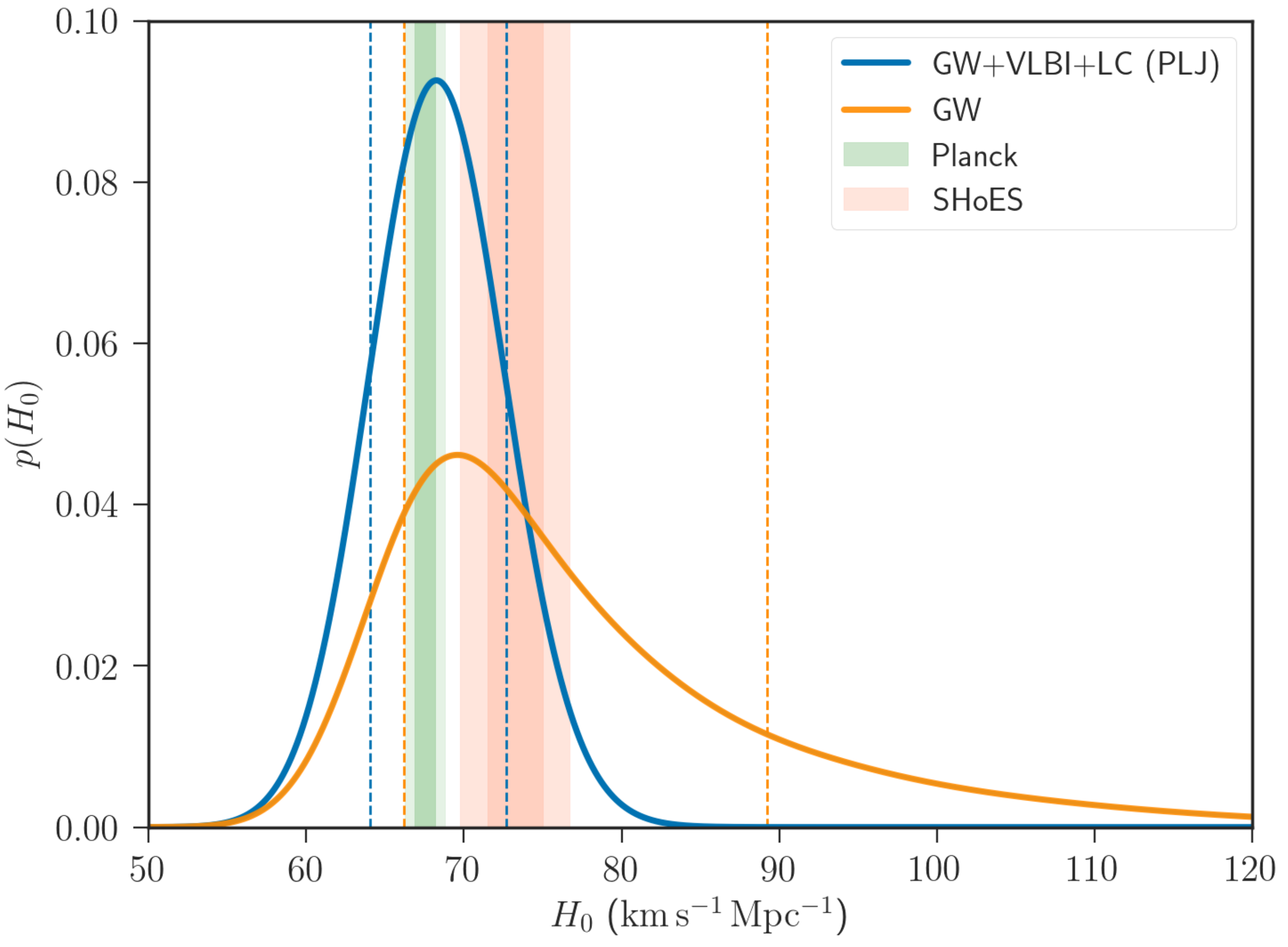}
	\caption{Posterior distributions for H0. The results of the GW-only analysis and the combined GW-EM analysis with a  power-law jet model are shown. The vertical dashed lines show symmetric 68\% confidence intervals for each model. The 1 and 2-$\sigma$ regions determined by Planck CMB (green) and SH0ES Cepheid-SN distance ladder surveys (orange) are
also depicted as vertical bands. From  \citet{hotokezaka2019}.}%
	\label{fig:H0}
\end{figure}

The constraints from GW170817 on $H_0$ were explored based on the GW signal alone and by combining the GW information with the  EM constraints on  the inclination angle. \cite{abbott2017H0} considered only the GW signal, to constrain $D_L$, obtaining  $H_0 = 74^{+16}_{-8} \, {\rm~ km~s^{-1}~Mpc^{-1}}$ (1$\sigma$ confidence interval)\footnote{This value is the symmetric interval (i.e., median $\pm 1\sigma$). Another interval that is often quoted from \cite{abbott2017H0} is $H_0 = 70^{+12}_{-10} \, {\rm~ km~s^{-1}~Mpc^{-1}}$  which is the maximum a posteriori value and the  smallest range enclosing 68.3\% of the posterior.}. The highly asymmetric error is due to the asymmetry of the $D_L$-$i$ degeneracy, which completely dominates the error in this case. \cite{hotokezaka2019} used the constraints that the afterglow modeling provides on the inclination, which are rather tight due to the superluminal motion of the radio image \citep{mooley2018b}, $0.88<\cos(i)<0.97$ ($0.25 {\rm~rad}<\theta_{obs}<0.5 {\rm~rad})$. These constraints reduce the uncertainty significantly, and \cite{hotokezaka2019} measure $H_0 = 70.3^{+5.3}_{-5} \, {\rm~ km~s^{-1}~Mpc^{-1}}$ at 1 $\sigma$ confidence interval, where the errors are dominated by the GW SNR and the host peculiar motion, and are therefore symmetric. Figure \ref{fig:H0} (taken from \citealt{hotokezaka2019}) shows the various constraints on $H_0$ obtained from GW170817.  

Based on GW170817, some 50 future GW-EM events are needed, to reduce the error on $H_0$ to the level of 2\%, based on GW data alone \citep{chen2018,feeney2018}. If EM information about the inclination is included and it is of the same quality as in GW170817, then \cite{hotokezaka2019} estimate that about 15 GW170817-like events may be enough to bring the error on $H_0$ to the level of 2\%. This can be achieved only if the systematic uncertainties in the modeling of the jet is properly understood.

\subsection{{\bf The sub-relativistic ejecta}}\label{sec:GW170817_SubRel}
Modeling the UV/optical/IR emission from GW170817 using macronova theory (\S\ref{sec:macronova}) provides constraints on the properties of the sub-relativistic ejecta.  A comparison to the predictions of mass ejection processes (\S\ref{sec:massEjection}), can help us learn about the physics of the merger process (e.g., NS EOS, the fate of the remnant, etc.). Below, I discuss, at first, the constraints on the sub-relativistic ejecta, and then the implications of these constraints to the physics of the merger.

\subsubsection{Mass, velocity and composition}\label{sec:GW170817_ejecta_property}
The macronova of GW170817 has almost always been modeled by assuming that its main power source is radioactive decay of \rp elements. The predictions of these models have provided a good representation of the data (although as discussed below, there is much freedom in the models). The properties of the sub-relativistic ejecta which I list below are the most likely ones under this assumption. Caveats and constraints given other types of assumptions are discussed further on.  

A partial list of papers that perform modeling of the optical/IR emission of GW170817 using a radioactively powered macronova includes \citet{buckley2018,cowperthwaite2017,drout2017,evans2017,kasen2017,kasliwal2017,kilpatrick2017,nicholl2017,pian2017,shappee2017,smartt2017,shibata2017,tanvir2017,villar2017,villar2018,kawaguchi2018,waxman2018,waxman2019} and \cite{kasliwal2019}. These models use different levels of approximations, and they all provide a reasonable fit to the data. The main uncertainty in the modeling is the radiative transfer, which is dictated by the uncertain \rpe line opacity, and therefore none of the models can provide a reliable fit to the entire series of spectra available for the event. This uncertainty, together with the fact that the observed broad spectral features are a blend of many lines, makes it difficult to obtain a direct identification of specific elements in the ejecta. Nevertheless, the general properties of the observed macronova, which do not resemble any previously seen transient, fit the expectations from the emission by $r$-process-rich ejecta (e.g., heat production and opacity). The conclusion that the sub-relativistic ejecta from GW170817 contains \rpe and that these elements play the central role in shaping the observed emission is therefore rather robust. This conclusion is supported by the possible identification of the 1$\mu \rm m$ feature, in the first-week spectra (see \S\ref{sec:GW170817_obsMN}), with a blend of lines from singly ionized Sr (Z=38; A=88) \citep{watson2019}, a light \rp element which is relatively abundant in the Sun.\\

\noindent{{\underline {\it Total mass, velocity distribution, and opacity}}}\\
Given the uncertainty in the outflow properties (e.g., composition, velocity distribution, and angular distribution), and the uncertainty in the theory (e.g. opacity), most models attempt to obtain quantitative fits only to the multi-wavelength photometry data, and there are enough degrees of freedom such that there is no unique solution (the fits to the spectra are mostly qualitative). Thus, there are many different models that provide reasonable fits. Nevertheless, there are some rather robust conclusions about some of the outflow properties that can be drawn from the data, if we assume that the emission is powered by radioactive decay of \rp elements. These conclusions are shared by most of the models cited above, and in the following, I summarize them and the observations on which they are based.

\begin{itemize} 
\item {\bf The total ejected mass is about $0.05\msun$.} This is the value, to within about a factor of 2, obtained in almost all the models, and it is constrained by the total energy radiated. The most robust method to constrain the mass is using equation \ref{eq:katz} (section \ref{sec:KatzIntegral}; \citealt{katz2013}), which is independent of radiative transfer. This method uses the observed bolometric luminosity to measure an integral over the heat deposition which, in turn, depends mostly on the total mass. 
%$t_{ph}$ in equation \ref{eq:katz}  is the photospheric time at which the optical depth drops to unity, which in our case is probably around $10$d, as indicated by the time at which the spectrum cannot be modeled by a blackbody anymore. Using bolometric luminosity from \cite{waxman2018} and \cite{arcavi2018} one obtains $\int_0^{10d} \dot{Q}~t~dt \approx \int_0^{10d} L_{bol}~t~dt = 4 \times 10^{52} {\rm~erg~s}$. For a given composition and velocity profile this value determines the total ejected mass. 
\cite{hotokezaka2019a} find that for solar abundances with $A_{\rm min}<A$ and a velocity profile that fits the photospheric velocity evolution, the ejected mass is $0.05\pm0.01 \msun$ for $85<A_{\rm min}<130$. The mass required for different realistic compositions can vary by a factor of order unity. A mass that is lower by a factor of a few is possible if there is a significant contribution of $\alpha$-decay or fission during the first 10 days.

\item {\bf About $0.02 \msun$ were ejected at a velocity of 0.2-0.3c. The opacity of this material to optical and near-UV light during the first day is relatively low.} The first observation after half a day showed a luminous blue signal,  $L_{\rm bol} \approx 10^{42} {\rm~erg~s^{-1}}$ and $T \approx 10,000$ K. The blackbody radius implies that the photosphere, at that time, moves at a velocity of 0.3c, and the luminosity requires a radioactive heat of about  $0.02 \msun$ of $r$-process-rich material, deposited above the trapping radius. Equation \ref{eq:mobs} shows that in order to have this amount of mass above the trapping radius at the required velocity, its effective opacity for optical and near-UV light must be small. An optical/near-UV effective opacity of $\kappa \approx 0.3 {\rm~cm^2~g^{-1}}$ provides a good fit, and it seems hard to find a model where $\kappa >0.5 {\rm~cm^2~g^{-1}}$. 

\item {\bf About $0.03 \msun$ were ejected at a velocity of 0.1-0.2c. The opacity to IR light of the material that dominates the emission a week after the merger is higher than the opacity seen during the first day.} The photospheric velocity (measured from the blackbody radius) drops from about 0.3c at half a day to about 0.15c after three days, and to about 0.1c after a week (after that, the spectrum cannot be modeled by a blackbody). The total bolometric luminosity during this time and the photospheric velocity implies that, in addition to the faster material seen during the first day, there is a comparable amount of mass moving more slowly, at 0.1-0.2c. After a week, the spectrum is still quasi-thermal and it peaks in the near-IR ($T \approx 2500$K). After about 10 days, it is clearly non-thermal, suggesting that this is roughly the time during which most of the energy is deposited  above the photosphere, i.e.,  $t_{\rm ph} \approx 10$d \citep[e.g.,][]{waxman2018}. Equation \ref{eq:tph} shows that this requires that the average effective opacity of the ejecta to IR light at 10 d is $\kappa \approx 1 {\rm~cm^2~g^{-1}}$. 

\end{itemize}

Note that, while the first conclusion (about the total mass) is largely independent of the radiative transfer, the velocity profile, or the geometry of the outflow, the two other conclusions (about the velocity and opacity distributions) uses a simplified radiative transfer approximation and a spherical geometry. Models using more complex radiative transfer and non-spherically symmetric geometries can obtain somewhat different results, although the general conclusions remain similar. Also, the constraints on the effective opacity are an average over the entire ejecta that contributes to the emission at the relevant time. Thus, if the ejecta has several components (see below), then the opacity of each component can differ from the average (e.g., some of the slowly moving material can have $\kappa > 1 {\rm~cm^2~g^{-1}}$).\\

\noindent{{\underline {\it How many ejecta components are there?}}}\\
One of the interesting debates about the interpretation of the data is whether the data can be explained by a single component, or are two or even three different components required. This is important because, if different components are identified, having different velocities, compositions and/or angular structure,  it might be possible to map them on to specific ejection processes, and improve the constraints that we can pose on issues such as the NS EOS or the fate of the remnant.

During this debate, various models have been in fundamental disagreement as to whether or not one component is sufficient. However, examination of the details of the different models show that they share many similarities, and they all agree with the three conclusions I listed above about the mass, the velocity profile, and the opacity. The major difference between different studies lies mainly in the interpretation of the modelling. Models with several components \citep[e.g.,][]{cowperthwaite2017,kilpatrick2017,villar2018} have at least one fast component ($v \approx 0.25c$) and one slow component ($v \approx 0.15$c), in order to fit the color-luminosity evolution. The masses of the those components are comparable and, in almost all models, the fast component has low opacity and the slower component has higher opacity. In contrast, \cite{waxman2018} presented a single-component model that fits both the luminosity and the temperature evolution. However, this single component has a power-law velocity profile with velocities between 0.1-0.3c, and a time-evolving opacity that grows during the first week from $0.3 {\rm~cm^2~g^{-1}}$  to $1 {\rm~cm^2~g^{-1}}$. In this model, the early colors are determined by the faster part of the ejecta, when the opacity is low, and the late-time emission is determined by the slower part of the ejecta when the opacity is higher. Thus, the multi-component models where each component has a single velocity and constant opacity are replaced by a model with a single component that has a range of velocities and a time varying opacity. The main difference between the models is in the interpretation of the varying opacity. While the multi-component models interpret this variation as an indication of varying composition, \cite{waxman2018} suggest that it may be a result of temporal evolution of the ejecta density and temperature, which may well-affect the opacity. As I discuss below, a comparison between early and late spectra suggests that the composition does vary, at least to some extent, between different parts of the ejecta.\\

\noindent{{\underline {\it Composition}}}\\ 
The main clue about the composition of the ejecta is the opacity, since even a small fraction of lanthanides should have a significant effect on the optical depth, both in the UV/optical and in the IR. The lanthanides are, in turn, tracers for heavy \rp elements (see section \ref{sec:rprocess}). Specifically, a low opacity indicates lanthanide-poor material that contains a small fraction of heavy elements (if any), while a high opacity suggests the presence of lanthanides, and thus of heavy \rp elements.

The low optical depth to UV/optical light during the first day of the event implies that the approximately $0.02\msun$ of material that dominates the emission, during that time, must be lanthanide-poor. For example, \cite{nicholl2017} use macronova models from \cite{kasen2017} to find that the blue emission in the spectrum at 1.5 d requires a lanthanide fraction $X_{\rm lan} \ll 10^{-4}$ (see also \citealt{pian2017,shappee2017}). While the large uncertainty in the opacity of lanthanide elements handicaps the accuracy with which  actual upper-limit on the lanthanide fraction can be determined, it is quite clear that a solar abundance is ruled out. Since the early emission was most likely dominated by fast material, it seems that, following the merger, about $ 0.02\msun$ of $\Ye \gtrsim 0.3$ were ejected at a velocity $v>0.2$c.

On day 7, the blackbody spectrum with temperature of about 2500K implies a significant opacity in the near-IR ($\kappa \gtrsim 1{\rm~cm^2~g^{-1}}$). Such a high opacity in the IR seems to be inconsistent with that of iron-peak or light \rp elements, and it is most likely dominated by lanthanides. Again, it is hard to obtain a robust lower limit on the fraction of lanthanides in the slow moving material, but it seems to be larger than $X_{\rm lan} \gtrsim 10^{-3}$, yet most likely lower than the solar abundance \citep[e.g.,][]{chornock2017,waxman2018}. 

The significant difference between the UV opacity during the first day, and the IR opacity after a week, strongly suggests that the composition is not uniform, and that the slower material contains more heavy \rpe than the faster material. However, it seems that there is no significant amount of mass with $\kappa \approx 10{\rm~cm^2~g^{-1}}$, as expected for low $\Ye \lesssim 0.1$ ejecta, so even the slow ejecta probably had a moderate $\Ye$. The most likely conclusion is that most of the ejecta had $\Ye \gtrsim 0.2$.

The radioactive heating rate can also provide some clues about the composition, if we assume that the composition of the ejecta resembles the solar abundance pattern over some atomic mass number range\footnote{This assumption is supported by the observations of \rp enhanced metal-poor stars, if BNS mergers are the sources of \rpe in those systems.} $A_{\rm min} \leq A \leq A_{\rm max}$, and the heat is dominated at all times by $\beta$-decay, i.e., the distribution of $A$ does not change during the radioactive decay  \citep[e.g.,][]{hotokezaka2019a}. First, when examining the heating rate by the various elements with solar abundance, one finds that if the ejecta contained all the \rpe ($A>69$), then the heat deposition during the first week would be completely dominated by the chain of elements with A=72, $^{72}$Ge $\rightarrow$ $^{72}$Ga $\rightarrow$ $^{72}$Zn (see figure \ref{fig:Qtot}). Now, if the total mass of A=72 elements is set to fit the observed heating rate of GW170817 during the first week, then there is not enough heat deposition to fit the observations at late times. Thus, the abundance of $^{72}$Zn in the ejecta was most likely much lower than Solar. This result is consistent with the finding of abundances in \rp enhanced metal-poor stars, and with the expectations from theoretical calculations for the composition of BNS merger ejecta (see \S\ref{sec:rprocess}). The first peak elements $73 \leq A \leq 84$ contain most of the mass of \rpe in the Sun, but they have no significant contribution to the heating rate. Thus, if $A_{\rm min}$ is in this range, then the total mass of the ejecta is about $0.1\msun$ \citep{hotokezaka2019a}, which is higher than expectations for the total mass ejection in BNS mergers (see \S\ref{sec:BNSejecta}). Hence, it is most likely that $A_{\rm min} \geq 85$. The next interesting decay chain is $^{88}$Kr $\rightarrow$ $^{88}$Rb $\rightarrow$ $^{88}$Sr. This chain contributes a significant amount of heat during the first hours, and including it enhances the peak luminosity. For example, when modeling the macronova of GW170817, the peak luminosity with $A_{\rm min}= 88$ is higher by a factor of $\sim 2$ than that with $A_{\rm min} = 89$ (for the same ejecta mass and velocity profile). The high observed peak luminosity of the macronova may indicate that this decay chain significantly contributes to the heating around the peak. This is also consistent with the identification of the spectral feature around 1$\mu \rm m$ as $^{88}$Sr lines \citep{watson2019}. Finally, the heavy \rpe (A>140) are not very efficient heat sources, and they account only for a small fraction of the ejecta mass, and therefore the heating rate per unit mass is insensitive to whether the ejecta contain heavy \rpe or not. To conclude, the heating rate seen in GW170817 supports a composition that does no contain first-peak elements, but does contain elements with $85 \lesssim A \lesssim 140$. The late IR spectra of GW170817 suggest that elements with $140<A$ are also present in the ejecta. However, in the Sun, the fraction of lanthanides from the entire \rp elements, A>69, is $X_{\rm lan}\approx 0.03$, and the fraction among the elements above the first peak, A>85, is $X_{\rm lan}\approx 0.1$. These values seem significantly larger than the lanthanide fraction in the ejecta of GW170817, as suggested by the IR spectra. Thus, it seems that the ratio of light to heavy \rpe in GW170817 was larger than seen in our Sun.\\

\noindent{{\underline {\it Caveats and non-radioactive energy sources}}}\\ 
The properties of the sub-relativistic ejecta derived above are the most likely ones, given the theoretical predictions and the set of observations. However, there are also alternative ejecta models that can explain the observations. One example is a model where the dominant energy source is a central engine. In this model the heating rate is decoupled from the ejecta, which now serve only as a screen that absorbs the energy from the source and re-emits it with the right spectrum. \cite{matsumoto2018} show that, in this scenario, the observations can be explained with a total ejecta mass as low as $\sim 0.005\msun$. Another possibility is that the early blue signal has a different origin than the rest of the emission, such as cooling emission following shocks that cross the ejecta at large radii $\gtrsim 10^{10}$ cm \citep{piro2018,chang2018,metzger2018}. In that case, the mass of the fast component can be significantly lower than $\sim 0.02\msun$. 

Finally, even if only radioactive heating is considered, one can devise an aspherical geometry that will alter the conclusions about the mass of the various components of the ejecta by a factor of a few. For example, motivated by the properties of the dynamical ejecta found in simulations, \cite{kawaguchi2018} suggest a scenario where the fast component of the ejecta has a relatively low mass ($0.009 \msun$) and low $\Ye$ (i.e, high opacity). It is highly anisotropic, and only small fraction of it (0.001) is ejected  towards the polar region. The slow component is more massive ($0.02 \msun$) but its opacity is low. In this model, the energy that powers the early blue emission is generated also by the slow component, while the fast component acts as a screen that reprocesses this energy to produce the color evolution seen in GW170817. The anisotrpoic outflow produces also an anisotropic emission, where an observer at a high latitude sees brighter emission than an observer at low latitude, leading to a bright signal for some of the observers but with a lower total heating rate. Finally, in this model the composition of the outflow is  taken from \cite{wanajo2014}, where the low-$\Ye$ material is rich with heavy, $A>209$, elements, so the contribution of $\alpha$-decay to the heating is significant. The required heating rate is thus obtained with a lower mass of material.

\subsubsection{Implications for mass ejection processes}
%\noindent{{\underline {\it Mass ejection processes}}}\\
Comparison of the ejecta properties (as inferred from the observations) with the theoretical predictions (summarized in table \ref{table:BNSsubRel}) shows general agreement between the models and the observations. First, the total mass of the ejecta, $\sim 0.05\msun$, is near the upper limit of the theoretical predictions. It requires the formation of a relatively massive disk, $0.1-0.3 \msun$, which in turn requires a HMNS or SMNS remnant that does not collapse promptly to a BH. Second, the range of observed velocities, $0.1c-0.3$c, is consistent with the predictions of the various components. However, a more detailed comparison shows that there are some discrepancies. While the slow component can be well-explained by the secular disk winds, the mass of the lanthanide-poor fast component, $\sim 0.02\msun$, is significantly higher than  theoretical predictions. The only process by which ejection of a significant fraction of the mass at $v>0.2$c is predicted is the dynamical ejecta. However, the typical mass of the dynamical ejecta is $\sim 10^{-3}\msun$, and there is no simulation where the dynamical mass exceeds $\sim 0.01\msun$. Moreover, the dynamical ejecta contains a large range of velocities and $\Ye$ values, so only a fraction of the dynamical ejecta can contribute to the fast component observed. Thus, based on the simulations available to date, it seems that, while the ejecta is expected to contain a fast low-opacity component, its mass is predicted  to be about one order of magnitude lower than the one inferred from observations (even under favorable conditions of a soft EOS and no prompt formation of a BH). Given that the discrepancy is only by a factor of $\sim 10$, it may simply be a result of the limitations of the numerical simulations (e.g., insufficient resolution or inaccurate approximations of physical processes). Alternatively, it may suggest that there are additional processes at work which are not captured by current theoretical models. Finally, it is also possible that the simple interpretation of the observations is wrong, and the fast low-opacity component in GW170817 was significantly less massive than $\sim 0.02\msun$, as may be the case, for example, if the early luminous and blue signal is not powered by radioactive $\beta$-decay and/or if a complex geometry  played a role in shaping the light curve (see the caveats in the previous section).

\subsection{{\bf The relativistic outflow}}\label{sec:GW170817relativistic}

\subsubsection{Afterglow}\label{sec:GW170817_Afterglow}
The afterglow of GW170817, and especially the direct VLBI radio imaging, provide unprecedented information about the structure of the relativistic outflow. The apparent superluminal motion of the image, when combined with the shape of the light curve, indicate that a narrowly collimated, highly energetic, jet was launch following the merger of GW170817, and that the jet broke out of the ejecta successfully, then driving a shock into the circum-merger medium and generating the observed afterglow emission \citep{mooley2018b,mooley2018c,lazzati2018,alexander2018,lyman2018,wu2018a,ghirlanda2019,hajela2019}. Below, I describe the specific properties of the jet and how those properties are deduced from the observations.

The afterglow was first detected 9 days after the merger. At that time, the emission is expected to be dominated by the synchrotron emission from the forward shock that the outflow drives into the circum-merger medium. A prediction of this model is that, at frequencies in the range $\nu_m,\nu_a<\nu<\nu_c$, the spectrum is a single power-law, $F_\nu \propto \nu^{-(p-1)/2}$, where the typical value of $p$ is in the range $2-2.5$. During the entire evolution of the afterglow, the radio and the X-rays (as well as the optical where observations were available) followed a single power-law, $F_\nu \propto \nu^{-0.584 \pm 0.06}$, which corresponds to $p=2.17 \pm 0.01$ \citep{margutti2018}. First of all, this observation lends strong support to the standard afterglow model. Second, it provides one of the the most accurate and reliable measurements of $p$ in a single event. Next, under the interpretation of the radio to X-ray emission as synchrotron radiation from a forward shock, the  slow rise as $t^{0.8}$ indicates that we are seeing a structured jet for which, at all times, we are within the beam of the  emission from the source, i.e., the source is on-axis. In other words, at all times $\theta(t) \lesssim 1/\Gamma(t)$ is obeyed, where $\Gamma$ and $\theta$ are the instantaneous source Lorentz factor, and the angle between the source and the line of sight, respectively. In this picture, the fast decay following the peak is consistent with a successful jet, seen away from the core. 

\begin{figure}
	\center
	\includegraphics[width=0.4\textwidth]{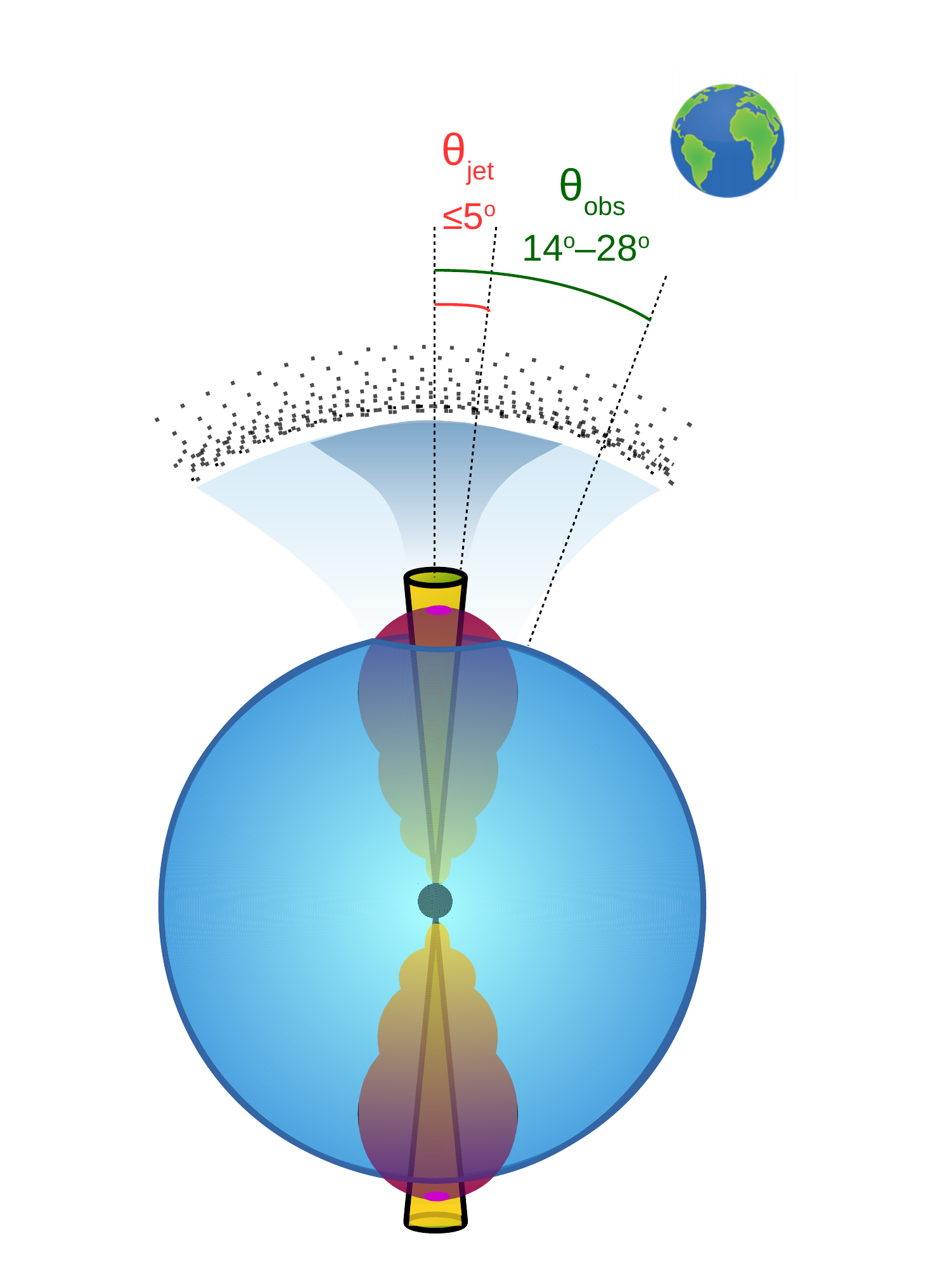}
	\caption{An illustration of the geometry of the relativistic jet that was launched following the merger of GW170817. The jet has a narrow core, $\theta_{\rm j} \lesssim 5^{\circ}$, with a total energy of $10^{49}-10^{50}$erg and isotropic equivalent energy $\gtrsim 10^{52}$ erg. The jet is surrounded by lower energy material, i.e., a structured jet, which is consistent with the expected structure of the cocoon and its interface with the jet (see section \ref{sec:JetCocoon}).
	The observer is at an angle of $\sim 20^\circ$. From \cite{mooley2018b}.}
	\label{fig:GW170817_jetSketch}
\end{figure}

The full geometry is revealed by the VLBI observations, as discussed in \cite{mooley2018b}. Around the time of the peak ($\sim 130$d), the angle between the observer and the core of the jet is roughly $\theta_{\rm obs}-\theta_{\rm j} \approx 1/\Gamma$, where $\theta_{\rm obs}$ is the angle between the jet axis and the line of sight and $\theta_{\rm j}$ is the opening angle of the jet core. There are several indications that the radio source is compact near the peak, namely that $\theta_{\rm obs} \gg \theta_{\rm j}$. These include the fact that the source in the image is unresolved, and the fast turnover of light curve at the peak, from rise to afast decay (see \citealt{mooley2018b,mooley2018c} for a full discussion). Given that the source is compact and that
the time of the VLBI observations (between 75d and 235d) is around the time of the peak, the radio source can be regarded as a point source at an angle $1/\Gamma$ with respect to the observer. This geometry implies that the apparent velocity of the image is $\beta_{\rm app} \approx \Gamma$. The VLBI images taken before and  after the  peak find $\beta_{\rm app} \approx 4$, which implies $\theta_{\rm obs}-\theta_{\rm j} \approx 0.25$rad. Numerical simulations find that $\theta_{\rm j} \lesssim 0.1$rad is required in order to explain the fast turn-over of the light curve when $\theta_{\rm obs}-\theta_{\rm j} \approx 0.25$ rad. The conclusion is that the jet is extremely narrow, $\theta_{\rm j} \lesssim 5^\circ$ and that that the observing angle is roughly 0.35 rad, namely $\theta_{\rm obs} \approx 20^\circ$. A set of numerical RHD simulations have shown that, indeed, jets with this geometry can simultaneously explain the light curve and the motion of the VLBI image. Taking into account the uncertainty of the measurement and  the modeling, \cite{mooley2018b} estimated that the observing angle is in the range $14^\circ<\theta_{\rm obs}<29^\circ$. Note that some papers find that even an angle as large as $\theta_{\rm obs} = 35^\circ$  is consistent with the light curve data \citep[e.g.,][]{lazzati2018,wu2018a,hajela2019}. However, these papers do not include in their fits any modeling of the VLBI data, which contains the most constraining information about $\theta_{\rm obs}$. A sketch that illustrates the geometry of the relativistic jet in GW170817 is presented in figure \ref{fig:GW170817_jetSketch}. 

Once the geometry is constrained, the time and the flux of the peak can be used to constrain the energy of the jet and the circum-merger density (equations \ref{eq:peak_time} \& \ref{eq:peak_mag};  \citealt{mooley2018b}). Equation \ref{eq:peak_time} is independent of the microphysical parameters, and since the peak is observed around day 130 and $\theta_{\rm obs} - \theta_{\rm j} \approx 15^\circ$, it implies $E \approx 10^{50} (n/10^{-3} {\rm~ cm^{-3}}) {\rm~ erg}$. Using the upper limit on the jet opening angle, $\theta_{\rm j} \lesssim 0.1$ rad, we can derive a lower limit on the isotropic equivalent energy, $E_{\rm iso} \gtrsim 10^{52} (n/10^{-3} {\rm~ cm^{-3}}) {\rm~ erg}$. At the location of the merger (projected distance of about 2 kpc from the nucleus of the host galaxy) the density is not expected to be much 
lower than $10^{-3} {\rm~ cm^{-3}}$. The isotropic equivalent energy of the jet in GW170817 was therefore very high, when compared to that of typical sGRBs (see section \ref{sec:sGRBs} for a discussion of the implications). Equation \ref{eq:peak_mag} depends also on the microphysical parameters, $\epsilon_e$ and $\epsilon_B$, and therefore, in order to obtain additional constraints, one needs to make assumptions. Typically, modeling of GRB afterglows suggest that $\epsilon_e \approx 0.1$ is a 
good approximation (to within an order of magnitude). Using this approximation and the constraint of equation \ref{eq:peak_time}, we obtain $n \sim 10^{-3} (\epsilon_B/10^{-4})^{0.44} {\rm~ cm^{-3}}$. The value of $\epsilon_B$ is poorly constrained. Allowing it to take a value somewhere within a relatively large range, $10^{-5}-10^{-1}$, corresponds to a density range of $5 \times 10^{-5} - 5 \times 10^{-3} {\rm~ cm^{-3}}$. If, instead, we use the information that the density within the ISM of the host galaxy is not expected to be much lower than $10^{-3} {\rm~ cm^{-3}}$, then the conclusion is that $\epsilon_B \lesssim 10^{-4}$.

\subsubsection{\grays}\label{sec:GW170817_grays}
The first reaction of many people to GRB 170817A, the burst of \grays  that followed the GW signal from GW170817, was that we had finally observed the long-predicted prompt emission of a short GRB that accompanies a BNS merger. However, the burst \grays from GW170817 does not look like any other sGRB seen before. Most notable is its extremely low luminosity, but also its softness and smooth light curve are atypical (although not unique). The explanation proposed for these properties was that we have observed a sGRB's emission off-axis. According to this interpretation, the \grays that we observe are generated by the core of the jet, which is off-axis, but the same emission, when seen on-axis, would be  observed as a regular sGRB.
%Namely, for us $\theta>1/\Gamma$ where $\Gamma$ is the Lorentz factor of the core of the jet, which is the source of the $\gamma$-rays, and $\theta$ is the angle between the velocity of the source and the line of sight. But, an alien observer with line of sight that is within the jet core, saw a regular sGRB. 
This explanation sounds promising at first, since due to the different Lorentz boosts, off-axis emission is fainter, softer, and less variable than on-axis emission. However, a closer look shows that this scenario can be ruled out by several lines of argument. First, if the emission we saw was a typical sGRB seen off-axis, then its isotropic equivalent energy, as seen by an on-axis observer, was $\sim 10^4-10^6$ times larger than the one we observed. This implies that the on-axis peak photon energy, $E_p$, was at least a few MeV, harder than typical sGRBs (see section \ref{sec:off_axis_prompt} and \citealt{matsumoto2019a}). 
Second, and more importantly, compactness arguments put an upper limit on the angle between the source of the \grays and the observer. \citealt{matsumoto2019a} show that if the sGRB, as it would have been seen on-axis,  was typical, with $E_{\rm iso} \sim 10^{50}$ erg, then it cannot be at an angle greater than $0.05$ rad away from our line of sight  \citep[c.f.,][]{eichler2018}. This angle is much smaller than the angle between the core of the jet and the line of sight found from the afterglow modeling, $0.25$ rad. Moreover, \citealt{matsumoto2019} find that the \gray source cannot be at angle that is larger than $0.15$ rad away, even if the on-axis emission was much fainter than that of a regular sGRB. 

The two remaining possible sources which are commonly discussed in the literature\footnote{A third possible source of the observed \grays is off-axis  scattering of photons from baryonic material, which is presumably the material caught at the front of the jet as it propagates in the ejecta \citep{eichler2018}. The baryonic material in this model is accelerating due to its interaction with the radiation, and at the time that the photons we observed were scattered, its Lorentz factor was moderate, $\Gamma \sim 20$. The scattered emission that we have seen was not observed by an on-axis observer as a sGRB. Only after the baryonic material accelerates significantly, an on-axis observer can see the sGRB. The degree of consistency of this model with the compactness limits derived by \citealt{matsumoto2019a} is still unclear.} are shock breakout (\S\ref{sec:breakout}) and emission from internal dissipation within regions of the outflow that are outside of the core of the jet (\S\ref{sec:high_inclination_grays}). In the latter scenario, we observed high-inclination emission from a structured jet, where the angle between us and the emitting region is small, $<0.15{\rm~rad}~(10^\circ)$, and its Lorentz factor and isotropic equivalent energy are much smaller than that of the jet-core \citep[e.g.,][]{ioka2019,kathirgamaraju2019}. Thus, we are either within, or not very far from, the $1/\Gamma$ beam of the observed source emission. In any case, the emission that we saw was emitted from regions different from those producing the emission that would have been seen by an on-axis observer as a sGRB. The high-inclination emission model does not specify the dissipation mechanism that is the source of the radiation, nor the emission mechanism that produces the \grays we observed. Thus, it has no specific predictions with regard to the luminosity, duration, temporal structure or spectrum of the observed emission. Hence, while this is a viable model, there is no way to test it or to make predictions regrding \gray emission from future BNS mergers. I therefore do not discuss this model further. The shock breakout model, however, does make specific predictions for the properties of the emission.  Below, I discuss the compatibility of shock breakout with the observations, and the constraints on the properties of the emitting region, if this is the correct model.\\

\noindent \underline{\textbf{Shock breakout}}\\
The model of a shock breakout from the ejecta of a binary merger is described in section \ref{sec:breakout}, and the following discussion is based on it.
A shock breakout was suggested by several authors as the source of the observed \grays \citep{kasliwal2017,gottlieb2018b,nakar2018,pozanenko2018,beloborodov2018}. The most plausible breakout is that of the shock driven by the cocoon out of the merger ejecta. A major advantage of this model is that, since the jet in GW170817 was successful, there must have been a shock breakout, and some of its general predictions are a low variability \gray flash that contains a minute fraction of the total explosion energy, and shows a hard-to-soft spectral evolution. All of these properties are seen in 
GRB 170817A. The model also predicts specific relations between the \gray signal isotropic equivalent energy, duration, photon peak energy, and delay from the time of the merger (see section \ref{sec:breakout}). Assuming that the breakout emission is dominated by the breakout layer, and under the assumptions described in section \ref{sec:breakout}, equations \ref{eq:Ebo}, \ref{eq:tbo} 
and \ref{eq:R_Dt} relate three observables: (i) the total observed isotropic equivalent energy, $E_{\rm bo}$; (ii) the duration $t_{\rm bo}$; and (iii) the time between the termination of the gravitational wave signal that marks the moment of collision and the onset of the \gray flash, $\delta t_{\scriptscriptstyle GW,\gamma}$, with three parameters of the breakout: (i) the shock Lorentz factor, $\gamma_{\rm s,bo}$; (ii) the ejecta Lorentz factor, $\gamma_{\rm e,bo}$ (both measured in the observer frame); and (iii) the breakout radius, $R_{\rm bo}$. In GRB 170817A the observables were $E_{\rm bo} \approx 3 \times 10^{46}$ erg, $t_{\rm bo} \approx 0.5$ s and $\delta t_{\scriptscriptstyle GW,\gamma} \approx 1.7$ s. The breakout layer parameters that produce these observables are
\begin{equation}\label{eq:Rbo_GW170817}
	R_{\rm bo} \approx 6 \times 10^{11} {\rm~ cm} ,
\end{equation}
\begin{equation}
	\gamma_{\rm s,bo} \approx 4 ~,
\end{equation}
and
\begin{equation}\label{eq:ge_bo_GW170817}
	\gamma_{\rm e,bo} \approx 2.5 ~.
\end{equation}
The corresponding shock velocity, as seen in the upstream frame, $\beta_{\rm s,bo}'$, is not relativistic, implying that the post shock acceleration is mild, so that $\gamma_{\rm f,bo} \approx 4.5$. The mass of the breakout layer is $\sim 3 \times 10^{-8} \msun$ and the density at the time of the breakout is $\sim 10^{-10} {\rm~g~cm}^{-3}$. The velocity of the shock in the upstream frame, $\beta_{\rm s,bo}'$, is sensitive to the exact values of $\gamma_{\rm s,bo}$ and $\gamma_{\rm e,bo}$ (see equation \ref{eq:beta_s_tag}), and those can vary slightly and still provide a theoretical prediction that is consistent with the three observables, $E_{\rm bo}$ , $t_{\rm bo}$  and $\delta t_{\scriptscriptstyle GW,\gamma}$, to within a factor of order unity. The corresponding range of the shock velocity, which is consistent with these observables, is\footnote{For example, according to equations \ref{eq:Ebo}, \ref{eq:tbo} 
and \ref{eq:R_Dt},  $R_{bo} = 5 \times 10^{11} {\rm~ cm}$, $\gamma_{s,bo}=5$ and $\gamma_{e,bo}=2.4$, which correspond to $\beta_{s,bo}'=0.65$,   predict $E_{bo} \approx 5 \times 10^{46}$ erg, $t_{bo} \approx 0.2$ s and $\delta t_{\scriptscriptstyle GW,\gamma} \approx 1.7$ s. On the other hand the breakout parameters  $R_{bo} = 8 \times 10^{11} {\rm~ cm}$, $\gamma_{s,bo}=3.7$ and $\gamma_{e,bo}=2.9$, which correspond to $\beta_{s,bo}'=0.25$,   predict $E_{bo} \approx 2 \times 10^{46}$ erg, $t_{bo} \approx 0.9$ s and $\delta t_{\scriptscriptstyle GW,\gamma} \approx 1.7$ s. } $\beta_{s,bo}' \approx 0.25-0.65$. 
As I show below, in this velocity range there is always a solution that fits the observed temperature, $T_{bo}$, which is a fourth observable.

The peak energy of the observed spectrum during the peak of the luminosity is $E_p=520^{+310}_{-290}$ keV, and when the spectrum is integrated over the entire initial pulse, its peak energy is $E_p=185\pm 62$ \citep{goldstein2017,veres2018}. Thus, the corresponding observed temperature is $T_{bo} \sim 100$ keV. For a given shock velocity and upstream density, the post shock temperature can be calculated using equations \ref{eq:dot_nff}-\ref{eq:Td}. For the range of velocities obtained above, $\beta_{s,bo}' \approx 0.25-0.65$, pair production is marginal and, therefore, the observed temperature is expected to depend sensitively on the details, particularly on the exact breakout velocity and ejecta composition\footnote{The bulk of the ejecta is \rp rich, but in the  fastest parts of the ejecta, where the breakout takes place,  the composition may be dominated by light elements and possibly also by free neutrons. Here I discuss only the effect of light elements on the temperature, and ignore the (unknown) possible effect of free neutron.} (see Fig 8 and the discussion in \S\ref{sec:breakout}). At the upper end of values for the shock velocity ($\beta_{s,bo}' \approx 0.65$), the downstream temperature is regulated by pair creation for any composition, and the radiation will be released only after the pairs annihilate as the plasma cools to a rest-frame temperature of 50 keV, that translates to an observed temperature  $T_{bo}\sim 200$ keV (after accounting for the Lorentz boost). At the lowest possible value of the shock velocity ($\beta_{s,bo}' \approx 0.25$), pairs are not produced and the temperature depends on the composition. For an upstream density of $10^{-10} {\rm~g~cm}^{-3}$, the restframe temperature is $\sim 1$keV for \rp rich material and $\sim 10$keV for H-rich ejecta, corresponding to observed temperatures of $T_{bo}\sim 4$ keV and $T_{bo}\sim 40$ keV, respectively. Thus, for every ejecta composition, there is a shock velocity in the range $\beta_{s,bo}' \approx 0.25-0.65$ that predicts an observed temperature that is consistent with the one that was observed. In particular, there are three shock breakout parameters ($R_{bo}$, $\gamma_{s,bo}$, $\gamma_{e,bo}$), that provide a theoretical prediction that is consistent with the four main observables ($E_{bo}$ , $t_{bo}$, $\delta t_{\scriptscriptstyle GW,\gamma}$, $T_{bo}$), to within a factor of order unity.

The agreement of the emission from the breakout layer with the observations implies that the contribution from deeper layers during the planar phase cannot be much larger than that of the breakout layer. A planar phase that contributes a factor of a few more than the breakout layer is still consistent with the observations. In that case, the Lorentz factor at the time of the breakout is $\gamma_{s,bo} \approx 3$, and the ejecta maximal Lorentz factor is $\gamma_{e,bo} \approx 1.5$.

To conclude, the shock breakout model has only limited flexibility (it depends on a small number of well defined physical parameters) and it makes specific predictions regarding the relations between the observables, the general light curve structure, and the spectral evolution. 
Considering this, the agreement of the shock-breakout model with the properties of GRB170817A is remarkable. The model, which is over-constrained, can explain the main properties of GRB170817A well and the observed \grays show all the characteristics of shock breakout emission, most of which were discussed already in \cite{nakar2012}. These include low energy compared to that of the explosion, a smooth light curve, hard-to-soft evolution, a delay between the GW signal and the $\gamma$-rays satisfying the closure relation, and emission over a wide angle (see the summary of section \ref{sec:breakout}).

\subsection{{\bf The fate of the remnant and the equation of state}}
%\noindent{{\underline {\it The fate of the remnant}}}\\
\subsubsection{The fate of the remnant}\label{sec:remnant}
We cannot be certain what was the evolution of the central compact object following the merger. The most popular scenario  is that the merger formed at first a HMNS (a hypermassive NS supported by differential rotation; see definitions at \S\ref{sec:massEjection}) that collapsed within $\sim 1-10$ s to a BH \citep{margalit2017,shibata2017,rezzolla2018,ruiz2018}. This scenario is based on the EM emission only, since the GW signal does not provide much constraining information on the post-merger evolution. Each stage in this chain of events is supported by a different property of the EM counterpart, and they are all limited by our limited understanding of the mass-ejection processes, as I briefly discuss below.
\begin{itemize}
\item The first, and most robust evidence is the large amount of sub-relativistic ejecta, which indicates that the central object did not collapse promptly to a BH. In all GRMHD simulations of BNS mergers, a prompt collapse to a BH (within $\sim 1$ ms) leads to an ejecta mass that is about two orders of magnitude smaller than the one inferred from the EM counterpart. It seems that the central NS must have survived for at least  $\sim 10$ ms in order for there to be significant dynamical ejecta and a disk massive enough to explain the observed ejecta mass (see \S\ref{sec:BNSejecta}). 

\item The second clue is that most of the ejecta was not lanthanide-rich. This is more certain for the fast component, but also the slow component seems to contain a large fraction with high $\Ye$ ejecta. The implication is that a significant part of the outflow was irradiated by neutrinos, which in turn, suggests a central NS at the time that this mass was ejected (see \S\ref{sec:BNSejecta}). Since about half of the disk wind mass is ejected within the first second, this argues that the central object did not collapse to a BH for at least some fraction of a second.

\item The total energy of the ejecta places another constraint on the remnant. The energy of the sub-relativistic ejecta is $\sim 10^{51}$ erg and that of the relativistic jet is lower. This sets a limit on the energy that the central object can deposit in the ejecta. Since the remnant carries a rotational energy of $\sim 10^{53}$ erg following the merger, it can deposit less than $1\%$ of it in the ejecta. A magnetized NS with a dipole magnetic field $B_d$ and a period $P$ loses its rotational energy via magnetized winds roughly at a luminosity of $L \sim 10^{50} (B_d/10^{15}G)^{-2} (P/1{\rm~ms})^{-4}~$erg s$^{-1}$ \citep{spitkovsky2006}. Since the energy deposited in the ejecta over a time scale of at least a few weeks is lower than $10^{51}$ erg, this implies that either the remnant collapsed to a BH before it was able to deposit its rotational energy, or that the magnetic field was lower than $\sim 10^{12}$ G. This upper limit on the energy released as a magnetized wind is  supported by the upper limit on the X-ray emission after 100 days, since on this time scale, a fraction of $\sim 10^{-3}$ of the magnetized wind energy is expected to be radiated in X-rays \citep{pooley2018}. The strength of the remnant dipole field before its collapses is unknown, but during the merger, small-scale fields with a strength of $\sim 10^{15}-10^{16}$ G must be formed. Since we know that magnetars with $B_d \sim 10^{15}$G are formed following the core collapse of some massive stars, it is reasonable to expect that a magnetar with similar dipole field is formed also following a merger, where the rotation is the fastest possible and the small scale field is extremely strong. If this was the case, then the central object must have collapsed to a BH within less than about 10s, to avoid injecting too much energy into the ejecta. In addition, since the NS collapsed before losing a significant fraction of its rotational energy, it was most likely a HMNS (supported by differential rotation) and not a SMNS (supported by rigid rotation).

\item The last argument favoring a collapse to a BH within 1s after the merger is the delayed gamma-ray signal. The gamma-ray emission requires the launch of a relativistic jet within less than 1.7 s. The two main models for the engine driving this jet are accretion onto a BH, or a magnetar with $B_d \sim 10^{15}$G. The former obviously requires a collapse within less than 1.7 s, while the latter must collapse within less than about 10 s (see above).
Note that the time delay in the observed gamma-ray signal does not necessarily reflect a delay between the merger and the jet launching, as the jet could have been launched at any time between the merger and the time at which the gamma-rays were observed (see \S\ref{sec:grays}). 

\end{itemize}

To conclude, it is almost certain that the merger formed, initially, a NS that did not collapse to a BH for at least 10 ms, and most likely for much longer. It is likely that the NS did collapse, eventually, to a BH, where the most natural (but highly uncertain) collapse time is $\sim 1-10$ s after the merger.

%\noindent{{\underline {\it Equation of state}}}\\
\subsubsection{Equation of state}\label{sec:GW170817_EOS}
The EOS is constrained both by direct limits set by the GW signal on the deformability, and by the indirect EM-counterpart constraints on the fate of the remnant (see above). These are typically used to constrain the NS radius and/or the mass at which a non-rotating NS collapses to a BH, $m_{TOV}$, which in turn sets constraints on the EOS.

The GW signal sets an upper limit on the deformability and thus on the binary members' radii. The initial limit on the deformability was $\tilde{\Lambda} \lesssim 800$ \citep{abbott2017GW}.  This sets an upper limit on the radius of about $13.5$ km \citep{raithel2018,annala2018,most2018,fattoyev2018}, implying that the EOS cannot be too stiff. A more detailed analysis of the GW signal constrained the deformability to $\tilde{\Lambda}=300^{+420}_{-230}$ (90\% highest posterior density interval) when low spins are assumed \citep{abbott2019Binary}. When this range is combined with  the lower limit on the mass of some Galactic NSs (about $2\msun$) the radius of the two NS is constrained to $R=11.9^{+1.4}_{-1.4}\msun$ \citep{abbott2018EOS}. 
 
The EM signal requires no direct collapse to a BH, which implies that the EOS cannot be too soft. This has been used by various authors  \citep{radice2018b,radice2019,kiuchi2019,margalit2017,shibata2017,bauswein2017,rezzolla2018,ruiz2018,coughlin2018} to put a lower limit on the NS deformability, $\tilde{\Lambda} \gtrsim 300$, on the NS radius, $R \gtrsim 11$ km, and on the maximal mass of a non-rotating NS. The most recent and tightest constraints were obtained by \cite{capano2019} who, by combining the GW and EM information, conclude that the radius of a  $1.4\msun$ NS should be $R=11^{+0.9}_{-0.6}\msun$.

\subsection{ {\bf  Implication for the origin of \rpe in the Universe}}\label{sec:GW170817_rpe}
Before GW170817 there was no direct evidence for any astrophysical object that produced \rp elements, and several different phenomena were discussed as possible \rp sources. In recent years, the idea that BNS and/or BH-NS mergers are the main \rp sources in the Universe has gained popularity, after many years that core-collapse SNe were the favored option. The main reason for the shift in opinion was the realization that \rp sources must be rare, and that the yield per event must be high (\S\ref{sec:rp_rate}). This source characterization fits compact binary mergers well but is inconsistent with normal ccSNe. The alternative to compact binary mergers was that the sources are rare types of SNe, such as magneto-rotational SNe or collapsars. The main advantage of BNS mergers was that their merger rate was constrained from Galactic binaries, and it was relatively certain that the ejecta must be composed almost entirely from \rp elements. In contrast, the rate of some SNe types is unknown (e.g., magneto-rotational) and for all SNe types the yield of \rpe is highly uncertain. The main argument against mergers was that the chemical evolution of \rpe in the Milky Way may require the \rp source to take place shortly after star formation, while most models of BNS mergers predict a significant delay between the formation of the binary and the merger. However, this argument was not conclusive for two reasons. First, Galactic chemical evolution is a complex process, and at least according to some models, the observations are fully consistent with the predicted delay-time distribution of BNS mergers (see \S\ref{sec:rp_rate} and for a more detailed discussion \citealt{cowan2019}). Second, while most theoretical models of binary mergers predict a significant delay, there is considerable theoretical uncertainty and no direct observational measurements of the delay-time distribution. Thus,
while observations of Galactic BNS systems show that there must be a population with a long delay, it is certainly possible that there is also a significant population of mergers with a short delay. Moreover, the observed Galactic population even seems to support the existence of such a population \citep{beniamini2019a}. Thus, two central questions, whose answers awaited the detection of simultaneous GW-EM emission from BNS mergers, were whether BNS mergers indeed produce enough \rpe to account for the abundances seen in the Universe, and whether they produce both light and heavy \rp elements, or rather just one of those components.

\begin{figure}
	\center
	\includegraphics[width=0.7\textwidth]{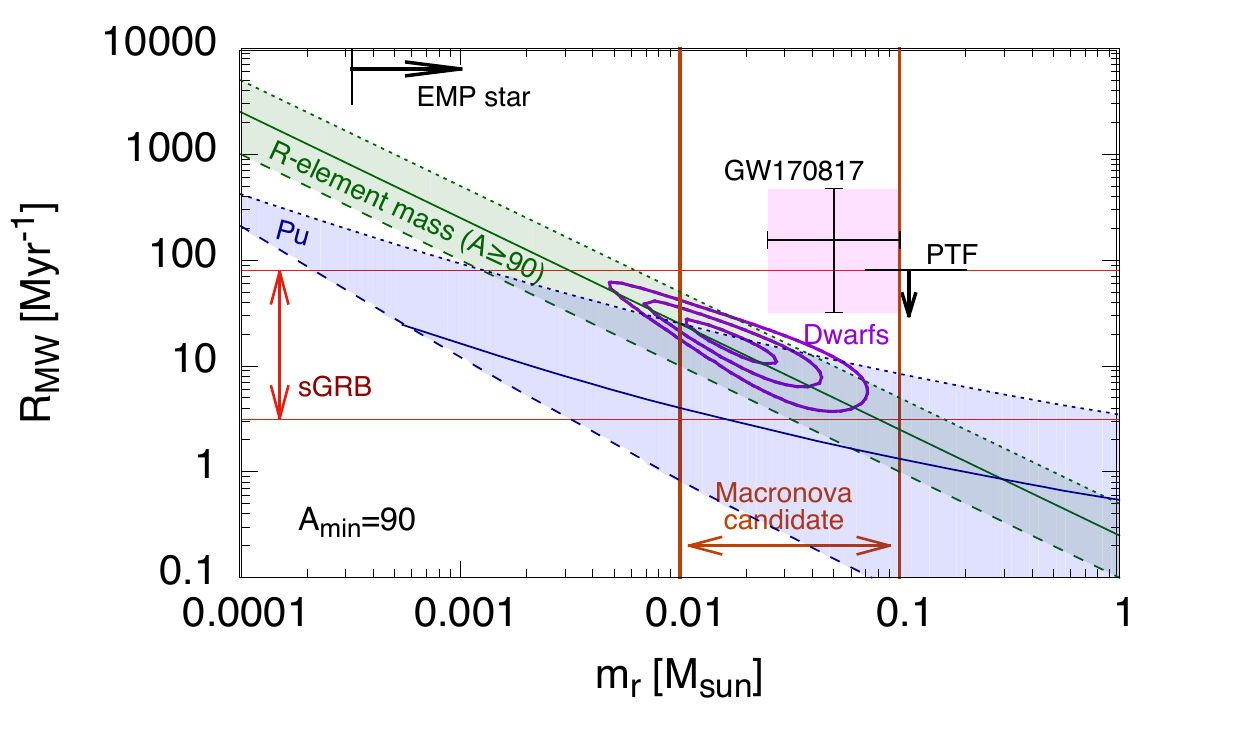}
	\caption{A comparison between the production of \rpe by BNS mergers, as inferred from GW170817, and constraints from various measurements on the Galactic rate of \rp events and on the mass produced per event.
Both GW170817 and Galactic sources are assumed to produce \rpe with the solar abundance pattern for $A \geq 90$.   From \citet{hotokezaka2018a}. }%
	\label{fig:rp_rate}
\end{figure}

GW170817 is the first astrophysical event where there is robust evidence for the production of \rp elements. Moreover, as I show below, its observations suggest that BNS mergers are major, and probably even the dominant, source of \rpe in the Universe \citep[e.g.][]{hotokezaka2018a,rosswog2018}.  The ejecta of GW170817 contained, most likely, both light and heavy \rp elements, although their ratio is not necessarily similar to the solar one (\S\ref{sec:GW170817_ejecta_property}). This single event can be translated (with considerable uncertainty) into 
a rate of total \rp production by BNS mergers in the local Universe. The mass of \rpe ejected in GW170817 is $\sim 0.05\msun$ and the volumetric rate of BNS mergers as implied by this single event is ${\cal R}_{BNS}=75-7100 \,{\rm Gpc^{-3}\,yr^{-1}}$ at 90\% confidence (\S\ref{sec:merger_rate}). Assuming that the ejecta from GW170817 is typical, this then translates to a volumetric \rp production rate of  $4-350 \,{\rm \msun~ Gpc^{-3}\,yr^{-1}}$. Assuming a Milky-Way-like galaxy density of $0.01~ {\rm Mpc^{-3}}$, the implied Galactic rate is $0.4-35 ~\msun/{\rm Myr}$. This can be compared to the estimated average \rp production rate in the Milky Way of $\sim 1 ~\msun/{\rm Myr}$ (\S\ref{sec:rp_rate}). The agreement is remarkable.

Observations provide not only an estimate of the total \rp production rate, but also constraints on the rate of individual events and their average \rp yield per event. This comes about from several different types of observations, including the abundance of radioactive \rp elements in meteorites and in the deep sea floor, and \rp abundances in extremely metal-poor Galactic stars and in ultra-faint dwarf galaxies. All of these various methods provide similar results. The rate of \rp sources is lower by a factor of $\sim 1000$ than the rate of ccSNe (a Galactic \rp event rate of $\sim 30 \,{\rm Myr^{-1}}$), and the yield of each event is $\sim 0.05\msun$ (\S\ref{sec:rp_rate}). The observations of GW170817 are fully consistent with these constraints. Figure \ref{fig:rp_rate} (from \citealt{hotokezaka2018a}) shows a comparison of the constraints obtained by various methods on the rate of \rp events, and on the mass of \rp material produced per event, with the estimates of BNS mergers based on GW170817. These results show that, even though there is considerable uncertainty in each of the steps above, it is highly likely that BNS mergers are major sources of \rpe in the Universe, and it is plausible that they are the dominant sources. In fact, these results suggest that unless the BNS rate is near the low end allowed by the Poisson uncertainty on the rate from a single event, then in order not to over-produce the \rp material in the universe, the total ejected mass in GW170817 must be higher than typical.

To conclude, following GW170817, BNS mergers are the only confirmed sources of \rp elements. They are almost certain to produce at least a significant fraction of the \rpe in the Universe, and are likely to be the dominant \rp sources. Significant contributions of other sources, such as various types of rare SNe \citep[e.g.,][]{nishimura2017,siegel2019}, are still possible, but at the current stage, their contributions are only postulated and their levels are highly uncertain.

\section{Compact binary mergers and short gamma-ray bursts}\label{sec:sGRBs}
\subsection{\textbf{ A brief historical account}}
The connection between GRBs and BNS and/or BH-NS mergers was first predicted\footnote{Several earlier papers briefly mentioned the possibility that neutron-star mergers are related to GRBs \citep{blinnikov1984,goodman1986,goodman1987}.} by \cite{eichler1989}. This paper, written at a time when most of the community believed that GRBs were Galactic, already included most of the ingredients that turned out to be relevant three decades later, such as a prediction that a significant fraction of the \rpe in the Universe originate from mergers, and that BNS mergers generate relativistic outflows that are the sources of a sub-class of the observed gamma-ray bursts. 
This idea gained popularity several years later, when BATSE\citep{meegan1992} and then BeppoSax data \citep{vanParadijs1997} established the cosmological nature of GRBs. The leading alternative GRB model, suggested by \cite{woosley1993}, was the collapsar model, in which the collapse of a massive star leads to the formation of an accretion disk around a newly formed BH (in the original model the GRB was a result of a failed SN but, as was learned from observations a decade later, the SNe that accompany long GRBs are actually quite successful). Soon after, the BATSE GRB duration distribution suggested the existence of (at least) two separate GRB sub-classes, "short" and "long"\footnote{Hints for the separation of GRBs into two distinct populations was found already in earlier data \citep{mazets1981,norris1984}.} \citep{kouveliotou1993}. This led to the theoretical expectation, based on the expected life-time of the accretion disk, that long GRBs (lGRBs) are associated with the collapse of massive stars, while short GRBs (sGRBs) are associated with compact binary mergers. The association of lGRBs with the collapse of massive stars was confirmed within a decade. At first, there were only inconclusive clues, such as the association of lGRBs with star-forming regions, and the appearance of a red bump in the afterglow on a time scale of a week or two, which suggested an underlying SN. The final confirmation came with the secure spectroscopic association of a broad-lined Type Ic SN (2003bh) with GRB 030329 \citep{hjorth2003,stanek2003}. Additional reading on the association of lGRBs with SNe can be found in\cite{woosley2006} and references therein.

Progress in understanding the origin of sGRBs required some more time. sGRBs are observed less frequently than lGRBs (1/4 in BATSE and 1/10 in {\it Swift}), and are much harder to localize, due to their lower fluence and fainter afterglow. It is therefore much harder to identify their afterglows, and without an afterglow there can be no redshift measurement and no host identification. The first sGRB afterglows were observed only in 2005 \citep{gehrels2005,fox2005,berger2005}, about 8 years after the first detection of lGRB afterglows. These observations confirmed that sGRBs and lGRBs are two separate astrophysical phenomena. The most evident property of sGRBs, that was revealed by the detection of their afterglows, was that sGRBs are not directly related to star formation. sGRBs were found in galaxies of all types, including blue galaxies with active star formation and red elliptical galaxies with a very low star-formation rate. In addition, sGRBs have shown no evidence of association with SNe of any known type, and their location within the host shows a wide range of offsets with respect to the galactic nucleus. All of these observations were fully consistent with the predicted properties of BNS and BH-NS mergers, and therefore supported the possible association. In addition, the volumetric rate of sGRBs was measured and found to be consistent with the (rather uncertain) estimates of BNS merger rates, which were based on Galactic BNS systems. The most recent support for the association of sGRBs with compact binary mergers was the identification of an IR excess in the late afterglow of GRB 130603B, which was interpreted as possible macronova emission \citep{tanvir2013,berger2013}. Following this report, there were several other candidates for macronova emission in the light curves of sGRB afterglows (see \S\ref{sec:sgrb_macronova}). However, all of these cases, while suggestive, can be considered as either circumstantial (e.g., host types, offsets, rates, etc.) or highly uncertain (e.g., the identification of a single IR measurement in a variable sGRB afterglow as a macronova is not conclusive). Thus, before GW170817, the origin of sGRBs was an open question with a favored, but uncertain, solution.
For more details on sGRBs and evidence of their association with compact binary mergers (before GW170817), I refer the reader to the comprehensive reviews on sGRBs by  \cite{nakar2007}, \cite{lee2007}, \cite{berger2014}, and references therein.

\subsection{\textbf{Macronova signatures in sGRB afterglows}}\label{sec:sgrb_macronova}
The afterglow emission from an on-axis sGRB jet is blue, and typically is already fading within minutes after the burst. The macronova emission, in contrast, is red, and rises in the IR on a time scale of a day to a week. The optical/IR luminosities of the afterglow and the macronova on a timescale of a week are comparable. Thus, if sGRBs are compact binary mergers, it may be possible to identify the macronova emission in their afterglows. This is not an easy task, as it requires a model for the jet emission that can be subtracted from the data to reveal the macronova contribution. However, many afterglows (of both short and long GRBs) show variability in the jet emission, which is unaccounted for by the standard model. Moreover, sometimes this variability is chromatic and seen in one band, but not in another. Thus, it is almost impossible to be certain, based on photometry in one or two bands alone, that a deviation from the model is due to a genuine macronova, rather than the result of some variable jet emission. Nevertheless, identification of macronova candidates in sGRB afterglows provides some support for the link between sGRBs and compact binary mergers. Moreover, it provides information (even if only  upper limits) on the mass ejection from a large sample of mergers, under the assumption that they are the progenitors of the sGRBs. Finally, the EM counterpart of GW170817 provided strong support, almost a proof, of the connection between BNS mergers and sGRBs (see below), suggesting that at least some of the macronova detections in sGRB afterglows are real. 

The first to suggest that irregularities in the light curve of a sGRB afterglow are associated with macronova emission were \cite{perley2009}, based on the peculiar afterglow of GRB080503. However, the afterglow of this burst showed also an X-ray flare, roughly at the same time that the optical light curve peaked, and therefore \cite{perley2009} concluded that jet emission is most consistent with the data, rather than \rpe radioactive decay, which predicts no X-ray emission. The next macronova candidate was identified in the afterglow of GRB130603B as an IR excess over the jet emission model \citep{tanvir2013,berger2013}. The IR excess was seen only in a single band (F160W) at a single epoch  (7 days after the burst, as measured in the GRB rest frame). Interestingly, this burst also exhibited a simultaneous X-ray excess, casting some doubt on the interpretation of the IR excess as a macronova powered by radioactive decay of r-process elements. Nevertheless, due to the significant reddening of the afterglow, which was predicted by the improved macronova models that were available at that time \citep{tanaka2013,kasen2013}, GRB130613B was considered a more-reliable macronova candidate. The IR emission from GRB130613B was brighter than that of GW170817 by about a factor of 3 at the corresponding age, in the same band. Modeling of the single IR data point suggested that, if it were powered by radioactive decay of \rpe, then the ejecta mass would be $\sim 0.1\msun$ \citep{tanvir2013,berger2013}.

Following GRB130613B, a re-examination of archival sGRB afterglows led to the identification of two additional macronova candidates, GRB050709 \citep{jin2016} and GRB060614 \citep{jin2015,yang2015}. The estimated ejecta mass, in both events, under the assumption that the IR excess is due to radioactive decay of \rpe, was $\sim 0.05-0.1\msun$. Most interestingly, 
GRB050709 also shows a clear X-ray flare, roughly simultaneous with the IR excess  \citep{fox2005,jin2015}. This flare, which is seen at day 16, shows variability on a time scale of hours and therefore cannot be explained by emission from the blast wave that the jet drives into the circum-burst material. The X-ray afterglow of GRB060614 shows an unexpected flattening after $10^6$s \citep{mangano2007}. This flattening is seen at the end of the XRT observation, when the flux is close to the detection threshold, and it may be the result of a constant, unrelated, underlying X-ray source such an AGN, although the host-galaxy spectrum shows no indication for AGN activity \citep{gal-yam2006}. Alternatively, it may be related to the afterglow emission, similarly to the three other sGRBs with macronova candidates, discussed above.
The X-ray excesses that are seen in at least three, and possibly all four, of these macronova candidates, elucidate how difficult it is to obtain a confident identification of the source of the IR emission. The X-ray excess and the IR excess seen simultaneously in all of these bursts may be unrelated coincidences, but they may also well be closely connected, in which case the IR emission is most likely not powered by radioactive decay. For example, an alternative explanation to the IR light, supported by the comparable energy seen in the IR and in the X-rays, is that the IR excess is powered by the enhanced X-ray emission. In this model, we see the central X-ray source directly through the opening in the ejecta that was cleared out by the GRB jet, while the emission in all other directions is absorbed by the ejecta and reprocessed, before being emitted quasi-isotropically in the optical/IR \citep{kisaka2016}.

Following GW170817, several papers compared its macronova emission to archival sGRB afterglows. This search revealed additional macronova candidates. One candidate is GRB160821B \citep{lamb2019a,troja2019}, which is fainter than GW170817 by a factor of about 3. The inferred ejecta mass, assuming a radioactively powered macronova, is $\sim 0.01\msun$. If the emission is not powered by radioactive decay, then this is an upper limit on the mass of \rp material ejected during the event. Similarly to the other macronova candidates, the X-ray emission in this burst shows deviations from a simple power-law decay, where near the time of the optical/IR excess there is a significant flattening, or even a re-brightening, of the X-ray emission \citep{lamb2019a}.  An additional candidate is GRB 150101B \citep{troja2018a}. The optical emission after 2 days is brighter than GW170817 by a factor of about 2, and is interpreted by \cite{troja2018a} as macronova emission. However, this identification is based on only two detections in the $r$-band, and almost no optica/IR color information, and it is therefore highly uncertain.
\cite{gompertz2018} and \cite{rossi2019} have compared the optical/IR emission from GW170817 with a large sample of sGRB afterglows. They find that many afterglows are brighter than GW170817. However, there are about half a dozen sGRBs with afterglows that are fainter than GW170817, some by at least a factor of 5. 

To conclude, there are currently six sGRB afterglows in which an optical/IR excess was interpreted as possible macronova emission. The ejecta mass inferred, assuming it is macronova emission, is in the range $0.01-0.1\msun$. The identification of these candidates as powered by radioactive decay of \rpe became much more credible following the observation of GW170817, but it is still not fully secure (and in some events, it is quite uncertain). This is especially true, given that most of these candidates exhibit an X-ray excess roughly at the same time that the optical/IR excess is seen. In addition to macronova candidates, there are also a handful of sGRB afterglows with deep limits on the optical/IR luminosity on a time scale of days. In these sGRBs, any macronova emission, if present, must be significantly fainter than that in GW170817. This implies that, if sGRBs are BNS mergers, then there is significant variance, of at least one order of magnitude, in the macronova luminosity, which most likely reflects a corresponding variance in the amount of mass ejected during the mergers.

\subsection{\textbf{ GW170817 and sGRBs}}\label{sec:GW170817_sGRB}
\begin{figure}
	\center
	\includegraphics[width=0.7\textwidth]{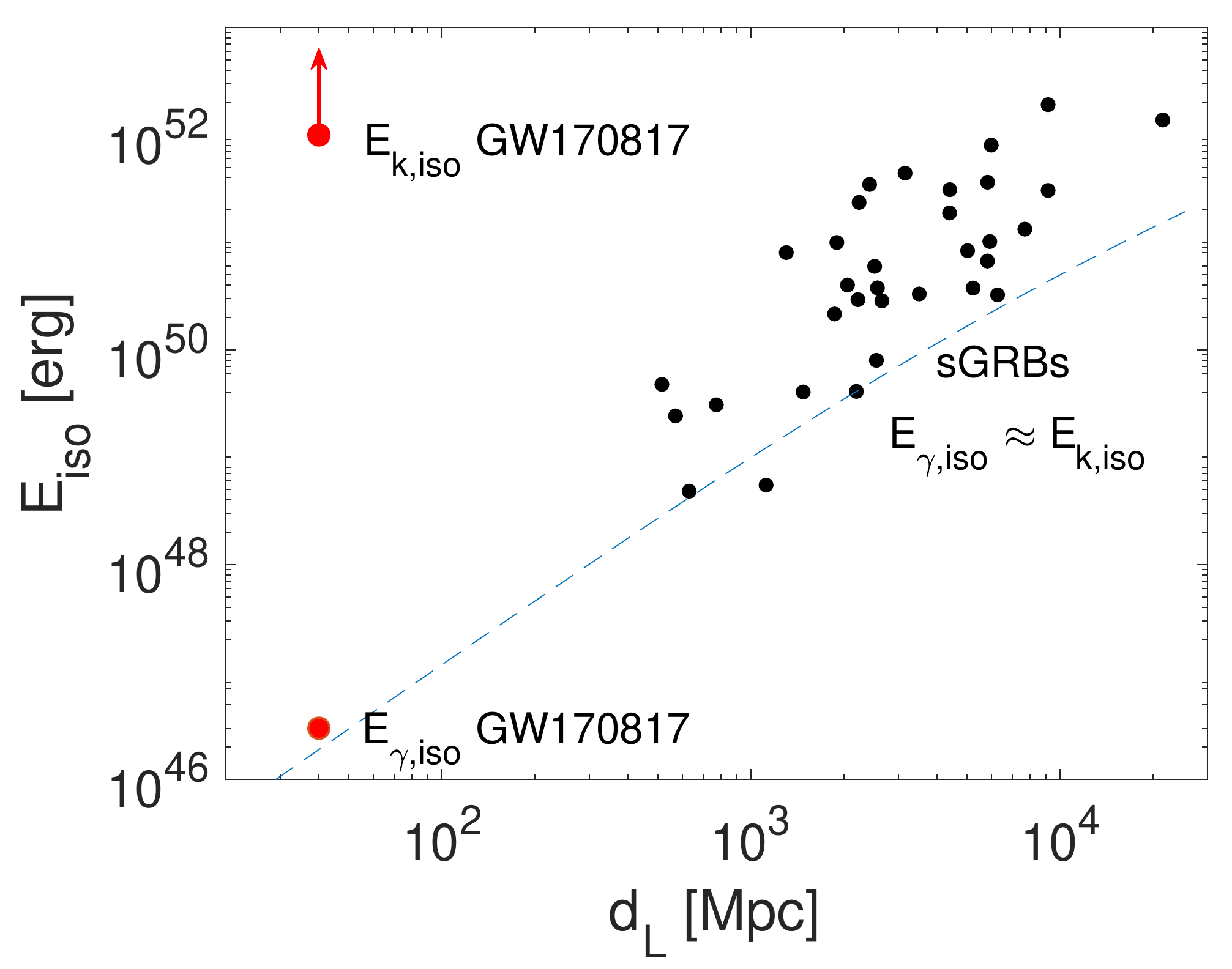}
	\caption{Isotropic equivalent \gray energy of sGRBs with known redshifts (black circles), compared to the isotropic equivalent \gray energy and the jet kinetic energy of GW170817 (red circles). The sGRB redshifts are taken from \cite{fong2015} (and references therein) and the fluences at $15-150$keV are from the Swift archive. I have applied a constant bolometric correction, $k_{bol}=5$, to account for emission outside of Swift's spectral window. For sGRBs, the isotropic equivalent \gray energy is typically comparable to the kinetic energy of the jet (to within an order of magnitude). The lower limit on the isotropic equivalent kinetic energy of the jet is based on the constraints obtained by the VLBI observations (\citealt{mooley2018b} and \S\ref{sec:GW170817relativistic}). The dashed line marks an observed fluence of $10^{-7} {\rm~erg~cm^{-2}}$, which is roughly the \gray detection threshold.
	}\label{fig:E_sGRB}
\end{figure}

Figure \ref{fig:E_sGRB} shows the isotropic equivalent \gray energy, $E_{\rm \gamma,iso}$, of sGRBs with known redshifts, as a function of their luminosity distance, $d_L$. The \gray efficiency of sGRBs is high, but not too high, so $E_{\rm \gamma,iso}$ is a good approximation to the isotropic equivalent kinetic energy of the jet as inferred from the afterglow, $E_{\rm k,iso}$ \citep{nakar2007,fong2015}. Figure \ref{fig:E_sGRB} shows also $E_{\rm \gamma,iso}$ of GRB170817A \citep{goldstein2017} and $E_{\rm k,iso}$ in the core of the jet of GW170817, as inferred from the afterglow observations \citep{mooley2018b}. This figure illustrates several interesting points on the nature of sGRBs, and about the relation between GW170817 and sGRBs, as I discuss below. 

The observed sGRB sample shows a clear relation between $E_{\rm \gamma,iso}$ and the distance to the burst, roughly $E_{\gamma,iso} \propto d_L^2$. As far as we can tell, there is no evidence for evolution with redshift of the intrinsic properties of sGRBs, and the observed relation is purely an observational effect, reflecting the fact that this is a flux limited sample. The dashed line in figure \ref{fig:E_sGRB} marks the energy of a burst with an observed fluence of $10^{-7} {\rm~erg~cm^{-2}}$.  This fluence is roughly (to within an orer of magnitude) at the detection threshold of GRBs by {\it Swift} and Fermi-GBM\footnote{The actual detection thresholds of Swift and Fermi are different by a factor of a few, and depends on the luminosity and the duration in a non-trivial way.}. Thus, bursts below this line cannot be detected, which explains why there are no observed low-energy bursts at large distances. The explanation for the lack of nearby high-energy bursts is also observational. The volumetric rate of sGRBs with $E_{\rm \gamma,iso} \sim 10^{49}$ erg is much higher than the rate of sGRBs with $E_{\rm \gamma,iso} \sim 10^{52}$ erg. Thus, at a distance $\lesssim 1$Gpc, where the volume is relatively small, the probability to see a burst with $E_{\rm \gamma,iso} \sim 10^{52}$ erg is too low. This property of sGRBs was found in a large number of papers that studied the sGRB luminosity function\footnote{There is a single paper, \cite{ghirlanda2016} , that argues that the luminosity function of sGRBs peaks around $10^{52} {\rm~ erg~s^{-1}}$, and that per unit volume, there are more sGRBs with peak luminosity of $\sim 10^{52}  {\rm~ erg~s^{-1}}$ than sGRBs with $\sim 10^{50} {\rm~ erg~s^{-1}}$. A clear prediction of this claim is that, also at low redshifts, the observed population will be dominated by bright bursts. This prediction is clearly inconsistent with the observations.}  \cite[e.g.,][]{nakar2006,guetta2005,guetta2006,dietz2011,petrillo2013,dAvanzo2014,wanderman2015}. All of these studies find that the volumetric rate of sGRBs decreases monotonically with the \gray peak luminosity, at least in bursts with peak luminosity\footnote{The papers that derive the luminosity function consider the  peak luminosity, since most of the triggering criteria of \gray detectors are based on peak luminosity.  The typical durations of  sGRBs are $0.1-1$s, and thus the peak luminosity of a given burst is larger by a factor of roughly 1-10 than the isotropic equivalent \gray energy, when both are measured in c.g.s. units.} larger than  $\sim 10^{49}-10^{50} {\rm~erg~s^{-1}}$.

Figure \ref{fig:E_sGRB} also shows how unique was the \gray emission that we observed from GW170817. This emission was discussed at length in section \ref{sec:GW170817relativistic}, where it was shown that the observed \grays were not simply from a sGRB seen off-axis, and that if the core of the jet in GW170817 did produce a sGRB, then the jet was pointed away from us, and what we saw was not the emission of the jet itself. Note that the fact that GW170817 lies on the same energy-distance relation as sGRBs, $E_{\rm \gamma,iso} \propto d_L^2$, is, again, an observational selection effect,  due to the \grays from GW170817 being not much brighter than the GBM detection limit. 
Thus, quite surprisingly, the observed \grays from GW170817 were not the conclusive evidence for the connection between BNS mergers and sGRBs. The strongest evidence for this connection came from the radio VLBI observations that imaged the emission from a  successful relativistic jet with energy and opening angle that appear to be fully consistent with those observed in sGRBs.  This is not a "smoking gun", since we do not know if this jet produced a sGRB that was pointed away from us, but it is the best evidence, by far, in support of BNS mergers as the origin of sGRBs.  

\subsubsection{Anti-correlation between $E_{\gamma,\rm iso}$ and $\theta_j$ in sGRBs}\label{sec:GW170817_sGRBjet}
Interestingly, if the core of the jet in GW170817 produced a typical sGRB, then its isotropic equivalent kinetic energy is surprisingly high\footnote{It is often claimed that the isotropic equivalent energy of the jet seen in GW170817 is similar to that of typical sGRBs. This claim is based on a comparison to the observed sample of sGRBs (see figure \ref{fig:E_sGRB}).  It does not take into consideration the fact that the observed sample is shaped mostly by selection effects and that there is a vast difference between the volume-limited sample and the flux-limited sample.}. The identification of the host and thus the derivation of the distance to GW170817 were triggered by its GW signal. Had the \gray emission been detected with no GW detection, the afterglow of GW170817 would most likely never have been detected\footnote{The macronova might have been detected by  a wide field-of-view optical surveys, such as ZTF, also without the GW alert. However, such surveys have been running only for the past several years. The afterglow, which was detectable for only ten days after the GRB, would not have been detected by the usual GRB-afterglow searches.}. On the other hand, the GW signal and the following macronova would have been detected,  regardless of the jet emission. Therefore, it is reasonable to expect that the energy in the jet of GW170817  \textit{is} typical for a volume-limited sample of BNS mergers. In sGRBs, $E_{\rm \gamma,iso} \sim E_{\rm k,iso}$, and therefore if GW170817 produced a typical sGRB that pointed away from us, then this GRB had  $E_{\gamma,\rm iso} \gtrsim 10^{52}$ erg, and a corresponding peak luminosity $L_{\rm p,iso} > 10^{52} {\rm~ erg~s^{-1}}$. However,   according to studies of the luminosity function, and as evident from figure \ref{fig:E_sGRB}, bursts with $E_{\rm \gamma,iso} > 10^{52}$ erg and  $L_{\rm p,iso} > 10^{52} {\rm~ erg~s^{-1}}$ are extremely rare in a volume-limited sample. Almost all the studies of the luminosity function  find that only fewer  than about $1\%$ of sGRBs have $L_{\rm p,iso} > 10^{52} {\rm~ erg~s^{-1}}$ \citep{nakar2006,guetta2005,guetta2006,dietz2011,petrillo2013,dAvanzo2014,wanderman2015}. How does this low fraction of energetic sGRB jets fit with the high-energy jet seen in GW170817?

There are several possible explanations for this apparent discrepancy. One possibility is that we have been lucky and the only merger that took place during O2 within the  LIGO/Virgo detection horizon had an extremely rare and powerful jet. Alternatively, it is possible that the \gray efficiency of sGRBs is much lower than our estimates, and $E_{\gamma,\rm iso} \ll E_{\rm k,iso}$. Yet another possibility, which I consider the most likely and also the most interesting, is that the luminosity function that we derived based on the observed sGRB sample is biased. An sGRB is observed only if its jet points towards us and therefore the luminosity function that we derive must be corrected by the beaming factor in order to represent the true volumetric rate. Now, if the beaming is correlated with $E_{\gamma,\rm iso}$, then the true, beaming-corrected, luminosity function differs from the one we derived based on the observed sample. This can explain the energetic jet of GW170817, if bursts with higher isotropic equivalent energy are also more narrowly beamed. In such a case, bursts with $L_{\rm p,iso} > 10^{52} {\rm~ erg~s^{-1}}$ are much more frequent than their representation among sGRBs that point towards us. This possibility may even be supported by the very narrow jet of GW170817 (a few degrees). Note, also, that since $E_{\rm iso} \approx 2E/\theta_j^{2}$, where $E$ is the total (beaming-corrected) energy in the jet, and $E_{\rm iso}$ is the jet-core isotropic equivalent energy, if $E$ and $\theta_j$ are not correlated, then there must be a strong anti-correlation between the isotropic equivalent energy of the jet-core and its opening angle.

\subsubsection{GRB 170817A-like events}\label{sec:GRB170817A_like}
GRB 170817A was detected by the Fermi-GBM and by INTEGRAL, independently of the GW detection by LIGO/Virgo. Therefore, it is reasonable to expect that there are similar sGRBs in archival data of detectors such as BATSE, {\it Swift}, GBM, INTEGRAL and Konus-Wind. \cite{burns2018} examined the closest known sGRB, 150101B, an event at a distance of about $550$ Mpc (z=0.1341). They re-analysed its prompt emission, finding many similarities with GRB 170817A. Most notably, the prompt emission of GRB 150101B began with a hard and bright non-thermal pulse, followed by fainter and softer emission that is consistent with a quasi-thermal spectrum. There are, however, also significant differences between GRB 150101B and GRB 170817A. First, GRB 150101B was much more luminous and energetic. Its peak \gray luminosity was  $L_{\rm p,iso} \approx 10^{51} {\rm~ erg~s^{-1}}$, and its total isotropic equivalent energy was $E_{\rm \gamma,iso} \approx 10^{49}$ erg.  Second, its duration was much shorter, with the intense hard pulse lasting $\sim 0.01$s. Finally, it was harder, with $E_p \sim 1300$ keV  at the time of peak luminosity. \cite{troja2018a} analyzed the afterglow of GRB 150101B which, like GW170817, peaked late (in this case, after a few days) and contained much more kinetic energy than the energy carried by the $\gamma$-rays. The similarities in the \gray signals and, even more, in the afterglows, suggest that GRB 150101B was similar to GW170817. The peak in the afterglow seen after a few days can be interpreted, within this picture, as a jet-core that points away from us, but the angle between us and the jet-core is smaller than in GW170817. One possibility is that the observed \grays are emitted from the jet-core and are less luminous than the kinetic energy in the afterglow because it is off-axis. However, \cite{matsumoto2019} show, based on compactness arguments, that the angle between the \gray source and the observer was smaller than $3^\circ$, which makes the \gray off-axis interpretation highly unlikely. Interestingly, the \gray emission is consistent with a shock breakout scenario, including the closure relation of equation \ref{eq:closure}, where $R_{\rm bo} \sim 10^{12}$ cm and $\gamma_{\rm f,bo} \approx 35$. Such parameters are reasonable for a cocoon breakout, when it is observed close to the jet-core (closer than GW170817).

\cite{von-kienlin2019} have searched the Fermi-GBM catalog, looking for GRBs with properties that are similar to GRB 170817A. They find 13 events that they consider GRB 170817A-like,  based on their prompt \gray emission showing an initial hard and bright non-thermal spike, followed by a fainter and softer emission that is consistent with a quasi-thermal spectrum. The main limitation of this approach is that the prompt emission of GRBs is very diverse, and there is a continuum in any measurable property that has been explored (e.g., duration, variability, spectrum, spectral evolution). Moreover, there is no clear way to know which of the properties of GRB 170817A are the characteristics that should be chosen in the quest for physically similar events, with a dependence of the search results on  the very-fluid definition of "similar". All 13 GRBs found by \cite{von-kienlin2019}, except for GRB 150101B, have no redshift information or afterglow detection. Therefore, we will likely never know whether the \gray emission processes in any of these events was similar to that of GRB 170817A,  or whether there are sGRBs in the GBM sample that are truly similar to GRB 170817A,  but were not among these 13 chosen bursts. 

Statistically, we can estimate how many events similar to GW170817 are hiding in archival sGRB samples, based on the following simple argument. During O2 there was a single BNS merger within a distance of about 80Mpc. This merger produced \gray emission that can be detected by BATSE and by the Fermi-GBM roughly out to that distance. Thus, we can expect that an all-sky survey of sGRBs with sensitivity similar to that of BATSE should detect, every year, \gray emission from, at most, of order unity number of BNS mergers at a distance of up to 80Mpc. The true number is probably lower by about an order of magnitude, since it is most likely that GRB 170817A would not have been detected if the viewing angle with respect to the jet axis had been significantly larger than in GW170817. Over the entire sky, there are about 170 sGRBs brighter than the BATSE threshold every year \citep[e.g.,][]{nakar2007}. Thus, about 0.1-1\% of the sGRB sample is from BNS mergers within a distance of $\sim 100$ Mpc. Given that the entire sample of sGRBs (observed by BATSE, {it Swift}, GBM, etc.) contains about 1,000 bursts, we can expect that 1-10 of the bursts in this sample are nearby, $\sim 100$ Mpc, BNS mergers. This reasoning also explains why we have never identified any such event in the past; there are only several dozen sGRBs with detected afterglows and the detection of a regular sGRB X-ray afterglow is easier than the detection of a delayed GW170817-like afterglow, so
it is not surprising that we have not identified any nearby BNS mergers in this sample.

\subsection{\textbf{ Rates}}\label{sec:sGRB_rate}
Almost all studies of the sGRB luminosity function find a local sGRB rate of $\sim 10 \,{\rm Gpc^{-3}\,yr^{-1}}$, to within an order of magnitude \citep[][ and references therein]{wanderman2015}. This rate accounts only for the events that point toward us, and therefore, in order to find the true rate, it need to be multiplied by the average beaming-correction factor. This factor is not well constrained, but analyses of sGRB afterglows suggest a typical jet opening angle of $5^\circ-10^\circ$ \citep[e.g.,][]{fong2015}, which corresponds to a beaming correction of $50-200$ (note that there are considerable uncertainties involved in the process of inferring a jet opening angle from an afterglow light curve). The best guess for the true (beaming -corrected) local rate of sGRBs is ${\cal R}_{sGRB} \sim 1000 \,{\rm Gpc^{-3}\,yr^{-1}}$. This rate is similar to the BNS rate, as estimated based on the single case of GW170817. 

This agreement is an additional support for the connection between sGRBs and BNS mergers, and it has some interesting implications. If BNS mergers are the main (or only) progenitors of sGRBs, then a significant fraction of these mergers produce sGRBs, i.e.  a significant fraction of BNS mergers launch relativistic jets that break out of the sub-relativistic ejecta successfully \citep{beniamini2019b}. Given the uncertainties, there may be a considerable, or even dominant, population of BNS mergers with choked jets, but it cannot be larger by orders of magnitude than the population of mergers with successful jets (as may, in fact, be the case for long GRBs).

\subsection{\textbf{ BH-NS mergers and sGRBs}}\label{sec:sGRB_BHNS}
Similar to the case of BNS mergers, BH-NS mergers are also prime candidates for sGRB progenitors. The accretion of the disk that forms during the merger on the BH is expected to launch a relativistic jet, and possibly to produce a sGRB. The detection of a successful jet in GW170817 supports this picture. Moreover, the path of the jet from a BH-NS merger is expected to be cleaner than in BNS mergers, and if there is a jet, it is not expected to be choked (see \S\ref{sec:relativistic}). It is therefore possible that both BNS and BH-NS mergers produce sGRBs. Moreover, although there is no clear evidence for sub-classes within the sGRB population, sGRBs are so diverse that the two types of mergers may still produce sGRBs with different characteristics. For example it is possible that the sub-group of sGRBs that show extended X-ray emission (see \citealt{nakar2007} and references therein) are generated by one type of merger and not the other.

Nonetheless, the recent LIGO/Virgo results suggest that BH-NS are perhaps not the main source of sGRBs. During O2 there were no clear BH-NS candidates, and while the results from the LIGO/VIRGO O3 run are not published yet, the public alerts during the first six months indicate that the detection rate of GW signals from BH-NS mergers is not higher than that from BNS mergers. However, the detection volume of BH-NS binaries is significantly larger than that of BNS mergers. This implies that the volumetric rate of BH-NS mergers is significantly lower than the rate of BNS mergers, and it may be too low to explain a significant fraction of the observed sGRB sample. If the final results of O3 confirm this result, a possible caveat to the above is that the opening angles of jets from BH-NS mergers could be considerably larger than those in BNS mergers. BH-NS mergers could then still constitute a significant fraction of the observed sGRB population, even if their true (beaming-corrected) local rate is low. 
%\section{Future prospects}	

\section*{Acknowledgements}
Special thanks to Dan Maoz for a careful and constructive review of the manuscript. His invaluable comments and help have improved it considerably. I thank Tsvi Piran for many enlightening discussions. I also thank Iair Arcavi, Omer  Bromberg, Francois Foucart, Ore Gottlieb, Kenta Hotokezaka, Mansi Kasliwal, Amir Levinson and Dovi Poznanski for valuable discussions and comments. This work was partially supported by an ERC grant (JetNS) and by Israel Science Foundation Grant 1114/17.

\bibliographystyle{apj}
%\bibliography{GW170817Review}

\end{document}